\documentclass[a4paper,11pt]{article}
\pdfoutput=1
\usepackage{jheppub}
\usepackage[mathscr]{eucal}
\usepackage{hhline}
\usepackage{relsize}
\usepackage{psfrag}
\usepackage{mathrsfs} 
\usepackage{hyperref}
\usepackage{bm}
\usepackage{array}
\usepackage{tikz}
\usepackage{subcaption}
\usepackage{textcomp}
\usepackage{hyperref}
\usepackage{mathrsfs} 
\usepackage{enumitem}
\usepackage{caption}
\usepackage{epic}
\usepackage{eepic}
\usepackage{gensymb}
\usepackage{pifont}
\DeclareMathAlphabet{\mathpzc}{OT1}{pzc}{m}{it}

\makeatletter
\@addtoreset{equation}{section}
\makeatother

\def\bea{\begin{eqnarray}}
\def\eea{\end{eqnarray}}
\def\be{\begin{equation}}
\def\ee{\end{equation}}

\def\be{\begin{equation}}
\def\ee{\end{equation}}
\def\bdm{\begin{displaymath}}
\def\edm{\end{displaymath}}
\def\bea{\begin{eqnarray}}
\def\eea{\end{eqnarray}}

\def\sgn{{\rm sgn}}

\def\ri{{\rm i}}
\def\half{\textstyle\frac{1}{2}}

\def\XXint#1#2#3{{\setbox0=\hbox{$#1{#2#3}{\int}$}
    \vcenter{\hbox{$#2#3$}}\kern-.5\wd0}}

\newcommand{\rd}{\mbox{d}}
\newcommand{\re}{\mbox{e}}

\newcommand{\Ecal}{{\mathcal E}}

\newcommand{\Hcal}{{\mathcal H}}

\DeclareMathAlphabet{\mathpzc}{OT1}{pzc}{m}{it}

\title{ODE/IQFT correspondence for the generalized
 affine $\mathfrak{ sl}(2)$  Gaudin model}

\author[a]{Gleb A. Kotousov,}
\author[b,c]{Sergei L. Lukyanov}

\affiliation[a]{DESY, Theory Group, Notkestrasse 85, Hamburg 22607, Germany}
\affiliation[b]{NHETC, Department of Physics and Astronomy,
     Rutgers University,\\
     Piscataway, NJ 08855-0849, USA}
\affiliation[c]{Kharkevich Institute for Information Transmission Problems,\\
Moscow, 127994, Russia}

\emailAdd{gleb.kotousov@desy.de}
\emailAdd{sergei@physics.rutgers.edu}

\abstract{An
integrable system is introduced, which is a generalization of the
$\mathfrak{sl}(2)$ quantum affine Gaudin model.
Among other things, the
Hamiltonians are constructed and their spectrum
is calculated
using the ODE/IQFT approach.
The model fits into the framework 
of Yang-Baxter integrability.
This opens a way for the
 systematic quantization of a large class of
integrable non-linear sigma models. 
There may also be some interest in terms of Condensed Matter applications,
as  the theory can be thought of as a multiparametric
generalization of the Kondo model.}

\arxivnumber{2106.01238}

\begin{document}
\captionsetup[figure]{labelfont={small},labelformat={default},labelsep=period,name={Fig.\!}}
\captionsetup[table]{labelfont={small},labelformat={default},labelsep=period,name={Tab.\!}}

\maketitle

\pagebreak

\section{Introduction}

Suppose we are a given a set  of quantum spins  $\vec{S}^{(a)}=
\big(S^{(a)}_1,S^{(a)}_2, S^{(a)}_3\big)$:
 \bea
 \big[S^{(a)}_A, S^{(b)}_B\,\big]=\ri\,  \delta_{ab}\ {\varepsilon_{ABC}}\  S^{(a)}_C\ ,\ \ \ \  \ \ \ \ \  
  \big(\vec{S}^{(a)}\big)^2= {\mathfrak j}_a({\mathfrak j}_a+1)
 \eea
with $a=1,\ldots,r$.
As was  pointed out  by Gaudin \cite{Gaudin,Gaudin1},
 the operators
 \bea\label{oias89129812}
{ \bf{H}}^{(a)}=2\ \sum_{b=1\atop
 b\not=a}^r\frac{\vec{S}^{(a)}\cdot\vec{S}^{(b)}}{z_a-z_b}\ : \ \  \ \ \ \ \ \ \ \  \  \ \ \ \sum_{a=1}^r{\bf H}^{(a)}=0
 \eea
mutually commute for an arbitrary choice of the parameters
$\{z_a\}_{a=1}^r$. They  
 also  commute with any projection
of the total spin operator ${\vec S}=\sum_{a=1}^r{\vec S}^{(a)}$ and, furthermore,
 ${\vec S}^2$ is linearly expressed through the Hamiltonians:
\bea\label{oias8912981}
{\vec S}^{\,2}=\sum_{a=1}^rz_a{\bf H}^{(a)}
+\sum_{a=1}^r{\mathfrak j}_a({\mathfrak j}_a+1)\ .
\eea
It turns out that
 the  problem of the simultaneous diagonalization of ${\bf H}^{(a)}$
can be ``solved'' within the framework of the
Bethe ansatz \cite{Gaudin,Gaudin1}.  In this approach the 
 energy $E_a$ of the $a$-th Hamiltonian  is expressed as 
\be\label{op2929292929}
E_a=\sum_{b= 1\atop
b\not=a}^r\frac{2{\mathfrak j}_a\mathfrak{j}_b}{z_a-z_b}
\,-\,\sum_{m=1}^{\tt {M}_+}\frac{2{\mathfrak j}_a}{z_a-x^{\scriptscriptstyle{(+)}}_m }\ ,
\ee
where the set of auxiliary parameters $\{x^{\scriptscriptstyle{(+)}}_m\}_{m=1}^{{\tt M}_+}$
is determined through 
 the solution of the system of algebraic equations
\be\label{apo0402321}
\sum_{a=1}^r\frac{\mathfrak{j}_a}{z_a-x^{\scriptscriptstyle{(+)}}_m}\,-\,
\sum_{n=1\atop
n\not=m}^{{\tt M}_+}\frac{1}{x^{\scriptscriptstyle{(+)}}_n-x^{\scriptscriptstyle{(+)}}_m}
=0\qquad\qquad
(m=1,2,\ldots,\tt{M}_+)\ .
\ee
The integer ${\tt M}_+$ takes all possible values from $0$ to $2\sum_{a=1}^r\mathfrak{j}_a$.

\bigskip
The Gaudin model admits an almost straightforward generalization to any
simple Lie algebra ${\mathfrak{g}}$ (see sec.13.2.2 in \cite{Gaudin1} and  ref.\cite{Gurco}).
The development of the mathematical apparatus of $2D$ Conformal Field Theory 
led  to the idea that there should be a
meaningful  generalization to the case when the finite-dimensional Lie  algebra
is replaced by an affine Kac-Moody algebra  $\widehat{\mathfrak{g}}$.
Then the Hilbert space would be built out of
 Verma modules for  an algebra of  extended conformal symmetry. 
According to general principles of integrability in CFT \cite{Zam},
the diagonalization problem would be
 formulated for an  infinite set $\{{\bf I}_s\}$ of so-called local
Integrals of Motion  (IM), which 
depend on the arbitrary parameters $\{z_a\}_{a=1}^r$.
These would mutually commute, while by local what is meant is that
\be
{\bf{I}}_s=\int_0^{2\pi} \frac{\rd u}{2\pi}\ T_{s+1}(u)\,,
\ee
where $T_{s+1}$ is a chiral local field of Lorentz spin $s+1$.

\bigskip
An important step in exploring the affine Gaudin model was made by Feigin and Frenkel in ref.\cite{Feigin:2007mr}.
The key idea came from  an interesting link between the  original Gaudin model described above
and a certain class of linear
differential equations of the form \cite{Gaudin1,Feigin:1994in,Frenkel:2004qy}
\bea\label{oias8918912}
\big(-\partial_z^2+t_0(z)\big)\,\Psi=0\ \ \ \ \qquad {\rm with}\ \ \ \qquad t_0(z)=\sum_{a=1}^r\bigg(\frac{\mathfrak{j}_a\,(\mathfrak{j}_a+1)}
{(z-z_a)^2}+\frac{E_a}{z-z_a}\bigg)\ .
\eea
The ODE possesses $r$ regular singular points at $z=z_a$.
If we further assume that
\bea
\sum_{a=1}^rE_a=0
\eea
then $z=\infty$ is also a regular  singularity so that \eqref{oias8918912}
is a Fuchsian differential equation. 
A remarkable phenomenon occurs when the residues
$\{E_a\}_{a=1}^r$ in the potential $t_0(z)$ 
coincide with the set of energies corresponding to some common eigenvector of the
Hamiltonians  ${\bf H}^{(a)}$.
In this case all the singular points at $z=z_a$
turn out to be apparent.\footnote{Recall that  a singularity $z_a$ is called apparent
if the ratio of any two solutions
of the ODE is single valued in the vicinity of that point.}
The roots $x^{\scriptscriptstyle{(+)}}_m$ of the algebraic system \eqref{apo0402321},
such that \eqref{op2929292929} is satisfied for the given set
$\{E_a\}_{a=1}^r$, have a simple interpretation.
They are the zeroes of the function 
\be
\Psi_+(z)=
            \frac{\prod_{m=1}^{{\tt M}_+}(z-x^{\scriptscriptstyle{(+)}}_m)}
            {\prod_{a=1}^r  
           (z-z_a)^{\mathfrak{j}_a}}\,,
\ee
which is a solution of the ODE \eqref{oias8918912}.
There is another linearly independent solution of the form
\be
\Psi_-(z)=
            z\ \frac{\prod_{m=1}^{\tt{ M}_-}(z-{x}^{\scriptscriptstyle{(-)}}_m)}
                   {\prod_{a=1}^r  
           (z-z_a)^{\mathfrak{j}_a}}
\ \qquad\qquad\ \qquad \ \  \Big(\,{\tt M}_++{\tt M}_-=2\sum_{a=1}^r\mathfrak{j}_a\,\Big)\,.
\ee
Here the set  $\{x^{\scriptscriptstyle{(-)}}_m\}_{m=1}^{{\tt M}_-}$ also solves the Bethe ansatz like equations,
\be\label{apo0402321AAdd}
\frac{1}{{x}^{\scriptscriptstyle{(-)}}_m}+\sum_{a=1}^r\frac{\mathfrak{j}_a}{z_a-{x}^{\scriptscriptstyle{(-)}}_m}-
\sum_{n=1\atop
n\not=m}^{{\tt M}_-}\frac{1}{{x}^{\scriptscriptstyle{(-)}}_n-{x}^{\scriptscriptstyle{(-)}}_m}
=0\qquad\qquad
(m=1,2,\ldots,\tt{M}_-)\ ,
\ee
while
\be
E_a=-\frac{2{\mathfrak j}_a}{z_a}+\sum_{b= 1\atop
b\not=a}^r\frac{2{\mathfrak j}_a\mathfrak{j}_b}{z_a-z_b}\,
-\,\sum_{m=1}^{{\tt M}_-}\frac{2{\mathfrak j}_a}{z_a-{x}^{\scriptscriptstyle{(-)}}_m }
\ .
\ee
Thus there is a link between the spectrum of the Gaudin Hamiltonians and a  class of 
differential equations possessing certain monodromy properties. 
This provides, perhaps, one of the simplest illustrations of a broad phenomena,
known as the ODE/IQFT correspondence  \cite{Voros:1994,Dorey:1998pt,Bazhanov:1998wj,Bazhanov:2003ni}.\footnote{%
The term ODE/IM correspondence, where
IM stands for integrable model,  is also used 
in the literature. However, we find that the latter
tends to be heavily focused on 
formal properties of ODEs at the expense 
of a clear study of the quantum theory which, in our opinion,
is the most interesting part of the relation. Recently, there has been a great
flux of works that introduce and discuss different sorts of
``correspondences'', which are essentially ODE/IM. 
In order to emphasize that our main motivation is the
study of integrable quantum field theory, rather than 
 abstract ``integrable models'',   we use the term ODE/IQFT.
}

\bigskip

In ref.\cite{Feigin:2007mr} Feigin and Frenkel 
introduce the Hamiltonians, which can be interpreted as an  ``affinization''
of ${\bf H}^{(a)}$ \eqref{oias89129812}.
They are built from   $r$ independent 
copies of the affine Kac-Moody $\widehat{\mathfrak{sl}}_{k_a}(2)$ algebra 
at levels $k_a=1,2,\ldots\ $. The currents would obey the operator
product expansions of the form
  \bea\label{aiissaias}
 J^{(a)}_{A}(u)J^{(b)}_{B}(0)=-\delta_{ab}\ \bigg(\,\frac{k_a}{2 u^2}\ \eta_{AB}+\frac{\ri}{u}\
{f_{AB}}^C\,  J_{C}^{(a)}\,\bigg)+O(1)\ .
 \eea
To each copy one can associate the Virasoro field,
 \bea\label{asod8912iod}
 G^{(a)}= \frac{\eta^{AB}J_A^{(a)}J_B^{(a)}}{k_a+2}=
 \frac{1}{4\,(k_a+2)}\ \Big(
J_0^{(a)}J_0^{(a)}+   2\, J_+^{(a)}J_-^{(a)}+2\, J_-^{(a)}J_+^{(a)}\Big)
 \eea
(here $\eta^{AB}$ stands for the Killing form, while $\eta^{AC}\,\eta_{CB}=\delta^A_B$).
Then the Hamiltonians of the affine Gaudin model are given by
\be\label{asio89128432}
{\bf H}_{{\rm\scriptscriptstyle G}}^{(a)}=\frac{1}{2}\int_{0}^{2\pi}\frac{\rd u}{2\pi}\
\sum_{b=1\atop
b\ne a}^{r}\, \frac{k_b\, G^{(a)}+k_a\, G^{(b)}-2 \eta^{AB}J_A^{(a)}J_B^{(b)}
}{z_a-z_b}\ .
\ee
Feigin and Frenkel put forward the
conjecture that the spectrum of these operators would be encoded in 
a class of differential equations that generalizes \eqref{oias8918912},
though they did not explain exactly how the spectrum would be extracted from the ODEs.
The last point was clarified in refs.\cite{Lacroix:2018fhf,Lacroix:2018itd}.
\bigskip

In this work we introduce and study the model, which possesses 
an infinite set of mutually commuting local integrals of motion. The simplest ones,
the ``Hamiltonians'', are expressed in terms of the $\mathfrak{sl}(2)$
Kac-Moody currents and Virasoro field \eqref{asod8912iod} as 
 \bea\label{oias8912}
{ \bf{H}}^{(a)}_{\rm gen} &=&
\int_{0}^{2\pi}\frac{\rd u}{2\pi}\,\bigg[\,\frac{\beta^2 K}{1-\beta^2}\ G^{(a)}
+\frac{1}{4K}\ \frac{1-\beta}{1+\beta}\ \Big(\, k_a\, \big(J_0^{(\rm tot)}\big)^2
 -K\, J_0^{(a)} J_0^{(\rm tot)}\,\Big)\\[0.3cm]
&-&  \sum_{b=1\atop
b\ne a}^{r}
\frac{1}{z_a-z_b}\, \Big(\, \frac{1}{4}\, (z_a+z_b)\, J^{(a)}_0J^{(b)}_0+
 z_a\, J_+^{(b)} J_-^{(a)}+
 z_b\, J_+^{(a)} J_-^{(b)}
 -k_a z_b\,G^{(b)}-k_bz_a\,G^{(a)}
  \Big)\,\bigg] ,\nonumber
 \eea
where
\be
J_0^{({\rm tot})}=\sum_{a=1}^r J_0^{(a)}\qquad\qquad
{\rm and}\qquad\qquad K=\sum_{a=1}^r k_a\ .
\ee
Notice that 
\be
\frac{1-\beta^2}{\beta^2 K}\ \sum_{a=1}^r\,{\bf H}^{(a)}_{\rm gen}\,=\,\sum_{a=1}^r \,
 \int_{0}^{2\pi}\frac{\rd u}{2\pi}\ 
G^{(a)}\,,
\ee
which can be thought of as the affine counterpart of eq.\,\eqref{oias8912981}.
\bigskip

The operators \eqref{oias8912} depend on the parameter $\beta$.
The  Hamiltonians of the affine Gaudin model \eqref{asio89128432} are obtained
through a certain limiting procedure, which includes taking $\beta\to 1^-$.
We formulate the ODE/IQFT correspondence for the model and explain
how the spectrum
of ${\bf H}^{(a)}_{\rm gen}$ for arbitrary $\beta\in(0,1)$ can be extracted from
the differential equations.
The theory will be referred to as the 
Generalized Affine Gaudin Model (GAGM). 
\bigskip

The GAGM fits within the framework of the standard
Yang-Baxter integrability. In particular, 
the Hamiltonians ${\bf H}^{(a)}_{\rm gen} $ are part of a large commuting family
which, as usual, involves the quantum transfer-matrices and Baxter
$Q$-operators. These are explicitly constructed along the lines
of the BLZ approach \cite{Bazhanov:1994ft,Bazhanov:1996dr,Bazhanov:1998dq}. 
It is proposed that the GAGM governs the critical
behaviour of a lattice system, which in the simplest case coincides with the
inhomogeneous six-vertex model introduced by Baxter \cite{Baxter:1971cs}. The 
local Boltzmann weights are contained in the $R$-matrix that is
 the trigonometric solution of the Yang-Baxter equation.
The anisotropy parameter entering into the $R$-matrix,
commonly referred to as $q$, is related to 
the  parameter $\beta$ in \eqref{oias8912} as
\be
q=-\,\re^{\frac{\ri\pi}{K}(\beta^2-1)}\, .
\ee
In the limit $\beta\to1^-$ the trigonometric $R$-matrix becomes the rational one.
\bigskip

The paper is organized as follows.
Sections \ref{sec2}-\ref{sec5} contain no
new material and their purpose is to illustrate
the ideas of the BLZ approach. Its main ingredient is
a realization of the Borel subalgebra of the quantum algebra
 $U_q\big(\widehat{\mathfrak{ sl}}(2)\big)$ in terms of the
vertex operators.
We use the example from ref.\cite{Lukyanov:2006gv} as it 
 contains the essential blueprints for the construction
of the commuting family of operators in the GAGM
 as well as the ODE/IQFT correspondence.
The realization of  the
Borel subalgebra of  $U_q\big(\widehat{\mathfrak{ sl}}(2)\big)$,
which gives rise to the commuting family for the GAGM,
is presented in sec.\,\ref{sec6}. The next section is 
a central one and contains a detailed discussion of
 the ODE/IQFT correspondence for the model. Some comments
concerning the  literature are also presented therein.
The way the spectrum of the local and  non-local integrals of motion
are extracted from the ODE is described in sec.\,\ref{sec8}.
Sections \ref{sec9} and \ref{sec10} deal with some
specific cases and provide an illustration of 
the rather abstract ideas that preceded them.
The Hamiltonians ${ \bf{H}}^{(a)}_{\rm gen} $
\eqref{oias8912}
are deduced in sec.\,\ref{sec11}.
Also considered are certain limits of the model such as the isotropic, classical 
as well as the limit,
which yields the Hamiltonians of the affine Gaudin model \eqref{asio89128432}.
Finally, we sketch how the GAGM appears in the scaling
limit of the Baxter-type statistical systems in sec.\,\ref{sec12}.

\section{Quantum transfer-matrices\label{sec2}}
 The   algebraic structure underlying
 the  Yang-Baxter relation
was clarified within the theory of quasi-triangular Hopf algebras by Drinfeld  \cite{Drinfeld1986}.
A basic example
is when the Hopf algebra is  $U_q(\widehat{\mathfrak{g}})$ 
 -- the
quantum deformation of the universal enveloping algebra of
the affine algebra \cite{Drinfeld1986,Jimbo:1985zk}. 
The central r\^{o}le is played  by the universal $R$-matrix, which lies in the
tensor product $U_q(\widehat{\mathfrak{g}})\otimes U_q(\widehat{\mathfrak{g}})$
and satisfies the relation  
\be\label{YB1}
{\cal R}^{12}\,{\cal R}^{13}\,{\cal R}^{23}\,=\,{\cal R}^{23}\,{\cal R}^{13}\,{\cal R}^{12}\ .
\ee
An important feature of ${\cal R}$ is that it is decomposed as
 ${\cal R}\in U_q(\widehat{{\mathfrak b}}_+)\otimes U_q(\widehat{{\mathfrak b}}_-)$, where 
$U_q(\widehat{{\mathfrak b}}_\pm)$  stand for the Borel subalgebras of $U_q(\widehat{\mathfrak{g}})$.
 In this paper we restrict to the case
 ${\mathfrak g}=\mathfrak{ sl}(2)$.

\bigskip

Let us consider  the evaluation homomorphism of $U_q(\widehat{{\mathfrak g}})$ to the loop algebra
$U_q({\mathfrak{g}})[\lambda,\lambda^{-1}]$ and specify a finite dimensional matrix representation $\pi$ of
 $U_q(\mathfrak{g})$. 
In the case  under consideration the  Borel subalgebra $U_q(\widehat{{\mathfrak b}}_+)$  is generated by four elements,
$\{y_0,y_1,h_0,h_1\}$,
 and its evaluation homomorphism is defined by
\be\label{evalhomo1}
y_0\mapsto \lambda q^{\frac{{\tt h}}{2}} \,{\tt e}_+\ , \qquad y_1\mapsto \lambda \,q^{-\frac{{\tt h}}{2}} \,
{\tt e}_-\ , \qquad h_0\mapsto {\tt h}\ , \qquad h_1\mapsto -{\tt h}\ .
\ee
Here ${\tt h},{\tt e}_\pm$ are the generators of $U_q\big({\mathfrak{sl}}(2)\big)$,
subject to the commutation relations
 \bea\label{iosdif090as}
[\,{\tt h}, {\tt e}_\pm\,]=\pm 2\, {\tt e}_{\pm}\ ,\ \ \ \ \ \ \ [\,{\tt e}_+,{\tt e}_-\,]=\frac{q^{\tt h}-q^{-{\tt h}}}{q-q^{-1}}\ .
\eea
Then 
 \bea\label{aissia}
{\boldsymbol L}_{\ell}(\lambda) = \big(\pi_{\ell}(\lambda)\otimes 1\big)[{\cal R}] \qquad\qquad
\qquad\big(\ell=\tfrac{1}{2},1,\tfrac{3}{2},\ldots\big)
\eea
 is a 
$U_q(\widehat{{\mathfrak b}}_-)$-valued 
$(2{\ell}+1)\times (2{\ell}+1)$ matrix
whose entries depend on an auxiliary parameter $\lambda$. 
In turn
the formal algebraic relation \eqref{YB1}  becomes the Yang-Baxter algebra
\be\label{YBeq1a}
 R_{{\ell},{\ell}'} \big(\lambda_1/\lambda_2\,\big)
 \big({\boldsymbol L}_{{\ell}}
 (\lambda_1)
 \otimes {\boldsymbol 1 }\big)\big({\boldsymbol 1}\otimes {\boldsymbol L}_{{\ell}'}
(\lambda_2)\big)=
 \big({\boldsymbol 1}\otimes {\boldsymbol L}_{{\ell}'}
 (\lambda_2)\big)\big({\boldsymbol L}_{{\ell}}
(\lambda_1)\otimes {\boldsymbol 1 }\,\big)
 R_{{\ell},{\ell}'}
 \big(\lambda_1/\lambda_2\,\big) 
\ee
with
\bea
{R}_{{\ell}_1,{\ell}_2}(\lambda_1/\lambda_2)=\big(\pi_{{\ell}_1}(\lambda_1)\otimes 
\pi_{{\ell}_2}(\lambda_2)\big)[{\cal R}]\, .\nonumber
\eea
As an immediate consequence the operators
\bea\label{asusuya}
{\boldsymbol{\tau}}_{\ell}(\lambda)={\rm Tr}_{\ell}\Big[ q^{\frac{1}{2}{\tt h}\, h_0}\,
{\boldsymbol L}_{{\ell}}(\lambda)\Big]\,,
\eea
usually referred to as the transfer-matrices,
obey the commutativity condition
\be
\big[{\boldsymbol{\tau}}_{\ell}(\lambda),\,{\boldsymbol{\tau}}_{{\ell}'}(\lambda')\big]=0\ .
\ee
\medskip

With the expression for the universal $R$-matrix
given in \cite{Khoroshkin:1994um}, 
one can obtain ${\boldsymbol L}_{\ell}(\lambda)$ as a formal series expansion in 
powers of the spectral parameter $\lambda$. The first few terms read as
\bea\label{series1}
{\boldsymbol L}_{\ell}(\lambda)&=&q^{\frac{1}{2}{\tt h}\, h_0}\ 
\bigg[1+\lambda\,(q-q^{-1})\,
(x_0\,q^{\frac{{\tt h}}{2}}\,{\tt e}_++x_1\,q^{-\frac{{\tt h}}{2}}\,{\tt e}_-)
\nonumber\\
&+&\lambda^2\  \frac{q-q^{-1}}{q^2\,[2]_q}\,
\bigg(
\,(q^{2}-1)\,x_0^2\ (q^{\frac{{\tt h}}{2}}\,{\tt e}_+)^2+
\,(q^{2}-1)\,x_1^2\ (q^{-\frac{{\tt h}}{2}}\,{\tt e}_-)^2\\
&+&
\,(q^2\,x_1x_0-x_0x_1)\ (q^{\frac{{\tt h}}{2}}\,{\tt e}_+)
(q^{-\frac{{\tt h}}{2}}\,{\tt e}_-)+
(q^{2}\,x_0 x_1-x_1x_0)\,(q^{-\frac{{\tt h}}{2}}\,{\tt e}_-)
(q^{\frac{{\tt h}}{2}}\,{\tt e}_+)\bigg)+\ldots\bigg]\, .
 \nonumber
\eea
Here   and below, abusing notation,
we do  not distinguish between the formal generators of $U_q({\mathfrak{sl}}_2)$ and their
$(2{\ell}+1)\times (2{\ell}+1)$ matrices in a finite
dimensional representation $\pi_{\ell}$. Also,
$[n]_q\equiv(q^n-q^{-n})/(q-q^{-1})$.
The expression in the square brackets contains the elements
$x_0,x_1\in U_q(\widehat{{\mathfrak b}}_-)$,
which obey the quantum
Serre relations
\be\label{Serre1}
x_a^3x_b-[3]_q\,x_a^2x_bx_a+[3]_q\,x_ax_bx_a^2-x_bx_a^3=0 \qquad (a,b=0,1) \, .
\ee
There are
two remaining generators  $h_0,h_1$ satisfying
\be\label{comm2}\arraycolsep=0.5cm
\begin{array}{lll}
  [h_0,x_0]=-[h_1,x_0]=-2x_0\, , & [h_0,x_1]=- [h_1,x_1]=2x_1\,, & [h_0,h_1]=0 \, .
\end{array}
\ee
Since $h_0+h_1$
is a central element, for our purposes and
without loss of generality we have set it to be zero.

\bigskip

Up to this point,
there was no need to specify a representation of  $U_q(\widehat{{\mathfrak b}}_-)$ -- 
the Yang-Baxter relation \eqref{YBeq1a} holds identically 
provided \eqref{Serre1},\,\eqref{comm2} are true. An important case is when
the generators $x_0$, $x_1$ are realized  as integrals over the vertex operators
\be\label{contourInt1}
x_0 = \frac{1}{q-q^{-1}}\, \int_0^{2\pi}{\rm d}u\, V_+(u) \ ,\qquad 
x_1 = \frac{1}{q-q^{-1}}\,\int_0^{2\pi}{\rm d}u\, V_-(u) \ .
\ee
The latter are required to satisfy the braiding relation
 \bea\label{jausus}
V_{\sigma_1}(u_1) V_{\sigma_2}(u_2)=q^{2\sigma_1\sigma_2}\ V_{\sigma_2}(u_2) V_{\sigma_1}(u_1)\ ,\ \ \qquad
 u_1>u_2
\eea
 along  with the  quasiperiodicity condition
 \bea\label{iisisaisa}
V_\pm(u+2\pi)=q^{-2}\ \Omega^{\pm 1}\ V_\pm(u)\ .
\eea
Here  the operator $\Omega$ obeys 
\bea\label{aisaisisaias}
\Omega\, V_\pm (u)\, \Omega^{-1}=q^{\pm 4}\ V_\pm(u)
\eea
and can be identified with
\bea\label{io9121}
\Omega=q^{2h_0}\ .
\eea
\bigskip

As was pointed out in the work \cite{Bazhanov:1998dq}, using the braiding relations \eqref{jausus}
it is possible to express  monomials built from the generators $x_0$ and $x_1$
in terms of the ordered integrals
\be\label{Iordered}
J(\sigma_1,\ldots,\sigma_m) = \!\!\!\!\!\!\!\!\int\limits_{2\pi> u_1> u_2>\ldots> u_m> 0}
\hspace{-1cm}{\rm d}u_1\ldots{\rm d}u_m\, V_{\sigma_1}(u_1)\ldots V_{\sigma_m}(u_m)\ .
\ee
This way, the formal power series \eqref{series1} can be brought to the form
\be\label{Mseries}
{\boldsymbol L}_{\ell}(\lambda)=\,\Omega^{\frac{1}{4}{\tt h}}\,
\sum\limits_{m=0}^\infty\ \lambda^{m}\!\!\!\sum\limits_{\sigma_1\ldots\sigma_m=\pm}
\big(q^{\frac{{\tt h}}{2}\sigma_1}{\tt e}_{\,\sigma_1}\big)\ldots 
\big(q^{\frac{{\tt h}}{2}\sigma_m}{\tt e}_{\,\sigma_m}\big)\ J(\sigma_1,\ldots,\sigma_m)\ .
\ee
The latter is recognized as the path ordered exponent
\bea\label{9sd090as}
\bm{{ L}}_{\ell}(\lambda)= 
\Omega^{\frac{1}{4}{\tt h}}\ \overset{\leftarrow}{{\cal P}}\exp\bigg(\lambda \int_0^{2\pi}\rd u\,
\Big(V_-(u)\ q^{+\frac{{\tt h}}{2}}\,{\tt e}_++
 V_+(u)\ q^{-\frac{{\tt h}}{2}}\,{\tt e}_-\Big)\bigg)\ .
\eea
\bigskip

 The first realization of the
generators $x_0$, $x_1$
in terms of the vertex operators was proposed in refs.\cite{Bazhanov:1994ft,Bazhanov:1996dr,Bazhanov:1998dq}.
It reads as
\bea\label{isaisiisa}
V_{\pm}=  \re^{\pm 2\ri\beta \phi}\,,
\eea
where 
$\phi=\phi(u)$ stands for the chiral Bose field satisfying the Operator
Product Expansion (OPE)
\bea\label{iiiasiasias}
\phi(u_1)\phi(u_2)=-\tfrac{1}{2}\, \log(u_1-u_2)+O(1)\ .
\eea
 The parameter $\beta$  is related to $q$, entering into the
braiding relation \eqref{jausus}, as $q=\re^{\ri\pi\beta^2}$.
It is  taken to lie in the domain
\bea
0<\beta<1\ .
\eea
If the field $\phi$ is assumed to be quasiperiodic
it can be expanded in a Fourier series
of the form\footnote{%
We usually employ the hat notation for 
the oscillator modes to emphasize that they are operators
rather than $c$\,-\,numbers. 
However in some cases, when the formulae become too cluttered,
the ``hat'' will be omitted (see, e.g., eq.\,\eqref{oiasd891221}).
}
\bea
\phi(u)=\hat{\phi}_0+\hat{a}_{0}u+\ri\ \sum_{m\not=0}\frac{\hat{a}_{m}}{m}\ \re^{-\ri mu}
\eea
so that
\bea
\phi(u+2\pi)=\phi(u)+2\pi \hat{a}_0\ .
\eea
In turn the vertices satisfy the quasiperiodicity condition \eqref{iisisaisa} with
$
\Omega=\re^{4\pi\ri \beta \hat{a}_0}\,.
$
It is easy to see that the operators ${\boldsymbol{\tau}}_{\ell}(\lambda)$ \eqref{asusuya}
commute with the zero mode momenta,
\bea
\hat{a}_0\ :\ \ \ \ \ [\hat{\phi}_0,\hat{a}_0]=\tfrac{\ri}{2}\ ,
\eea
 and hence
act invariantly in the Fock space ${\cal F}_P$.
The latter is generated by the action of the creation operators $\hat{a}_n\ (n=-1,-2,\ldots)$ on the Fock 
vacuum
\bea
|\,P\,\rangle\ :\ \ \  \ \ \hat{a}_{n}\,|\,P\,\rangle=0\ \ \ 
\qquad(n=1,2,\ldots)\ ,\ \ \ \qquad \hat{a}_{0}\,|\,P\,\rangle={P}\, |\,P\,\rangle\ .
\eea

\section{Parafermionic realization of $U_q(\widehat{{\mathfrak b}}_-)$}

In ref.\cite{Lukyanov:2006gv} a
 generalization of \eqref{isaisiisa} was proposed,
which goes along the following lines. 
Suppose we are given the algebra $\widehat{\mathfrak {sl}}_k(2)$ with 
central charge $k=1,2,\ldots\ $.
The Kac-Moody currents obey the OPEs
\bea
&&J_+(u) J_-(0)=-\frac{k}{u^2}-\frac{\ri}{u}\ J_0(0)+O(1)\ ,\ \ \   
 J_0(u) J_\pm (0)=\mp\frac{2\ri}{u}\  J_\pm (0)+O(1)\nonumber\\[0.2cm]
 && J_0(u)  J_0(0)=-\frac{2k}{u^2}+O(1)\ .
\eea
As is well known \cite{Fateev:1985mm}, they admit a realization in terms of the $\mathbb{Z}_k$
parafermionic fields $\psi_\pm$ and the chiral Bose field $\phi$ \eqref{iiiasiasias}:
\bea\label{aisisai}
J_{\pm}=\sqrt{k}\ \psi_\pm\ \re^{\pm \frac{2\ri\phi}{\sqrt{k}}}\ ,\ \ \ \ \ \ \ \ 
J_0=2\sqrt{k}\ \partial\phi\ .
\eea
The  parafermionic fields are quasiperiodic
\bea
\psi_\pm(u+2\pi)=\re^{\frac{2\pi\ri}{k}}\ \big(\hat{\Omega}_k\big)^{\pm 1 }\, 
\psi_\pm(u)
\eea
with  $\hat{\Omega}_k$ being  the operator  of the ${\mathbb Z}_k$ charge:
 \bea\label{oisaoisisa}
 \hat{\Omega}_k\, \psi_{\pm}\,  \big(\hat{\Omega}_k\big)^{-1}=\omega^{\pm 2}\,  \psi_{\pm}\,,\qquad
\ \ \ \ \ \ \ \omega=\re^{-\frac{2\pi\ri}{ k}}\ .
 \eea
Then
\bea\label{iaisisa}
V_{\pm}=\sqrt{k}\, \psi_\pm\ \re^{\pm \frac{2\ri\beta \phi}{\sqrt{k}}}\,,
\eea
where again $\beta\in(0,1)$. 
Note that  the vertex operators
 can be interpreted as a one-parameter  deformation of the currents $J_\pm$
from the case $\beta=1$.
The braiding relations \eqref{jausus} and quasiperiodicity condition \eqref{iisisaisa} 
are satisfied with 
\bea\label{oias90129012}
q=-\re^{\frac{\ri\pi}{k}(\beta^2-1)}\,,\qquad\qquad \Omega=\re^{\frac{4\pi\ri \beta \hat{a}_0}{\sqrt{k}}}\ \hat{\Omega}_k\ .
\eea

\bigskip
Let us now briefly discuss the diagonalization problem 
for the transfer matrices ${\boldsymbol{\tau}}_{\ell}(\lambda)$  \eqref{asusuya}
with the vertex operators as in  \eqref{iaisisa}.
To this end  some basic facts concerning 
 representations  of the algebra of parafermionic currents are needed.
The theory was developed  by Fateev and Zamolodchikov in ref.\cite{Fateev:1985mm},
in the construction of the 
CFTs
 describing the multicritical points of the  ${\mathbb Z}_k$ statistical systems \cite{Fateev:1982wi} (certain
 generalizations
of the ${\mathbb Z}_2$  invariant Ising model).
The algebra 
 contains a set of non-local currents $\{\psi_n\}_{n=1}^{k-1}$
with conformal dimensions
\bea
\Delta_{n}=\frac{n\,(k-n)}{k}\ .
\eea
Their defining OPEs  are invariant w.r.t. ${\mathbb Z}_k$  transformations
$\psi_m\mapsto \omega^{2m a}\,  \psi_m$ with $a=0,1,\ldots,k-1$ and $ \omega=\re^{-\frac{2\pi\ri}{ k}}$.
All the parafermionic currents can be generated through the OPE of the 
currents with the lowest conformal dimensions
\bea
\psi_+\equiv \psi_1\ ,\ \ \ \ \ \ \psi_-\equiv \psi_{k-1}\ ,
\eea
which carry  the 
 ${\mathbb Z}_k$-charges $+2$ and $-2$, respectively (see eq.\eqref{oisaoisisa}).

\bigskip

The OPE of the fundamental parafermions is of special interest. It has the form
\bea\label{aoisaisais}
\psi_+(u)\psi_-(v)&=&-\re^{\frac{\ri\pi}{k}}\ (u-v)^{-2+\frac{2}{k}}\, \Big[ 1+\tfrac{k+2}{2k}\ (u-v)^2\ \big(W_2(u)+W_2(v)\big)\nonumber\\[0.2cm]
&+&
\tfrac{1}{2}\, k^{-\frac{3}{2}}\  (u-v)^3\  \big(W_3(u)+W_3(v)\big)+\ldots\Big]\ .
\eea
All the fields occuring in the expansion are ${\mathbb Z}_k$ neutral. 
The field  $W_2$, having Lorentz spin $2$,  generates the Virasoro algebra with  central charge
\bea\label{oias91212}
c_k=\frac{2(k-1)}{k+2}\ .
\eea
Further terms in the OPE \eqref{aoisaisais} 
involve  a set of local fields $W_3,\,W_4,\ldots\,$, labeled by 
their value of the Lorentz spin.
They form the
$W\!A_{k-1}$ algebra introduced in 
refs.\cite{Zamolodchikov:1985wn,Fateev:1987zh} with the  special value of the
central charge $c=c_k$ \eqref{oias91212}.

\bigskip

The chiral component of the Hilbert space
 of the ${\mathbb Z}_k$ CFT
 can be decomposed into  irreps ${\cal V}^{(k)}_{\mathfrak j}$  of the chiral algebra.  Here  the subscript ${\mathfrak j}$  stands 
 for the highest weight of the irrep with highest weight vector $\big|\,\sigma^{(k)}_{\mathfrak j}\,\big\rangle$
 having  
 conformal dimension
 \bea\label{hssaysasy}
 \Delta_{\mathfrak j}=\frac{{\mathfrak j}\,(k-2 {\mathfrak j})}{k\,(k+2)}\ ,
\eea 
 while
\bea
\hat{ \Omega}_k\ \big|\,\sigma^{(k)}_{\mathfrak j}\,\big\rangle=\omega^{2{\mathfrak j}}\, 
 \big|\,\sigma^{(k)}_{\mathfrak j}\,\big\rangle\ .
 \eea
It will be further assumed that
 \bea
 {\mathfrak j}=0,\tfrac{1}{2},1,\ldots,\tfrac{k}{2}\ .
 \eea
If ${\mathfrak j}\not=0,\tfrac{k}{2}$
the states   $|\,\sigma^{(k)}_{\mathfrak j}\,\rangle$ and $|\,\sigma^{(k)}_{\frac{k}{2}-{\mathfrak j}}\,\rangle$
possess the same non-vanishing
 conformal dimension. However they are  distinguished  by their value of the ${\mathbb Z}_k$ charge.
 In fact the whole space
   ${\cal V}_{\mathfrak j}$ is  naturally splitted   on the invariant subspaces of the operator  $\hat{\Omega}_k$\cite{Fateev:1985mm}:
 \bea
{\cal V}_{\mathfrak j}=\Big[\oplus_{s=0}^{2{\mathfrak j}}\, 
{\cal V}^{(k)}_{{\mathfrak j},2{\mathfrak j}-2s}\,\Big]\oplus
\Big[\oplus_{s=1}^{k-2{\mathfrak j}-1}\, 
{\cal V}^{(k)}_{{\mathfrak j},2{\mathfrak j}+2s}\Big]
\ :\ \ \ \quad
\hat{ \Omega}_k\, {\cal V}_{{\mathfrak j}, {\mathfrak m}}^{(k)}=
 \omega^{{\mathfrak m}}\  {\cal V}_{{\mathfrak j}, {\mathfrak m}}^{(k)}\ .
 \eea
 The lowest possible  conformal dimension in  the subspace ${\cal V}_{{\mathfrak j},{\mathfrak m}}^{(k)}$ is given
 by 
\be\label{oias0910912}
\Delta_{{\mathfrak j}, {\mathfrak m}}=\frac{{\mathfrak j}({\mathfrak j}+1)}{k+2}-\frac{{\mathfrak m} ^2}{4k}
\ee
for  ${\mathfrak m}=-2{\mathfrak j},\, -2{\mathfrak j}+2,\,\ldots, 2{\mathfrak j}$, and
 $
\Delta_{\frac{k}{2}-{\mathfrak j},k-{\mathfrak m}}
 $
 when  ${\mathfrak m}=2{\mathfrak j}+2,\ldots,2 k-2{\mathfrak j}-2$.
\bigskip

Each of the ${\mathbb Z}_k$-invariant   subspaces
 ${\cal V}_{{\mathfrak j},{\mathfrak m}}^{(k)}\subset {\cal V}_{\mathfrak j}^{(k)}$ possess a structure
of the highest weight irrep of the  $W\!A_{k-1}$ algebra.
From the formal algebraic point of view  such 
an irrep  is obtained by factorizing a highest weight module over submodules of 
``null-vectors''. In the case under consideration the highest weight can be
thought of as a pair of eigenvalues $(\Delta,w)$ of the mutually commuting operators
\bea
\frac{c_k}{24}+\int_0^{2\pi}\frac{{\rd u}}{2\pi}\ W_2(u), \ \  \ \ \ \ \int_0^{2\pi}\frac{{\rd u}}{2\pi}\ W_3(u)
\eea
corresponding to the highest weight vector. In the case of   ${\cal V}_{{\mathfrak j},{\mathfrak m}}^{(k)}$
 with ${\mathfrak m}=2{\mathfrak j},\, 2{\mathfrak j}-2,\,\ldots,- 2{\mathfrak j}$
the conformal dimension $\Delta$ is given by \eqref{oias0910912},
 while $w=w_{{\mathfrak j},{\mathfrak m}}$ with
\bea
w_{{\mathfrak j},{\mathfrak m}}=\frac{\mathfrak m}{6\sqrt{k}}\ \bigg(\, \frac{3k+4}{2k}\, {\mathfrak m}^2-
6\, {\mathfrak j}({\mathfrak j}+1)+k\,\bigg)\ .
\eea
For ${\mathfrak m}=2{\mathfrak j}+2,\ldots,2 k-2{\mathfrak j}-2$,
the highest weight is 
 $\big(\Delta_{\frac{k}{2}-{\mathfrak j},k-{\mathfrak m}},w_{\frac{k}{2}-{\mathfrak j},k-{\mathfrak m}}\big)$. 
The primary state (the highest weight vector)  w.r.t.  the $W\!A_{k-1}$ algebra will be 
denoted by $\big|\,\sigma^{(k)}_{{\mathfrak j},{\mathfrak m}} \,\big\rangle\in{\cal V}_{{\mathfrak j},{\mathfrak m}}^{(k)}$.

\bigskip
Let us return to  the transfer-matrices ${\boldsymbol{\tau}}_{\ell}(\lambda)$
built from the vertex operators $V_{\pm}=\sqrt{k}\,
 \psi_\pm\ \re^{\pm \frac{2\ri\beta \phi}{\sqrt{k}}}$.
First of all it is straightforward to show that they  commute with $\hat{a}_0$ and  $\hat{\Omega}_k$:
\bea\label{isaususa}
[{\boldsymbol{\tau}}_{\ell}(\lambda), \hat{a}_0]=[{\boldsymbol{\tau}}_{\ell}(\lambda), \hat{\Omega}_k]=0\ .
\eea
This implies that ${\boldsymbol{\tau}}_{\ell}(\lambda)$ acts invariantly
in the space ${\cal V}_{{\mathfrak j},{\mathfrak m}}^{(k)}\otimes {\cal F}_P$.
Furthermore, it turns out that
\bea\label{isaususAAAA}
[{\boldsymbol{\tau}}_{\ell}(\lambda),{\bf I}_1]=0\ ,
\eea
where 
\bea\label{isaisisa}
{\bf I}_1=\int_0^{2\pi}\frac{\rd u}{2\pi}\ \big(\,(\partial\phi)^2+W_2\,\big)\ .
\eea
The eigenvalues of this
operator
 are given by 
\bea
P^2+\Delta_{{\mathfrak j}, {\mathfrak m}}-\frac{c_k+1}{24}+{\tt L}
\eea
with  ${\tt L}$ a non-negative integer. 
As such ${\bf I}_1$ naturally  introduces a grading in 
${\cal V}_{{\mathfrak j},{\mathfrak m}}^{(k)}\otimes {\cal F}_P$.
The dimensions of  each level eigenspaces characterized by  given ${\tt L}=0,1,\ldots$
is finite. Due to the commutativity condition \eqref{isaususAAAA} the transfer-matrix acts invariantly in
 each level subspace
of ${\cal V}_{{\mathfrak j},{\mathfrak m}}^{(k)}\otimes {\cal F}_P$. An immediate consequence is that
 the primary state $
\big|\,\sigma^{(k)}_{{\mathfrak j},{\mathfrak m}} \,\big\rangle\otimes|P\rangle$ 
is an eigenvector of the  transfer-matrix.

\bigskip

One can address  the diagonalization problem 
of ${\boldsymbol{\tau}}_{\ell}(\lambda)$ in the level subspaces of  
${\cal V}_{{\mathfrak j},{\mathfrak m}}^{(k)}\otimes {\cal F}_P$.
It turns out that the transfer-matrices are part 
of a larger commuting family. The latter includes the operators
 \be
{\mathlarger{\mathlarger{\mathlarger {\boldsymbol a}}}}_\pm(\lambda)\in {\rm End}\big(
{\cal V}_{{\mathfrak j},{\mathfrak m}}^{(k)}\otimes {\cal F}_P
\big)\
   :\ \ \  \ \ \ \ \ \ \ \big[ \boldsymbol{\tau}_{\ell}(\lambda),{\mathlarger{\mathlarger{\mathlarger {\boldsymbol  a}}}}_\pm(\lambda)\big]=
 \big[{\mathlarger{\mathlarger{\mathlarger { \boldsymbol a}}}}_+
 (\lambda), {\mathlarger{\mathlarger{\mathlarger {\boldsymbol  a}}}}_-(\lambda)\big]= 
   0\,,
 \ee
 which
 satisfy the  Baxter type  relation
  \bea\label{iaisaisa}
  {\boldsymbol \tau}_{\scriptscriptstyle \frac{1}{2}}(\lambda)\,
  {\mathlarger{\mathlarger{\mathlarger {\boldsymbol  a}}}}_\pm(\lambda)=
   \Omega^{\pm \frac{1}{2}}
\,
{\mathlarger{\mathlarger{\mathlarger {\boldsymbol  a}}}}_\pm(q^{+1}\lambda)  + \Omega^{\mp \frac{1}{2}}
 \,
{\mathlarger{\mathlarger{\mathlarger {\boldsymbol  a}}}}_\pm(q^{-1}\lambda)
\ .
  \eea
The  definition  of ${\mathlarger{\mathlarger{\mathlarger {\boldsymbol  a}}}}_\pm(\lambda) $
is similar to that given by  eqs.\,\eqref{9sd090as},\,\eqref{asusuya}
and involves the vertex operators $V_\pm$ \eqref{iaisisa} and $\Omega$ \eqref{oias90129012}.
However, the path-ordered exponent now
contains the generators of the
$q$-oscillator algebra ${\cal E}_\pm$ and ${\cal H}$:
\be\label{qosc}
[\Hcal,\Ecal_\pm]=\pm2\,\Ecal_\pm\,,\qquad\qquad q\, \Ecal_+\Ecal_--
q^{-1}\,\Ecal_-\Ecal_+=\frac{1}{q-q^{-1}} \ .
\ee
Let $\rho_\pm$  be representations of this algebra such that the traces 
\be\label{tr1}
{{\rm Tr}}_{\rho_\pm}\big[
\re^{\pm 2\ri\pi\beta P{\cal H}}\big]\ne 0,\, \infty \qquad {\rm with} \qquad {\Im} m(P)<0
\ee
exist and are non-vanishing.
Then
one  may introduce the operators ${\mathlarger{\mathlarger{\mathlarger{\mathlarger {\boldsymbol  a}}}}}_\pm(\lambda)$
as 
\bea\label{soso1a}
{\mathlarger{\mathlarger{\mathlarger {\boldsymbol  a}}}}_\pm(\lambda)=
\frac{{\rm Tr}_{\rho_\pm}\big[\,\re^{\pm\ri\pi\beta \hat{ a}_0{\cal H}}
{{\boldsymbol L}}_\pm(\lambda)\,\big]}{{\rm Tr}_{\rho_\pm}\big[\,\re^{\pm2\ri\pi \beta \hat{a}_0{\cal H}}\,\big]}\,,
\eea
where
\bea\label{Lop2a}
{\boldsymbol L}_\pm(\lambda)=\Omega^{\pm \frac{1}{4}{\cal H}}\ 
 \overset{\leftarrow}{{\cal P}}\exp\bigg(\lambda \int_0^{2\pi}\rd u\,
\Big(V_-(u)\ q^{\pm\frac{\cal H}{2}}\,{\cal E}_\pm+ V_+(u)\ q^{\mp\frac{\cal H}{2}}\,{\cal E}_\mp\Big)\bigg)\ .
\eea
It turns out \cite{Bazhanov:1998dq} that all the transfer-matrices are expressed 
through  ${\mathlarger{\mathlarger{\mathlarger{\mathlarger {\boldsymbol a}}}}}_\pm(\lambda)$ as 
\be\label{aisisaisa}
\big(\Omega^\frac{1}{2}-\Omega^{-\frac{1}{2}}\big)\ \mathlarger{{\boldsymbol{\tau}}}_{\ell}(\lambda)=\Omega^{\frac{2\ell+1}{2}}\,
{\mathlarger{\mathlarger{\mathlarger{ {\boldsymbol  a}}}}}_+\big(q^{\ell+\frac{1}{2}}\lambda\big)
{\mathlarger{\mathlarger{\mathlarger{ {\boldsymbol  a}}}}}_-\big(q^{-\ell-\frac{1}{2}}\lambda\big)-
\Omega^{-\frac{2\ell+1}{2}}\, {\mathlarger{\mathlarger{\mathlarger{ {\boldsymbol  a}}}}}_+\big(q^{-\ell- \frac{1}{2}}\lambda\big)
{\mathlarger{\mathlarger{\mathlarger{ {\boldsymbol  a}}}}}_-\big(q^{\ell+ \frac{1}{2}}\lambda\big)
\ee
with $\ell=0,\frac{1}{2},1\ldots$ and ${\boldsymbol{\tau}}_{0}(\lambda)\equiv\bf{1}$.

\section{Local IM\label{sec4}}

Among the operators which commute with the transfer-matrix, a special r${\hat{\rm o}}$le belongs to the  
local Integrals of Motion (IM) \cite{Zam}. 
These are a set of mutually commuting operators which 
can be written  in the form
\bea\label{laasss}
{\bf I}_s=\int_0^{2\pi}\frac{\rd u}{2\pi}\ T_{s+1}(u)
\eea
with $T_{s+1}$ being a chiral local density of integer  Lorentz spin $s+1$.
Remarkably, for a given choice of vertex operators, there exists a purely algebraic procedure
which, in principle,  allows one 
 to explicitly build  the local IM. 
Later, a generalization of the vertex operators
$V_{\pm}= \sqrt{k}\,\psi_\pm\, \re^{\pm \frac{2\ri\beta \phi}{\sqrt{k}}}$
will be proposed, which
gives rise  to new commuting families  involving  ${\boldsymbol{\tau}}_{\ell}(\lambda)$, 
${\mathlarger{\mathlarger{\mathlarger {\boldsymbol a}}}}_\pm(\lambda)$ as well as the corresponding sets of local IM.
Here, for future references, we illustrate the construction of $\{{\bf I}_s\}$  
for  the basic case.

\bigskip
Let ${\cal L}^{(s)}$ be the  linear space of chiral local fields of
Lorentz spin $s$ built out of the Bose field  $\phi$ and the fundamental parafermions  $\psi_\pm$.
For given 
positive integer $s$ it is a finite dimensional space.
We choose one of the vertices, say $V_+$, and  consider the linear subspace
${\cal W}^{(s)}\subset{\cal L}^{(s)}$ made up of  local fields $X_{s}$
such that
the singular part of the OPE $X_{s}(u)V_+(v)$ is a total derivative in $v$:
\bea\label{gasststa}
X_{s}(u)V_+(v)=\partial_v(\ldots)+O(1)\ .
\eea
In   the physical slang,  one says that the fields $X_{s}$ commute
with the ``screening charge''
\bea\label{aiissa}
Q=\oint\rd v\ V_+(v)\ .
\eea
Suppose $X_{s}$ and $Y_{s'}$ both commute with  $Q$.
By construction any    local field which appears in the OPE
$X_{s}(u)Y_{s'}(v)$  also commutes with  the  screening charge. 
Hence the direct sum  $\oplus_{s\geq 1}{\cal W}^{(s)}$ possesses
the structure of an operator  algebra.
Starting from the seminal  work of A.B. Zamolodchikov \cite{Zamolodchikov:1985wn},  
algebras of such type are referred to as
the $W$-algebras.

\bigskip
In the case under consideration 
the space ${\cal L}^{(1)}$ contains only the field $\partial \phi$, while ${\cal W}^{(1)}=\emptyset$.
The space ${\cal L}^{(2)}$ is  spanned by  $(\partial\phi)^2,\,\partial^2\phi$ and the
$W_2$ current occuring  in the OPE of the fundamental parafermionic fields $\psi_\pm$\ \eqref{aoisaisais}.
It is easy to check  that, up to an overall multiplicative factor,
 there is a single  spin 2 field commuting with the screening charge,
\bea\label{X2jad12}
X_2=(\partial\phi)^2+\tfrac{{\ri }}{\sqrt{k}}\ \big(\beta^{-1}-\beta\big)\, \partial^2\phi
+W_2\ .
\eea
Further,  ${\cal L}^{(3)}={\rm span}\big((\partial\phi)^3,\,\partial^2\phi\partial\phi,\ \partial^3\phi, 
\,W_3,\,\partial\phi W_2,\,\partial W_2\big)$ while
any local field from ${\cal W}^{(3)}$ is  proportional to the derivative $\partial X_2$.
The space ${\cal W}^{(4)}$ contains  $\partial^2 X_2$ and $(X_2)^2$. The latter is a spin 4 local field
that is the first  regular term in the OPE:
\bea\label{iiiasiasis}
X_2(u)X_2(v)=\frac{c}{2(u-v)^2}-\frac{X_2(u)+X_2(v)}{(u-v)^2}+(X_2)^2(v)+O(u-v)\ ,
\eea
where $c= \frac{3 k}{k+2}-\frac{6}{k}(\beta^{-1}-\beta)^2$.
It turns out that $\dim\big({\cal W}^{(4)}\big)=3$, i.e., together with the descendants of $X_2$ there is
an extra field $X_4$. Its construction   can be simplified if one  takes advantage of the
bosonization formulae for the parafermion currents \cite{Zam1986,Wakimoto,Gerasimov:1989mz}
\bea\label{isiasis}
\psi_\pm=\Big(\partial\alpha\pm\ri\, \sqrt{\tfrac{k+2}{k}}\ \partial\gamma\Big)\ \re^{\pm\frac{2\alpha}{\sqrt{k}}}\, .
\eea
This  involves two chiral Bose fields  with
 $\alpha(u)\alpha(v)=-\tfrac{1}{2}\log(u-v)+O(1)$ and similarly for $\gamma$.
Then the field $X_4$ is a  certain differential polynomial built  from
 $\partial\phi$, $\partial\alpha$, $\partial\gamma$.
In principal one can proceed further and explicitly describe the higher spin components ${\cal W}^{(s)}$.
The original definition of the parafermion algebra requires that
$k$ is a positive integer.
Nevertheless it  can be treated as an
arbitrary complex number
in the bosonization formulae \eqref{isiasis}.
This makes it possible  to introduce
the $W$-algebra associated with the screening charge $Q$ \eqref{aiissa},
 for  arbitrary  $k$ as  was done in the  unpublished work of V.A. Fateev as well as
 in refs.\cite{Feigin:2001yq,Lukyanov:2012wq}. In  the last paper   $\oplus_{s=1}^\infty{\cal W}^{(s)}$
 was referred to as the corner-brane $W$-algebra.

\bigskip
 The second vertex operator $V_-$ comes in to play when
obtaining the local IM from the fields
in the $W$-algebra. One searches for local 
fields $T_{s+1}\in {\cal W}^{(s+1)}$ such that the OPE of $T_{s+1}$  and  $V_-$ possesses the following structure
\bea\label{aiisaisasia}
T_{s+1}(u)V_-(v)=\sum_{m=2}^{s+1}\frac{{R}_{-m}(v)}{(u-v)^m}+\frac{{R}_{-1}(v)}{u-v}+O(1)\ \ \ \ \ 
{\rm with}\ \ \ \  \ {R}_{-1}=\partial {\cal O}(v)\ .
\eea
Note that, if $T_{s+1}$ exists,  it is  defined up to the addition of a total derivative
$\partial X_s$  $(\forall X_s \in {\cal W}^{(s)})$.
This ambiguity has no affect on  ${\bf I}_s$, which is expressed through an integral as in \eqref{laasss}.
There is also an ambiguity in the overall multiplicative 
normalization of $T_{s+1}$,
which is carried over to the local IM.
Following the arguments of ref.\cite{Bazhanov:1998dq}, it is expected that ${\bf I}_{s}$
commutes with the transfer-matrices and ${\mathlarger{\mathlarger{\mathlarger {\boldsymbol  a}}}}_\pm(\lambda)$:
 \bea
\big[ \boldsymbol{\tau}_{\ell}(\lambda), {\bf I}_{s}\big]=
 \big[{\mathlarger{\mathlarger{\mathlarger {\boldsymbol a}}}}_\pm (\lambda),{\bf I}_{s} \big]= 
   0\ .
 \eea
Also, assuming that the transfer-matrices resolve all the degeneracies in
$ {\cal V}_{{\mathfrak j},{\mathfrak m}}^{(k)}\otimes{\cal F}_P$, one arrives at
the mutual commutativity condition
\bea
\big[\,{\bf I}_{s} , {\bf I}_{s'}\big]=0\ .
\eea
In the case under consideration 
the  operator   ${\bf I}_{s}$ exists and is unique   (up to overall  normalization)
 only for odd  $s=1,3,5,\ldots\ $. For example, for $s=1$ the density $T_2=X_2 $ from \eqref{X2jad12}, while
the explicit formula for $ {\bf I}_{3}$ was originally obtained in refs.\cite{Fateev:1995ht,Fateev:1996ea}.
 Further facts on this commuting family
of local IM can be found in \cite{Bazhanov:2013cua}.

\section{Prototype  example of the ODE/IQFT correspondence\label{sec5}}

The most effective way for computing the spectrum of 
the operators ${\boldsymbol{\tau}}_{\ell}(\lambda)$,
 ${\mathlarger{\mathlarger{\mathlarger {\boldsymbol  a}}}}_\pm(\lambda)$ and ${\bf I}_{2n-1}$
in the space ${\cal V}_{{\mathfrak j},{\mathfrak m}}^{(k)}\otimes{\cal F}_P $
is provided
by the ODE/IQFT correspondence. 
As part of our study of the new commuting family, a class of ODEs
for the eigenstates will be proposed.
In order to prepare for that discussion it would be useful to 
demonstrate the approach on the
eigenvalues of  ${\mathlarger{\mathlarger{\mathlarger {\boldsymbol  a}}}}_+(\lambda)$ 
corresponding to the primary state 
$|\,\sigma^{(k)}_{{\mathfrak j},{\mathfrak m}} \,\big\rangle\otimes\big|P\rangle$.

\bigskip
Let us start with the simplest case when ${\mathfrak j}={\mathfrak m}=0$.
According to the work \cite{Lukyanov:2006gv}, one should consider the linear  differential equation
\bea\label{asoosao}
\bigg[-\partial_x^2+\kappa^2\ \big(\,\re^{\xi x }+\re^{(1+\xi) x}\,\big)^k-A^2\
\bigg]\ \Theta=0\, .
\eea
Here $\xi>0$ while $k$  is  assumed to be  a positive integer.
For $\Im m (A)\geq 0$ one can introduce the Jost
solution, which asymptotically approaches to a plane wave 
\bea
{\Theta}^{\scriptscriptstyle{(\leftarrow)}}_{A}(x)\asymp \re^{-\ri A x}\ \ \ \ \ \ \ \ {\rm as}\ \ \ \ x\to-\infty\ .
\eea
It turns out that as a function of $A$ it is meromorphic. 
This allows one to 
unambiguously define  $\Theta^{\scriptscriptstyle{(\leftarrow)}}_{A}(x)$ 
for any complex $A$ except for a discrete, pure imaginary  set of values,
where it develops simple poles. 
The function
 $\Theta^{\scriptscriptstyle{(\leftarrow)}}_{-A}(x)$
 is another solution to the ODE \eqref{asoosao}, 
which is linearly independent from  $\Theta^{\scriptscriptstyle{(\leftarrow)}}_A(x)$ when $A\ne 0$.
\bigskip

Since the potential in \eqref{asoosao} grows rapidly for large positive $x$, the ODE admits a solution which 
 decays at $x\to+\infty$.
We denote it by $\Theta^{\scriptscriptstyle{(\rightarrow)}}$ and specify its overall  normalization through the asymptotic condition
\bea\label{iasisias}
\Theta^{\scriptscriptstyle{(\rightarrow)}}\asymp \re^{-F(y)+o(1)}\ \ \  \ \ {\rm as}\ \ \ y=x+\log(\mu)\to+\infty\ ,
\eea
where
\be
F(y)=\tfrac{1}{4}\,  (1+\xi)k\,   y+\tfrac{2 }{(1+\xi) k}\  \re^{\frac{1 }{2}(1+\xi)k\,y}\,
{}_{2}F_1\big(-\tfrac{1}{2}\, (1+\xi)k,-\tfrac{1}{2}\,k,
1-\tfrac{1}{2}\, (1+\xi) k\,;\,-\mu\re^{-y}\,\big)
\ee
(${}_2F_1$ stands for the Gauss  hypergeometric function)
and
\bea
\mu=\kappa^{\frac{2}{(1+\xi)k}}\ .
\eea
The solution $\Theta^{\scriptscriptstyle{(\rightarrow)}}$ 
may be introduced for any complex value of $A^2$ and is  an entire function
of this parameter.

\bigskip
Consider  the connection coefficients for the ODE \eqref{asoosao},  which are expressed
via the Wronskian, ${\rm Wron}[f,g]=f\partial_xg-g\partial_xf $, as
\bea\label{iissisiaBB}
W(\mu)=\mu^{- \ri A}\ {\rm Wron}\big[\,\Theta^{\scriptscriptstyle{(\rightarrow)}}, \Theta^{\scriptscriptstyle{(\leftarrow)}}_{ A}\,\big]\ .
\eea
Here the overall factor has been chosen to ensure
the existence of the limit
\bea\label{ajsusus}
\lim_{\mu\to 0}W(\mu)=W(0)\ .
\eea
The connection coefficient $W$  is a meromorphic function of $A$.
For given $A$ it 
turns out to be  an entire function of $\mu$
and, therefore,  admits a convergent Taylor series expansion.
This is related to the fact that
in the variable $y=x+\log(\mu)$,
  the ODE takes the form
\bea\label{asoosaoA}
\big(-\partial_y^2+U(y)-A^2\,\big)\Theta=0\ \ \ \ \ \qquad {\rm with}\ \ \  \ \  \ \qquad U(y)=\re^{(1+\xi)k y}+ \delta U\ .
\eea
The term $\delta U=(\re^{(1+\xi) y }+\mu\, \re^{\xi y})^k-\re^{(1+\xi)k y}$
 analytically depends on $\mu$ and   can be treated perturbatively so long as $\xi>0$, $k=1,2,\ldots\ $.
 In ref.\cite{Lukyanov:2006gv} it was shown that  the ratio $W(\mu)/W(0)$ coincides with the eigenvalue 
of
${\mathlarger{\mathlarger{\mathlarger {\boldsymbol a}}}}_+(\lambda)$
 corresponding to the primary state   $
\big|\,\sigma^{(k)}_{{\mathfrak j},{\mathfrak m}} \,\big\rangle\otimes|P\rangle $ with $ {\mathfrak j}={\mathfrak m}=0$,
provided the following identification of the parameters is made:
\bea\label{siisaiasd}
\xi=\tfrac{\beta^2}{1-\beta^2}\ ,\ \ \ \ \ \ \  \ \mu=-\lambda^2\ \Gamma^2\big(\tfrac{1-\beta^2}{k}\big)
\big(\tfrac{k}{1-\beta^2}\big)^{\frac{2}{k}(1-\beta^2)}
\eea
and also
\bea
A=\tfrac{\ri \sqrt{k}}{\beta^{-1}-\beta}\ P\ .
\eea
Recall  that
the vertices $V_\pm$ are  non-local fields with fractional conformal dimensions.
In writing the $\mu-\lambda$  relation \eqref{siisaiasd} we  assume that the phase ambiguity of $V_\pm$ 
is specified   through the following condition imposed on the OPE
 \bea\label{iissauas}
q\ V_\pm(u_1)V_\mp(u_2)\big|_{u_1-u_2\to 0^+}\to k\times  (u_1-u_2)^{-\frac{2}{k}(k-1+\beta^2)}\, \big(1+o(1)\,\big)>0\ .
\eea

We now turn to  the eigenvalues corresponding to the states
$
\big|\,\sigma^{(k)}_{{\mathfrak j},{\mathfrak m}} \,\big\rangle\otimes|P\rangle $
with ${\mathfrak j}$ and ${\mathfrak m}$ not necessarily zero. 
To the best of our knowledge the ODE/IQFT correspondence for this case
 has not yet been discussed in the literature
despite that it follows immediately from the results of the works 
\cite{Lukyanov:2012wq,Bazhanov:2013cua}.
The generalization of the ODE \eqref{asoosao} reads as
\bea\label{aiisaisa}
\bigg[-\partial_x^2+\kappa^2\ 
\big(\re^{(1+\xi)x}+\re^{\xi x}\,\big)^k-\frac{ A^2+B^2\re^{x}}{1+\re^x}-
(C^2-\tfrac{1}{4}\big)\ \frac{\re^{x}}{(1+\re^x)^2}
\bigg]\ \Theta=0\,.
\eea
Together with $\xi$ and $k$ this equation involves  three extra parameters. With the specialization
$B^2=A^2$ and $C^2=\tfrac{1}{4}$, it boils down  to \eqref{asoosao}.
As with the previous case one can consider the Wronskian  $W(\mu)$ \eqref{iissisiaBB}. 
However an important difference now  is that $W(\mu)$
 turns out to be  an entire function of $\log(\mu)$ rather than $\mu$.  
This is related to the fact that  when the  ODE is  rewritten in the form \eqref{asoosaoA},
the term $\delta U$  is no longer an entire function of $\mu$.

\bigskip
For the analysis of eq.\eqref{aiisaisa} it is convenient to perform the change of variables
\bea\label{iasisisaswq}
z=\re^x\ ,\ \ \ \ \  \ \ { \Psi}(z)=\re^{\frac{x}{2}}\, \Theta(x)\ ,
\eea
bringing the ODE to the form
\bea\label{isaiisaisa}
\bigg[\!\!\!&-& \partial_z^2+\kappa^2\ z^{-2+\xi k  }\,
(1+z)^k\\[0.2cm]
&-&\frac{A^2+\frac{1}{4}}{z^2}+\frac{A^2-B^2-C^2+\frac{1}{4}}{z}
+\frac{{ B}^2+{ C}^2-{A}^2-\frac{1}{4}}{1+z}+
\frac{{C}^2-\frac{1}{4}}{ (1+z)^2}
\,\bigg]\ {\Psi}=0\, .\nonumber
\eea
For generic values of the parameters it possesses three singular points at $z=0,-1,\infty$.
Let us explore the condition when the singularity
 at $z=-1$ becomes apparent.
First we consider the case $\kappa=0$ when \eqref{isaiisaisa} belongs to  the Fuchsian class. The test for the apparent
singularity is well known and can be found, for example, in ref.\cite{Gao}. 
Suppose we are given the equation
\bea
\big(-\partial_z^2+V(z)\big)\, \Psi=0
\eea
and $V(z)$ admits a  Laurent  expansion of the form
\bea
V(z)=\frac{1}{(z-w)^2}\ \bigg( v_0+\sum_{m=1}^\infty v_m (z-w)^m\bigg)\, .
\eea
The singularity at $z=w$ is apparent if and only if
\bea\label{iasiisisaa}
v_0={\mathfrak j}({\mathfrak j}+1)\ \ \ \ \ \ {\rm with}\ \ \ \ {\mathfrak j}=0,\tfrac{1}{2},1,\ldots
\eea
and
\bea\label{asuasisuaiu}
F_{{\mathfrak j}}(v_1,\ldots, v_{2{\mathfrak j}+1})=0\ .
\eea
Here the polynomials  $F_{{\mathfrak j}}$ are defined through the determinant
\bea\label{hasgsats}
F_{{\mathfrak j}}(v_1,\ldots, v_{2{\mathfrak j}+1})=
\det\begin{pmatrix}
&v_1& 1\cdot( 2{\mathfrak j}) &     0    & \ldots  &0\\
&v_2& v_1&2\cdot (2{\mathfrak j}-1)&\ldots  &0\\
&\vdots &\vdots   &\vdots&\ddots &\vdots\\
&v_{2{\mathfrak j}}&v_{2{\mathfrak j}-1}&v_{2{\mathfrak j}-2}&\ldots &2 {\mathfrak j}\cdot (1)\\
&v_{2{\mathfrak j}+1}&v_{2{\mathfrak j}}&v_{2{\mathfrak j}-1}&\ldots & v_1
\end{pmatrix}.
\eea
In the simplest cases \eqref{asuasisuaiu} reads explicitly as
\bea
{\mathfrak j}&=&\tfrac{1}{2}\ :\ \ \ \ \ v_1^2-v_2=0\\[0.3cm]
{\mathfrak j}&=&1\ :\ \ \ \ \  v_1\,\big(v_1^2-4 v_2\,\big)+4 v_3=0\ .\nonumber
\eea
Applying the above condition to  \eqref{isaiisaisa} with $\kappa=0$
 one concludes that the singularity at $z=-1$ is  apparent provided the parameters
 $A$, $B$ and $C$ satisfy the  conditions
 \be\arraycolsep=0.4cm
\begin{array}{lll}
 C^2=\big({\mathfrak j}+\tfrac{1}{2}\big)^2  & {\rm with} & {\mathfrak j}=0,\tfrac{1}{2},1,\ldots\\[0.3cm]
B^2=\big(A-
 \tfrac{\ri}{2}\,{\mathfrak m}\big)^2  & {\rm with} & 
{\mathfrak m}=-2{\mathfrak j},\, -2{\mathfrak j}+2,\ldots, {\mathfrak j}-2,2{\mathfrak j}\, .
\end{array}
 \ee
\bigskip
Suppose now that $\kappa\not=0$. 
For
\bea
k=1,2,3,\ldots
\eea
the term $\propto \kappa$ has a zero of order $k$ at $z=-1$.
Hence its presence will not affect the conditions \eqref{iasiisisaa} and \eqref{asuasisuaiu}
as long as  
\bea
{\mathfrak j}=0,\tfrac{1}{2},1,\ldots,\tfrac{k}{2}\ .
\eea

\bigskip
When $z=-1$  is an apparent  singularity 
we define 
 the normalized connection coefficients in  the frame $(z,\Psi)$ \eqref{iasisisaswq} as
\be\label{iissisiao}
D_{{\mathfrak j},{\mathfrak  m},A}(\mu)=\frac{W_{{\mathfrak j},{\mathfrak m},A}(\mu)}
{W_{{\mathfrak j},{\mathfrak m},A}(0)}\ ,\ \ \ \  \quad {\rm where}\ \ \ 
W_{{\mathfrak j},{\mathfrak m},A}(\mu)=\mu^{-\frac{1}{2}\mathfrak{m}- \ri A}\
{\rm Wron} \big[\,\Psi^{\scriptscriptstyle{(\rightarrow)}}, \Psi^{\scriptscriptstyle{(\leftarrow)}}_{A}\,\big]\, .
\ee
Then 
$W_{{\mathfrak j},{\mathfrak m},A}$ admits a power series expansion in $\mu$.\footnote{In fact the phenomena 
was already discussed in ref.\cite{Bazhanov:2017nzh} for the case $\xi=0$.}
It is possible to show that
\bea
W_{{\mathfrak j},{\mathfrak m},A}(0)=\frac{1}{\sqrt{\pi}}\  \big({(1+\xi) k}\big)^{\frac{1}{2}-\frac{\mathfrak{m}+ 2\ri A}{(1+\xi) k}}\ 
\frac{\Gamma\big(1- \tfrac{\mathfrak{m}+ 2\ri A}{(1+\xi) k}\big)
\Gamma(1- 2\ri A)\, \Gamma(-\mathfrak{m}- 2\ri A)}
{ \Gamma\big( \mathfrak{j}+1-\tfrac{\mathfrak{m}}{2} - 2\ri A\big)\,
\Gamma\big(-\mathfrak{j}-\tfrac{\mathfrak{m}}{2} - 2\ri A \big)}\ ,
\eea
while 
\bea\label{iaiissiaasi}
D_{{\mathfrak j},{\mathfrak  m},A}(\mu)&=&1+
\bigg(\frac{2}{(1+\xi) k}\bigg)^{\frac{2}{(1+\xi)k}}\ 
\frac{\Gamma\big(-\frac{1}{(1+\xi) k}\big)\Gamma\big(\frac{1}{2}+\frac{1}{(1+\xi) k}\big)}{4\sqrt{\pi}}\ 
\frac{\Gamma\big(1-\frac{1+\mathfrak{m}+ 2\ri A}{(1+\xi) k}\big)}
{\Gamma\big(\frac{1-\mathfrak{m} - 2\ri A}{(1+\xi) k}\big)}
\nonumber\\[0.3cm]
&\times& 
 \bigg(\,\frac{4\mathfrak{j}(\mathfrak{j}+1)+
{\mathfrak m}^2
+  4\ri\mathfrak{m} A}{(1-{\mathfrak m}- 2\ri A)(1+{\mathfrak m}+ 2\ri A)}
+
\frac{2k}{(1+\xi) k -2}\,\bigg)\ \mu+O(\mu^2)\ .
\eea

\bigskip
A comparison of the first nontrivial expansion coefficient of 
the eigenvalue  of ${\mathlarger{\mathlarger{\mathlarger {\boldsymbol a}}}}_+(\lambda)$
 corresponding to the state
$
\big|\,\sigma^{(k)}_{{\mathfrak j},{\mathfrak m}} \,\big\rangle\otimes|P\rangle$ 
with formula \eqref{iaiissiaasi} results in the relations
\be\arraycolsep=0.7cm
\begin{array}{ll}
A=\ri\, \big(\,\sqrt{\xi (1+\xi) k}\, P-\frac{1}{2}\,
\xi\, {\mathfrak m}\big)\, ,&
   {C}^2=\big({\mathfrak j}+\frac{1}{2}\big)^2
 \\[0.2cm]
  \!\! \!{B}^2=- \big(\,\sqrt{\xi (1+\xi) k} \,P-\frac{1}{2}
 (1+\xi)\, {\mathfrak m}\big)^2
\ .&
\end{array}
\ee
It also confirms   eq.\,\eqref{siisaiasd}.
This way one identifies
 $D_{{\mathfrak j},{\mathfrak  m},A}(\mu)$   with the eigenvalue  of ${\mathlarger{\mathlarger{\mathlarger { \boldsymbol a}}}}_+(\lambda)$
 corresponding to the state
$
\big|\,\sigma^{(k)}_{{\mathfrak j},{\mathfrak m}} \,\big\rangle\otimes|P\rangle$ in the case
${\mathfrak m}=-2{\mathfrak j},\, -2{\mathfrak j}+2,\ldots, 2{\mathfrak j}-2,2{\mathfrak j}$.

 \bigskip
Regarding the other members of the commuting family, the
eigenvalue of ${\mathlarger{\mathlarger{\mathlarger {\boldsymbol  a}}}}_-(\lambda)$
for the state $
\big|\,\sigma^{(k)}_{{\mathfrak j},{\mathfrak m}} \,\big\rangle\otimes|P\rangle$
coincides with $D_{{\mathfrak j},-{\mathfrak  m},-A}(\mu)$.
The eigenvalues of the transfer-matrices ${\boldsymbol{\tau}}_{\ell}(\lambda)$ 
are also given by certain connection 
 coefficients for the ODE  \eqref{isaiisaisa}. Those of the
local IM can be extracted from the 
  large $\mu$ behaviour of
$D_{{\mathfrak j},{\mathfrak  m},A}$ within the standard WKB technique.
All this is well known and widely discussed in the literature and will
not be elaborated on here.

 \section{New realization of $U_q(\widehat{{\mathfrak b}}_-)$\label{sec6}}
Introduce the $r$ non-local fields 
\bea\label{iuaususa}
V_{\pm}^{(a)}=\sqrt{k_a}\ \psi_\pm^{(a)}\ \exp\bigg[\pm 2\ri\, \Big(\,\tfrac{\beta-1}{K} \sum_{b=1\atop b\not=a}^r\sqrt{k_b}\ \phi_b+
\big(\tfrac{\beta-1}{K}\ \sqrt{ k_a}+\tfrac{1}{\sqrt{k_a}}\big)\, \phi_{a}\,\bigg)\bigg]\,,
\eea
where $\phi_a$ and $\psi_\pm^{(a)}$ are 
$r$ independent copies of the chiral Bose  
 and ${\mathbb Z}_{k_a}$ parafermionic  fields, respectively.
Also  the positive integer $K$ stands for
\bea
K=\sum_{a=1}^rk_a\, .
\eea
The definition \eqref{iuaususa} has been arranged so that
the braiding relations
 \bea\label{oiass90121}
V^{(a)}_{\sigma_1}(u_1) V^{(b)}_{\sigma_2}(u_2)=q^{2\sigma_1\sigma_2}\ V^{(b)}_{\sigma_2}(u_2) V^{(a)}_{\sigma_1}(u_1)\ ,\ \ \ u_1>u_2
\eea
with $q=-\re^{\frac{\ri\pi}{K}(\beta^2-1)}$ are satisfied.
A simple way to see this is to rewrite $V^{(a)}_\pm$ in the form
\bea
V_{\pm}^{(a)}=J^{(a)}_{\pm}\ \re^{\pm \frac{2\ri(\beta-1)}{\sqrt{K}} \varphi}\ ,
\eea
where $J^{(a)}_\pm$ is given by eq.\,\eqref{aisisai} with 
$k$, $\psi_\pm$, $\phi$ swapped by $k_a$, $\psi_\pm^{(a)}$ and $\phi_a$, respectively,
while
\bea\label{iasisiasass}
\varphi=\frac{1}{\sqrt{K}}\sum_{a=1}^r\sqrt{k_a}\,\phi_a\ :\ \ \ \ J_0^{\rm (tot)}\equiv \sum_{a=1}^rJ_0^{(a)}=2\sqrt{K}\,
 \partial \varphi\ .
\eea
Then the braiding relations follow from the fact that the
 Kac-Moody currents $J^{(a)}_{\pm}$, $J^{(a)}_{0}$ are mutually local fields.
Notice that the field $\varphi$ satisfies the
OPEs
\be
 \varphi(u_1)\varphi(u_2)=-\tfrac{1}{2}\ \log(u_1-u_2)+O(1)\ ,\ \ \  \ 
  \phi_a(u_1)\varphi(u_2)=-\sqrt{\tfrac{k_a}{4K}}\ \log(u_1-u_2)+O(1)
\ee
We will  assume that    the non-local fields $V^{(a)}_\pm$
are normalized by means of  the condition 
  \bea
q\ V^{(a)}_\pm(u)\,V^{(b)}_\mp(0)\big|_{u\to 0^+}\to k_a\, u^{-\frac{2}{k_a}(k_a-1+\beta^2)}\,
\big(\delta_{ab}+o(1)\big)>0\ ,
\eea
which is similar to \eqref{iissauas}.

\bigskip
Let ${\cal V}_{{\mathfrak j}_a}$ be the space of representation of the 
parafermionic  algebra generated by the fundamental parafermions $\psi_\pm^{(a)}$.
 As before the subspace denoted  by  ${\cal V}_{{\mathfrak j}_a,m_a}^{(k_a)}$  is the one
with fixed ${\mathbb Z}_{k_a}$ charge $m_a$, so that
 \bea
 {\cal V}_{{\mathfrak j}_a,m_a}^{(k_a)}=\emptyset\ \ \ \  \ \ {\rm as}\ \ \  \ m_a-2{\mathfrak j}_a\not=2\,{\mathbb Z}
 \eea
 and 
 \bea
 {\cal V}_{{\mathfrak j}_a,m_a}^{(k_a)}={\cal V}_{{\mathfrak j}_a,m'_a}^{(k_a)}\ \ \ \ \ \ \ \ {\rm for}\ \ \ \ m_a=m_{a}'\ \ 
 ({\rm mod}\  k_a)\, .
 \eea
Also  ${\cal F}^{(a)}_{P_a}$ is the space of  representation for the Heisenberg algebra
 generated of the Fourier modes of  the
field $\partial\phi_a$ with $P_a$ being the value of the corresponding zero mode momentum.
Introduce the space
 \bea\label{oi981981212}
 {\cal H}\Big[
 \small{
 \begin{matrix}
 m_1,\ldots m_r\\
 {\mathfrak j}_1,\ldots ,{\mathfrak j}_r
 \end{matrix}
 }
 \Big|\, s \,\Big]= \bigotimes_{b=1}^r\Big(\, 
 {\cal V}_{{\mathfrak j}_b,{m}_b}^{(k_b)}\otimes{\cal F}^{(b)}_{P_{b}({ m}_b|s)}\, \Big)\ ,
 \eea
 where
\bea
P_b({ m}_b|s)=
\tfrac{1}{2}\ \big(\,\tfrac{{m}_b}{\sqrt{k_b}}+
{s}\, \sqrt{k_b}\,\big)
\eea
and
$s$ is an arbitrary number.
 The  operators $V^{(a)}_\pm$   act as the intertwiners
 \bea
V^{(a)}_\pm\ :\ \ \ \
 {\cal H}\bigg[
 \begin{matrix}
 m_1,\ldots, m_r\\
 {\mathfrak j}_1,\ldots ,{\mathfrak j}_r
 \end{matrix}
 }
 \bigg|\, s \,\bigg]
 \mapsto
 {\cal H}\bigg[
 \small{
 \begin{matrix}
 m_1,\ldots, m_a\pm 2,\ldots, m_r\\
 {\mathfrak j}_1,\ldots ,\phantom{\pm2}{\mathfrak j}_a\phantom{\pm2},\ldots,{\mathfrak j}_r
 \end{matrix}
 \bigg|\, s\pm\tfrac{2}{K}\, (\beta-1) \,\bigg]
 \eea
and hence   invariantly   within  the direct sum
\bea\label{aisisi}
{\cal H}_{ {\mathfrak j}_1,\ldots ,{\mathfrak j}_r}=
  \bigoplus_{{m}_b\in 2{\mathfrak j}_b+2{\mathbb Z}}
  {\cal H}\bigg[
 \begin{matrix}
 m_1,\ldots, m_r\\
 {\mathfrak j}_1,\ldots ,{\mathfrak j}_r
 \end{matrix}
 \bigg|\, \tfrac{2P_0}{\sqrt{K}}+\tfrac{\beta-1}{K} \sum_{c=1}^r m_c \,\bigg]\ .
 \eea
In other words, for any given $P_0$ and ${\mathfrak j}_1,\ldots ,{\mathfrak j}_r$,
\bea
  V^{(a)}_\pm\in {\rm End} \big( {\cal H}_{\bm{\mathfrak{j}}}\big)\ ,
\eea
where we use the shortcut notation
\be
{\boldsymbol {\mathfrak j}}=({\mathfrak j}_1,\ldots,{\mathfrak j}_r)\,.
\ee

\bigskip
 Each of the components in the  linear decomposition \eqref{aisisi} is an eigenspace  for the zero mode momentum of the field
 $\varphi$ \eqref{iasisiasass}: 
 \bea
\hat{a}_0=\int_{0}^{2\pi}\frac{\rd u}{2\pi}\, \partial\varphi=\frac{1}{2\sqrt{K}}\ 
 \int_{0}^{2\pi}\frac{\rd u}{2\pi}\, J_0^{(\rm tot)}
 \eea
with the corresponding eigenvalue
 \bea\label{oasisis}
 P=P_0+\tfrac{\beta}{2\sqrt{K}}\ \sum_{c=1}^rm_c\ .
 \eea
 Together with $\hat{a}_0$ introduce  $\hat{U}\in {\rm End}\big( {\cal H}_{\bm{\mathfrak{j}}}\big)$
 such that
 \bea\label{aiiaasyy}
\hat{ U}\ {\cal H}\Big[
 \begin{matrix}
 m_1,\ldots m_r\\
 {\mathfrak j}_1,\ldots ,{\mathfrak j}_r
 \end{matrix}
 \Big|\, s \,\Big]=\re^{-\frac{\ri \pi}{K}\sum_{c=1}^r m_c}\ 
 {\cal H}\Big[
 \begin{matrix}
 m_1,\ldots, m_r\\
 {\mathfrak j}_1,\ldots ,{\mathfrak j}_r
 \end{matrix}
 \Big|\, s \,\Big]\ 
 \ \ \ \  \ \ \ \  (\,\forall\ s\,)\ .
 \eea
 An important property is that
the  non-local fields
 $V^{(a)}_\pm\in {\rm End}\big( {\cal H}_{\bm{\mathfrak{j}}}\big)$
 obey the quasiperiodicity condition
 \bea\label{oias09120912}
 V^{(a)}_\pm(u+2\pi)=q^{-2}\ \Big(\re^{\frac{2\pi \ri\beta}{\sqrt{K}} \hat{a}_0}\, \hat{U}\Big)^{\pm 2}\  V^{(a)}_\pm(u)
 \eea
with  the operator valued factor being independent of  $a=1,2,\ldots,r$.
 The relations \eqref{oiass90121} and \eqref{oias09120912} imply  that the 
 vertex operators
\bea\label{isaiiasias}
V_\pm\in{\rm End} \big( {\cal H}_{\bm{\mathfrak{j}}}\big)\ :\ \ \ 
V_+= \sum_{a=1}^rV_+^{(a)}\ ,\ \ \ \ \ \ \ 
V_-= \sum_{a=1}^r z_aV_-^{(a)}\ ,
\eea
 satisfy  the conditions \eqref{jausus}-\eqref{aisaisisaias}
with
\bea\label{oias8o21}
q=-\re^{\frac{\ri\pi}{K}(\beta^2-1)}\ , \ \ \ \ \Omega^{\frac{1}{2}}=\re^{\frac{2\pi \ri\beta}{\sqrt{K}} \hat{a}_0}\, \hat{U}
\in
{\rm End} \big( {\cal H}_{\bm{\mathfrak{j}}}\big)\ .
\eea
The parameters $\{z_a\}_{a=1}^r$ entering into the definition of $V_-$
may be arbitrary.
\bigskip

The
operators 
\be\label{contourInt1a}
x_0 = \frac{1}{q-q^{-1}}\, \int_0^{2\pi}{\rm d}u\,  \sum_{a=1}^rV_+^{(a)}\ ,\qquad 
x_1 = \frac{1}{q-q^{-1}}\,\int_0^{2\pi}{\rm d}u\,  \sum_{a=1}^r z_a\,V_-^{(a)}
\ee
are expected to  obey  the quantum  Serre
relations \eqref{Serre1}.
Together with 
\bea\label{oias90121A}
q^{h_0}=\re^{\frac{2\pi \ri\beta}{\sqrt{K}} \hat{a}_0}\, \hat{U}
\eea
they provide a
 realization of the generators of  the  Borel subalgebra 
$U_q(\widehat{{\mathfrak b}}_-)$.\footnote{It deserves to be mentioned that the construction of the transfer matrices 
 ${\boldsymbol{\tau}}_\ell(\lambda)$
 given by eqs.\eqref{asusuya},\,\eqref{9sd090as}  involves the operator
  $\Omega^{\frac{1}{2}}=q^{h_0}$ in the case of half-integer $\ell$.}
 In all likelihood, it can be 
understood as coming from $r$ 
consecutive  applications of the co-multiplication
\bea
\delta(x_i)=x_i\otimes 1+q^{-h_i}\otimes x_i\ ,\ \ \ \  \delta(q^{h_i})=q^{h_i}\otimes 1+1\otimes q^{h_i}
\eea
to the operators $x_0$, $x_1$ with $r=1$.

\bigskip
As soon as a representation of the Borel subalgebra $U_q(\widehat{{\mathfrak b}}_-)$ 
is specified, an infinite family of commuting operators can be constructed
via the general definitions \eqref{asusuya}  for the transfer-matrices  and
 \eqref{soso1a}  for ${\mathlarger{\mathlarger{\mathlarger {\boldsymbol a}}}}_\pm(\lambda)$.
They commute
with the zero-mode momentum of the field $\varphi$ and the operator $\hat{U}$ \eqref{aiiaasyy}:
\bea
\big[\hat{a}_0,{\boldsymbol{\tau}}_\ell(\lambda)\big]=
\big[\hat{a}_0,{\mathlarger{\mathlarger{\mathlarger {\boldsymbol a}}}}_\pm(\lambda)\big]=0\ ,\ \ \ 
\big[\hat{U},{\boldsymbol{\tau}}_\ell(\lambda)\big]=\big[\hat{U},
{\mathlarger{\mathlarger{\mathlarger {\boldsymbol a}}}}_\pm(\lambda)\big]=0\, .
\eea
As a result, they act invariantly in the subspaces ${\cal H}_{ \mathfrak{j}}$ \eqref{aisisi}
with fixed value of these operators. The latter will be denoted by 
${\cal H}_{\bm{\mathfrak{j}}, {\mathfrak m},P}\subset {\cal H}_{ \mathfrak{j}}$:
 \bea\label{oias90121B}
&& \hat{a}_0\, {\cal H}_{\bm{\mathfrak{j}}, {\mathfrak m},P}
=P\ 
{\cal H}_{\bm{\mathfrak{j}}, {\mathfrak m},P}
\\[0.3cm]
&& \hat{U}\, {\cal H}_{\bm{\mathfrak{j}}, {\mathfrak m},P}\,
=\re^{-\frac{\ri\pi }{K} \mathfrak{m}}\
{\cal H}_{\bm{\mathfrak{j}}, {\mathfrak m},P}
 \, .\nonumber
 \eea
Here the integer $\mathfrak{m}$ is defined modulo $2K$. 
We employ the conventions that
\bea
\mathfrak{m}= -2\mathfrak{J}, -2\mathfrak{J}-2,\ldots,   2K-2\mathfrak{J}-2\ ,\ \ \ \ \ \ {\rm where}\ \ \ 
\mathfrak{J}=\sum_{a=1}^r{\mathfrak j}_a\ ,\ \ \  \ K=\sum_{a=1}^rk_a\, .
\eea
As follows from \eqref{oasisis}, the value of the zero-mode momentum $P$ must obey the relation
 \bea\label{iiasiasias}
 \re^{\frac{2\pi\ri}{\beta\sqrt{K}}(P-P_0)}=\re^{\frac{\ri\pi}{K} {\mathfrak m}}\ .
 \eea
Explicitly, the decomposition of ${\cal H}_{\bm{\mathfrak{j}}, {\mathfrak m},P}$ is given by
 \bea\label{aisisidd}
{\cal H}_{\bm{\mathfrak{j}}, {\mathfrak m},P}=\sum_{N=-\infty}^\infty\ 
  \bigoplus_{(m_1,\ldots m_{r})\in \Sigma_{\bm{\mathfrak{j}},{\mathfrak m},N}}
  {\cal H}\bigg[
 \begin{matrix}
 m_1,\ldots, m_r\\
 {\mathfrak j}_1,\ldots ,{\mathfrak j}_r
 \end{matrix}
 \bigg|\, \tfrac{2P}{\sqrt{K}}-\tfrac{{\mathfrak m}}{K} -2 N \,\bigg]\ ,
 \eea
where each of the components in the direct sum is described by formula \eqref{oi981981212}.
The summation  is taken over the set
\bea
\Sigma_{\bm{\mathfrak{j}},\mathfrak{m}, N}=\Big\{\, (m_1,\ldots,m_r)\,|\, {m}_a\in 2{\mathfrak j}_a+2{\mathbb Z},\ 
\sum_{a=1}^r m_a={\mathfrak m}+2NK\,\Big\}\ .
\eea
In view of eqs.\eqref{oasisis},\,\eqref{aiiaasyy} and \eqref{iiasiasias}, 
${\cal H}_{\bm{\mathfrak{j}}, {\mathfrak m},P}$ is a common eigenspace for the operators $ \hat{a}_0$ and $\hat{U}$.
In turn
\bea
{\boldsymbol{\tau}}_\ell(\lambda)\, ,\ {\mathlarger{\mathlarger{\mathlarger { \boldsymbol a}}}}_\pm(\lambda)\,\in\,{\rm End}
\big({\cal H}_{\bm{\mathfrak{j}}, {\mathfrak m},P}\big)\ .
\eea
Also  ${\cal H}_{\bm{\mathfrak{j}}, {\mathfrak m},P}$ is an eigenspace
of  ${\boldsymbol{\tau}}_\ell(0)$ with eigenvalue
\be\label{ioasd90812}
{{\tau}}_\ell(0)=\frac{\sin\big(\frac{\pi}{K}(2\ell+1)\,\eta\big)}
{\sin\big(\frac{\pi}{K}\eta\big)}\,,\qquad\quad
{\rm where} \qquad\quad
\eta=2\sqrt{K}\,\beta P-\mathfrak{m}\, .
\ee

 \bigskip
 Similar to the case $r=1$ it is expected that
 the commuting family of  operators contains an infinite  set of local IM for any positive integer $r$.
 As was already discussed,  their construction starts by exploring the $W$-algebra built out
of  the local fields commuting with the screening charge $\oint\rd u\, V_+$.
 It is easy to check  that the  algebra includes the field
 \bea\label{oias81aaa212}
 T_2= \sum_{b=1}^r\Big(\,(\partial\phi_b)^2+\tfrac{\ri \sqrt{k_b}}{K}\ \big(\beta^{-1}-\beta\big) \ 
 \partial^2\phi_b+W^{(b)}_2\,\Big)\ ,
 \eea
 where the spin 2 local fields $W^{(b)}_2$ appear in the first non-trivial term
 in the OPE $\psi_+^{(b)}(u_1)\psi_-^{(b)}(u_2)$, see eq.\,\eqref{aoisaisais}.
 The field $T_2$  forms a  closed subalgebra, similar to  \eqref{iiiasiasis},
with the central charge
\bea
c=\sum_{a=1}^r\frac{3 k_a}{k_a+2}-\frac{6}{K}\ \big(\beta^{-1}-\beta\big)^2 \ .
\eea 
 The study of this $W$-algebra for $r>1$ is a considerably more difficult task
 than that for the 
corner-brane $W$-algebra corresponding to  $r=1$.
Some results concerning the case when all $k_a=1$ are presented  in secs.\,\ref{sec9}
and \ref{sec10}.
Having considered in detail various
 particular cases, we conclude that there exists a non-trivial $W$-algebra for 
any $r=1,2,\ldots\ $.
It will be denoted by
 ${W}^{(c,r)}_{{\boldsymbol k}}$, where the multiindex 
 ${\boldsymbol k}=(k_1,\ldots,k_r)$ is used and $c$ stands for the
 central charge of the Virasoro subalgebra.\footnote{%
Similar as for the corner-brane $W$-algebra, ${ W}^{(c,r)}_{{\boldsymbol k}}$
can be defined for arbitrary values of $k_a$ if one uses the bosonization formulae \eqref{isiasis} for each copy
of the parafermionic currents $\psi_\pm^{(a)}\ (a=1,\ldots, r)$.
}
We conjecture that,
out of the fields from the $W$-algebra,  one 
can construct a set of local IM  such that
 \bea
 {\bf I}_{2n-1}^{(a)}\ \ \ \ {\rm with}\ \  n=1,2,\ldots\ ;\ \ a=1,\ldots, r\ .
 \eea
A certain linear combination of ${\bf I}_1^{(a)}$ gives the integral
 \bea\label{oi89129821}
{ \bf I}_1=\sum_{a=1}^rC_a\,{\bf I}_1^{(a)}=\int_0^{2\pi}\frac{\rd u}{2\pi}\  T_2\ ,
 \eea
where $T_2$ is the local density defined in eq.\,\eqref{oias81aaa212}.
\bigskip

The space ${\cal H}_{\bm{\mathfrak{j}}, {\mathfrak m},P}$ is 
a highest weight module of ${ W}^{(c,r)}_{{\boldsymbol k}}$.
As usual, a grading is introduced by the operator ${ \bf I}_1$.
Its spectrum is bounded from below,
while the eigenvalues are given by ${  I}^{({\rm min})}_1+{\tt L}$ with
non-negative integer 
${\tt L}=0,1,2,\ldots\ $. 
This way
${\cal H}_{\bm{\mathfrak{j}}, {\mathfrak m},P}$ decomposes into 
finite-dimensional level subspaces  ${\cal H}_{\bm{\mathfrak{j}}, {\mathfrak m},P}^{({\tt L})}$.
The component corresponding to ${\tt L}=0$ would be formed by the 
  $W$-primary states.
\bigskip

The representations  ${\cal H}_{\mathfrak{j},\mathfrak{m},P}$ are labeled by
the set $\bm{\mathfrak{j}}=(\mathfrak{j}_1,\ldots,\mathfrak{j}_r)$ with
$\mathfrak{j}_a=0,\frac{1}{2},1,\ldots,\frac{1}{2}\,k_a$ and 
 the integer $\mathfrak{m}\sim\mathfrak{m}+2K$, which we take to lie in the interval
 $-2\mathfrak{J},-2\mathfrak{J}+2,\ldots
2K-2\mathfrak{J}-2$. The latter can be further subdivided into two
subsets
\begin{subequations}\label{casei}
\bea\label{caseiaaaaa}
{\rm case} \ (i) \quad && \mathfrak{m} = -2\mathfrak{J},\,  -2\mathfrak{J}-2,\ldots, \, 2\mathfrak{J}-2,  2\mathfrak{J}
\\[0.3cm]
\label{caseibbbbb}
{\rm case} \ (ii) \quad && \mathfrak{m} =  2\mathfrak{J}+2,\,2\mathfrak{J}+4,\ldots,\,2K-2\mathfrak{J}-2\, .
\eea
\end{subequations}
Notice that the transformation $\mathfrak{j}_a\mapsto\check{\mathfrak{j}}_a$ and 
$\mathfrak{m}\mapsto\check{\mathfrak{m}}$ with
\be
\check{\mathfrak{j}}_a=\tfrac{1}{2}\,k_a-\mathfrak{j}_a\,,\qquad
\qquad\qquad\check{\mathfrak{m}}=\mathfrak{m}+K\sim\mathfrak{m}-K
\ee
maps $\mathfrak{m}$ from the interval $(ii)$ to the interval $(i)$ with
$\mathfrak{J}$ replaced by $\check{\mathfrak{J}}=\sum_{a=1}^r\check{\mathfrak{j}}_a$.
In the next section it will be proposed that the spaces
${\cal H}_{\mathfrak{j},\mathfrak{m},P}$ and ${\cal H}_{\check{\mathfrak{j}},\check{\mathfrak{m}},P}$  
should be treated as equivalent representations of the algebra
 ${ W}^{(c,r)}_{{\boldsymbol k}}$:
\be\label{ioasioioasioas12}
{\cal H}_{\mathfrak{j},\mathfrak{m},P}\,\cong\, {\cal H}_{\check{\mathfrak{j}},\check{\mathfrak{m}},P}\, .
\ee
We'll mainly focus  on case $(i)$, and comment on case $(ii)$ as required.

 \section{ODE/IQFT correspondence
\label{sec7}}
 \subsection{ODE for the  $W$-primary states}
Provided that the integer $\mathfrak{m}$ is restricted as in \eqref{caseiaaaaa},
the    spectrum of the operator ${ \bf I}_1$ in  ${\cal H}_{\mathfrak{j},\mathfrak{m},P}$
is given by
\be\label{isisiasias}
I_1=P^2-\frac{{\mathfrak m}^2}{4K}+\sum_{a=1}^r\frac{{\mathfrak j}_a ({\mathfrak j}_a+1)}{k_a+2}-\frac{1}{24}\,
\sum_{a=1}^r\frac{3k_a}{k_a+2}+{\tt L}\qquad\qquad ({\tt L}=0,1,2,\ldots) \, .
\ee
The dimensions of the subspace of primary states, ${\cal H}_{\bm{\mathfrak{j}}, {\mathfrak m},P}^{(0)}$,
can be read off from the formula
\bea\label{oias891298129821}
\sum_{{\mathfrak m}}
\dim\big[{\cal H}^{(0)}_{\bm{\mathfrak{j}}, {\mathfrak m},P}\big]\ {\tt q}^{\mathfrak{m}}
=
\prod_{a=1}^{r}\,[2\mathfrak{j}_a+1]_{{\tt q}}
\eea
with $[ n]_{{\tt q}}=({\tt q}^{n}-{\tt q}^{-n})/({\tt q}-{\tt q}^{-1})$.
When ${\mathfrak m}=\pm 2\sum_{a=1}^r{\mathfrak j}_a$
the corresponding eigenspaces 
are one-dimensional and spanned by the  ground  states
\be\label{ioas9812}
{\boldsymbol  e}_{\boldsymbol{\mathfrak j},\pm 2 {\mathfrak J},P}=
 \bigotimes_{a=1}^r\Big(\, 
\big|\,\sigma^{(k_a)}_{{\mathfrak j}_a,\pm2{\mathfrak j}_a} \,\big\rangle
\otimes\big|P^{( {\rm vac},\pm)}_a\big\rangle
\Big)\, ,
\qquad
P^{ ({\rm vac},\pm)}_a=
P\,\sqrt{\frac{k_a}{K}}\pm \sum_{b=1}^r
\frac{{\mathfrak j}_ak_b-{\mathfrak j}_bk_a}{K\sqrt{k_a}}
\ee

\bigskip

 The  path-ordered integral formulae for
${\boldsymbol{\tau}}_\ell(\lambda)$ and
${\mathlarger{\mathlarger{\mathlarger {\boldsymbol a}}}}_\pm(\lambda)$
allow one to express their  matrix elements 
as a convergent power series in $\lambda^2$. The
corresponding expansion coefficients are given by
the well defined multifold contour integrals (for further details see \cite{Bazhanov:1998dq}).
In the simplest set-ups, it is possible to calculate them directly. For example 
for the  vacuum states \eqref{ioas9812}, one can show that
\bea
{\mathlarger{\mathlarger{\mathlarger {\boldsymbol  a}}}}_+(\lambda)\, {\boldsymbol  e}_{\boldsymbol{\mathfrak j},\pm 2 {\mathfrak J},P}=
{\mathlarger{\mathlarger{\mathlarger { \it a}}}}^{( {\rm vac},\pm)}_+ (\lambda)\, 
{\boldsymbol  e}_{\boldsymbol{\mathfrak j},\pm 2{\mathfrak J},P}
\ :\ \ \  \ 
\log{\mathlarger{\mathlarger{\mathlarger { \it a}}}}^{({\rm vac},\pm)}_+(\lambda)=-\sum_{n=1}^\infty H^{( {\rm vac},\pm)}_{n}\,\lambda^{2n}
\eea
with
\bea\label{oias90121121}
H_{1}^{({\rm vac},\pm)}&=&-\,
 \frac{\pi\Gamma\big(-1+\frac{2}{K}(1-\beta^2)\big)}{\sin(\frac{\pi}{K}(1-\beta^2))}\ 
\frac{\Gamma\big(1-\frac{1}{K}(1-\beta^2)+\frac{1}{K}\,
\eta^{\rm (vac,\pm)}\big)}
 {\Gamma\big(\frac{1}{K}(1-\beta^2)+\frac{1}{K}\,
\eta^{\rm (vac,\pm)}\big)}\ \ \sum_{a=1}^{r}k_a z_a \nonumber\\[0.2cm]
&\pm&
 \frac{\pi\Gamma\big(\frac{2}{K}(1-\beta^2)\big)}{\sin(\frac{\pi}{K}(1-\beta^2))}\ 
\frac{\Gamma\big(\frac{1\pm 1}{2}-\frac{1}{K}(1-\beta^2)+\frac{1}{K}\,
\eta^{\rm (vac,\pm)}\big)}
 {\Gamma\big(\frac{1\pm 1}{2}+\frac{1}{K}(1-\beta^2)+\frac{1}{K}\, 
\eta^{\rm (vac,\pm)}\big)}\ \ \sum_{a=1}^r2\mathfrak{j}_a z_a
\eea
and
\bea
\eta^{\rm (vac,\pm)}
=2\sqrt{K}\beta P \mp \sum_{a=1}^r2\mathfrak{j}_a\ .
 \eea
However, the computation of the higher order coefficients $H_{2},\,H_{3},\ldots$, even for the ground states,
 turns out to be highly cumbersome. 
A  practical way of studying the spectrum of ${\boldsymbol{\tau}}_\ell(\lambda)$ and
${\mathlarger{\mathlarger{\mathlarger {\boldsymbol  a}}}}_\pm(\lambda)$ is through the ODE/IQFT correspondence
for the commuting family, which is proposed below.

\bigskip
Generalizing eq.\eqref{isaiisaisa} we   consider the ODE
\bea\label{aois90121}
\bigg[\!\!\!&-& \partial_z^2+\kappa^2\, z^{-2+\xi\, { \sum_{a=1}^r }k_a }\,
\prod_{a=1}^r(z-z_a)^{k_a}\\[0.2cm]
&-&\frac{{A}^2+\frac{1}{4}}{z^2}+
\sum_{a=1}^r\bigg(\,\frac{{\mathfrak j}_a({\mathfrak j}_a+1)}{ (z-z_a)^2}+\frac{z_a\gamma_a}{z(z-z_a)}\,\bigg)
\,\bigg]\ \Psi=0\,,\nonumber
\eea
where $\xi>0$ and ${\boldsymbol k}=(k_1,\ldots k_r)$ is a set of  positive integers.
As before  the singularities at $z=z_a\ (a=1,\ldots,r)$ are required to be apparent. Repeating the same
line of arguments as for the case $r=1$ one  arrives at the following conditions imposed on the parameters.
An immediate one is that
\bea\label{aiasiiasias}
{\mathfrak j}_a\in\big\{0,\tfrac{1}{2},1,\ldots,\tfrac{k_a}{2}\big\}\ \ \ \ \ \ \ \  \ \ (a=1,\ldots, r)\ .
\eea
Then for some given (half-)integers $({\mathfrak j}_1,\ldots,{\mathfrak j}_r)$ and fixed value of $A$,
 the set ${\boldsymbol\gamma}=(\gamma_1,\ldots,\gamma_r)$ 
should solve the system of algebraic equations
\bea\label{aosisisa}
F_{{\mathfrak j}_a}\big(v^{(a)}_1,\ldots, v^{(a)}_{2{\mathfrak j}_a+1}\big)=0\ \ \  \ \ \ \ \  (a=1,\ldots,r)\ .
\eea
Here $v^{(a)}_{m}=v^{(a)}_{m}({\boldsymbol\gamma})\ \ 
(m=1,\ldots, 2{\mathfrak j}_a+1)$ are defined though the Laurent expansion of
\bea\label{aoasosaisa}
t_0(z)=-\frac{{A}^2+\frac{1}{4}}{z^2}+
\sum_{a=1}^r\bigg(\,\frac{{\mathfrak j}_a({\mathfrak j}_a+1)}{ (z-z_a)^2}+\frac{z_a\gamma_a}{z(z-z_a)}\,\bigg)
\eea
 in the vicinity of  $z=z_a$:
\bea
t_0(z)=\frac{1}{(z-z_a)^2}\ \Big(v^{(a)}_0+\sum_{m=1}^\infty v_{m}^{(a)}\, (z-z_a)^m\,\Big)\,.
\eea
The polynomials 
$F_{{\mathfrak j}}(v_1,\ldots, v_{2{\mathfrak j}+1})$ are given by the determinant \eqref{hasgsats}.

\bigskip
Numerical work shows that, if $A$ is generic,
  the algebraic system \eqref{aosisisa} possesses $\prod_{a=1}^r(2{\mathfrak j}_a+1)$ solutions.
These
  are splitted on the classes labeled by an integer
${\mathfrak m}$ 
such that
\bea\label{iaisaisaisaias}
\sum_{a=1}^r z_a\gamma_a=
\tfrac{1}{2}\, {\mathfrak{m}}\,\big(
\tfrac{1}{2}\, {\mathfrak{m}}+2\ri A\big)-\sum_{a=1}^r{\mathfrak j}_a({\mathfrak j}_a+1)\,,
\eea
where ${\mathfrak m}$ is restricted as in eq.\eqref{caseiaaaaa}, i.e.,
$\mathfrak{m}= -2\mathfrak{J},\,  -2\mathfrak{J}-2,\ldots, \, 2\mathfrak{J}-2,  2\mathfrak{J}$.
The connection coefficients are defined similar to the case $r=1$. Namely,
\be\label{iissisia}
D_{\boldsymbol{\mathfrak j},\pm \mathfrak{ m},\pm A}
=\frac{W_{\boldsymbol{\mathfrak j},\pm \mathfrak{ m},\pm A}(\mu)}{W_{\boldsymbol{\mathfrak j},\pm \mathfrak{ m},\pm A}(0)}
\ \ \ \  \quad {\rm with}\ \ \ \ \ 
W_{\boldsymbol{\mathfrak j},\pm \mathfrak{ m},\pm A}(\mu)=\mu^{\mp(\frac{1}{2}\mathfrak{m}+ \ri A)}\
W\big[\,\Psi^{\scriptscriptstyle{(\rightarrow)}}, \Psi^{\scriptscriptstyle{(\leftarrow)}}_{\pm A}\,\big]
\ee
They are treated as a function of
\be
\mu\,:\ \ \ \kappa^2=\mu^{(1+\xi)K}\ .
\ee
\bigskip

Let $\bm{\gamma}$ be a solution of \eqref{aosisisa}.
We expect that there exists a 
state  labeled by this set
\bea\label{osososaoa}
{ \boldsymbol e}_{\boldsymbol {\mathfrak j},\mathfrak{m},P}({\boldsymbol\gamma})
\in{\cal H}^{(0)}_{\bm{\mathfrak{j}},\mathfrak{m},P}
\eea
such that it is a simultaneous eigenstate for
the operators of the commuting family
with
\bea\label{jasusauasu}
{\mathlarger{\mathlarger{\mathlarger {\boldsymbol  a}}}}_\pm (\lambda)\, { \boldsymbol e}_{\boldsymbol{\mathfrak{j}},
{\mathfrak m},P}({\boldsymbol \gamma})
=
D_{\boldsymbol{\mathfrak j},\pm \mathfrak{ m},\pm A}
(\mu |{\boldsymbol\gamma})\,
{ \boldsymbol e}_{\boldsymbol {\mathfrak j}, \mathfrak{m},P}({\boldsymbol \gamma})\ .
\eea
The parameters on the ODE and the field theory side are related as
\bea\label{siisaias11a}
 \xi=\tfrac{\beta^2}{1-\beta^2}\ ,\ \ \ \ \ \ \ \ \ \ \ \ \ \ \mu=-\lambda^2\ \Gamma^2\big(\tfrac{1-\beta^2}{K}\big)\,
\big(\tfrac{K}{1-\beta^2}\big)^{\frac{2}{K}(1-\beta^2)}
\ ,
\eea
while
\bea\label{siisaiasfff}
{A}=
\tfrac{\ri }{\beta^{-1}-\beta}\ \big(\,\sqrt{K}\, P-\tfrac{1}{2}\, \beta\, \mathfrak{m}\,\big)\ .
\eea
The states ${ \boldsymbol e}_{\boldsymbol {\mathfrak j},\mathfrak{m},P}({\boldsymbol\gamma})$
form a basis in ${\cal H}^{(0)}_{\bm{\mathfrak{j}},\mathfrak{m},P}$.
\bigskip

The support for the  proposed ODE/IQFT correspondence 
is based on the arguments that were
 originally developed in the works 
\cite{Bazhanov:1998wj,Bazhanov:2003ni}. Following ref.\cite{Bazhanov:1998wj}, it is 
straightforward to derive a
set of functional relations for various connection coefficients. The most fundamental of these is the 
so-called quantum Wronskian relation 
\be\label{jasasasasay}
\big(q^2\big)^{2\ri A+{\mathfrak m}}\, D_{\boldsymbol{\mathfrak j},
\mathfrak{ m},A}\big(q^2\mu\big)D_{\boldsymbol{\mathfrak j},-\mathfrak{ m},-A}(\mu)
-
D_{\boldsymbol{\mathfrak j},\mathfrak{ m},A}(\mu)D_{\boldsymbol{\mathfrak j},-\mathfrak{ m},-A}\big(q^2\mu\big)
=\big(q^2\big)^{2\ri A+{\mathfrak m}}-1
\ee
with $q^2=\re^{-\frac{2\pi\ri}{(1+\xi)K}}$. It holds  true as long as all the
singularities 
at $z=z_a$ are apparent. Furthermore in this case
the connection
coefficients turn out to be
 entire functions of $\mu$. On the other hand,
all the eigenvalues of ${\mathlarger{\mathlarger{\mathlarger {\boldsymbol a}}}}_\pm(\lambda)$ 
obey the quantum Wronskian relation,
which comes from the operator valued relation \eqref{aisisaisa} with $\ell=0$.
This coincides with \eqref{jasasasasay} upon making the identification
\eqref{siisaiasfff} as well as
$\xi=\frac{\beta^2}{1-\beta^2}$ and
$\mu\propto \lambda^2$.
The precise  $\mu-\lambda$ relation   \eqref{siisaias11a} can be established via a first order
perturbative calculation in $\mu$ of  $D_{\boldsymbol{\mathfrak j},\mathfrak{ m},A}$ for \eqref{aois90121}
with all $\mathfrak{j}_a=\mathfrak{ m}=\gamma_a=0$ and comparing the result with the corresponding
specialization of eq.\,\eqref{oias90121121}.

\bigskip
For given $A$ and  ${\boldsymbol {\mathfrak j}}=({\mathfrak j}_1,\ldots,{\mathfrak j}_r)$,
there exists two 
solutions of the algebraic system \eqref{aosisisa}
which are particularly simple:
\bea\label{jausuas}
\gamma^{{( {\rm vac},\pm)}}_a=\frac{2{\mathfrak j}_0\mathfrak{j}_a  }{z_a-z_0}+
\sum_{b=1\atop b\not=a}^{r}\frac{2\mathfrak{j}_a \mathfrak{j}_b}{z_a-z_b}\ 
\ \  \  \ \ \ \  \ (a=1,\ldots,r)\ ,
\eea
where
\bea\label{jaussauaus}
z_0=0\ ,\ \ \ \ \ \qquad {\mathfrak j}_0=\pm\ri A-\tfrac{1}{2}\ .
\eea
The ODE \eqref{aois90121} in this case  takes the form 
\be
\bigg[\!- \partial_z^2+\kappa^2
\prod_{a=0}^r(z-z_a)^{k_a}+
\sum_{a=0}^r\frac{{\mathfrak j}_a({\mathfrak j}_a+1)}{ (z-z_a)^2}
+\sum_{ 0\leq a<b\leq r}
\frac{2\,\mathfrak{j}_a \mathfrak{j}_b}{(z-z_a)(z-z_b)}
\,\bigg] \Psi=0\ .
\ee
Here, together with the notations  \eqref{jaussauaus}, we use
\bea
k_0=-2+\xi\, \sum_{a=1}^rk_a\ .
\eea
The integer ${\mathfrak m}$ \eqref{iaisaisaisaias} for the solutions $\gamma^{{( {\rm vac},\pm)}}_a$
 coincides with $\pm 2{\mathfrak J}$. Numerical work shows 
that they  correspond to the ground  states 
${ \boldsymbol e}_{\boldsymbol {\mathfrak j},\pm 2\mathfrak{J},P}$
 \eqref{ioas9812}. Note that, in  view of eq.\eqref{siisaiasfff},
 \bea
 {\mathfrak j}_0= \mp \frac{ \sqrt{K}\, P}{\beta^{-1}-\beta}-\frac{1}{2}
 +
 \frac{\beta }{\beta^{-1}-\beta}\
 \sum_{a=1}^r{\mathfrak j}_a\ .
 \eea
\bigskip

It is important to keep in mind that the connection coefficients
$D_{\boldsymbol{\mathfrak j},\pm\mathfrak{ m},\pm A}$ are defined
with $\mathfrak{m}$  ranging from $-2\mathfrak{J}$ to $+2\mathfrak{J}$
as in eq.\,\eqref{caseiaaaaa}. In order to cover the case \eqref{caseibbbbb},
we introduce $\check{\mathfrak{j}}_a$, $\check{\mathfrak{m}}$  and $\check{A}$ via the formulae
\be\label{iowewewew9889}
\check{\mathfrak{j}}_a=\tfrac{1}{2}\,k_a-\mathfrak{j}_a\,,\qquad\qquad 
\check{\mathfrak{m}}=\mathfrak{m}-K\,,\qquad\qquad
\check{A}=\tfrac{\ri }{\beta^{-1}-\beta}\ \big(\,\sqrt{K}\, P-\tfrac{1}{2}\, \beta\, (\mathfrak{m}-K)\,\big)
\ee
and consider the ODE  
\bea\label{isaisasisaisaias} 
\bigg[\!&-& \partial_z^2+\kappa^2\, z^{-2+\xi\, { \sum_{a=1}^r }k_a }\,
\prod_{a=1}^r(z-z_a)^{k_a}\\[0.2cm]
&-&\frac{\check{{A}}^2+\frac{1}{4}}{z^2}+
\sum_{a=1}^r\bigg(\,\frac{\check{{\mathfrak j}}_a\,(\,\check{{\mathfrak j}}_a+1)}{ (z-z_a)^2}+
\frac{z_a\check{\gamma}_a}{z(z-z_a)}\,\bigg)
\,\bigg]\ \Psi=0\,.\nonumber
\eea
Here the set $\{\check{\gamma}_a\}$ solves the system of equations \eqref{aosisisa}
with $A$ and ${{\mathfrak j}}_a$ replaced by their ``checked'' counterparts.
In turn, eq.\,\eqref{iaisaisaisaias} is swapped for
\be
\sum_{a=1}^r z_a\check{\gamma}_a=
\tfrac{1}{4}\, \check{\mathfrak{m}}\,\big(
\check{\mathfrak{m}}+2\ri \check{A}\big)-\sum_{a=1}^r\check{{\mathfrak j}}_a(\check{{\mathfrak j}}_a+1)\, .
\ee
Then the ODE/IQFT correspondence for the primary states
\bea\label{osososaoas}
{ \boldsymbol e}_{\boldsymbol {\mathfrak j},\mathfrak{m},P}(\check{\boldsymbol\gamma})
\in{\cal H}^{(0)}_{\bm{\mathfrak{j}},\mathfrak{m},P}\qquad \qquad {\rm with}\qquad\qquad
\mathfrak{m}=2\mathfrak{J}+2,\,2\mathfrak{J}+4,\ldots,\,2K-2\mathfrak{J}-2
\eea
is formulated as
\bea\label{jasusauasur}
{\mathlarger{\mathlarger{\mathlarger {\bf \it a}}}}_\pm (\lambda)\, { \boldsymbol e}_{\boldsymbol{\mathfrak{j}},
{\mathfrak m},P}(\check{\boldsymbol \gamma})
=
D_{\check{\boldsymbol{\mathfrak j}},\pm \check{\mathfrak{ m}},\pm \check{A}}
(\mu |\check{{\boldsymbol\gamma}})\,
{ \boldsymbol e}_{\boldsymbol {\mathfrak j}, \mathfrak{m},P}(\check{{\boldsymbol \gamma}})\ .
\eea
Notice that 
\be
\big(q^2\big)^{2\ri A+{\mathfrak m}}\,=\,
\re^{-\frac{4\pi\ri}{\sqrt{K}}\,\beta P-\frac{2\pi\ri\mathfrak{m}}{K}}\,=\,
\re^{-\frac{4\pi\ri}{\sqrt{K}}\,\beta P-\frac{2\pi\ri\check{\mathfrak{m}}}{K}}\,=\,
\big(q^2\big)^{2\ri \check{A}+\check{\mathfrak m}}
\ee
and therefore the quantum Wronskian relation
for the 
 connection coefficients $D_{\check{\boldsymbol{\mathfrak j}},
\pm\check{\mathfrak{ m}},\pm\check {A}}$ 
is  identical to \eqref{jasasasasay}  as long as the parameter $P$ is kept fixed.
This is the first piece of support for the isomorphism \eqref{ioasioioasioas12}.

\subsection{Relation to the  Bethe ansatz equations for the $\mathfrak{ sl}(2)$  Gaudin model\label{sec72}}
There is an alternative description of the sets
$\bm{\gamma}$, which solve the algebraic equations 
\eqref{aosisisa} going back to the works \cite{Gaudin1,Feigin:2007mr,Feigin:1994in,Frenkel:2004qy}.
Assuming that $t_0$ \eqref{aoasosaisa} is given, consider the Riccati equation for the unknown function  $f_0=f_0(z)$,
\bea
t_0=f^2_0-\partial_z f_0\,,
\eea
so that $t_0$ is the Miura transform  of $f_0$.
One can search for a solution using the ansatz
\bea
f_0(z)=\frac{\ri A-\frac{1}{2}}{z}+
\sum_{a=1}^r\frac{\mathfrak{j}_a}{z-z_a}-\sum_{m=1}^{{\tt M}_+}\frac{1}{z-x^{\scriptscriptstyle{(+)}}_m}\,,
\eea
which involves  a set of  parameters $\bm{x}=(x^{\scriptscriptstyle{(+)}}_1,\ldots,x^{\scriptscriptstyle{(+)}}_{{\tt M}_+})$. 
Substituting $f_0$   into the Riccati equation gives
\bea
t_0(z)=-\frac{{A}^2+\frac{1}{4}}{z^2}+
\sum_{a=1}^r\bigg(\,\frac{{\mathfrak j}_a({\mathfrak j}_a+1)}{ (z-z_a)^2}+\frac{\gamma_a}{z(z-z_a)}\bigg)+
\sum_{m=1}^{{\tt M}_+}\frac{r_m}{z-x^{\scriptscriptstyle{(+)}}_m}\ ,
\eea
where
\bea\label{iosaiisa}
\gamma_a=\mathfrak{j}_a \ \bigg(\,
\frac{2\ri A-1}{z_a}+
\sum_{b=1\atop b\not=a}^{r}\frac{2\mathfrak{j}_b}{z_a-z_b}-
\sum_{m=1}^{{\tt M}_+}\frac{2}{z_a-x^{\scriptscriptstyle{(+)}}_m}\,\bigg)
\eea
and
\bea
r_m=\sum_{n=1\atop n\not=m}^{{\tt M}_+}\frac{2}{x^{\scriptscriptstyle{(+)}}_m-
x^{\scriptscriptstyle{(+)}}_n}-
\frac{2\ri A-1}{x^{\scriptscriptstyle{(+)}}_m}
-\sum_{a=1}^r
\frac{2\mathfrak{j}_a}{x^{\scriptscriptstyle{(+)}}_m-z_a}\, .
\eea
The requirement that $r_m=0$ yields an algebraic system  for the auxiliary parameters $x^{\scriptscriptstyle{(+)}}_m$:
\bea\label{aisisaiias}
\sum_{n=1\atop n\not=m}^{{\tt M}_+}\frac{2}{x^{\scriptscriptstyle{(+)}}_m-
x^{\scriptscriptstyle{(+)}}_n}
\,-\,\frac{2\ri A-1}{x^{\scriptscriptstyle{(+)}}_m}
\,-\,\sum_{a=1}^r
\frac{2\mathfrak{j}_a}{x^{\scriptscriptstyle{(+)}}_m-z_a}
=0\ \ \ \ \ \ \ \ \ \ (m=1,\ldots,{{\tt M}_+})\ .
\eea
Note that for ${\tt M}_+=0$, formula \eqref{iosaiisa}  
gives back the vacuum solution $\gamma_a^{({\rm vac},+)}$  \eqref{jausuas}.
Moreover it  is straightforward,  using eqs.\,\eqref{aisisaiias} 
and 
\eqref{iosaiisa}, to calculate the sum $\sum_{a=1}^rz_a\gamma_a$. This yields \eqref{iaisaisaisaias} 
with ${\mathfrak m}$  being related to
the non-negative integer ${\tt M}_+$  as
\bea
{\tt{M}}_+=\sum_{a=1}^r {\mathfrak{j}}_a-\tfrac{1}{2}\, \mathfrak{m}\ .
\eea
\bigskip

Having at hand a solution of eq.\,\eqref{aisisaiias}, the set 
$\bm{\gamma}$ obtained via \eqref{iosaiisa} would automatically
obey the conditions \eqref{aosisisa}, which guarantee  that the singularities
at $z=z_a$ of the ODE are apparent. Our numerical work suggests that for \emph{generic} $A$
there is a one-to-one correspondence between $\bm{x}^{\scriptscriptstyle{(+)}}$ solving \eqref{aisisaiias}
and $\bm{\gamma}$. However, for some specific values of $A$ the correspondence breaks down.
For example upon taking
 $A=-\frac{\ri}{2}$,  eqs.\eqref{iosaiisa},\,\eqref{aisisaiias} imply that 
 $\sum_{a=1}^r\gamma_a=0$. On the other hand, for $A^2=-\frac{1}{4}$,
the system \eqref{aosisisa} admits solutions with $\sum_{a=1}^r\gamma_a\ne 0$.

\bigskip

The ODE depends  on $A^2$ rather than $A$.
However eqs.\,\eqref{iosaiisa} and \eqref{aisisaiias}
are not invariant if the sign of $A$ is flipped.
For this reason 
the set $\bm{\gamma}$ solving 
\eqref{aosisisa} may be alternatively expressed as
\bea\label{ioaoi901290812}
\gamma_a=\mathfrak{j}_a\ \bigg(\!
-\frac{2\ri A+1}{z_a}+
\sum_{b=0\atop b\not=a}^{r}\frac{2\mathfrak{j}_b}{z_a-z_b}-
\sum_{m=1}^{{\tt M}_-}\frac{2}{z_a-{x}^{\scriptscriptstyle{(-)}}_m}\,\bigg)\, .
\eea
Here ${x}^{\scriptscriptstyle{(-)}}_m$ are subject to the equations
\bea\label{oiais891891212aaa}
\sum_{n=1\atop n\not=m}^{{\tt M}_-}\frac{2}{{x}^{\scriptscriptstyle{(-)}}_m-
{x}^{\scriptscriptstyle{(-)}}_n}\,+\,\frac{2\ri A+1}{{x}^{\scriptscriptstyle{(-)}}_m}
\,-\,
\sum_{a=1}^r
\frac{2\mathfrak{j}_a}{{x}^{\scriptscriptstyle{(-)}}_m-z_a}
=0\ \ \ \ \ \ \ \ \ \ (m=1,\ldots,{{\tt M}_-})\ ,
\eea
while 
\be
{\tt{M}}_-=\sum_{a=1}^r {\mathfrak{j}}_a+\tfrac{1}{2}\, \mathfrak{m}\ .
\ee
\bigskip

When  $A=-\frac{\ri}{2}$ the algebraic systems \eqref{aisisaiias} and \eqref{oiais891891212aaa}
are  identical to the
Bethe ansatz equations for the Gaudin model eqs.\,\eqref{apo0402321} and \eqref{apo0402321AAdd},
respectively.
As follows from a comparison of  eqs.\,\eqref{op2929292929} 
and \eqref{iosaiisa}, the set of energies  $\{E_a\}_{a=1}^r$
coincides with $\{\gamma_a\}_{a=1}^r$.
An immediate question arises as to whether there exists a generalization of the Hamiltonians
\eqref{oias89129812},  which results in the Bethe ansatz equations \eqref{aisisaiias} 
with arbitrary $A$. An evident candidate is
\be\label{ois89128912}
{\bf H}^{(a)}=\frac{2}{z_a}\,\vec{S}^{(0)}\cdot\vec{S}^{(a)}+2\ \sum_{b=1\atop
 b\not=a}^r\frac{\vec{S}^{(a)}\cdot\vec{S}^{(b)}}{z_a-z_b}\ ,
\ee
where $\vec{S}^{(0)}$ is a spin operator that acts in a highest weight
 infinite dimensional representation (Verma module) of $\mathfrak{sl}(2)$.
The corresponding Casimir would be given by
\be
\big(\vec{S}^{(0)}\big)^2=
\mathfrak{j}_0\,(\mathfrak{j}_0+1)\ \qquad\qquad\qquad \big(2\mathfrak{j}_0\notin\mathbb{Z}\big)\,, 
\ee
while the value of $S_3^{(0)}$ on the highest weight is $\mathfrak{j}_0=\ri A-\frac{1}{2}$.
The diagonalization problem can be considered in the 
finite dimensional eigenspaces of the operator $S_3^{(0)}+\sum_{a=1}^r S_3^{(a)}$,
which commutes with the Hamiltonians.
If instead of the highest weight, one takes the lowest weight
representation for $\vec{S}^{(0)}$ with lowest weight $-1-\mathfrak{j}_0$,
the spectral problem for the Hamiltonians \eqref{ois89128912}
would lead to the Bethe ansatz equations \eqref{oiais891891212aaa}.

\subsection{Excited states ODE}

In  ref.\cite{Bazhanov:2013cua}, developing the ideas from \cite{Bazhanov:2003ni},
the ODE/IQFT correspondence was proposed for  the highest state irreps of  the pillow brane $W$-algebra.
Here, following the construction from that work,  we   extend the
 correspondence to 
all the eigenstates of the commuting family of operators.
These    form
 a basis in the highest weight irrep of ${ W}^{(c,r)}_{{\boldsymbol k}}$.
Such  basic  states   
\bea\label{ososossssaoa}
{ \boldsymbol e}_{\boldsymbol {\mathfrak j},\mathfrak{m},P}({\boldsymbol\gamma};\bm{w})
\in{\cal H}^{({\tt L})}_{\bm{\mathfrak{j}},\mathfrak{m},P}
\eea
would be
 distinguished by the sets $(\bm{\gamma};\bm{w})=(\gamma_1,\ldots,\gamma_{r},w_1,\ldots, w_{\tt L})$,
which are solutions of a certain algebraic system.
The latter
has already been discussed for ${\tt L}=0$.
The  system of equations imposed on
$({\boldsymbol \gamma};\bm{w})$ 
for general ${\tt L}=0,1,2,\ldots$ is described as follows.

\bigskip
Consider the meromorphic  function
\be\label{aiasaisias}
t_{\tt L}(z)=-\frac{{A}^2+\frac{1}{4}}{z^2}+
\sum_{a=1}^r\bigg(\,\frac{{\mathfrak j}_a({\mathfrak j}_a+1)}{ (z-z_a)^2}+\frac{z_a\gamma_a}{z(z-z_a)}\bigg)+
\sum_{\alpha=1}^{\tt L}
\bigg(\,\frac{2}{ (z-w_\alpha)^2}+\frac{w_\alpha\Gamma_\alpha}{z(z-w_\alpha)}\bigg)
\ee
which possesses second order poles at $z=z_a\  (a=1,\ldots,r)$ and $z=w_\alpha\ (\alpha=1,\ldots,{\tt L})$.
The corresponding residues are given by the sets
${\boldsymbol\gamma}=(\gamma_1,\ldots\gamma_r)$ and
${\boldsymbol\Gamma}=(\Gamma_1,\ldots,\Gamma_{\tt L})$.
Introduce 
\bea\label{sisisa}
v^{(a)}_{m}\ \ \ \ 
(m=1,\ldots, 2{\mathfrak j}_a+1)\ \ \ \ \  {\rm as\ well\ as }\ \ \ \ \ 
t^{(\alpha)}_1,\, t^{(\alpha)}_2, \,t^{(\alpha)}_3\ \ \ \ \ \ (\alpha=1,\ldots,{\tt L})
\eea
  through the Laurent expansion of $t_{{\tt L}}(z)$ 
in the vicinity of  $z=z_a$ and $z=w_\alpha$, respectively:
\bea\label{isaiisa}
t_{\tt L}(z)&=&\frac{1}{(z-z_a)^2}\ \Big(v^{(a)}_0+\sum_{m=1}^{2{\mathfrak j}_a+1} v_{m}^{(a)}\, (z-z_a)^m+\ldots\,\Big)\\
&=&\frac{1}{(z-w_\alpha)^2}\ \Big(2+t_1^{(\alpha)} (z-w_\alpha)+t_2^{(\alpha)} (z-w_\alpha)^2+
t_3^{(\alpha)} (z-w_\alpha)^3+\ldots\,  \Big)\ .\nonumber
\eea
Assuming that  $(z_1,\ldots,z_r)$ and $(w_1,\ldots,w_{\tt L})$ are  given, the residues
${\boldsymbol\gamma}$ 
 and
${\boldsymbol\Gamma}$ are determined through the solution of the coupled system of $r+{\tt L}$ equations:
\bea\label{aiasiisasia}
&&F_{{\mathfrak j}_a}\big(v^{(a)}_1,\ldots, v^{(a)}_{2{\mathfrak j}_a+1}\big)=0\ \ \  \ \ \ \ \  \ \  \ \ \ \ \ \
 \ \ \  \ \  \ \ \ \ \  \   (a=1,\ldots,r)\\[0.2cm]
&&t^{(\alpha)}_1\,\Big(\,\big(t^{(\alpha)}_1\big)^2-4 t^{(\alpha)}_2\,\Big)+4 t^{(\alpha)}_3=0\ \ \  \ \ \ \ \  \ \ \ \ \ \ \ 
(\alpha=1,\ldots,{\tt L})\ .\nonumber
\eea
Here  the polynomials 
$F_{{\mathfrak j}}(v_1,\ldots, v_{2{\mathfrak j}+1})$ are defined via the determinant \eqref{hasgsats}.
The meaning of these equations should be obvious at this point: if
all $2{\mathfrak j}_a+1$  are positive integers
they form the full set of   conditions that  all  the singularities
 of  the Fuchsian differential equation $\big(-\partial_z^2+t_{{\tt L}}(z)\big)\Psi=0$ are apparent
except for $z=0$ and $z=\infty$.

\bigskip
Consider now the ODE
\bea\label{aois90121AA}
\big(\, - \partial_z^2+t_{\tt L}(z)+\kappa^2\, {\cal P}(z)\,\big)\ \Psi=0\,,
\eea
where $t_{\tt L}(z)$ is given by \eqref{aiasaisias}, while
\bea\label{jasuasuas}
{\cal P}(z)= z^{-2+\xi\, { \sum_{a=1}^r }k_a }\,
\prod_{a=1}^r(z-z_a)^{k_a}\ .
\eea
We impose that all the singularities except $z=0,\infty$
for {\it  any}  value of the parameter $\kappa$ are apparent.
Eqs.\eqref{aiasiisasia}  guarantee this property for $\kappa=0$ so that
 they  are necessary conditions. Furthermore,
as was already discussed,  for generic $\kappa$  the  admissible values of ${\mathfrak j}_a$ 
must be restricted as in  \eqref{aiasiiasias}:
\bea
{\mathfrak j}_a\in\big\{0,\tfrac{1}{2},1,\ldots,\tfrac{k_a}{2}\big\}\ \ \ \ \ \ \ \  \ \ (a=1,\ldots, r)\ .
\eea
 Also it turns out that the positions
 of the apparent singularities  $w_\alpha$
 may not be chosen at will.  Instead they
should satisfy ${\tt L}$ extra  conditions  (for details see \cite{Bazhanov:2013cua})
\bea\label{aois901828912}
\Gamma_\alpha=\partial_z\log{\cal P}(z)|_{z=w_\alpha}\ ,
\eea
or explicitly
\bea\label{oisaisai}
\Gamma_\alpha=-\Big(2-\xi\, { \sum_{b=1}^r }k_b\Big)\ \frac{1}{w_\alpha}
+\sum_{b=1}^r\frac{k_b}{w_\alpha-z_b}\ \ \ \ \ \ \ \ \  \ \ \ \  (\alpha=1,\ldots, {\tt L})\ .
\eea
This  expresses  the set of residues $\bm{\Gamma}$ through $\bm{w}$. 
Thus \eqref{aiasiisasia} becomes
a system of $r+{\tt L}$ equations imposed on  the $r+{\tt L}$ variables 
 $(\gamma_1,\ldots,\gamma_r;w_1,\ldots,w_{\tt L})$.
 Similar to   the case of ${\tt L}=0$,
its solutions   
  are splitted on the classes labeled by an integer
${\mathfrak M}$ 
such that
\bea\label{kausy}
\sum_{a=1}^r z_a\gamma_a+\sum_{\alpha=1}^{\tt L} w_\alpha\Gamma_\alpha=
\tfrac{1}{2}\, { \mathfrak M}\,\big(
\tfrac{1}{2}\, {{\mathfrak M}}+2\ri A\big)-2\,{\tt L}-\sum_{a=1}^r{\mathfrak j}_a({\mathfrak j}_a+1)\ .
\eea
This integer takes the values
\be\label{oasisisaasi}
\mathfrak{M}=-\mathfrak{M}_{\rm max}, -\mathfrak{M}_{\rm max}+2,\ldots,\mathfrak{M}_{\rm max}-2,
\mathfrak{M}_{\rm max}
\ee
with some $\mathfrak{M}_{\rm max}$.  Numerical work suggests that
\be\label{oi8921891221}
2\mathfrak{J}\le \mathfrak{M}_{\rm max}\le 2\mathfrak{J}+2{\tt L}  \qquad
\qquad\qquad\qquad \Big({\mathfrak J}=\sum_{a=1}^r{\mathfrak j}_a\Big)\,.
\ee
\bigskip

An alternative description   is provided if one considers,
rather than $t_{\tt L}(z)$, the solution of the Riccati equation
$t_{\tt L}=f_{\tt L}^2-\partial_z f_{\tt L}$. 
As explained in sec.\,\ref{sec72}, it leads one to introduce
the auxiliary sets  $\bm{x}^{\scriptscriptstyle{(\pm)}}
=(x^{\scriptscriptstyle{(\pm)}}_1,\ldots,x^{\scriptscriptstyle{(\pm)}}_{{\tt M}_\pm})$.
These would parameterize the residues $\gamma_a$ as
\bea\label{oi09120912}
\gamma_a=\mathfrak{j}_a\,\bigg(\,
\frac{\pm2\ri A-1}{z_a}+
\sum_{b=1\atop
b\ne a}^r\frac{2\mathfrak{j}_b}{z_a-z_b}-
\sum_{n=1}^{{\tt M}_\pm}\,\frac{2}{z_a-x^{\scriptscriptstyle{(\pm)}}_n}+\sum_{\beta=1}^{{\tt L}}\frac{2}{z_a-w_\beta}\,\bigg)\ ,
\eea
where
\bea\label{ioas19829A}
{\tt M}_\pm=\mathfrak{J}+{\tt L}\mp \tfrac{1}{2}\,\mathfrak{M}\ .
\eea
Then
\bea 
\label{aiiias}
\frac{\pm 2\ri A-1}{x^{\scriptscriptstyle{(\pm)}}_m}+
 \sum_{a=1}^{r}\frac{2\mathfrak{j}_a}{x^{\scriptscriptstyle{(\pm)}}_m-z_a} -
\sum_{n=1\atop
n\ne m}^{{\tt M}_\pm}
\frac{2}{x^{\scriptscriptstyle{(\pm)}}_m-x^{\scriptscriptstyle{(\pm)}}_n}+
 \sum_{\alpha=1}^{\tt L}\frac{2}{x^{\scriptscriptstyle{(\pm)}}_m-w_\alpha}=0\,
\qquad \ \ ( m=1,2,\ldots,{\tt M}_\pm)\nonumber\\[-0.1cm]
\\
\Gamma_\alpha-
\frac{\pm 2\ri A-1}{w_\alpha}-
 \sum_{b=1}^{r}\frac{2\mathfrak{j}_b}{w_\alpha-z_b}+
\sum_{m=1}^{{\tt M}_\pm}\frac{2}{w_\alpha-x^{\scriptscriptstyle{(\pm)}}_m}-\sum_{\beta=1\atop
\beta \ne \alpha}^{{\tt L}}\frac{2}{w_\alpha-w_\beta}=0\,\ \ \ \ \quad
 (\alpha=1,2,\ldots, {\tt L})\nonumber
\eea
together with \eqref{oisaisai} forms a closed system of ${\tt L}+{\tt M}_\pm$
equations, which determine the sets
$\bm{x}^{\scriptscriptstyle{(\pm)}}
=(x^{\scriptscriptstyle{(\pm)}}_1,\ldots,x^{\scriptscriptstyle{(\pm)}}_{{\tt M}_+})$ and $\bm{w}=
(w_1,\ldots,w_{\tt L})$. Finally   $\bm{\gamma}=(\gamma_1,\ldots\gamma_r)$ is obtained via eq.\,\eqref{oi09120912}.
\bigskip

In order to formulate a precise  conjecture regarding the ODE/IQFT correspondence,
there is an important issue that needs to be addressed. According to eqs.\,\eqref{kausy}-\eqref{oi8921891221},
the algebraic system \eqref{aiasiisasia} admits 
solutions $(\bm{\gamma};\bm{w})$ with $|\mathfrak{M}|>2\mathfrak{J}$.
However, there exists a map   
$(\bm{\gamma};\bm{w})\mapsto (\tilde{\bm{\gamma}},\tilde{\bm{w}})$ such that the 
transformed set
satisfies the equations similar to \eqref{aiasiisasia}  with the parameters $\mathfrak{j}_a$, ${\tt L}$ 
and $A$ replaced by
\be\label{ioasoi89219812}
\tilde{\mathfrak{j}}_a=\tfrac{1}{2}\,k_a-\mathfrak{j}_a\,,\qquad\qquad
\tilde{{\tt L}}={\tt L}-\tfrac{1}{2}\,\big(\,|\mathfrak{M}|-2\mathfrak{J}\,\big)\ge 0\,,
\qquad\qquad
\tilde{A}=A+\tfrac{\sigma\ri}{2}\,\xi K\,,
\ee
$\sigma=\sgn(\mathfrak{M})$,
along with the conditions
\bea\label{aios891298121}
\sum_{a=1}^r z_a\tilde{\gamma}_a+\sum_{\alpha=1}^{\tilde{\tt L}} \tilde{w}_\alpha\tilde{\Gamma}_\alpha=
\tfrac{1}{2}\, \widetilde{ \mathfrak M}\,\big(
\tfrac{1}{2}\, {\widetilde{\mathfrak M}}+2\ri \tilde{A}\big)-2\,\tilde{{\tt L}}-\sum_{a=1}^r
\tilde{{\mathfrak j}}_a(\tilde{{\mathfrak j}}_a+1)\,,
\eea
where
\be\label{oias8921212121}
\widetilde{ \mathfrak M}=\begin{cases}
\mathfrak{M}-K\, &\qquad {\rm for}\ \ \ \  \ \mathfrak{M}>+2\mathfrak{J} \\[0.2cm]
\mathfrak{M}+K\, &\qquad {\rm for}\ \ \ \ \ \mathfrak{M}<-2\mathfrak{J}
\end{cases}\ .
\ee
For $\mathfrak{M}>2\mathfrak{J}$ the transformation is described as follows.
Suppose that $w_\alpha$ and $\gamma_a$ are given.
The set $\tilde{\bm{w}}_\alpha$ is the  solution of the 
equations:
\be\label{oi98219812}
\frac{2\ri A-1}{\tilde{w}_\alpha}+
 \sum_{b=1}^{r}\frac{2\mathfrak{j}_b}{\tilde{w}_\alpha-z_b} -
\sum_{\beta=1\atop
\beta\ne \alpha}^{\tilde{{\tt L}}}
\frac{2}{\tilde{w}_\alpha-\tilde{w}_\beta}+
 \sum_{\beta=1}^{\tt L}\frac{2}{\tilde{w}_\alpha-w_\beta}=0\, \qquad\quad
(\alpha=1,2,\ldots,\tilde{{\tt L}})\,,
\ee
which is uniquely specified by the extra conditions
\be
\frac{2\ri A-1}{z_a}+
\sum_{b=1\atop
b\ne a}^r\frac{2\mathfrak{j}_b}{z_a-z_b}-
\sum_{\beta=1}^{\tilde{{\tt L}}}\,\frac{2}{z_a-\tilde{w}_\beta}+\sum_{\beta=1}^{{\tt L}}\frac{2}{z_a-w_\beta}
=\frac{\gamma_a}{\mathfrak{j}_a} \qquad\qquad\ (a=1,2,\ldots,r)\, .
\ee
Once $\tilde{\bm{w}}$  is found, the residues $\tilde{\bm{\gamma}}$
are obtained from
\bea\label{chekiuas8912fgf}
\tilde{\gamma}_a&=&\tilde{\mathfrak{j}}_a\,\bigg(\frac{-2\ri \tilde{A}-1}{z_a}+
\sum_{b\ne a}\frac{2\tilde{\mathfrak{j}}_b}{z_a-z_b}
-\sum_{\beta=1}^{{\tt L}}\frac{2}{z_a-w_\beta}+
\sum_{\beta=1}^{\tilde{\tt L}}\frac{2}{z_a-\tilde{w}_{\beta}}\bigg)\,,
\eea
while $\tilde{\Gamma}_\alpha$ is expressed  in terms of $\tilde{w}_\alpha$ as
\bea\label{oisa98219821}
\tilde{\Gamma}_\alpha=\partial_z\log{\cal P}(z)|_{z=\tilde{w}_\alpha}\ \ \ \ \ \ \ \ \  \ \ \ \  
\qquad (\alpha=1,\ldots, \tilde{{\tt L}})\ .
\eea
If one now considers the change of variables
\be\label{iu872187128721}
\widetilde{\Psi}=\big({\cal P}(z)\big)^{-\frac{1}{2}}\,\bigg(
\partial_z+\frac{2\ri A-1}{2z}+
\sum_{a=1}^r\frac{\mathfrak{j}_a}{z-z_a}+
\sum_{\alpha=1}^{{\tt L}}\frac{1}{z-w_\alpha}-\sum_{\alpha=1}^{\tilde{\tt L}}\frac{1}{z-\tilde{w}_\alpha}\bigg)\,\Psi\,,
\ee
the ODE \eqref{aois90121AA},\,\eqref{aiasaisias}
becomes
\bea\label{ioas89218921}
\bigg[&-&\partial_z^2+\kappa^2\, {\cal P}(z)-\frac{\tilde{A}^2+\frac{1}{4}}{z^2}+
\sum_{a=1}^r\bigg(\,\frac{\tilde{{\mathfrak j}}_a(\tilde{{\mathfrak j}}_a+1)}{ (z-z_a)^2}+
\frac{z_a\tilde{\gamma}_a}{z(z-z_a)}\bigg)\nonumber
\\[0.2cm]
&+&
\sum_{\alpha=1}^{\tilde{\tt L}}
\bigg(\,\frac{2}{ (z-\tilde{w}_\alpha)^2}+\frac{\tilde{w}_\alpha\tilde{\Gamma}_\alpha}{
z(z-\tilde{w}_\alpha)}\bigg)\bigg]\widetilde{\Psi}=0\,.
\eea
For the other case with $\mathfrak{M}<-2\mathfrak{J}$,
the sign of $A$ and $\tilde{A}$ in eqs.\,\eqref{oi98219812}-\eqref{iu872187128721}
needs to be flipped. 
\bigskip

An important point is that 
since $\Psi$ and $\widetilde{\Psi}$ are related through the action of a 
first order differential operator, the monodromic properties of the ODEs
\eqref{aois90121AA} and \eqref{ioas89218921} are the same.
In consequence, the corresponding  connections coefficients for these  differential equations 
coincide.
Thus, although there exists the solutions sets 
 $(\bm{\gamma};\bm{w})$ with $|\mathfrak{M}|>2\mathfrak{J}$,
by  performing the above transformation, 
the value of $|\mathfrak{M}|$ can be reduced. 
By successive  applications if necessary, we expect that
the integer $\mathfrak{M}$ may always be brought to the interval
\be
\mathfrak{M}=-2\mathfrak{J},-2\mathfrak{J}+2,\ldots,2\mathfrak{J}\ .
\ee

\bigskip

\bigskip

Like for the primary states, it is necessary to distinguish  the cases $(i)$ 
and $(ii)$ from \eqref{casei}.
For the former, one makes the identification
\be\label{ioas19829BA}
{\rm case}\ (i)\ \ \ \ \ \qquad\qquad \mathfrak{M}=\mathfrak{m}=-2\mathfrak{J},\,-2\mathfrak{J}+2,\,\ldots,2\mathfrak{J}\ .
\ee
The connection coefficients 
$D_{\boldsymbol {\mathfrak j},\pm {\mathfrak  m},\pm A}(\mu\,|\,\bm{\gamma},\bm{w})$  
are introduced similarly as in eq.\,\eqref{iissisia}.
 They turn out to be entire
 functions of 
 \bea\label{ioas19829B}
 \mu=\kappa^{\frac{2}{(1+\xi)K}}
 \eea
  for any given choice
 of  $({\boldsymbol \gamma},\bm{w})$.
Since the singularities $w_\alpha$ that were introduced are apparent,
the quantum Wronskian
relation \eqref{jasasasasay} is still satisfied.
Interpreting $D_{\boldsymbol {\mathfrak j},\pm\mathfrak{ m},\pm A}(\mu\,|\,{\boldsymbol\gamma},\bm{w})$ 
as an eigenvalue of the operator
${\mathlarger{\mathlarger{\mathlarger {\boldsymbol  a}}}}_\pm(\lambda)$ for a certain state
in the level subspace ${\cal H}^{({\tt L})}_{\bm{\mathfrak{j}},\mathfrak{m},P}$\,,
the zero-mode momentum $P$ would be related to $A$  as in eq.\,\eqref{siisaiasfff}.
With this set-up, we conjecture that the connection coefficients
coincide with certain  eigenvalues of  ${\mathlarger{\mathlarger{\mathlarger {\boldsymbol  a}}}}_\pm(\lambda)$:
 \be
{\rm case}\ (i)\qquad\qquad
{\mathlarger{\mathlarger{\mathlarger {\boldsymbol  a}}}}_\pm(\lambda)\, 
{ \boldsymbol e}_{\boldsymbol {\mathfrak j},\mathfrak{m},P}({\boldsymbol\gamma},\bm{w})
=D_{\boldsymbol {\mathfrak j},\pm\mathfrak{ m},\pm A}(\mu\,|\,{\boldsymbol\gamma},\bm{w})\,
{ \boldsymbol e}_{\boldsymbol {\mathfrak j},\mathfrak{m},P}({\boldsymbol\gamma},\bm{w})\,
\in\,
{\cal H}^{({\tt L})}_{\bm{\mathfrak{j}},\mathfrak{m},P}\  .
\ee
Moreover,  all such   eigenstates 
 ${ \boldsymbol e}_{\boldsymbol {\mathfrak j},\mathfrak{m},P}({\boldsymbol\gamma},\bm{w})$
  form a basis in ${\cal H}^{({\tt L})}_{\bm{\mathfrak{j}},\mathfrak{m},P}\,$.
Of course, the  $\xi-\beta$ and $\mu-\lambda$ relations remain unchanged  
and are given by \eqref{siisaias11a}.

\bigskip
For the second case  in \eqref{casei} we make the identification
\be
{\rm case}\ (ii)\qquad\qquad \mathfrak{M}=\check{\mathfrak{m}}=-2\check{\mathfrak{J}},
\,-2\check{\mathfrak{J}}+2,\,\ldots,2\check{\mathfrak{J}}
\ee
with $\check{\mathfrak{m}}=\mathfrak{m}-K$ and
$\check{\mathfrak{J}}=\frac{1}{2}\,K-\mathfrak{J}$.
The conjectured  ODE/IQFT correspondence is described as
\be\label{oiasio982121}
{\rm case}\ (ii)\qquad\qquad
{\mathlarger{\mathlarger{\mathlarger {\boldsymbol  a}}}}_\pm(\lambda)\, 
{ \boldsymbol e}_{\boldsymbol {\mathfrak j},\mathfrak{m},P}({\check{\boldsymbol\gamma}},\check{\bm{w}})
=
D_{\check{\boldsymbol {\mathfrak j}},\pm\check{\mathfrak{ m}},\pm \check{A}}
(\mu\,|\,\check{{\boldsymbol\gamma}},\check{\bm{w}})\,
{ \boldsymbol e}_{\boldsymbol {\mathfrak j},\mathfrak{m},P}(\check{{\boldsymbol\gamma}},\check{\bm{w}})\,
\in\,
{\cal H}^{({\tt L})}_{\bm{\mathfrak{j}},\mathfrak{m},P}\ .
\ee
Here $\check{\boldsymbol {\mathfrak j}},\check{\mathfrak{ m}}$ and  $ \check{A}$ are defined  in 
\eqref{iowewewew9889}. It is important to keep in mind that $(\check{{\boldsymbol\gamma}},\check{\bm{w}})$,
which label the states in ${\cal H}^{({\tt L})}_{\bm{\mathfrak{j}},\mathfrak{m},P}$, are solutions of the algebraic system
\eqref{aiasiisasia} and satisfy  \eqref{kausy} with $A$, $\mathfrak{j}_a$  replaced by their
 ``checked'' counterparts and $\mathfrak{M}=\check{\mathfrak{m}}$.
\bigskip

It is expected that the eigenstates of ${\mathlarger{\mathlarger{\mathlarger { \boldsymbol a}}}}_+(\lambda)$ 
 form a basis in the representation
${\cal H}_{\bm{\mathfrak{j}},\mathfrak{m},P}$,
which is fully specified by its eigenvalues.
The conjecture \eqref{oiasio982121} implies that the spectrum of 
${\mathlarger{\mathlarger{\mathlarger {\boldsymbol  a}}}}_+(\lambda)$ in the
space ${\cal H}_{\bm{\mathfrak{j}},\mathfrak{m},P}$ with
$\mathfrak{m}=2\mathfrak{J},\ldots,2K-2\mathfrak{J}-2$  coincides with its spectrum in
 ${\cal H}_{\check{\bm{\mathfrak{j}}},\check{\mathfrak{m}},P}$.
This leads us to propose the isomorphism
of these two representations of the ${ W}^{(c,r)}_{{\boldsymbol k}}$ algebra, see eq.\,\eqref{ioasioioasioas12}.

\subsection{Some comments concerning the literature\label{sec74}}

 \subsubsection*{Isotropic limit}

 In ref.\cite{Bazhanov:1994ft} an expression for the
 vacuum eigenvalue  of the transfer-matrix 
${\boldsymbol \tau}_{\scriptscriptstyle \frac{1}{2}}(\lambda)$
was presented for the case $r=1$.
It coincides with the grand canonical partition function of a
1D plasma  of alternating charges. The latter, since the seminal work of Anderson and  Yuval \cite{Anderson},  is well
known to be the partition function of the
one-channel anisotropic  Kondo model, where the impurity has spin $\frac{1}{2}$ \cite{Fendley:1995kj}.
This allowed the techniques of ref.\cite{Bazhanov:1994ft} and its further developments, including the ODE/IQFT correspondence,
to be transferred to different variants of the
quantum  impurity problem such as the quantum dot \cite{Lukyanov:2003rt} and the Coqblin-Schrieffer
model \cite{Bazhanov:2003ua}.
In  the work \cite{Lukyanov:2006gv}, the partition function
of the multichannel anisotropic  Kondo model was expressed in terms of the solution of the ODE \eqref{asoosao}.
Renewed  interest in the  Kondo  problem 
came from refs.\cite{Gaiotto:2020fdr,Gaiotto:2020dhf}, where
some  previously obtained facts concerning
the isotropic case were  rediscovered. Here we explain the relation
between the differential equation \eqref{aois90121} and the  one that was 
considered in those works.

\bigskip
The isotropic limit corresponds to taking  $\beta\to1^-$. It is a  complicated problem
to perform the limit starting  with  the path-ordered exponent \eqref{9sd090as},
 which in the context of the impurity 
problem is interpreted as an imaginary time evolution operator in the interaction picture.
One of the main advantages of the ODE/IQFT correspondence  is that it is well adapted for
exploring such a limit.
The corresponding  technique was developed in the works \cite{Lukyanov:2003rt,Bazhanov:2003ua} and 
can be straightforwardly applied to the  ODE \eqref{aois90121AA}.

\bigskip
Let us  make a uniform shift of the variable $z$ and the  positions of the apparent singularities
 \bea\label{ausausau}
z\mapsto z+\tfrac{1}{\varepsilon}\ ,\ \ \ z_a\mapsto z_a+\tfrac{1}{\varepsilon}\ ,\ \ \ \ \ 
w_\alpha\mapsto  w_\alpha+\tfrac{1}{\varepsilon}
\eea
and rewrite the ODE    \eqref{aois90121AA} using the parameters
\bea
\widetilde{\kappa}^2=\kappa^2\,\varepsilon^{2-\xi K}\ ,\ \ \qquad \widetilde{A}=\varepsilon\,A\, .
\eea
Of course such an innocent
change of variables and redefinition of the parameters should not affect  the properties of the
ODE. 
We set
\bea
\varepsilon=\tfrac{2}{\xi K}
\eea
 and consider the limit $\xi\to+\infty$ keeping $\widetilde{\kappa} $ and $\widetilde{A}$  fixed. 
This  corresponds  to $\beta=\sqrt{\frac{\xi}{1+\xi}}\to 1^-$.
Taking the isotropic limit  in the differential equation   \eqref{aois90121AA}, one obtains
 \bea\label{aois90121a}
\Big(\, - \partial_z^2+\widetilde{\kappa}^2\, \re^{2z}\ \prod_{a=1}^r(z-z_a)^{k_a}+
{t}_{\tt L}^{({\rm iso})}(z)\,\Big)\ \Psi=0\ ,
\eea
 where
 \bea\label{aois90121b}
{t}_{\tt L}^{({\rm iso})}(z)=-\widetilde{A}^2+
\sum_{a=1}^r\bigg(\,\frac{{\mathfrak j}_a({\mathfrak j}_a+1)}{ (z-z_a)^2}+\frac{\gamma_a}{z-z_a}\bigg)+
\sum_{\alpha=1}^{\tt L}
\bigg(\,\frac{2}{ (z-w_\alpha)^2}+\frac{\Gamma_\alpha}{z-w_\alpha}\bigg)
\eea
and  the limiting form of \eqref{oisaisai} reads as
\bea
\Gamma_\alpha=2
+\sum_{b=1}^r\frac{k_b}{w_\alpha-z_b}\ \ \ \ \ \ \ \ \  \ \ \ \  (\alpha=1,\ldots, {\tt L})\ .
\eea
The sets  $\boldsymbol{\gamma}$ and  $\boldsymbol{w}$ still satisfy the algebraic system \eqref{aiasiisasia}.
Of course, now   $v^{(a)}_{m}$  and $t^{(\alpha)}_{i}$ \eqref{sisisa} are defined by 
 \eqref{isaiisa} with ${t}_{\tt L}(z)$  substituted by ${t}_{\tt L}^{({\rm iso})}(z)$.
 In the isotropic  limit  the relation \eqref{siisaiasfff}  becomes
 \be\label{oias90121ss}
\widetilde{A} =
\tfrac{\ri}{K}\ \big(\,2\sqrt{K}\,P- \mathfrak{m}\,\big)\ .
\ee

 \bigskip
For the ``vacuum'' case
 corresponding  to the states ${\boldsymbol  e}_{\boldsymbol{\mathfrak j},\pm 2 {\mathfrak J},P}$
  \eqref{ioas9812}, one should set  ${\tt L}=0$ and choose $\gamma_a$ to be
 \bea
\gamma^{{({\rm vac},\pm)}}_a=-\ri \widetilde{A}^{{({\rm vac},\pm)}}\  \frac{2\mathfrak{j}_a  }{z_a}+
\sum_{b=1\atop b\not=a}^{r}\frac{2\mathfrak{j}_a \mathfrak{j}_b}{z_a-z_b}\  \ \  \  \ \ \ \  \ (a=1,\ldots,r)\,,
\eea 
where
\bea\label{aiaiisa}
\widetilde{A}^{{({\rm vac},\pm)}}=
\frac{2\ri}{K}\ \Big(\,\sqrt{K}\,P\mp \sum_{a=1}^r{\mathfrak j}_a\,\Big)\ .
\eea
 The ODE \eqref{aois90121a} boils down to
 \bea\label{aois90121ab}
\bigg(\!\!\!\! &-&\! \partial_z^2+\tilde{\kappa}^2\, \re^{2z}\ \prod_{a=1}^r(z-z_a)^{k_a}-\big(\widetilde{A}^{{({\rm vac},\pm)}}\big)^2-
\ri\widetilde{A} ^{{({\rm vac},\pm)}}\ \sum_{a=1}^r \frac{2\mathfrak{j}_a  }{z_a(z-z_a)}
\nonumber
\\[0.3cm]
&+&\!
\sum_{a=1}^r\frac{{\mathfrak j}_a({\mathfrak j}_a+1)}{ (z-z_a)^2}
+\sum_{ 1\leq a<b\leq r}
\frac{2\,\mathfrak{j}_a \mathfrak{j}_b}{(z-z_a)(z-z_b)}
\,\bigg)\, \Psi=0\ .
\eea
 This equation  with  $r=k=1$ and ${\mathfrak j}_1=0$ originally appeared in refs.\cite{Lukyanov:2003rt,Bazhanov:2003ua} in the
description of the isotropic Kondo model. 
It is worth mentioning that   the parameter $2\ri P$ can be identified
with $H/T$, where $H$ is an external local magnetic field acting on the impurity spin, while $T$ stands for the temperature.
Also $\widetilde{\kappa}=T_{K}/T$ with $T_K$ being the Kondo temperature.

\bigskip
The ODE \eqref{aois90121ab} with $r> 1$ and arbitrary $k_a$
was introduced  in the recent work \cite{Gaiotto:2020fdr}.
Note that in the case   $\widetilde{A}^{{({\rm vac},\pm)}}=0$ it corresponds to 
the chiral primary states for  the $\otimes_{a=1}^r \widehat{\mathfrak { sl}}_{k_a}(2)$ WZW model.
Indeed, as  follows
 from \eqref{isisiasias} with $P=\frac{\mathfrak{m}}{2\sqrt{K}}=\pm\frac{1}{\sqrt{K}}\sum_{a=1}^r{\mathfrak j}_a$,
  the value of the local IM  ${\bf I}_1$ is given by
  \bea\label{iasiisaiasdd}
I_1=\sum_{a=1}^r\frac{{\mathfrak j}_a ({\mathfrak j}_a+1)}{k_a+2}-\frac{c_{\tt wzw}}{24}\ ,\ \ \  \ \ \ \qquad
\ {\rm where}\ \ \ \ \ \ \qquad
c_{\tt wzw}=\sum_{a=1}^r\frac{3k_a}{k_a+2}\ .
\eea

\subsubsection*{Gaudin limit}
The ordinary differential equation for the isotropic case \eqref{aois90121a}-\eqref{aois90121b}
still admits an interesting limit, which can be interpreted as a double scaling limit
of the original system.\footnote{%
The authors thank S.~Lacroix  and B.~Vicedo for pointing out this possibility.
} One performs the change of variables 
\be
z=\delta\  y
\ee
 and takes $\delta\to0$, keeping fixed the combinations
\be
z_a=\delta\ y_a\,,\qquad\qquad 
\kappa_{\rm \scriptscriptstyle G} =\delta^{1+\frac{K}{2}}\ 
\widetilde{\kappa} \ .
\ee
Further assuming that $\widetilde{A}=0$, i.e., 
\be
P=\frac{\mathfrak{m}}{2\sqrt{K}}
\ee
the differential equation becomes
 \be\label{ioas8919823}
\bigg[\, - \partial_y^2+\kappa_{\rm \scriptscriptstyle G}^2\,
\prod_{a=1}^r(y-y_a)^{k_a}+
\sum_{a=1}^r\bigg(\,\frac{{\mathfrak j}_a({\mathfrak j}_a+1)}{ (y-y_a)^2}+
\frac{\gamma_a^{(\rm \scriptscriptstyle G)}}{y-y_a}\bigg)+
\sum_{\alpha=1}^{\tt L}
\bigg(\,\frac{2}{ (y-v_\alpha)^2}+
\frac{\Gamma_\alpha^{(\rm \scriptscriptstyle G)}}{y-v_\alpha}\bigg)\bigg]\ \Psi=0\ .
\ee
Here
\be
\Gamma_\alpha^{(\rm\scriptscriptstyle G)}=
\sum_{b=1}^r\frac{k_b}{v_\alpha-y_b}\ \ \ \ \ \ \ \ \  \ \ \ \  (\alpha=1,\ldots, {\tt L})\ ,
\ee
while the value of $\gamma_a^{(\rm\scriptscriptstyle G)}$ and 
$v_\alpha$ are solutions of an algebraic system, which ensures that the points $y=y_a,v_\alpha$ are apparent
singularities of the ODE. Of course, this system can be obtained via a limit of \eqref{aiasiisasia}.
The differential equation \eqref{ioas8919823} was originally
proposed by Feigin and Frenkel \cite{Feigin:2007mr} in their study of the affine Gaudin model.
\bigskip

Finally we note that the simplest variants 
 of the ODE \eqref{aiasaisias}-\eqref{oisaisai} were studied in 
 refs.\cite{Voros:1999,Ito:2020ueb}.

\section{ 
\label{sec8}
Large $\mu$ asymptotic expansion   of \texorpdfstring{%
$D_{\bm{ {\mathfrak j}},\mathfrak{ m}, A}$}{}}

\subsection{Leading behaviour}
The operators ${\mathlarger{\mathlarger{\mathlarger {\boldsymbol  a}}}}_\pm(\lambda)$,
which depend continuously on the spectral parameter $\lambda$, are  generating functions
of the family of commuting operators. In this regard, of special interest is the large $\lambda$
expansion of ${\mathlarger{\mathlarger{\mathlarger {\boldsymbol  a}}}}_\pm(\lambda)$ \cite{Bazhanov:1996dr}.
With the ODE/IQFT correspondence at hand, it can be explored 
via the study of the connection coefficients $D_{\boldsymbol {\mathfrak j},\pm \mathfrak{m},\pm A}(\mu)$
at large $\mu$. Below we restrict our attention to $D_{\boldsymbol {\mathfrak j}, \mathfrak{m}, A}$.

\bigskip
A brief inspection of the ODE shows that
the connection coefficients 
depend on $\mu$
and  the parameters $\{z_a\}_{a=1}^r$  only through the combinations $z_a\mu$.
Thus a multiplication of all the $z_a$ by the same factor can be absorbed into a redefinition of the spectral
parameter. Focusing on the case  when  
$z_a\not=0$ for any $a$, without loss of generality, one   can set
\bea\label{concj128921}
\prod_{a=1}^r(-z_a)^{k_a}=1\ .
\eea

Some
 details of the derivation of the 
leading  terms in the asymptotic expansion of $D_{\boldsymbol {\mathfrak j}, \mathfrak{m}, A}(\mu)$
at $\mu\to +\infty$ have been relegated to Appendix \ref{AppA}. 
Assuming the convention \eqref{concj128921} and also that none of the
 $z_a$ are positive real numbers,
the final result reads as
\be\label{oia38923ss1A}
D_{\boldsymbol {\mathfrak j}, \mathfrak{m}, A}(\mu)\asymp
R_{\boldsymbol {\mathfrak j}, \mathfrak{m}, A}
\ \mu^{\frac{\ri A}{\xi}-\frac{1}{2}\mathfrak{m}}\ 
\ \exp\bigg(\mu^{\frac{1}{2}(1+\xi) K}\ q_{-1}+
o(1)\bigg)\ \ \ \ \ \ \ \qquad\big(\mu\to+\infty)\, .
\ee
 The asymptotic coefficients are given by
  \bea\label{aisisaasias}
q_{-1}=\frac{1}{(1-\re^{-\ri \pi (1+\xi) K})\ 
(1-\re^{\ri \pi \xi K})}\ 
\oint_{{\cal C}}\rd z\, \sqrt{{\cal P}(z)}
\eea
and 
\be\label{usaussuasa}
R_{\boldsymbol {\mathfrak j}, \mathfrak{m}, A}=
(-1)^{\frac{1}{2} \mathfrak{m}}\,\big((1+\xi) K\big)^{\frac{2\ri A+ \mathfrak{m}}{(1+\xi)K}-\frac{1}{2}}\ 
\big(\xi K)^{\frac{1}{2}-\frac{2\ri A}{\xi K}}
\ 
 \frac{\Gamma\big(1-\tfrac{2\ri  A}{ \xi K}\big)}{
 \Gamma\big(1-\tfrac{2\ri  A+\mathfrak{m}}{(1+\xi)K})}\ \frac{\prod_{m=1}^{{\tt M}_+}x^{\scriptscriptstyle{(+)}}_m}{\prod_{a=1}^r  
           (z_a)^{\mathfrak{j}_a}\prod_{\alpha=1}^{\tt L}w_\alpha}\ .
          \ee
The branch of the multivalued function  in the integrand
in  \eqref{aisisaasias} is chosen in such a way that
\bea\label{siusauasusa}
\Im m\big(\sqrt{{\cal P}(z)}\,\big)\to 0\,,\ \ \ \qquad
 \Re e\big(\sqrt{{\cal P}(z)}\,\big)>0\ \ \ \ \ \ \ {\rm as}\ \ \ z\to+\infty\ .
\eea
To describe  the closed path of integration
${\cal C}$,
let us  define
the domain of the complex plane
\bea
{\cal{D}}=\big\{z\, :\ -\delta<\arg(z)<+\delta\,\big\}\!
\begin{array}{c} \\[-0.4cm] {\displaystyle{\mathlarger{\mathlarger \cup}}}
\end{array}\!
\big\{z\,: \ |z|<\varepsilon\,\big\}
\!
\begin{array}{c} \\[-0.4cm] {\displaystyle{\mathlarger{\mathlarger \cup}}}
\end{array}\!\big\{z\,: \ |z|>\varepsilon^{-1}\,\big\}\ .
\eea
Here the parameters  $\varepsilon$ and  $\delta$ should be taken to be sufficiently
small, such that  $z_a\notin{\cal D}$ for all $a=1,\ldots ,r$.
 The  contour  ${\cal C}$ lies inside $\cal{D}$ and
is the image of
a  Pochhammer  loop on the Riemann sphere under stereographic projection.
The homotopy class of the   loop is schematically shown in fig.\ref{figa2}.
It winds around the  south and north poles of the sphere,
 which  are mapped to   $z=0$ and $z=\infty$,  respectively.
\begin{figure}
\centering
\begin{tikzpicture}
\node at (-0.1,0){
\includegraphics[width=8.cm,trim = 0cm 2.5cm 0cm 3cm,clip]{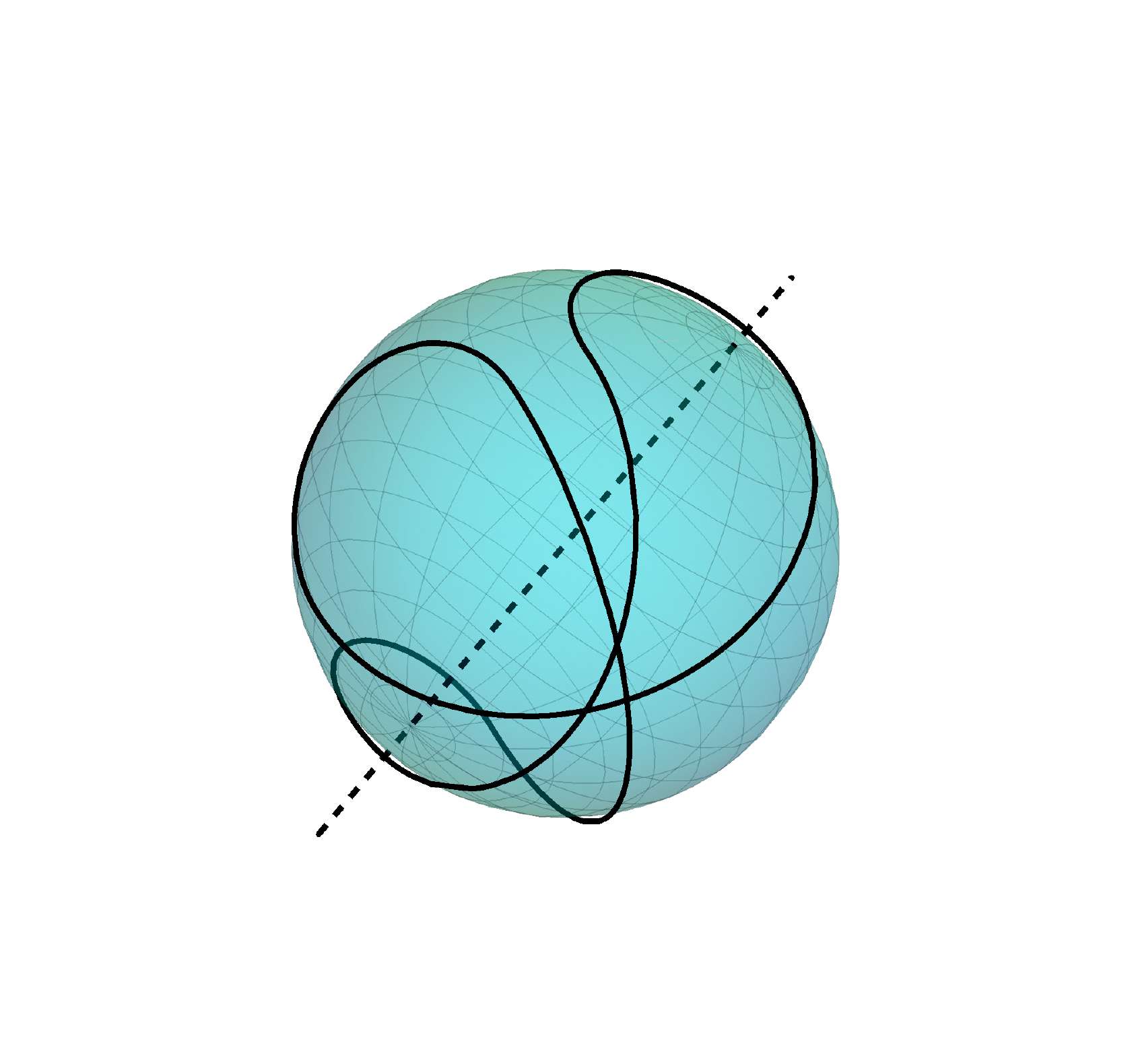}};
\node at (2.1,2.1) {\large $N$};
\node at (-2.15,-2.05) {\large $S$};
\draw [->,line width = 0.4mm] (1.4,1.8) -> (1.55,2);
\draw [->,line width = 0.5mm] (-0.9,-1.7) -> (-0.8,-1.7);
\draw [->,line width = 0.5mm] (0.2,1.15) -> (0.1,1.35);
\draw [->,line width = 0.5mm] (1.5,1.15) -> (1.58,1.05);
\draw [->,line width = 0.5mm] (1.3,-0.42) -> (1.2,-0.52);
\draw [->,line width = 0.5mm] (-1.87,-0.42) -> (-1.95,-0.3);
\draw [->,line width = 0.5mm] (-0.38,1) -> (-0.33,0.9);
\draw [->,line width = 0.5mm] (0.28,-1.8) -> (0.23,-1.9);
\end{tikzpicture}
\caption{\label{iaosi1212}\small  The Pochhammer loop on the Riemann sphere.
\label{figa2}}
 \end{figure}
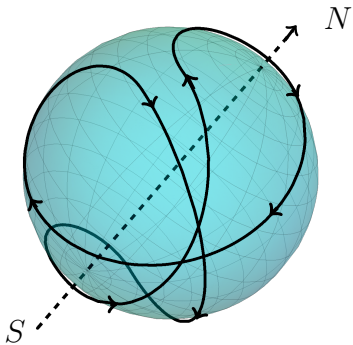

\bigskip

The asymptotic coefficient $R_{\boldsymbol {\mathfrak j}, \mathfrak{m}, A}$ 
\eqref{usaussuasa} is interpreted as an eigenvalue of the so-called
reflection operator.\footnote{%
The asymptotic formula \eqref{oia38923ss1A},
when combined with  the quantum Wronskian relation \eqref{jasasasasay},
yields an interesting identity
\bea
 \frac{\prod_{m=1}^{{\tt M}_+}x^{\scriptscriptstyle{(+)}}_m\,
 \prod_{m=1}^{{\tt M}_-}x^{\scriptscriptstyle{(-)}}_m}
 {\prod_{a=1}^r  
           (z_a)^{2\mathfrak{j}_a}\prod_{\alpha=1}^{\tt L}w^2_\alpha}=
           1+\ri\ \frac{{\tt M}_+-{\tt M}_-}{2 A}\qquad\qquad\qquad\qquad
\big(\,{\tt M}_\pm=\mathfrak{J}+{\tt L}\mp \tfrac{1}{2}\,\mathfrak{M}\,\big)\, .
\nonumber
           \eea
} For the case $r=1$ and $k=1$ it was
discussed in refs.\cite{Zamolodchikov:1995aa,Kotousov:2019ygw,Kotousov:2019nvt,Litvinov:2020zeq}.\footnote{%
In ref.\cite{Kotousov:2019nvt},
the auxiliary variables $x^{\scriptscriptstyle{(\pm)}}_m$ are not employed and the
eigenvalues of the reflection operator are expressed solely in terms of
the apparent singularities $w_\alpha$. This results in
 more cumbersome formulae.} It also appears
in various applications of the ODE/IQFT correspondence.

\subsection{Eigenvalues  of the  local IM in the GAGM\label{sec82}}

\bigskip
The omitted terms in \eqref{oia38923ss1A},
denoted by $o(1)$, encode the eigenvalues of the
local and so-called dual non-local integrals of motion.
It is expected that $D_{\boldsymbol {\mathfrak j}, \mathfrak{m}, A}(\mu)$  admits
a  systematic asymptotic  expansion of the form
\be\label{oia38923ss1AB}
D_{{\boldsymbol {\mathfrak j}}, \mathfrak{m}, A}(\mu)\asymp
R_{{\boldsymbol {\mathfrak j}}, \mathfrak{m}, A}
\ \mu^{\frac{\ri A}{\xi}-\frac{1}{2}\mathfrak{m}}\ 
\ \exp\Big(\mu^{\frac{1}{2}(1+\xi) K}\ q_{-1}\Big)\ \ 
B(\mu)\ X(\mu)
\ \ \ \qquad\big(\mu\to+\infty\big)\,,
\ee
where the eigenvalues of the local and dual non-local IM
appear in the  coefficients 
of the formal power series  $\log(B)=o(1)$ and $\log(X)=o(1)$, respectively.
The series  $\log(B)$ can be studied within the standard  WKB technique. 
Carrying over the analysis in refs.\cite{Lukyanov:2013wra,Bazhanov:2013cua}, one finds
\bea\label{isusuas}
\log(B)= \sum_{n=1}^\infty 
(-1)^{n-1}\ \frac{\Gamma(n-\frac{1}{2})}{2n!
\sqrt{\pi}} \ \ q_{2n-1}\ \mu^{\frac{1}{2}(1-2n)(1+\xi) K}
\eea
with
\bea\label{aiasisia}
 q_{2n-1}=\big[\,\big(1-\re^{\ri \pi  (2n-1)(1+\xi) K}\big)\ 
\big(1-\re^{-\ri \pi  (2n-1)\xi K}\big)\,\big]^{-1}\ 
  \oint_{{\cal C}}\rd z\ U_{2n}\, .
\eea
 There is a simple recursion procedure to generate the densities $U_{2n}$ for any $n$.
For instance,
 \bea
U_2&=&\big({\cal P}(z)\big)^{-\frac{1}{2}}\,  \bigg(\,t_{\tt L}+
\frac{4\,{\cal P}\partial^2_z{\cal P}-5\, (\partial_z {\cal P})^2}{16\,
{\cal P}^2}\, \bigg)\nonumber\\[0.3cm]
U_4&=&\big({\cal P}(z)\big)^{-\frac{3}{2}}\,  \bigg(\,t_{\tt L}+
\frac{4\,{\cal P}\partial^2_z{\cal P}-5\, (\partial_z {\cal P})^2}{16\,
{\cal P}^2}\, \bigg)^2\ .
\eea
Also
in writing  the numerical coefficients in the sum \eqref{isusuas},
 it was assumed that the densities are normalized by the condition
\bea
U_{2n}=\big({\cal P}(z)\big)^{\frac{1}{2}-n}\ \big(\,(\,t_{\tt L}\,)^n+\ldots\,\big)\ ,
\eea
 where the omitted terms contain derivatives and 
lower powers of $t_{\tt L}$ \eqref{aiasaisias}.

\bigskip
Let us focus on the first nontrivial integral $q_{1}$. It turns out
that the particular choice of the integration contour ${\cal C}$ is not essential. For this reason we consider
\bea\label{ysstops}
q_1^{\scriptscriptstyle{({\cal C})}}=\oint_{{\cal C}}\frac{\rd z}{\sqrt{{\cal P}}}\, 
\ \bigg(\,t_{\tt L}+
\frac{4\,{\cal P}\partial^2_z{\cal P}-5\, (\partial_z {\cal P})^2}{16\,
{\cal P}^2}\, \bigg)\, ,
\eea
where ${\cal C}$ is an arbitrary \emph{closed} contour.
 Independently on the choice of ${\cal C}$,
the following relation holds true
  \bea\label{jassjaasuasus}
q_1^{\scriptscriptstyle{({\cal C})}}=\Big(\sum_{a=1}^rI^{(a)}_1\, z_a\partial_{z_a}\,\Big)
 f^{\scriptscriptstyle{({\cal C})}}_1
\qquad\qquad
{\rm with}
\qquad\qquad
  f^{\scriptscriptstyle{({\cal C})}}_1=\oint_{{\cal C}}\frac{\rd z}{\sqrt{\cal P}}\,\frac{1}{z^2}\ .
 \eea
 The coefficients in the sum read explicitly as
\be\label{oias8912891}
I^{(a)}_1=\frac{2z_a{\gamma}_a}{k_a}-2z_a\,\sum_{\beta=1}^{{\tt L}}\frac{1}{z_a-w_\beta}
-\frac{2z_a}{k_a}\,\sum_{b=1\atop
b\not=a}^r\frac{d_ak_b+d_bk_a}{z_a-z_b}-\frac{2d_a}{k_a}\, \big(\xi K-2\big)-2d_0\ .
\ee
Here we use the notations
\bea\label{oi98129821}
d_a=\frac{\mathfrak{j_a}( \mathfrak{j_a}+1) }{k_a+2}-\frac{1}{24}\ \frac{3k_a}{k_a+2}\ ,\ \ \ \ \
\ \ \ \ \  K=\sum_{a=1}^rk_a\ ,
\eea
while
 \be\label{haysaysat}
d_0=-\frac{1}{8}-{\tt L}-\sum_{a=1}^rd_a+
\frac{1}{(1+\xi) K}\ \Big(2{\tt L}-A^2+
\sum_{a=1}^r\big(\,\mathfrak{j_a}( \mathfrak{j_a}+1)
+z_a{\gamma}_a\big)+
\sum_{\alpha=1}^{\tt L}w_\alpha\Gamma_\alpha
\Big)
\ee
 In spite of the somewhat  cumbersome expressions, the proof of  the identity \eqref{jassjaasuasus} is elementary.
 Namely,  taking into account eq.\eqref{aois901828912}, it is straightforward to show that
 \bea\label{kasjsaj}
\frac{1}{\sqrt{\cal P}}\, \bigg(\, t_{\tt L}+
\frac{4\,{\cal P}\partial^2_z{\cal P}-5\, (\partial_z {\cal P})^2}{16\,
{\cal P}^2}\,\bigg)=\partial_z V_1\,+\,
\sum_{a=1}^rI^{(a)}_1\   z_a\partial_{z_a}\bigg(\frac{1}{z^2\sqrt{\cal P}}\bigg)
\eea
with
\bea
{V}_1(z)=-\frac{2}{\sqrt{\cal P}}\, \bigg(\,
\frac{1}{z}\ \big(d_0-\tfrac{\xi K-2}{16}\,\big)
+\sum_{a=1}^r\frac{d_a-\frac{k_a}{16}}{z-z_a}+  \sum_{\alpha=1}^{\tt L}\frac{1}{z-w_\alpha}\,\bigg)\ .
\eea
Integrating  both sides of \eqref{kasjsaj} over the closed contour  yields \eqref{jassjaasuasus}.
 \bigskip

In the context of the ODE/IQFT correspondence, $q_1^{\scriptscriptstyle{({\cal C})}}$ 
\eqref{ysstops} should be interpreted as the eigenvalue of a certain local IM. 
Formula \eqref{jassjaasuasus}  shows that this quantity is a linear 
combination of $I^{(a)}_1$ with $a=1,2,\ldots r$, where the 
coefficients
\be
z_a\partial_{z_a} f^{\scriptscriptstyle{({\cal C})}}_1=\frac{1}{2}\ \oint_{{\cal C}}\frac{\rd z}{\sqrt{\cal P}}\,
\frac{z_ak_a}{z^2\,(z-z_a)}
\ee
encode all of the dependence on the integration contour. Thus $I^{(a)}_1$ would be expected to coincide with
the eigenvalues of a suitably chosen
set of IM 
\be
{\bf{I}}^{(a)}_1=\int_0^{2\pi} \frac{\rd u}{2\pi}\ T^{(a)}_{2}(u)\,\  \ \ \ \ \ \ \qquad (a=1,\ldots,r)\ .
\ee
As follows from the expression \eqref{oias8912891} and \eqref{kausy}, with $\mathfrak{M}=\mathfrak{m}$,
 the sum $\sum_{a=1}^r k_aI^{(a)}_1$ simplifies and can be brought
to the form
 \bea\label{iisisaias}
-\frac{1}{2\xi K}\, \sum_{a=1}^r k_aI^{(a)}_1=\sum_{a=1}^rd_a-\frac{\mathfrak{m}^2}{4K}+P^2
+{\tt L}\, ,
\eea
 where $A$ was swapped in favour of $P$ and the integer $\mathfrak{m}$ according to formula  \eqref{siisaiasfff}.
Keeping in mind the definition of $d_a$ \eqref{oi98129821},
this coincides   with the eigenvalue \eqref{isisiasias}
of  the local IM  ${\bf I}_1$
\eqref{oi89129821}.

\bigskip
For future references we include the expressions for the eigenvalues of the local IM through the sets
$\bm{x}^{\scriptscriptstyle{(\pm)}}
=(x^{\scriptscriptstyle{(\pm)}}_1,\ldots,x^{\scriptscriptstyle{(\pm)}}_{{\tt M}_\pm})$,
$\bm{w}=
(w_1,\ldots,w_{\tt L})$
with ${\tt M}_\pm=\mathfrak{J}+{\tt L}\mp \tfrac{1}{2}\,\mathfrak{m}$ 
solving the
algebraic system \eqref{aiiias}:
 \be\label{jhsasysay}
 \frac{k_a}{2z_a}\,  I^{(a)}_1=
 \sum_{b=0\atop
b\ne a}^r\, \frac{1}{z_a-z_b}\, \big(2\mathfrak{j}_a\mathfrak{j}_b
-d_ak_b-d_bk_a\big)
\,-\,
\sum_{m=1}^{{\tt M}_\pm}\,\frac{2\mathfrak{j}_a}{z_a-x^{\scriptscriptstyle{(\pm)}}_m}
\,+\,\sum_{\beta=1}^{{\tt L}}\frac{2\mathfrak{j}_a-k_a}{z_a-w_\beta}\ .
\ee
 Here $a=1,\ldots,r$ but the summation index ``$b$'' in the first sum runs from $0$ to $r$ and
  \bea
 k_0\equiv\xi K-2\ ,\ \ \ \  \ \qquad \ {\mathfrak j }_0\equiv \pm\ri A-\tfrac{1}{2}\ ,\qquad \ \ \ \ \ z_0=0\, .
\eea
Also $d_0$, defined by eq.\eqref{haysaysat}, can be rewritten as
 \bea
 d_0=-\sum_{a=1}^rd_a
 +\frac{1}{ (1+\xi) K}\, \bigg(\sum_{a=0}^r\mathfrak{j}_a
+\tfrac{1}{2} +{\tt L}-{\tt M}_\pm\bigg)^2-
{\tt L}- \frac{1}{8} \ .
 \eea
It is instructive to compare  \eqref{jhsasysay} with the expression
 \eqref{op2929292929}  for the energies $E_a$ in the original Gaudin model.

\bigskip
In principle, similar computations can be performed for the higher local IM.
Upon obtaining the expression for the density $U_{2n}$, one should
establish the identity
 \bea\label{isaisusaua}
 U_{2n}=\partial_z V_{2n-1}+\sum_{a=1}^rI^{(a)}_{2n-1}\   z_a\partial_{z_a}F_{2n-1}
 \eea
which is the analogue of \eqref{kasjsaj}. The eigenvalues of the local IM
would be read off from the expansion coefficients in the sum.
The calculations are purely algebraic and straightforward, but 
rather cumbersome. For the integral $I_3^{(a)}$, they were performed
only for the case $r=1$ in ref.\cite{Bazhanov:2013cua}.

 \subsection{Some comments on the eigenvalues  of the dual   nonlocal IM \label{sec83}}
The factor  $X$ in the asymptotic expansion \eqref{oia38923ss1AB} is a 
formal power series of the form
 \bea\label{isisaisa}
 \log X(\mu)\,\asymp\, -\sum_{n=1}^\infty X_n\, {\mu}^{-\frac{1+\xi}{\xi} n}\ .
  \eea
The coefficients $X_n$ 
are interpreted as the  eigenvalues of certain integrals of motion. However, contrary to ${\bf{I}}_{2n-1}^{(a)}$
the operators can not be represented as integrals over the local chiral densities.
Following the terminology of ref.\cite{Bazhanov:1996dr} we refer to these operators as  the ``dual nonlocal IM''.

\bigskip
The appearance of  $X(\mu)$  in the large $\mu$  expansion of the
connection coefficient is related to the fact that  the applicability of the WKB approximation  to the 
ODE   \eqref{aiasaisias}-\eqref{oisaisai} breaks down
in the vicinity of $z=0$. This makes it difficult to provide a
  mathematically rigorous justification of the 
asymptotic series  \eqref{oia38923ss1AB} and, in turn,  to calculate the expansion coefficients
$X_n$. Nevertheless,  the presence  of  $X(\mu)$ 
in \eqref{oia38923ss1AB} can be argued for   similarly as it was done  in ref.\cite{Bazhanov:1998wj} for the case $r=k=1$.

\bigskip
Consider the simplest variant of the ODE  \eqref{aois90121AA},\,\eqref{aiasaisias}:
\bea\label{hasysay}
\bigg(- \partial_z^2+\kappa^2\, z^{-2+\xi K }\,
\prod_{a=1}^r(z-z_a)^{k_a}
-\frac{{A}^2+\frac{1}{4}}{z^2}
\,\bigg)\ \Psi=0\,.
\eea
Taking into account the convention \eqref{concj128921}, 
the change  of variables $z=\zeta^{-1},\ \Psi(z)=\zeta^{-1}\, \tilde{\Psi} (\zeta)$ brings the 
differential equation to the form 
\bea
\bigg(- \partial_\zeta^2+\kappa^2\, \zeta^{-2-(1+\xi) K }\,
\prod_{a=1}^r\big( \zeta-z_{a}^{-1} \big)^{k_a}
-\frac{{A}^2+\frac{1}{4}}{\zeta^2}
\,\bigg)\ \tilde{\Psi}=0\,.
\eea
The latter looks similar to the original ODE with the
parameters   substituted as $\xi\mapsto -1-\xi$, $z_a\mapsto z^{-1}_a$.
 In  the parametric domain
 $-1<\xi<0$,
this formal substitution   can be interpreted as a duality transformation,
which allows one to relate the properties of the solutions  in the vicinity of $z=\infty$  and $z=0$.

 \bigskip
 The subject of our current interest is the domain  with $\xi>0$.
It is mapped to  $\xi<-1$ under the reflection 
 $\xi\mapsto -1-\xi$, which thus
can not be interpreted as a duality transformation. However,
since it relates  the   basic solutions $\Theta^{\scriptscriptstyle{(\leftarrow)}}_{ \pm A}$ ,
 defined through their behaviour 
 at  the regular singular point  $z=0$,
and  $\Theta^{\scriptscriptstyle{(\rightarrow)}}$, which  is the fast decaying
 solution at the irregular singularity $z=\infty$, it leads to an important property for
the connection coefficients.
 Namely, while
$D_{{\boldsymbol {\mathfrak j}}, \mathfrak{m}, A}(\mu)$  possesses the convergent Taylor series expansion in
$\mu$, its large $\mu$ asymptotic  involves  the  formal power series $X(\mu)$  in the ``dual'' parameter $\widetilde{\mu}$:
 \bea\label{suusauasu}
 \kappa^2=\mu^{(1+\xi) K}={\widetilde{\mu}}^{\,-\xi K}\ .
 \eea
 In this work, we forgo a systematic study of  the eigenvalues of the dual nonlocal IM. 
An illustration for a specific example with $r=2$ and $k_1=k_2=1$ is provided in sec.\,\ref{sec93}.
 
  \bigskip
  Finally we note  that  the eigenvalues of the transfer-matrices ${\boldsymbol{\tau}}_{\ell}(\lambda)$
can be expressed through $D_{{\boldsymbol {\mathfrak j}}, \mathfrak{m}, A}(\mu)$ by using the
set of well known functional relations. Hence, provided that the large  $\mu$ asymptotic expansion  of
the connection coefficients is available in the whole complex $\mu$ plane, one can straightforwardly
derive the large $\lambda$ expansion  of the eigenvalues of ${\boldsymbol{\tau}}_{\ell}(\lambda)$.
The latter are of special interest in the context
of the quantum impurity problem, where it permits a systematic study  of the infrared fixed points
in the theory.

 \section{Example with $r=2$ and $k_1=k_2=1$\label{sec9}}
The local IM for  the GAGM
have not been sufficiently studied in the general setup. 
The case $r=k=1$  has been explored in detail in the context of the quantum KdV theory 
\cite{Bazhanov:1994ft,Bazhanov:1996dr,Bazhanov:1998dq}, while
some results are known for $r=1$ and $k>1$.
The simplest case beyond the 
   quantum KdV model is when $r$ is a positive integer but with all $k_a=1$.
Then the representation of 
  $U_q(\widehat{{\mathfrak b}}_-)$  discussed  in sec.\,\ref{sec6}
 does  not involve the parafermionic fields, leading to significant simplifications. 
This section is devoted to an analysis of the case $r=2$ and $k_1=k_2=1$.

  \subsection{$W$-algebra
and   the first local IMs\label{sec91}}
 
 Taking $r=2$ and $k_1=k_2=1$ in  eqs.\eqref{isaiiasias} and \eqref{iuaususa}, the vertex operators $V_\pm$
become
  \bea\label{hasgssay}
 V_+=\big(\re^{\ri \sqrt{2} \theta}+\re^{-\ri \sqrt{2}\theta}\big)\ \re^{+\ri\sqrt{2}\beta \varphi}\ ,\ \  \ \  \ \ 
 V_-=\big( z_1\,\re^{-\ri \sqrt{2} \theta}+z_2\,\re^{+\ri \sqrt{2}\theta}\big)\ \re^{-\ri\sqrt{2}\beta \varphi}
 \eea
with
 \bea\label{oisaoioias9012}
 \varphi=\tfrac{1}{\sqrt{2}}\, \big(\phi_1+\phi_2\big)\ ,\ \ \ \  \ \ \ \ \ \ \ \ \ \ \ \ \ \ \ \  \theta=\tfrac{1}{\sqrt{2}}\, \big(\phi_1-\phi_2\big)\ .
 \eea 
The exponent involving the Bose field $\theta$  can be ``fermionized''
as
\be
\re^{\pm \ri \sqrt{2} \theta}=\tfrac{1}{\sqrt{2}}\, \big(\chi_1\pm \ri\,\chi_2\big)\ ,
\ee
where $\chi_{1,2}$ are  a pair  of chiral  Majorana fermion fields,
 satisfying the OPE
\bea\label{aiasisaisa}
\chi_a(u)\chi_b (0)=- \frac{\ri }{u}\ \delta_{ab}+O(1)\ .
\eea
Then formula \eqref{hasgssay} becomes
\bea\label{oias8912988921}
V_+=\sqrt{2}\ \chi_1\, \re^{+\ri\sqrt{2}\beta \varphi}\ ,\ \ \ \ \ \ \ \ \ \ \ \ \ \ \ 
V_-=\tfrac{1}{\sqrt{2}}\ \big(\,(z_1+z_2)\, \chi_1-\ri\,(z_1-z_2)\, \chi_2\, \big)\re^{-\ri\sqrt{2}\beta \varphi}\ .
\eea

 \bigskip
The construction of the local IM 
 follows the procedure outlined in  sec.\ref{sec4}. First  
 we note that the OPEs of $V_+$ with the spin $\tfrac{3}{2}$  and  spin 2 fields,
 \bea\label{ias891298321}
 {S}&=&\big(\rho\partial-\ri \partial\varphi\,\big)\, \chi_1\\[0.3cm]
 {G}&=&(\partial\varphi)^2+\ri \rho\ \partial^2\varphi+\tfrac{\ri}{2}\, \chi_1\partial \chi_1\,,\nonumber
 \eea
obey the condition \eqref{gasststa} provided the parameter $\rho$ is taken to be
 \bea
 \rho=\tfrac{1}{\sqrt{2}}\, \big(\beta^{-1}-\beta\big)\ .
 \eea
 The fields $S$ and $G$ form the    ${\cal N}=1$ super Virasoro algebra,
 whose commutation relations are encoded in the singular part of the operator product expansions
\bea\label{ajasusausa}
{{S}}(u){ S}(v)&=&-\frac{\ri\, c_{\scriptscriptstyle{sV}}}{3\,  (u-v)^3}+\frac{\ri \, ({{G}}(u)+
{{G}}(v)\big)}{2(u-v)}
+(u-v)\, 
\partial { S} { S}(v)+
O\big((u-v)^2\big)
\nonumber\\[0.2cm]
{{G}}(u){G}(v)&=&\frac{c_{\scriptscriptstyle{sV}}}{2\, (u-v)^4}-\frac{{ G}(u)+{{G}}(v)}{(u-v)^2}+
{G}^2(v)
+O (u-v)
\\[0.2cm]
{{G}}(u){ S}(v)&=&-\frac{3}{2}\ \frac{ {{S}}(v)}{(u-v)^2}-\frac{\partial {{S}}(v)}{u-v}+{{G}}{{S}}(v)+
O(u-v)\nonumber
\eea
with  the central charge  $c_{\scriptscriptstyle{sV}}=\tfrac{3}{2}-6\rho^2$.
The Majorana fermion $\chi_2$ also obeys the condition \eqref{gasststa} simply because the OPE
$\chi_2(u)V_+(v)$ is not singular at $u= v$.
The space of local fields satisfying \eqref{gasststa}  can be generated by the
spin $\frac{3}{2}$ field $S$ and spin $\frac{1}{2}$ field  $\chi_2$. 
Among such   fields we will focus on those having  integer Lorentz spin, which form a closed operator algebra.
The latter is a $W$-algebra that will be denoted by
 $W^{(c,2)}_{\boldsymbol 1}\equiv{ W}^{(c,2)}_{\bm{ k}}\big|_{\bm{ k}=(1,1)}$.

\bigskip
In sec.\,\ref{sec4}, the  linear subspace of
 chiral local fields of 
Lorentz spin $s$   commuting  with  the  screening charge
$\oint \rd u \,V_+$ was denoted by ${\cal W}^{(s)}$. In the case at hand
${\cal W}^{(2)}$ is a linear span of  three fields $G,\,\chi_2 S$ and $\chi_2\partial\chi_2$.
Out of these
\bea\label{aisisaisai}
T^{(e)}_{2}={{G}}+\tfrac{\ri}{2}\, \chi_2\partial \chi_2\ ,\ \ \ \ \ \ \ \ 
T^{(o)}_{2}=\chi_2 {S}+\ri \tau\, \chi_2\partial \chi_2
\eea
with
\bea
 \tau=-\frac{1}{\sqrt{2}}\ \big(\beta^{-1}-\beta\big)\ \frac{z_1+z_2}{z_1-z_2} 
 \eea
 satisfy the second requirement \eqref{aiisaisasia} involving the 
vertex operator
$V_-$. 
There is a shortcut way of checking the last statement.
 In the construction of the  densities  \eqref{aisisaisai}, one can 
alternatively start by considering the
space of local fields satisfying the condition \eqref{gasststa}, where
 $V_+$ is replaced by $V_-$.
Parameterizing $z_{1,2}$  via 
 the complex numbers $s$ and $\alpha$ as
\bea
z_1=\ri s\ \re^{-\ri \alpha}\ ,\ \ \ \ \  z_2=-\ri s\, \re^{\ri \alpha}
\eea
(if one adopts the convention \eqref{concj128921}, then $s^2=1$),
the vertices take the form
\bea
V_+=\sqrt{2}\,
 \chi_1\, \re^{+\ri\sqrt{2}\beta \varphi}\ ,\ \ \ \ \ \ \ \ \ \ \ \ \ \ \ \ \ V_-=\sqrt{2}\,s\,  \tilde{\chi}_2\, \re^{-\ri\sqrt{2}\beta \varphi}\,,
\eea
where
\bea
 \tilde{\chi}_2=\cos(\alpha)\,  \chi_2+\sin(\alpha)\,  \chi_1\ .
\eea
Then such local fields would be generated by the 
 $\frac{3}{2}$  and $\frac{1}{2}$ spin fields
\bea
 \tilde{S}&=&\big(\rho\partial+\ri \partial\varphi\,\big)\, \tilde{\chi}_2\ ,\ \ \ \  \ \ \ \ \ \ \ \ 
  \tilde{\chi}_1=\cos(\alpha)\,  \chi_1-\sin(\alpha)\,  \chi_2\ .
\eea
Again we introduce the $W$-algebra, $\widetilde{W}^{(c,2)}_{\boldsymbol 1}$, 
formed by the integer Lorentz spin fields
that commute with the screening charge $\oint \rd u \,V_-$. From the formal 
algebraic point of view, it is equivalent to $W^{(c,2)}_{\boldsymbol 1}$,
but the fields are expressed differently in terms of the current $\partial\varphi$
and the original  Majorana fermions $\chi_{1,2}$.
The two $W$-algebras are related through a so-called reflection operator.
It acts on the fields as
 \bea\label{oiasio98129812}
{\bf R}\, S\, {\bf R}^{-1}= {\tilde S}\ ,\ \ \ \ \ \ \
{\bf R}\,\chi_2 \,{\bf R}^{-1} = {\tilde\chi}_1\ .
\eea
The characteristic property of the densities entering into the local IM is that
they are invariant under the reflection, up to a total derivative:
 \bea\label{iaiaisiasias}
{\bf R}\,T_s\, {\bf R}^{-1}= T_s+\partial{\cal O}_{s-1}\ .
 \eea
Note that this immediately implies that the local IM
commute with the reflection operator:\footnote{%
The reflection operator is part of the commuting family 
and its eigenvalue already appeared in eq.\,\eqref{oia38923ss1A} as the subleading asymptotic
coefficient $R_{\boldsymbol {\mathfrak j}, \mathfrak{m}, A}$.}
\be
{\bf I}_s=\int_0^{2\pi}\frac{\rd u}{2\pi}\, T_{s+1}(u)\, :\ \ \ \ \ \  \qquad[\,{\bf I}_s,{\bf R}\,]=0\, .
\ee
It is easy to check that the fields \eqref{aisisaisai} possess the property \eqref{iaiaisiasias}.
\bigskip

Below we'll argue that the local IM
\bea\label{aoisi912898291}
{\bf I}^{(e)}_1&=&\int_0^{2\pi}\frac{\rd u}{2\pi}\, \big(\, {{G}}+\tfrac{\ri}{2}\,
 \chi_2\partial \chi_2\,\big)=\int_0^{2\pi}\frac{\rd u}{2\pi}\, \big(\, {\tilde{G}}+\tfrac{\ri}{2}\,
\tilde{ \chi}_1\partial \tilde{\chi}_1\,\big)
\\[0.3cm]
 {\bf I}^{(o)}_1&=&\int_0^{2\pi}\frac{\rd u}{2\pi}\ \Big( \chi_2 {S}+\ri\tau\, \chi_2\partial\chi_2\Big)
= \int_0^{2\pi}\frac{\rd u}{2\pi}\,
\Big(  \tilde{\chi}_1\tilde{S}+\ri\tau\, \tilde{\chi}_1\partial\tilde{\chi}_1\Big)
\nonumber
 \eea
are 
 linearly expressed in terms of ${\bf I}^{(1)}_1$ and  ${\bf I}^{(2)}_1$, whose eigenvalues 
are given  by eq.\,\eqref{oias8912891} with $r=2$, $k_1=k_2=1$. Namely,
 \bea\label{aiuususausa}
 {\bf I}^{(e)}_1&=&-\tfrac{1}{4\xi}\ \big({\bf I}^{(1)}_1+{\bf I}^{(2)}_1\big)\nonumber\\[0.2cm]
{\bf I}^{(o)}_1&=& \tfrac{1}{4\sqrt{2\xi(1+\xi)}}\, \big({\bf I}^{(1)}_1-{\bf I}^{(2)}_1\big)\ .
  \eea
 In fact the first relation follows immediately.  It was already mentioned (see eq.\eqref{iisisaias} and below) 
that its  r.h.s. coincides with 
  the local IM  ${\bf I}_1$  defined by \eqref{oi89129821}  with 
  the local density $T_2$ in  \eqref{oias81aaa212}. The latter, specialized to the case $r=2,\ k_1=k_2=1$,
  should be compared with the integrand in the first line of eq.\,\eqref{aoisi912898291},
taking into account that
  \bea\label{ssauasu}
  (\partial\theta)^2= G_{{\chi_1}}+{G}_{{\chi_2}}
\  , \ \ \ \ \ {\rm where}\ \ \ \ 
  G_{{\chi_a}}=
  \tfrac{\ri}{2}\,  \chi_{a}\partial \chi_{a}\ .
  \eea

  \bigskip
  One can proceed further and construct the higher spin densities $T_{s+1}$.
  It is easy to see that the linear subspace ${\cal W}^{(3)}\subset W^{(c,2)}_{\boldsymbol 1}$ includes, together with
  the derivatives of the fields   from ${\cal W}^{(2)}$,   a  single local  field  $\partial{ \chi}_2 S$. The latter 
  does not  satisfy the condition \eqref{iaiaisiasias}. The space ${\cal W}^{(4)}$, apart from total derivatives,
  contains  five composite fields
   $\partial S S$, $G^2$, $\chi_2\,GS$, $ \partial^2 \chi_2\, S$ and ${G}_{{\chi_2}} ^2$.
The first two of them, together with the spin $\frac{7}{2}$ field $GS$ are 
 defined through  the  regular terms in the OPE \eqref{ajasusausa}.
 The  spin 4 local field  ${G}_{{\chi_2}} ^2$ is 
the first regular term in the OPE ${G}_{{\chi_2}}(u){G}_{{\chi_2}}(v)$ as $u\to v$.
The explicit expressions in terms of the Bose field $\partial\varphi$ and
the Majorana fermions are quoted in Appendix \ref{hagf} (see \eqref{iaassususa}-\eqref{iaassususaf}).
Using these formulae and  the reflection condition \eqref{iaiaisiasias},
one obtains the local IM
  \bea\label{isisaias}
{\bf I}^{(e)}_3&=&\int_0^{2\pi}\frac{\rd u}{2\pi}\, \Big(
 G^2+2\ri\, \partial S S +6\, {G}_{{\chi_2}} G+\tfrac{3}{7}\, 
  \big(3-8\rho^2+24\tau^2\big)\,  G_{\chi_2}^2-6\tau\, 
   \partial^2 \chi_2\, S
  \Big)
\nonumber  \\[0.2cm]
{\bf I}^{(o)}_3&=&
\int_0^{2\pi}\frac{\rd u}{2\pi}\, \Big(
\chi_{2}\, GS -\tfrac{1}{2}\,\tau\, G^2+
\tau\,G_{\chi_2} G-
\tfrac{3}{14}\,\tau\,\big(8\rho^2-8\tau^2-7\big)\,G_{\chi_2}^2
\\[0.2cm]
&+&\big(\rho^2-\tau^2-\tfrac{3}{4}\big)\,\partial^2\chi_2\,S
\Big)\nonumber
 \eea
Again, the eigenvalues of certain linear combinations of these operators are expected to 
 coincide with the coefficients $I_{3}^{(a)}\ (a=1,2)$ from eq.\eqref{isaisusaua} specialized to $n=2$.
  
  \subsection{Irreps of the $W$-algebra}
The space   ${\cal H}_{\bm{\mathfrak{j}}, {\mathfrak m},P}$ was introduced in sec.\,\ref{sec6}
and conjectured to be a highest weight representation of the algebra of local fields
${ W}^{(c,r)}_{\bm{ k}}$. It is labeled by   $P$, which is the value of $\hat{a}_0$ -- the
zero mode momentum of $\varphi$ (see eq.\,\eqref{oias90121B}),
the set $\bm{\mathfrak{j}}=(\mathfrak{j}_1,\ldots,\mathfrak{j}_r)$ 
with $\mathfrak{j}_a=0,\frac{1}{2},\ldots,\frac{1}{2}\,k_a$
and the integer $\mathfrak{m}\sim\mathfrak{m}+2K$ such that
$$
\mathfrak{m}=-2\mathfrak{J},-2\mathfrak{J}+2,\ldots,2\mathfrak{J}-2,2\mathfrak{J}
\qquad\qquad\qquad \Big(\mathfrak{J}=\sum_{a=1}^r\mathfrak{j}_a\Big)\, .
$$ 
Moreover, the representations
 ${\cal H}_{\bm{\mathfrak{j}}, {\mathfrak m},P}$ and
 ${\cal H}_{\check{\bm{\mathfrak{j}}}, \check{{\mathfrak m}},P}$
with
$$
\check{\mathfrak{j}}_a=\tfrac{1}{2}\,k_a-\mathfrak{j}_a\,,\qquad\qquad
\check{\mathfrak{m}}=\mathfrak{m} - K\sim \mathfrak{m} + K\qquad\qquad\qquad
\Big(K=\sum_{a=1}^r k_a\Big)
$$
are equivalent. 
In the case when  $r=2$, $k_1=k_2=1$ we are left with four possibilities.
The irrep ${\cal H}_{\bm{\mathfrak{j}}, {\mathfrak m},P}$ corresponding to 
$\mathfrak{j}_1=\mathfrak{j}_2=\frac{1}{2}$ and $\mathfrak{m}=\pm 2$ 
is the same as the one with $\mathfrak{j}_1=\mathfrak{j}_2=\mathfrak{m}=0$.
As a shortcut, we'll denote it by ${\cal H}_{0,0}$.
Similarly the case with $\bm{\mathfrak{j}}=(\frac{1}{2},0)$ and $\mathfrak{m}=+ 1$
is equivalent to the one where 
$\bm{\mathfrak{j}}=(0,\frac{1}{2})$ and $\mathfrak{m}=- 1$. The corresponding
space will be denoted by ${\cal H}_{1,-1}$. The notation ${\cal H}_{1,+1}$
will stand for the irrep ${\cal H}_{\bm{\mathfrak{j}}, {\mathfrak m},P}$ with
$\bm{\mathfrak{j}}=(\frac{1}{2},0)$ and $\mathfrak{m}=-1$, which is the same
as $\bm{\mathfrak{j}}=(0,\frac{1}{2})$ and $\mathfrak{m}=+1$.
Finally ${\cal H}_{\bm{\mathfrak{j}}, {\mathfrak m},P}$ 
with $\mathfrak{j}_1=\mathfrak{j}_2=\frac{1}{2}$ and $\mathfrak{m}=0$
will be referred to as ${\cal H}_{2,0}$. 

\bigskip
Speaking in general, the  $W$-primary states are  the ones which are annihilated by  
$\int_0^{2\pi}\rd u \, {\cal O}(u)\,\re^{\ri n u}$ for any local field ${\cal O}\in { W}^{(c,r)}_{\bm{ k}}$ and any
positive integer $n=1,2,\ldots\ $.
They  can be chosen to be  eigenstates of the mutually commuting operators
 ${\bf I}^{(a)}_1\ (a=1,\ldots,r)$.
As follows from eq.\,\eqref{oias891298129821} the
 representations ${\cal H}_{0,0}$, ${\cal H}_{1,-1}$ and ${\cal H}_{1,+1}$ 
 all contain only one linearly independent state, which we denote by 
${\boldsymbol  e}_{0,0}$, ${\boldsymbol  e}_{1,-1}$ and ${\boldsymbol  e}_{1,+1}$,
respectively. 
As for ${\cal H}_{2,0}$, there are two 
   primary states ${\boldsymbol e}_{2,0}^{(\pm)}$.  
A summary of the notation, together with the corresponding eigenvalues of the local IM ${\bf I}_1^{(1)}$
and ${\bf I}_1^{(2)}$ obtained from the ODE prediction \eqref{oias8912891}
are provided in tab.\,\ref{tab1}.

\begin{table}
\centering
\scalebox{0.83}{
\begin{tabular}{|c|l|c|c|c|}
\hline
& & & & \\[-0.4cm]
{ Irrep} &\ \  \ \ \ \ \ \ \ \ $\bm{\mathfrak{j}},\mathfrak{m}$
& Primary states & $ -\tfrac{1}{4\xi}\ \big(I_1^{(1)}+I_1^{(2)}\big)$ &$\tfrac{1}{2}\, \big( I_1^{(2)}-I_1^{(1)}\big)$ \\[0.1cm]
\hline
& & & & \\[-0.3cm]
${\cal H}_{0,0}$ & 
$
\begin{array}{cc}
\bm{\mathfrak{j}}=(0,0)\,,&\!\!\!\mathfrak{m}=0 \\[0.2cm]
\bm{\mathfrak{j}}=(\frac{1}{2},\frac{1}{2})\,,&\mathfrak{m}=\pm 2 
\end{array}$  &
${\boldsymbol  e}_{0,0}$  &$P^2-\frac{1}{12}$ & $-\tfrac{1}{12}\,  \frac{z_1+z_2}{z_1-z_2}$ \\[0.5cm]
\hline
& & & & \\[-0.3cm]
${\cal H}_{1,-1}$ &
$
\begin{array}{cc}
\bm{\mathfrak{j}}=(\frac{1}{2},0)\,, & \mathfrak{m}=+1 \\[0.2cm]
\bm{\mathfrak{j}}=(0,\frac{1}{2})\, ,& \mathfrak{m}=-1 
\end{array}$ 
 &
${\boldsymbol  e}_{1,-1} $ &$P^2+\frac{1}{24}$   & $ \tfrac{1}{6}\,  \frac{z_1+z_2}{z_1-z_2}+ \sqrt{2\xi(1+\xi)}\, P$ \\[0.5cm]
\hline
& & & & \\[-0.3cm]
${\cal H}_{1,+1}$ &
$
\begin{array}{cc}
\bm{\mathfrak{j}}=(0,\frac{1}{2}) \,, & \mathfrak{m}=+1 \\[0.2cm]
\bm{\mathfrak{j}}=(\frac{1}{2},0)\, , & \mathfrak{m}=-1 
\end{array}$ 
 &
${\boldsymbol  e}_{1,+1}$ &$P^2+\frac{1}{24}$ & $ \tfrac{1}{6}\, \frac{z_1+z_2}{z_1-z_2}- 
 \sqrt{2\xi(1+\xi)}\, P$ \\[0.5cm]
\hline
& & & & \\[-0.4cm]
${\cal H}_{2,0}$ &
$
\begin{array}{cc}
\bm{\mathfrak{j}}=(\frac{1}{2},\frac{1}{2})\,, &
 \mathfrak{m}=0
\end{array}
$
&
$
\begin{array}{cc}
{\boldsymbol  e}_{2,0}^{(+)}
\\[0.3cm]
{\boldsymbol  e}_{2,0}^{(-)}
\end{array}
$
&
 $P^2+\frac{5}{12}$  & 
$
\begin{array}{cc}
\tfrac{11}{12}\, \frac{z_1+z_2}{z_1-z_2}
+\sqrt{8\xi(1+\xi)P^2+
\tfrac{ 4z_1z_2}{(z_1-z_2)^2}} \\[0.4cm]
\tfrac{11}{12}\, \frac{z_1+z_2}{z_1-z_2}
-\sqrt{{8\xi(1+\xi)P^2}+
\tfrac{ 4 z_1z_2}{(z_1-z_2)^2}}
\end{array}
$ \\[0.1cm]
\hline
\end{tabular}
}
\caption{\small
The algebra
${ W}^{(c,2)}_{\bm{ 1}}$ possesses four
inequivalent representations, whose corresponding
values of $\bm{\mathfrak{j}}=(\mathfrak{j}_1,\mathfrak{j}_2)$ and $\mathfrak{m}$
are listed in the second column. Given in the table are
 the eigenvalues of the symmetric
and anti-symmetric combinations of the local IM ${\bf I}_1^{(1)}$, ${\bf I}_1^{(2)}$ 
for the primary states from these representations. The eigenvalues were computed using
eq.\,\eqref{jhsasysay}, which itself comes from the analysis of the ODE \eqref{aois90121AA},
\eqref{aiasaisias}
specialized to the case $r=2$ and $k_1=k_2=1$. The realization of the primary states
in the bosonic and fermionic modules is described by eqs.\,\eqref{isisisasi} and \eqref{oias9812899812}.
\label{tab1}}
\end{table}

\bigskip

 The algebra $W^{(c,2)}_{\boldsymbol 1}$ was described in terms
of bosonic and fermionic fields. 
In turn, the irreps
${\cal H}_{0,0}$, ${\cal H}_{1,\pm1}$ and ${\cal H}_{2,0}$ can be identified
with certain subspaces of the   bosonic and fermionic modules.
Taking into account that the eigenvalues of $\hat{U}^2$ coincide with 
$ \re^{\frac{2\pi\ri}{K}{\mathfrak m}}$,
the quasiperiodicity condition  \eqref{oias09120912} for the vertices
\eqref{oias8912988921} implies that for the Majorana fermion fields
\be
\chi_a(u+2\pi)=(-1)^{\mathfrak{m}+1}\,\chi_a(u)\,.
\ee
 As a result, the spaces ${\cal H}_{0,0}$ and
${\cal H}_{2,0}$ would correspond to the anti-periodic boundary conditions, 
while for ${\cal H}_{1,\pm1}$ the Majorana fermions would be periodic fields.

 \subsubsection*{Neveu-Schwarz sector}
In this sector the fermion fields are antiperiodic,
so that their Fourier mode expansions are given by
\bea
 \chi_a=\sum_{\nu\in\frac{1}{2}+{\mathbb Z}} f^{(a)}_{\nu}\, \re^{-\ri \nu u}\ ,\ \ \ \ \ \ \ \ \ \ 
\qquad
\big\{f^{(a)}_{\nu}, f_{\mu}^{(b)}\big\}= \delta_{ab}\ \delta_{\nu+\mu,0}\ .
\eea
 The NS fermionic module (Fock space) ${\cal F}_{\rm NS}$ is generated  by the
action of the ``creation'' operators $f_{-\nu}^{(a)}$ 
on the NS vacuum, which itself is defined by the condition 
$f_{\nu}^{(a)}\,|0\rangle_{\rm NS}=0$  (here $\nu=\tfrac{1}{2},\tfrac{3}{2},\ldots$ and $a=1,2$).
 It is splitted  onto the eigenspaces ${\cal F}_{\rm NS}^{(+)}$ and
  ${\cal F}_{\rm NS}^{(-)}$ of the  fermion number operator
   \bea
 (-1)^{\tt F}=\exp\Big( \ri\pi\sum_{\nu>0} f^{{{(1)}}}_{{-\nu}}\,
 f_{{\vphantom{-}\nu}}^{{{(1)}}}+
 f^{{{(2)}}}_{{-\nu}}\,
 f_{{\vphantom{-}\nu}}^{{{(2)}}}
\Big)\ .
 \eea
Since all the fields of the $W$-algebra $W^{(c,2)}_{\boldsymbol 1}$
commute with $(-1)^{\tt F}$
and the bosonic zero mode momentum $\hat{a}_0$,
the following identifications can be made
 \bea\label{oiasio9821891289}
 {\cal H}_{0,0}={\cal F}_{\rm NS}^{(+)}\otimes{\cal F}_P\ ,\ \ \ \ \ \ \ {\cal H}_{2,0}={\cal F}_{\rm NS}^{(-)}\otimes{\cal F}_P\ .
  \eea
 The primary state
  ${\boldsymbol  e}_{0,0}$ coincides with $|0\rangle_{\rm NS}\otimes|P\rangle$, while 
 ${\boldsymbol  e}^{(\pm)}_{2,0}$  are certain linear combinations of
  $f^{\scriptscriptstyle{(1)}}_{\scriptscriptstyle{{-\frac{1}{2}}}}\,|0\rangle_{\rm NS}\otimes|P\rangle$ and
   $f^{\scriptscriptstyle{(2)}}_{\scriptscriptstyle{{-\frac{1}{2}}}}\,|0\rangle_{\rm NS}\otimes|P\rangle$
that make them  eigenstates of the local IM ${\bf I}^{(o)}_1$ \eqref{aoisi912898291}.

 \bigskip
 In terms of the oscillator modes ${\bf I}^{(e)}_1$  \eqref{aoisi912898291} reads as
 \bea\label{oiasd891221}
&&{\bf I}^{(e)}_1\,=\,-\tfrac{1}{12}+a_0^2+2\sum_{m=1}^\infty a_{-m}a_m+\sum_{\nu=\frac{1}{2},
\frac{3}{2},+\ldots}^\infty \nu\, \big(\, f^{(1)}_{-\nu}f^{(1)}_{\nu}+
f^{(2)}_{-\nu}f^{(2)}_{\nu}\,\big)\,,
\eea
while 
\bea
{\bf I}^{(o)}_1&=&-\tfrac{1}{24}\, \tau+ \sum_{\nu=\frac{1}{2},\frac{3}{2},\ldots}\nu\,
\big(\, 
2\tau\,
f^{(2)}_{-\nu}f^{(2)}_{\nu}
-\ri\rho\, (f^{(1)}_{-\nu}f^{(2)}_{\nu}+f^{(2)}_{-\nu}f^{(1)}_{\nu}) \big) \nonumber\\[0.2cm]
&-&
\sqrt{2}\, a_0b_0-\sqrt{2}\,  \sum_{n=1}^\infty \big(a_{-n}b_{n}+b_{-n}a_{n}\,\big)\,.
\eea
Here we use the notation 
\bea
&&b_n=-\tfrac{\ri}{\sqrt{2}} \sum_{\nu+\mu=n\atop
\nu,\mu\in{\mathbb Z}+\frac{1}{2}}f^{(1)}_{\mu}f^{(2)}_{\nu}\ \ \ \ \ \ \ \ \ (n=\pm1,\pm2,\ldots)\nonumber
\\[0.3cm]
&&b_0=-\tfrac{\ri}{\sqrt{2}}\ \sum_{\nu=\frac{1}{2},\frac{3}{2},\ldots}
\big(f^{(1)}_{-\nu}f^{(2)}_{\nu}-f^{(2)}_{-\nu}f^{(1)}_{\nu}\big)\ .
\eea
The latter coincide with the Fourier coefficients of the current 
 \bea
 \partial\theta=-\tfrac{\ri}{\sqrt{2}}\ \chi_1\chi_2=
 \sum_{n=-\infty}^\infty b_n\ \re^{-\ri n u}\ :\ \  \ \ \ \  [b_m,b_n]=\tfrac{m}{2}\ \delta_{m+n,0}\  ,
 \eea
where the bosonic field $\theta$ was introduced in \eqref{oisaoioias9012}.
Then  a simple calculation yields
 \bea\label{isisisasi}
 {\boldsymbol  e}_{0,0}&=&|0\rangle_{\rm NS}\otimes|P\rangle\ :\ \ \  I_1^{(e)}=P^2-\tfrac{1}{12}\ ,\ \ \ I_1^{(o)}=-\tfrac{1}{24}\, \tau\,,
 \\[0.4cm]
 {\boldsymbol  e}^{(\pm)}_{2,0}&=&\big(\alpha^{(\pm)}_1\,
 f^{{(1)}}_{\scriptscriptstyle{{-\frac{1}{2}}}}+
 \alpha_2^{(\pm)}\,
 f^{{(2)}}_{\scriptscriptstyle{{-\frac{1}{2}}}}\,\big)
 \,|0\rangle_{\rm NS}\otimes|P\rangle\ :\ \ \ \ I_1^{(e)}=P^2+\tfrac{5}{12} \ ,\ \ \ \ I_1^{(o)}=\tfrac{11}{24}\, \tau\mp
\tfrac{1}{2}\, g\nonumber
 \eea
with
\bea\label{gconstA}
\alpha_{1}^{(\pm)}=\tau\pm g\, ,\ \ \ \  \qquad
\alpha_{2}^{(\pm)}=\ri(\rho+2P)\ ,\ \ \ \qquad
g=\sqrt{4P^2-\rho^2+\tau^2}\ .
\eea

\subsubsection*{Ramond sector}
In the Ramond sector the Majorana fermions are periodic fields,
 so that
\bea
 \chi_a=\sum_{n\in{\mathbb Z}}f^{(a)}_n\, \re^{-\ri n u}\ ,\ \ \ \ \ \ \ \ \ \ 
\big\{f^{(a)}_n,f^{(b)}_{m}\big\}= \delta_{ab}\ \delta_{n+m,0}\ .
\eea
The Ramond vacuum states $|\pm\rangle_{\rm R}$ are annihilated by $f^{(a)}_n$ with $n>0$ and form a two 
dimensional representation for
the algebra of  zero modes
\bea
\big(f^{(1)}_{0}\big)^2=\big(f^{(2)}_{0}\big)^2=\tfrac{1}{2}\ ,\ \ \qquad
 \ \big\{f^{(1)}_{0},f^{(2)}_{0}\big\}=0\ .
\eea
Let  $\sigma^A$ $(A=x,y,z)$ be the Pauli matrices, such that $\sigma^z\,|\pm\rangle_{\rm R}=
\pm\,|\pm\rangle_{\rm R}$.
One can set
\bea
f_0^{(1)}=\tfrac{1}{\sqrt{2}}\ \sigma^y\ (-1)^{\tt F}\ ,\ \ \ \ f_0^{(2)}=\tfrac{1}{\sqrt{2}}\ \sigma^x\ (-1)^{\tt F}\ ,\ \ \ \
-\tfrac{\ri}{\sqrt{2}}\  f_0^{(1)}f_0^{(2)}=\tfrac{1}{\sqrt{8}}\ \sigma^z
\eea
with
 \bea
 (-1)^{\tt F}= \sigma^z\,\exp\Big( \ri\pi\sum_{n>0} f^{{{(1)}}}_{{-n}}\,
 f_{{\vphantom{-}n}}^{{{(1)}}}+
 f^{{{(2)}}}_{{-n}}\,
 f_{{\vphantom{-}n}}^{{{(2)}}}
\Big)\, .
 \eea
  The Ramond  fermionic module ${\cal F}_{\rm R}$, generated by the 
action of the creation operators $f_{-n}^{(a)}$   $(n>0)$ on the vacua,
can be  splitted  into ${\cal F}_{\rm R}^{(+)}$ and
  ${\cal F}_{\rm R}^{(-)}$ according to the value of $(-1)^{\tt F}$.
   The  fields from $W^{(c,2)}_{\boldsymbol 1}$ commute with
 $(-1)^{\tt F}$, which allows one to identify
 \bea\label{ioas98128912}
 {\cal H}_{1,\pm1}={\cal F}_{\rm R}^{(\pm)}\otimes{\cal F}_P\ .
  \eea
 In turn, for the primary states
 \bea
 {\boldsymbol  e}_{1,\pm 1}=|\pm\rangle_{\rm R}\otimes | P\,\rangle \, .
  \eea
The local IM   \eqref{aoisi912898291}
  in the Ramond sector, written in terms of the creation/annihilation operators, read as
  \bea
{\bf I}^{(e)}_{1}&=&-\tfrac{1}{12}+a_0^2+2\sum_{n=1}^\infty a_{-n}a_n+
\tfrac{1}{8}+\sum_{n=1}^\infty n\, \big(\,f^{(1)}_{-n}f^{(1)}_{\vphantom{-}n}+
f^{(2)}_{-n}f^{(2)}_{\vphantom{-}n}\,\big)
\eea
and
\bea
{\bf I}^{(o)}_1&=&\tfrac{1}{12}\, \tau+\sum_{n=1}^\infty n\,
\big(\,2\tau\,
 f^{(2)}_{-n}f^{(2)}_{\vphantom{-}n} -
\ri\rho\,(f^{(1)}_{-n}f^{(2)}_{\vphantom{-}n}+f^{(2)}_{-n}f^{(1)}_{\vphantom{-}n}) \big) 
\nonumber \\[0.2cm]
&-&
\sqrt{2}\, a_0b_0- \sqrt{2}\, \sum_{n=1}^\infty \big(a_{-n}b_{n}+b_{-n}a_{n}\big)\, .
\eea
 Now
\bea
&&b_n=-\tfrac{\ri}{\sqrt{2}} \sum_{m+s=n\atop
m,l\in {\mathbb Z}}f^{(1)}_{m}f^{(2)}_{s}\ \ \ \ \ \ \ \ \ \ \ (n=\pm1,\pm2,\ldots)\\[0.3cm]
&&b_0=-\tfrac{\ri}{\sqrt{2}}\  \Big(\,f^{(1)}_{0}f^{(2)}_{0}+
\sum_{m=1}^\infty \big(\,f^{(1)}_{-m}f^{(2)}_{\vphantom{-}m}-f^{(2)}_{-m}f^{(1)}_{\vphantom{-}m}\,\big)\,\Big)\ .\nonumber
\eea
 With these expressions it is easy to see that
 \bea\label{oias9812899812}
 {\boldsymbol  e}_{1,\pm 1}=|\pm\rangle_{\rm R}\otimes|P\rangle\ :\ \ \ \qquad
 I_1^{(e)}=P^2+\tfrac{1}{24}\ ,\ \ \ \qquad I_1^{(o)}=
 \tfrac{1}{12}\,\tau\mp\tfrac{1}{2} P\ .
 \eea
 The relations \eqref{aiuususausa} between the local IM 
$ {\bf I}_1^{(e)}$,\,${\bf I}_1^{(o)}$ and  $ {\bf I}_1^{(1)},\,{\bf I}_1^{(2)}$
come from a comparison of
 the eigenvalues  quoted in the above formula and eq.\eqref{isisisasi}
with the last two columns of tab.\ref{tab1}. 
One needs to also take into account that
$\tau=-\rho\, \frac{z_1+z_2}{z_1-z_2}$ and
$\rho=\frac{1}{\sqrt{2\xi(1+\xi)}}$.
For completeness, 
we present in tab.\ref{tab2}
the eigenvalues of the operators
${\bf I}_3^{(e)}$ and ${\bf I}_3^{(o)}$  \eqref{isisaias}
for the primary states.

\begin{table}
\centering
\scalebox{0.82}{
\begin{tabular}{|c|l|l|}
\hline
& & \\[-0.3cm]
 { Primary state } &\qquad\qquad
\qquad \qquad $I_3^{(e)}$  & 
\qquad\qquad\qquad\qquad \ $-\frac{2}{\tau}\,I_3^{(o)}$ \\[0.2cm]
\hline
 & & \\[-0.3cm]
$ {\boldsymbol  e}_{0,0}$ & $P^4-\frac{1}{2}\,P^2+\frac{7}{160}\,\tau^2-\frac{1}{30}\,\rho^2+\frac{3}{80}$   &
$P^4-\frac{1}{4}\,P^2-\frac{7}{480}\,\tau^2+\frac{1}{96}\,\rho^2$  \\[0.2cm]
\hline
& & \\[-0.2cm]
 ${\boldsymbol  e}_{1,\pm 1}$ &
$P^4+\frac{1}{4}\,P^2-\frac{1}{20}\,\tau^2+\frac{7}{240}\,\rho^2-\frac{3}{320}$
 & 
$
P^4-\frac{1}{4}\,P^2+\frac{1}{60}\,\tau^2-\frac{1}{48}\,\rho^2
+\frac{1}{64}\mp \frac{P}{\tau}\,\big(\frac{1}{8}-P^2\big) 
 $ \\[0.3cm]
\hline
& & \\[-0.2cm]
$ {\boldsymbol  e}^{(\pm)}_{2,0}$ &
$
P^4+\frac{5}{2}P^2+\frac{127}{160}\,\tau^2-\frac{8}{15}\,\rho^2+\frac{3}{80}\mp \frac{3}{4}\,\tau g
$
& 
$
 P^4-\frac{1}{4}\,P^2-\frac{127}{480}\,\tau^2+\frac{25}{96}\,\rho^2-\frac{1}{8}
\pm\frac{1}{4\tau}\big(1+g^2\big)\,g
$
\\[0.3cm]
\hline
\end{tabular}
}
\caption{\small
The eigenvalues of the local integrals of motion ${\bf I}_3^{(e)}$ and ${\bf I}_3^{(o)}$ 
for the primary states of the $W^{(c,2)}_{\boldsymbol 1}$ algebra. These were computed
directly using the expression \eqref{isisaias} for the local IM as well as the formulae 
\eqref{isisisasi},\,\eqref{gconstA}
for the states $ {\boldsymbol  e}_{0,0}$, $ {\boldsymbol  e}_{2,0}^{(\pm)}$ and 
 \eqref{oias9812899812}
for $ {\boldsymbol  e}_{1,\pm 1}$. 
Here we use the notation $g=\sqrt{4P^2-\rho^2+\tau^2}$.
\label{tab2}}
\end{table}

\bigskip

The identifications
\be
 {\cal H}_{0,0}={\cal F}_{\rm NS}^{(+)}\otimes{\cal F}_P\ ,\qquad\qquad
{\cal H}_{2,0}={\cal F}_{\rm NS}^{(-)}\otimes{\cal F}_P\,, \qquad \qquad
 {\cal H}_{1,\pm1}={\cal F}_{\rm R}^{(\pm)}\otimes{\cal F}_P
\ee
immediately yield formulae for the characters
\bea
{\rm{ch}}_{A,B}({\tt q})={\rm{Tr}}_{{\cal H}_{A,B}}\big({\tt q}^{{\bf I}_1^{(e)}}\big)\ .
\eea
Namely
\bea
{\rm{ch}}_{0,0}({\tt q})\pm{\rm{ch}}_{2,0}({\tt q})={\tt q}^{P^2-\frac{1}{12}}\ 
\frac{\prod_{\nu=\frac{1}{2},\frac{3}{2},\ldots}^\infty\big(1\pm{\tt q}^\nu)^2}{\prod_{n=1}^\infty(1-{\tt q}^n)}\ ,
\eea
while
\bea
{\rm{ch}}_{1,\pm 1}({\tt q})={\tt q}^{P^2+\frac{1}{24}}\,\prod_{n=1}^\infty\frac{(1+{\tt q}^n)^2}{1-{\tt q}^n}\ .
\eea 
The characters are the generating functions for the dimensions of the level subspaces.
In view of the ODE/IQFT correspondence discussed in sec.\,\ref{sec7},
these formulae provide highly nontrivial predictions
 concerning the number of  solutions of the algebraic system
\eqref{aiasiisasia}-\eqref{aois901828912}
 on the apparent singularities of the ODE for $r=2$ and $k_1=k_2=1$.

\subsection{Dual nonlocal IM\label{sec93}}
In sec.\,\ref{sec83}, we
 briefly mentioned a certain ``duality transformation'' 
 and traced its appearance at the level of the ODE.
On the quantum field theory side, the simplest instance where it
was originally observed was in the  work of Schmid \cite{Schmid1} on the instanton calculus
in dissipative quantum mechanics. Around ten years later, the duality was
used in the computation of the conductance in a fractional quantum Hall system \cite{Fendley}
and also appeared in the study of the quantum KdV theory \cite{Bazhanov:1996dr}. The latter
is the special case of the generalized affine Gaudin model with $r=k=1$.
For $r=2$ and $k_1=k_2=1$ the duality is manifest in essentially the same way,
as will be discussed below.\footnote{%
Some hints of the appearance of this duality were observed in the study
of the ${\cal Z}_2$ invariant inhomogeneous six-vertex model in ref.\cite{IJS2}.
The latter corresponds to the case $z_1=-z_2=\ri$.
The relation between the GAGM and the lattice system
is discussed in sec.\ref{sec12} of this work. 
}
\bigskip

An examination of the explicit formulae for the local IM \eqref{aoisi912898291} and \eqref{isisaias}
shows that they remain unchanged under the transformation
\bea\label{duality11a}
\varphi\mapsto -\varphi\ ,\ \ \ \ \chi_1\mapsto-\chi_1\ ,\ \  \ \chi_2\mapsto\chi_2
\eea
so long as one simultaneously swaps the parameters
\bea\label{duality11b}
\beta\mapsto\beta^{-1}\ ,\ \ \ \  z_a\mapsto z_a^{-1}\ .
\eea
The defining property of the corresponding local densities $T_{2n}^{(a)}$ is that they satisfy  the OPEs
of the form
\bea
T_{2n}^{(a)}
(u)\,V_\pm(v)=\sum_{m=2}^{2n+1}\frac{{R}^{(\pm)}_{-m}(v)}{(u-v)^m}+\frac{\partial {\cal O}_{\pm}(v)}{u-v}+
O(1)
\eea
with the vertices from eq.\,\eqref{oias8912988921}. The invariance of the local IM
under the ``duality'' transformation  immediately implies that
the similar OPEs hold true with $V_\pm$ replaced by the ``dual'' vertices
\bea\label{hasgssasssy}
 \widetilde{V}_+&=&-\sqrt{2}\ \chi_1\, \re^{-\frac{\ri\sqrt{2}}{\beta} \varphi}\\[0.2cm]
\widetilde{V}_-&=&-\tfrac{1}{\sqrt{2}}\ \big(\,(z^{-1}_1+z^{-1}_2)\, \chi_1+\ri\,\,(z^{-1}_1-z^{-1}_2)\, \chi_2\, \big)
\ \re^{+\frac{\ri\sqrt{2}}{\beta} \varphi}\ ,\nonumber
 \eea
which are obtained from the original ones via the substitutions \eqref{duality11a},\,\eqref{duality11b}.

\bigskip

Introduce the operators
\be\label{contourInt1r}
\widetilde{x}_0 = \frac{1}{\widetilde{q}-\widetilde{q}^{-1}}\, \int_0^{2\pi}{\rm d}u\, \widetilde{V}_+(u) \ ,\qquad 
\widetilde{x}_1 = \frac{1}{\widetilde{q}-\widetilde{q}^{-1}}\,\int_0^{2\pi}{\rm d}u\, \widetilde{V}_-(u) \ .
\ee
These would satisfy the Serre relations \eqref{Serre1}, but with $q$ replaced by
\bea
\widetilde{q}=-\re^{-\frac{\ri\pi}{2}(\beta^{-2}-1)}\, .
\eea
A realization of the Borel subalgebra $U_{\widetilde{q}}(\widehat{{\mathfrak b}}_-)$ is
provided by $\widetilde{x}_0$, $\widetilde{x}_1$, as well as $\widetilde{h}_0$,
which is defined such that
\bea\label{oias90121AN}
\tilde{q}^{\tilde{h}_0}=\re^{-\ri\pi\sqrt{2}\,\beta^{-1}\hat{a}_0} \ \hat{U}\equiv \widetilde{\Omega}^{\frac{1}{2}}
\eea
with  $\hat{U}$ from eq.\eqref{oias90121B}
 (${U}^2=+1$ and ${U}^2=-1$ in the Neveu-Schwarz and Ramond sector,
respectively). Consequently, one can construct the dual operator
$\widetilde{{\mathlarger{\mathlarger{\mathlarger {\boldsymbol  a}}}}}_\pm(\widetilde{\lambda})$
from the vertices $\widetilde{V}_\pm$ and $\widetilde{\Omega}^{\frac{1}{2}}$
via the formulae, which are only notationally different to \eqref{soso1a} and \eqref{Lop2a}.
Notice that the transformation \eqref{duality11a} flips the sign of $\varphi$ and,
  therefore, its zero mode momentum $\hat{a}_0$. As a result,
the operators $\widetilde{{\mathlarger{\mathlarger{\mathlarger { \boldsymbol a}}}}}_\pm$
would be given by the trace over a representation $\widetilde{\rho}_\pm$ 
of the $\widetilde{q}$ - oscillator algebra,
  \be\label{qoschtsr}
[\widetilde{\Hcal},\widetilde{\Ecal}_\pm]=\pm2\,\widetilde{\Ecal}_\pm\,,\qquad \qquad\widetilde{q}\, \widetilde{\Ecal}_+
\widetilde{\Ecal}_--
\widetilde{q}^{\,-1}\,\widetilde{\Ecal}_-\widetilde{\Ecal}_+=\frac{1}{\widetilde{q}-{\widetilde{q}}^{\,-1}} \ ,
\ee
subject to the requirement
\be
{{\rm Tr}}_{\widetilde{\rho}_\pm}\big[
\re^{\mp 2\ri\pi\beta^{-1} P\widetilde{\cal H}}\big]\ne 0,\, \infty \qquad {\rm with} \qquad {\Im} m(P)<0\, .
\ee
A comparison with the similar  condition \eqref{tr1} for the representation ${\rho}_\pm$,
appearing in the construction of ${\mathlarger{\mathlarger{\mathlarger {\boldsymbol  a}}}}_\pm$,
suggests that under the duality transformation  
${\rho}_\pm\mapsto \widetilde{\rho}_\mp$.
\bigskip

 The  operator $\widetilde{{\mathlarger{\mathlarger{\mathlarger {\boldsymbol a}}}}}_\pm(\widetilde{\lambda})$ possesses
the formal power series expansion in the ``dual'' spectral parameter:
\be
\log\widetilde{{\mathlarger{\mathlarger{\mathlarger {\boldsymbol  a}}}}}_\pm(\widetilde{\lambda})=
-\sum_{n=1}^\infty \widetilde{{\bf H}}_n^{(\pm)}\ \big(\,\widetilde{\lambda}^{2}\big)^n\ .
\ee
Based on the fact that the residue in the OPE 
$T_{2n}^{(a)}(u)\,\widetilde{V}_\pm(v)$ is a total derivative,
it is possible to argue, following the lines of \cite{Bazhanov:1998dq}, that 
the operators $\widetilde{{\bf H}}_n^{(\pm)}$ commute with the local IM.
They are referred to as the dual nonlocal IM.
\bigskip

Similar as in the case of the quantum KdV, 
one expects the eigenvalues of the dual nonlocal IM
to appear in the large $\mu$ asymptotics  
of the connection coefficients. 
For instance, the
formal series $X(\mu)$, which enters into the asymptotic formula \eqref{oia38923ss1AB}
for $D_{\boldsymbol {\mathfrak j},+\mathfrak{m},+ A}(\mu)$,  coincides with
the eigenvalue of $\widetilde{{{\mathlarger{\mathlarger{\mathlarger {\boldsymbol  a}}}}}}_-(\widetilde{\lambda})$,
provided that the expansion parameters $\mu$ and $\widetilde{\lambda}$ are properly related.
This way, formula \eqref{oia38923ss1AB} may be promoted to the operator relation:
\be\label{aiisisaisa}
{{\mathlarger{\mathlarger{\mathlarger {\boldsymbol  a}}}}}_+({\lambda})\,\asymp\,{\bf R}\,
\mu^{-\frac{\sqrt{2}}{\beta}\,\hat{a}_0}\ 
\exp\Big(\mu^{\frac{1}{1-\beta^2}}\ q_{-1}\Big)\ {\bf B}(\mu)\,
\widetilde{{\mathlarger{\mathlarger{\mathlarger {\boldsymbol  a}}}}}_-(\widetilde{\lambda})\qquad\qquad\qquad
(\mu\to+\infty)\, .
\ee
Here ${\bf R}$ is the reflection operator \eqref{oiasio98129812}, whose eigenvalues coincide with
$R_{\boldsymbol {\mathfrak j}, \mathfrak{m}, A}$ \eqref{usaussuasa},
while ${\bf B}(\mu)$, with eigenvalue
$B(\mu)$ \eqref{isusuas}, involves only the local integrals of motion.
 Recall that 
$\mu\propto\lambda^2$  according to eq.\eqref{siisaias11a} which, specialized
to the case at hand, reads as
\bea\label{asususuaas}
\mu=-\lambda^2\ \Gamma^2\big(\tfrac{1}{2}\,(1-\beta^2)\big)\,
\big(\tfrac{2}{1-\beta^2}\big)^{1-\beta^2}
\ .
\eea
The relation between $\widetilde{\lambda}$   and $\lambda$ can be 
 established in the following way. Introduce  $\widetilde{\mu}$ through the formula
\bea\label{asususu}
\widetilde{\mu}=-\widetilde{\lambda}^2\ \Gamma^2\big(\tfrac{1}{2}\,(1-\beta^{-2})\big)\,
\big(\tfrac{2}{1-\beta^{-2}}\big)^{1-\beta^{-2}}
\ ,
\eea
which is just the ``dual'' version of \eqref{asususuaas}.
On the other hand, the  analysis of the ODE performed in  sec.\,\ref{sec83} 
suggests that $\widetilde{\mu}=\mu^{-\beta^{-2}}$ (see \eqref{suusauasu}).
Combining this with the above equations one finds
  \bea\label{aysysa}
\widetilde{\lambda}^2=\big(\beta^2\big)^{\beta^{-2}-1}\, 
\Big( \Gamma\big(\tfrac{1}{2}\,(1-\beta^{-2})\big)\Big)^{-2}\,
\Big({\lambda}^2\, \Gamma^2\big(\tfrac{1}{2}\,(1-\beta^2)\big)\Big)^{-\frac{1}{\beta^2}}\ .
\eea
It should be pointed out 
that the relation $\widetilde{\mu}=\mu^{-\beta^{-2}}$ assumes
the convention \eqref{concj128921} so that  \eqref{aysysa}  is applicable for   $z_1z_2=1$.
Otherwise it would require a simple modification.

 \bigskip

 If  the eigenvalues of the coefficients in the Taylor  series expansion
\be\label{asoioi128921}
\log{{\mathlarger{\mathlarger{\mathlarger {\boldsymbol  a}}}}}_\pm({\lambda})=
-\sum_{n=1}^\infty\, {\bf H}_n^{(\pm)}\ \big({\lambda}^{2}\big)^n
\ee
are known in analytical form, then the duality relation allows one to 
determine the eigenvalues of the dual non-local IM $\widetilde{{\bf H}}_n^{(\mp)}$.
For instance, it follows from formula \eqref{oias90121121} that
the eigenvalue of ${\bf H}_1^{(+)}$ for the singlet primary states 
${\boldsymbol  e}_{0,0}$, ${\boldsymbol  e}_{1,-1}$ and ${\boldsymbol  e}_{1,+1}$
are given by
\be\label{oais89czA}
H_1^{(+)}({\boldsymbol  e}_{0,0})=
H^{(1)}\big(\sqrt{2}\,\beta P,\beta^2\,|\, Z\big)\ ,\ \ \ \ \ \ 
H_1^{(+)}({\boldsymbol  e}_{1,\pm 1})=
H^{(2)}\big(\sqrt{2}\,\beta P,\beta^2\,|\, Z^{\mp 1}\big)\ ,
\ee
where $Z=z_1=z_2^{-1}$ and
\bea\label{oais89czB}
H^{(1)}(p,g\,|\, Z)&=&-\,
 \frac{\pi\Gamma\big(-g\big)}{\cos(\frac{\pi g}{2})}\ 
\frac{\Gamma\big(\frac{1}{2}+ \frac{g}{2}+p\big)}
 {\Gamma\big(\frac{1}{2}-\tfrac{g}{2}+p\big)}\ \big(Z+Z^{-1}\big)\\[0.3cm]
 H^{(2)}(p,g\,|\, Z)&=& -\,
 \frac{\pi\Gamma\big(-g\big)}{\cos(\frac{\pi g}{2})}\ 
\frac{\Gamma\big( \frac{g}{2}+p\big)}
 {\Gamma\big(1-\frac{g}{2}+p\big)}\ \Big(\, p\, \big(Z+Z^{-1}\big)+\tfrac{g}{2}\ \big(Z-Z^{-1}\big)\,\Big)\ .\nonumber
 \eea
In turn, the eigenvalues of the corresponding dual non-local IM read as
\bea
\widetilde{H}_1^{(-)}({\boldsymbol  e}_{0,0})
=H^{(1)}\big(\sqrt{2}\,\beta^{-1} P,\beta^{-2}\,|\, Z\big)\ ,\ \ \  \ \ \
\ \ \   \widetilde{H}_1^{(-)}({\boldsymbol  e}_{1,\pm 1})
=H^{(2)}\big(\sqrt{2}\,\beta^{-1} P,\beta^{-2}\,|\, Z^{\pm1}\big)\ .
\nonumber
\eea

  \section{ $W$-algebra and   local IM ${\bf I}_{1}^{(a)}$ for $k_a=1$ \label{sec10}}
 For further illustration, we describe the lowest spin local fields for the 
$W_{\bm{k}}^{(c,r)}$ algebra with $\bm{k}=(1,1,\ldots,1)$ 
and arbitrary $r$. In turn,  explicit formulae are derived for the densities
$T_2^{(a)}$ corresponding to the local IM ${\bf I}_{1}^{(a)}$.
 \subsection{$W$-currents of Lorentz spin 2 and 3 }

 When all $k_a=1$, the vertex $V_+$  defined by eqs.\,\eqref{isaiiasias} and 
\eqref{iuaususa}      takes the form
 \bea
 V_+=
\sum_{a=1}^r\re^{2\ri \phi_a}\ \re^{\frac{2\ri(\beta-1)}{\sqrt{r}}\varphi}\ .
 \eea
Here $\{\phi_a\}_{a=1}^r$  is a set of independent chiral Bose fields subject to the OPE 
\bea\label{hsaysysa}
\phi_a(u) \phi_b(v)=-\tfrac{1}{2}\  \delta_{ab}\ \log(u-v)+O(1)\,,
\eea
while
\bea
\varphi=\frac{1}{\sqrt{r}}\,\sum_{a=1}^r\phi_a\ .
\eea
In what follows ${\cal O}_{ab}$ will stand  for  the exponential  operators
\bea
{\cal O}_{ab}&\equiv &\re^{2\ri(\phi_a-\phi_b)}\ \ \ \ \ \ \ \ \ (a\not=b)\, ,
\eea
which are local fields of Lorentz spin $2$.

\bigskip
Consider  the $r$ fields
\bea\label{hasyasy}
X_b=\beta^{-1}\,
R_{b}-(\beta^{-1}-\beta)\, \sum_{a=1\atop a\not=b}^r{\cal O}_{ab}\ \ \ \  \qquad \ \ \ \ (b=1,\ldots,r)
\eea
with
\bea\label{iasissau}
R_b&\equiv&
r\, (\partial\phi_b)^2
-2\sqrt{r}\, (1-\beta)\ \partial\varphi\,\partial\phi_b
+(1-\beta)^2\ (\partial\varphi)^2
\nonumber\\[0.3cm]
&+&\ri\, (1-\beta^2)\, \partial^2\phi_b-\frac{\ri}{ \sqrt{r}}\ (1-\beta)^2(1+\beta)\,\partial^2\varphi\,,
\eea
as well as the $\frac{1}{2}\, (r-1)\,r$ fields
\bea\label{hasyasyss}
Y_{ab}=Y_{ba}={\cal O}_{ab}+{\cal O}_{ba}
-\big(\partial\phi_a-\partial\phi_b\big)^2\ \ \ \ \ \ (a\not=b)\ .
\eea
One can check that 
  they commute with the screening charge $\oint\rd u\,V_+(u)$.
In other words, $X_a$, $Y_{ab}$   belong
 to the space ${\cal W}^{(2)}$.
We argue  that any spin two $W$-current can be expressed as  a linear combination
of the  fields $X_a$ and $Y_{ab}$. This way they
form a basis in ${\cal W}^{(2)}$.
\bigskip

The last statement
 is supported by the observation that the 
 OPE of $X_a$ and $Y_{ab}$ generates no spin 2
fields, which are linearly independent from the basic ones.
For example, a straightforward calculation yields the following result
\be\label{ksaisiasi}
X_{a}(u)X_{b}(v)=\frac{C_{ab}}{(u-v)^4}+
\frac{2X_{ab}(v)}{(u-v)^2}+\frac{1}{u-v}\, \Big(\,\partial X_{ab}(v)+\big(\beta^{-1}-\beta\big)\,Z_{ab}(v)\,\Big)+O(1)
\ee
where
\bea
X_{ab}&=&-\frac{r}{\beta}\ X_a\ \delta_{ab}+ (\beta^{-1}-\beta)\ \tfrac{1}{2}\,\big( X_a+X_b\big)\\[0.3cm]
&+&
(\beta^{-1}-\beta)\big(\beta^{-1}r+2\beta^{-1}-2\beta\big)\ \tfrac{1}{2}\ Y_{ab}
 \nonumber
 \ (1-\delta_{ab})
\eea
and $C_{ab}$ are some constants whose explicit form is unessential.
This  shows that the spin 2 fields generated in the OPE  $X_{a}(u)X_{b}(v)$
are linearly expressed in terms of the basic ones. The less singular term in \eqref{ksaisiasi} involves
the spin 3 local fields
\bea
Z_{ab}\ :\ \ \ \ \ \ \ Z_{ab}=-Z_{ba}\ \qquad\qquad (a\ne b)\, .
\eea
They also commute with $\oint\rd u\,V_+(u)$.
The explicit formula for $Z_{ab}$ reads as
\bea
Z_{ab}&=&(\beta^{-1}-\beta)\, \Big[\, \tfrac{1}{2}\ \partial\, \Big(\displaystyle{\sum_{c=1\atop
c\not=a}^r}{\cal O}_{ca}-\sum_{c=1\atop
c\not=b}^r{\cal O}_{cb}\,\Big)+
2\ri\,  \big(\partial\phi_a-\partial\phi_b\big)\, \Big(\displaystyle{\sum_{c=1\atop
c\not=a}^r}
{\cal O}_{ca}+\sum_{c=1\atop
c\not=b}^r
{\cal O}_{cb}\Big)\,\Big]\nonumber\\[0.2cm]
&+&\,2\ri r\beta^{-1}\ \big(\partial\phi_a{\cal O}_{ab}-\partial\phi_b{\cal O}_{ba}\big)-2\ri\sqrt{r}\beta^{-1}\ (1-\beta) \ \partial\varphi\, \big({\cal O}_{ab}-{\cal O}_{ba}\big)
\\[0.6cm]
&-&\,
\tfrac{1}{2} \big(\beta^{-1} r+2\beta^{-1}-2\beta\big)
 \partial({\cal O}_{ab}-{\cal O}_{ba})+(\beta^{-1}-\beta)
 \big(\,\tfrac{4\ri }{3}\, (\partial\phi_a-\partial\phi_b)^3
+\tfrac{\ri }{6}\, (\partial^3\phi_a-\partial^3\phi_b)\,\big)\nonumber
\\[0.3cm]
&+&r\beta^{-1}\big(\partial^2\phi_a\partial\phi_b-\partial^2\phi_b\partial\phi_a\big)
+\sqrt{r}\beta^{-1} (1-\beta) 
\big(\partial^2\varphi\ (\partial\phi_a-\partial\phi_b)-
(\partial^2\phi_a-\partial^2\phi_b)\,\partial\varphi\big)\,.\nonumber
\eea
\bigskip

The singular part of the OPE of $X_{a}$ and  $Y_{bc}$ is given by
\bea\label{ksaisiasidd}
X_{a}(u)Y_{bc}(v)&=&-\frac{\delta_{ab}+\delta_{ac}}{2 (u-v)^4}\ \big(\beta^{-1} r+2\beta^{-1}-2\beta\big)
+
\frac{2X_{a|bc}(v)}{(u-v)^2}\\[0.3cm]
&+&\frac{1}{u-v}\, \Big(\,\partial X_{a|bc}(v)
+
\big(\delta_{ac}-\delta_{ab}\big)\ Z_{bc}(v)\, \Big)+O(1)\ \ \ \ \ \ (n\not=j)\nonumber
\eea
with
\bea
2X_{a|bc}&=&\delta_{ac}\, (X_a- X_b)+\delta_{ab}\,
 (X_a-X_c)\\[0.3cm]
 &-&(\beta^{-1}r+2\beta^{-1}-2\beta\big)\ (\delta_{ab}+\delta_{ac})\ Y_{bc}\ .\nonumber
\eea
All the fields appearing   in the  singular parts of this OPE are linearly expressed in terms of the basic spin 
2 fields $X_a,\,Y_{ab}$
as well as the spin 3 fields $Z_{ab}$.

\bigskip
Finally one should consider the OPE involving the $Y$-fields only.
Its singular part possesses  the following structure
\be
Y_{ab}(u)Y_{cd}(v)=\frac{Y_{ab|cd}^{(4)}}{(u-v)^4}+
\frac{Y_{ab|cd}^{(2)}(v)}{(u-v)^2}+\frac{1}{u-v}\ \Big(\,\frac{1}{2}\ \partial Y_{ab|cd}^{(2)}(v)+
 Y_{ab|cd}^{(1)}(v)\,\Big)+O(1)
\ee
where $a\not=b,\ c\not=d$.
A calculation results in the explicit formulae
\bea
Y_{ab|cd}^{(4)}&=&\tfrac{1}{2}\ 
(\delta_{ac} +\delta_{ad} 
+\delta_{bc} 
+\delta_{bd}+      6\,   \delta_{ac} \delta_{bd}+      6\,      \delta_{ad} \delta_{bc}     )
\\[0.3cm]
Y^{(2)}_{ab|cd}&=&\big(\delta_{ac}+\delta_{bd}+\delta_{bc}+\delta_{bd}+
2\,\delta_{ac}\delta_{bd}+2\,\delta_{ad}\delta_{bc}\big)\ (Y_{ab}+Y_{cd})\nonumber\\[0.3cm]
&-&\big(1-\delta_{bc}\big)\,\delta_{ad}\ Y_{bc}
-\big(1-\delta_{ad}\big)\,\delta_{bc}\ Y_{ad}
-\big(1-\delta_{ac}\big)\,\delta_{bd}\ Y_{ac}
-\big(1-\delta_{bd}\big)\,\delta_{ac}\ Y_{bd}\ ,\nonumber
\eea
while
\bea
Y^{(1)}_{ab|cd}=\delta_{ac} \, {Z}_{bdc}+
\delta_{ad} \, {Z}_{bcd}+
\delta_{bc} \, {Z}_{adc} +\delta_{bd} \, {Z}_{acd}
\eea
with
\bea
{ Z}_{bdc}&=&\partial{\cal O}_{cd}-\partial {\cal O}_{cb}
+\ri\, (2\, \partial\phi_c-\partial\phi_b-\partial\phi_d)\, \big({\cal O}_{bd}-{\cal O}_{db}\big)
\\[0.3cm]
&+&
\ri\, 
\big(2\,\partial\phi_d-\partial\phi_c-\partial\phi_b\big)\,\big({\cal O}_{cb}+{\cal O}_{bc}\big)
-
\ri\, 
\big(2\,\partial\phi_b-\partial\phi_d-\partial\phi_c\big)\,\big({\cal O}_{dc}+{\cal O}_{cd}\big)\nonumber\\[0.3cm]
&+&\tfrac{1}{2}\,\big(2\, \partial\phi_c-\partial\phi_b-\partial\phi_d\big)\big(\partial^2\phi_b-\partial^2\phi_d\big)
-\tfrac{1}{2}\,\big(2\, \partial^2\phi_c-\partial^2\phi_b-\partial^2\phi_d\big)\big(\partial\phi_b-\partial\phi_d\big)\, .
\nonumber
\eea
Notice that
\bea
{Z}_{bdc}=-{Z}_{dbc}\  \ \ \ \ \  \ (b\not= d,\ \ d\not=c,\ c\not= b)
\eea
so that  the number of independent components of ${Z}_{bdc}$ is
equal to  $\frac{1}{2}\, r(r-1)(r-2)$.

\bigskip
The above analysis suggests that $X_a$, $Y_{ab}$ form
a basis in ${\cal W}^{(2)}$, while
 any spin 3 field from ${\cal W}^{(3)}$
may be represented as a linear combination of $Z_{ab}$, ${Z}_{abc}$ as well as the derivatives
$\partial X_a$ and $\partial Y_{ab}$.
One can continue the process of generating the higher Lorentz spin $s=4,5,\ldots$ fields,
which would belong to the
linear spaces ${\cal W}^{(s)}$.
An important property of the operator algebra $W_{\bm 1}^{(c,r)}=\oplus_{s=2}^\infty{\cal W}^{(s)}$ is that it contains the
spin 2 field
\bea\label{isiisias}
T_2=\frac{1}{\beta r}\, \bigg(\sum_{a=1}^rX_a+(\beta^{-1}-\beta)\, \sum_{b>a}Y_{ab}\bigg)\ ,
\eea
which forms the  Virasoro subalgebra 
\bea
T_2(u)T_2(v)=\frac{c}{2(u-v)^4}-\frac{T_2(u)+T_2(v)}{(u-v)^2}+O(1)
\eea
with central charge
\bea
c=r-\frac{6}{r}\ \big(\beta^{-1}-\beta\big)^2\ .
\eea

 \subsection{Local IM}
The  integrals of motion ${\bf I}_{1}^{(a)}$ are built from the densities
belonging to the linear space ${\cal W}^{(2)}$. They can be expressed as 
a linear combination of the basic fields with some numerical coefficients:
\bea
T^{(a)}_2=\sum_{b=1}^rA^{(a)}_{b} X_b+\sum_{b\not=c}B^{(a)}_{bc}\, Y_{bc}\ .
\eea
The coefficients may be fixed using the method discussed in sec.\,\ref{sec91}, which is based on
the notion of the reflection operator.
Introduce  ${\bf R}$, defined  via  its action on the
$W$-currents
 \bea\label{oiasio98129812aaa}
{\bf R}\, X_a\, {\bf R}^{-1}= {\tilde X}_a\ ,\ \ \ \ \qquad
{\bf R}\, Y_{ab}\, {\bf R}^{-1}= {\tilde Y}_{ab}\ ,
\eea
where  ${\tilde X}_a$ and  ${\tilde Y}_{ab}$ are obtained from the fields $X_a$ \eqref{hasyasy} and  $Y_{ab}$ \eqref{hasyasyss}
by means of the formal substitution
 \bea\label{oiasio98129812ss}
\phi_a \,\mapsto\, -\phi_a-\tfrac{\ri}{2}\,C_a\ .
\eea
The constants $C_a$ are defined as the  solution of the linear system,
\bea
\log(z_a)=C_a+ \frac{\beta-1}{r}\ \sum_{b=1}^rC_b\ ,
\eea
so  that under the replacement \eqref{oiasio98129812ss},
\bea
 V_+=
\sum_{a=1}^r\re^{2\ri \phi_a}\ \re^{\frac{2\ri(\beta-1)}{\sqrt{r}}\varphi}\ \ \ \ \ \ \ \ \ \mapsto\ \ \ \ \ \ \ \ 
 V_-=
\sum_{a=1}^rz_a\ \re^{-2\ri \phi_a}\ \re^{-\frac{2\ri(\beta-1)}{\sqrt{r}}\varphi}\ .
 \eea
Then the densities entering into the local IM are those, which are invariant under the reflection up to a total derivative:
 \bea\label{iaiaisiasiasAA}
{\bf R}\,T_2^{(a)}\, {\bf R}^{-1}= T_2^{(a)}+\partial{\cal O}_1\ .
 \eea

 \bigskip
 The following  $r$  spin 2 local fields 
 \bea\label{uyasyysayas}
 T_2^{(a)}=-\frac{2}{\beta^{-1}-\beta}\ X_a+
\sum_{b=1\atop
b\ne a}^{r}\, \frac{2 z_b}{z_a-z_b}\, Y_{ab}
 \eea
obey the condition \eqref{iaiaisiasiasAA}.
This becomes evident if  $T^{(a)}_2$ are rewritten using the differential polynomials  $R_a$ \eqref{iasissau}
and  the exponential fields
${\cal O}_{ab}=\re^{2\ri(\phi_a-\phi_b)}$:
\bea\label{isisiasis}
T^{(a)}_2&=&-\frac{2}{1-\beta^2}\ \Big(\,
  r\, (\partial\phi_a)^2
-2\sqrt{r}\, (1-\beta)\ \partial\varphi\,\partial\phi_a
+(1-\beta)^2\ (\partial\varphi)^2\,\Big)\\[0.3cm]
&-&
  \sum_{b=1\atop
b\ne a}^{r}\, \frac{2 z_b}{z_a-z_b}\,(\partial\phi_a-\partial\phi_b)^2
\,+\,
\sum_{b=1\atop
b\ne a}^{r}
\frac{2}{z_a-z_b}\ \Big(\,z_a\, \re^{2\ri(\phi_b-\phi_a)}+z_b\,\re^{2\ri(\phi_a-\phi_b)}\,\Big)+\partial\big(\ldots\big)\ .\nonumber
 \eea

 \bigskip
The prediction \eqref{oias8912891} from the ODE/IQFT correspondence concerns the eigenvalues
of the local IM, which a priori would be linearly expressed in terms of 
 $\oint\frac{{\rm d }u}{2\pi}\, T_{2}^{(a)}$.
To establish the precise relation  it is sufficient to consider the
 eigenvalue  for  the simplest $W$-primary states \eqref{ioas9812}.
 Introduce the short cut notation 
 \bea
{\boldsymbol  e}_{\mathfrak{m}}={\boldsymbol  e}_{\boldsymbol{\mathfrak j}, 2{\mathfrak J},P}\ :\ \ \ \ 
 \boldsymbol{\mathfrak j}=\big(\underbrace{\tfrac{1}{2},\ldots,\tfrac{1}{2}}_{\mathfrak{m}},
 \underbrace{0,\ldots, 0}_{r-\mathfrak{m}}\big)\ \ \ \  \ \ \ \ \ \ \  (\mathfrak{m}=0,1,\ldots, r-1)\, .
 \eea
As follows from \eqref{oias8912891} the corresponding eigenvalue  is given by
\bea\label{iisisaiasAA}
&&I_1^{(a)}({\boldsymbol  e}_{\mathfrak{m}})=\frac{1}{24}\ 
\sum_{b=1\atop
b\ne a}^{r}\, \frac{z_a+z_b}{z_a-z_b}\, \Big(\,3\,\epsilon(\mathfrak{m}-a)\,\epsilon(\mathfrak{m}-b)-1\,\Big)\\[0.3cm]
&&-\, \frac{\epsilon(\mathfrak{m}-a)}{\beta^{-1}-\beta}\  \Big(\sqrt{r}P+\tfrac{1}{8}\ \big(\beta+\beta^{-1}\big)\ 
(r-2\,\mathfrak{m})\,\Big)
-\frac{2\beta}{\beta^{-1}-\beta} \Bigg[\,
\frac{r }{48}+\bigg(P+\frac{r-2\,\mathfrak{m}}{4\beta\sqrt{r}}\bigg)^2\,\Bigg]\, ,\nonumber
\eea
where 
\bea
\epsilon(n)=\begin{cases}
+1\ \ \ \ &{\rm for}\ \ \ \ n\geq 0\\
-1\ \ \ \ &{\rm for}\ \ \ \ n< 0
\end{cases}\ .
\eea
On the other hand, for such primary states
\bea
\int_0^{2\pi}\rd u\ \re^{2\ri(\phi_a-\phi_b)}(u)\,{\boldsymbol  e}_\mathfrak{m}=0\ \ \ \  \ \ \ \ \ \ \ \ \ \ (\forall a,b,\mathfrak{m})\ .
\eea
This makes the
 calculation of  the eigenvalues of $\oint\frac{{\rm d} u}{2\pi}\, T_{2}^{(a)}$
 with the density  given by  \eqref{isisiasis}
elementary  and shows that they exactly coincide with $I_1^{(a)}({\boldsymbol  e}_{\mathfrak{m}})$
\eqref{iisisaiasAA}.

 \section{Hamiltonians for the GAGM\label{sec11}}
As we saw, the study 
 of the  algebra
$W_{\bm{k}}^{(c,r)}$ and the subsequent 
construction of the local IM is
rather involved.
Nevertheless using the
expression for the spectrum \eqref{jhsasysay},
as well as the $k_a=1$ result for the densities  \eqref{isisiasis}, 
it is possible to deduce the form of ${\bf I}_1^{(a)}$
 for arbitrary values of the parameters. Let's illustrate this
first in the isotropic limit.
 \subsection{The isotropic limit}
The  limit  was already discussed in the
context of the ODE 
 in sec.\eqref{sec74}.
As prescribed by
 eq.\eqref{ausausau}
one must first perform the substitution
 \bea\label{oiasio89129812}
 z_a\mapsto z_a+\tfrac{\beta^2 K}{2(1-\beta^2)}
 \eea
and  then take $\beta\to 1^-$. 
For the local IM ${\bf I}_1^{(a)}$,  we define
\be\label{aois981298}
{\bf I}_{1}^{(a,{\rm iso})}= \frac{1}{2K}\ \lim_{\beta\to 1^-}\big(\beta^{-2}-1\big) \,{\bf I}^{(a)}_1\, .
\ee
\bigskip

Consider the local densities \eqref{isisiasis}. 
Following the above procedure
with  $K=r$ in eqs.\eqref{oiasio89129812},\,\eqref{aois981298},
 one obtains
\be\label{usuusauadddds}
{\bf I}_{1}^{(a,{\rm iso})} =-
\int_{0}^{2\pi}\frac{\rd u}{2\pi}\,\bigg(
   (\partial\phi_a)^2
+\frac{1}{2}\,
  \sum_{b=1\atop
b\ne a}^{r}
\frac{1}{z_a-z_b}\ \Big(\,(\partial\phi_a-\partial\phi_b)^2- \re^{2\ri(\phi_b-\phi_a)}-\re^{2\ri(\phi_a-\phi_b)} \Big)\bigg)\, .
 \ee
Recall that these operators act invariantly in ${\cal H}_{\bm{\mathfrak{j}}, {\mathfrak m},P}$ \eqref{aisisidd}.
Provided that 
\be
P=\frac{\mathfrak{m}}{2\sqrt{r}}\,,
\ee
the latter can be realized as a subspace of the tensor product of $r$  integrable representations
of the  Kac-Moody algebra at level one. It is easy to see that the
 densities for the local integrals of motion
can be expressed in terms of the currents 
$J_\pm^{(a)}=\re^{\pm 2\ri\phi_a}$ and $J_0^{(a)}=2\partial\phi_a$ (see eq.\eqref{aisisai}) 
so that
 \be\label{ioas8912132}
 {\bf I}_{1}^{(a,{\rm iso})}=-
 \int_{0}^{2\pi}\frac{\rd u}{2\pi}\,\bigg(G^{(a)}+\frac{1}{2}\,\sum_{b=1\atop
b\ne a}^{r}\frac{ G^{(a)}+G^{(b)}-2\eta^{AB}J_A^{(a)}J_B^{(b)}}{z_a-z_b}\,\bigg)\ \ \ \ \ \ (k_a=1)\ ,
 \ee
 where
 \bea
2 \eta^{AB}J_A^{(a)}J_B^{(b)}=
 \tfrac{1}{2}\ J_0^{(a)}J_0^{(b)}+           J_+^{(a)}J_-^{(b)}+J_-^{(a)}J_+^{(b)}\ .
 \eea
The notation $G^{(a)}=(\partial\phi_a)^2$ stands for   the Virasoro field associated with
the  $k_a=1$ Kac-Moody algebra. 
\bigskip

Formula \eqref{ioas8912132} admits an immediate
 generalization for arbitrary positive
integers $k_a$.
One assumes
 that the local densities are  expressed as a quadratic combination of the
Kac-Moody currents, subject to the OPE\ \eqref{aiissaias}
 The local IM should be invariant w.r.t. the  global $\mathfrak{sl}(2)$ symmetry
that occurs in the isotropic limit.
Also, taking into account 
the expression \eqref{jhsasysay} for the eigenvalues of the IM,
specialized to $\xi\to+\infty$, one arrives at
the formula
\bea\label{aoisd89128912AA}
(-k_a)\,{\bf I}_{1}^{(a,{\rm iso})}=
 \int_{0}^{2\pi}\frac{\rd u}{2\pi}\,\bigg(G^{(a)}
  +\frac{1}{2}\,\sum_{b=1\atop
b\ne a}^{r}\frac{k_b\, G^{(a)}+k_a\, G^{(b)}-2 \eta^{AB}J_A^{(a)}J_B^{(b)}
}{z_a-z_b}\,\bigg)\,,
 \eea
 where 
 \bea
 G^{(a)}= \frac{\eta^{AB}J_A^{(a)}J_B^{(a)}}{k_a+2}\, .
 \eea
Notice  that the overall normalization of ${\bf I}_{1}^{(a,{\rm iso})}$ is such that
\bea
\sum_{a=1}^r(-k_a)\,{\bf I}_{1}^{(a,{\rm iso})}=\sum_{a=1}^r \,
 \int_{0}^{2\pi}\frac{\rd u}{2\pi}\ G^{(a)}\ .
\eea
\bigskip

It is instructive to consider the  limit
when all the $k_a$ tend to infinity.
One writes $k_a=\nu_a\,K$ and takes $K\to\infty $ while keeping
$\nu_a$ fixed. The latter automatically satisfy 
\be
\sum_{a=1}^r \nu _a= 1\ .
\ee
This  can be interpreted as the classical limit, with $K$ being identified with
the inverse Planck constant:
\be
K=\hbar^{-1}\ .
\ee
The currents 
\be
j^{(a)}_A=\hbar\,J^{(a)}_A
\ee
become classical fields obeying
the equal time Poisson bracket relations
\be\label{oiasoi1982AAABBB}
\big\{j^{(a)}_A(u),j^{(b)}_B(v)\big\}=\delta_{ab}\,\Big(\,
-\tfrac{1}{2}\,\nu_a\,
\eta_{AB}\,\delta'(u-v)+\ri\,{f_{AB}}^C\,j^{(a)}_C(u)\,\delta(u-v)\,\Big)\,.
\ee
In turn, for the integrals of motion \eqref{aoisd89128912AA},
 \bea\label{ioas98128932a}
\lim_{\hbar\to 0} \hbar^2\,(-{k_a})\,
{\bf I}_{1}^{(a,{\rm iso})}=\frac{1}{2}\ \sum_{b=1\atop
b\ne a}^{r}
 \int_{0}^{2\pi}\frac{\rd u}{2\pi}\
\eta^{AB}\ 
\frac{\big(\nu_a j^{(b)}_A-\nu_b j^{(a)}_A\big)\,\big(\nu_a j^{(b)}_B-\nu_b j^{(a)}_B\big)
}{\nu_a\nu_b\,\big(z_a-z_b\big)}\  .
 \eea
Notice that the term $G^{(a)}$ outside the sum in \eqref{aoisd89128912AA}
is absent from the classical expression. It can be interpreted as an effect of renormalization (quantum counterterm)
which appears already at the first perturbative order.

 \subsection{The Gaudin limit}
To take the Gaudin limit, following the discussion in  sec.\,\ref{sec74},
the parameters $z_a$ should be rescaled as $z_a=\delta\, y_a$
with $\delta\to 0$. Defining the Gaudin Hamiltonians as
\be
{\bf H}_{\rm \scriptscriptstyle G}^{(a)}=\lim_{\delta\to 0}\delta\, (-k_a)\,{\bf I}_{1}^{(a,{\rm iso})}\,,
\ee
one obtains
\be
{\bf H}_{\rm \scriptscriptstyle G}^{(a)}= 
\frac{1}{2}\int_{0}^{2\pi}\frac{\rd u}{2\pi}\
\sum_{b=1\atop
b\ne a}^{r}\, \frac{k_b\, G^{(a)}+k_a\, G^{(b)}-2 \eta^{AB}J_A^{(a)}J_B^{(b)}
}{y_a-y_b}\ .
\ee
Note that the spectrum of the first few lowest integrals of motion for
the affine Gaudin model associated to
$\mathfrak{sl}(N)$ was studied in the work
\cite{Lacroix:2018itd}, while for the case of an arbitrary Lie algebra
some conjectures are formulated in ref.\cite{Lacroix:2018fhf}.
 Also, the classical limit of 
${\bf H}_{\rm \scriptscriptstyle G}^{(a)}$ results in a similar expression
as in the r.h.s. of  \eqref{ioas98128932a} with $z_a$ replaced by $y_a$.

 \subsection{General case}
 For arbitrary positive integer $k_a$ and $\beta\in(0,1)$
we propose the following formula for the 
Hamiltonians of the GAGM,
\be
{ \bf{H}}^{(a)}_{\rm gen}\equiv(-\tfrac{k_a}{2})\,{\bf I}_{1}^{(a)}\,,
\ee
expressed in terms of the
set of parafermionic $\{\psi_\pm^{(a)}\}$ and bosonic $\{\phi_a\}$ fields:
\bea\label{usuadddds}
 { \bf{H}}^{(a)}_{\rm gen}&=&
\int_{0}^{2\pi}\frac{\rd u}{2\pi}\,\bigg[\, \frac{\beta^2 K}{1-\beta^2}\, G^{(a)}
+\frac{1-\beta}{1+\beta}\ \big(\,k_a\,(\partial \varphi)^2-\sqrt{K k_a}\ \partial\phi_a\partial\varphi\,\big)\nonumber\\[0.3cm]
&+&\sum_{b=1\atop
b\ne a}^{r}
\frac{1}{z_a-z_b} \ \Big(\, k_a z_b\,G^{(b)}+k_bz_a\,G^{(a)}\,\Big)
\\[0.3cm]
&-&  \sum_{b=1\atop
b\ne a}^{r}
\frac{\sqrt{k_ak_b}}{z_a-z_b}\, \Big(\, (z_a+z_b)\, \partial\phi_a\partial\phi_b+
 z_a\,\psi^{(b)}_+\psi^{(a)}_-
 \re^{2\ri(\phi_b-\phi_a)}+
 z_b\,\psi^{(a)}_+\psi^{(b)}_-
 \re^{2\ri(\phi_a-\phi_b)} \Big)\,\bigg]\,,\nonumber
 \eea
 where
 \bea
 G^{(a)}=(\partial\phi_a)^2+W_2^{(a)}\ ,\ \ \ \ \ \ \ \ \ \  \partial\varphi=\sum_{a=1}^r\sqrt{\frac{k_a}{K}}\ \partial\phi_a\ .
 \eea
 Recall that the spin 2 field $W^{(a)}_2$ occurs
  in the OPE of the fundamental parafermions $\psi^{(a)}_\pm$ as in eq.\,\eqref{aoisaisais}.
Formula \eqref{usuadddds}
 can be rewritten in terms of the Kac-Moody currents  $J_\pm^{(a)}=\sqrt{k_a}\,\psi_\pm^{(a)}\,
\re^{\pm\frac{2\ri\phi_a}{\sqrt{k}_a}}$,
$J_0^{(a)}=2\sqrt{k_a}\,\partial\phi_a$. The result  was already presented in the introduction, 
see eq.\eqref{oias8912}.
Taking the limit $\beta\to 1^-$, the combination $\frac{1}{K}(\beta^{-2}-1) \,{ \bf{H}}^{(a)}_{\rm gen}$ coincides 
with the r.h.s of \eqref{aoisd89128912AA}. Another consistency check is that the eigenvalue of 
the operators \eqref{usuadddds}, computed on the primary states
${\boldsymbol  e}_{\boldsymbol{\mathfrak j},\pm 2 {\mathfrak J},P}$ \eqref{ioas9812}, agrees with the 
prediction \eqref{jhsasysay} that comes from the ODE side. One can show by direct
computation that ${\bf H}_{\rm gen}^{(a)}$ 
with $a=1,2,\ldots,r$ mutually commute.

 \bigskip
The generalized affine Gaudin model,
being a multiparametric integrable system, admits
various interesting limits. In the previous subsection the
isotropic limit was discussed as well as the classical one. Let us emphasize that
the latter was performed only after taking 
$\beta\to 1^-$. It turns out that a meaningful classical limit
exists for any fixed value of $\beta$ and $z_a$.
Indeed, setting 
$
 k_a=\nu_a\,K$
and $j^{(a)}_A=J^{(a)}_A/\hbar $ in
eq.\,\eqref{oias8912} and taking
$K=\hbar^{-1}\to\infty$, 
one arrives at
 \bea
&&\lim_{\hbar\to 0} \hbar^{-2}\,{\bf H}_{\rm gen}^{(a)} =
\int_{0}^{2\pi}\frac{\rd u}{2\pi}\,\bigg[\, \frac{\beta^2 }{1-\beta^2}\ \frac{g^{(a)}}{\nu_a}
+\frac{1}{4}\ \frac{1-\beta}{1+\beta}\ \Big(\,\nu_a\, \big(j_0^{(\rm tot)}\big)^2
 -\, j_0^{(a)} j_0^{(\rm tot)}\,\Big)\\[0.3cm]
&&-  \sum_{b=1\atop
b\ne a}^{r}
\frac{1}{z_a-z_b}\, \Big(\, \frac{1}{4}\, (z_a+z_b)\  j^{(a)}_0j^{(b)}_0+
 z_a\  j_+^{(b)} j_-^{(a)}+
 z_b\  j_+^{(a)} j_-^{(b)}
 - \frac{\nu_a\,z_b}{\nu_b}\ g^{(b)}-\frac{z_a\,\nu_b}{\nu_a}\ g^{(a)}
  \Big)\,\bigg]\nonumber
 \eea
with
\be
g^{(a)}=\eta^{AB}j_A^{(a)}j_B^{(a)}\,,
\qquad\qquad\qquad
j_0^{({\rm tot})}=\sum_{a=1}^r j_0^{(a)}
\ee
and $\nu_1+\nu_2+\ldots+\nu_r=1$. Needless to say that these classical observables
mutually Poisson commute w.r.t. the Poisson bracket \eqref{oiasoi1982AAABBB}.

 \section{Baxter statistical systems in the scaling limit and
 the GAGM\label{sec12}}
\subsection{Baxter-type statistical systems}
Behind the integrability of the generalized affine
Gaudin model are the algebraic structures, which are inherited from the
quasi-triangular Hopf algebra $U_q\big(\widehat{\mathfrak{sl}}(2)\big) $
and the universal $R$-matrix ${\cal R}$. The  development  of quantum groups,
that encompasses these notions, 
was inspired by the works of Baxter on exactly soluble  lattice
models.  The $R$-matrix, encoding the Boltzmann weights of 
the statistical system, comes from taking a particular realization
of ${\cal R}$. Namely, the finite dimensional matrix
 \bea\label{poasop21}
{R}_{{\ell}_1,{\ell}_2}(\lambda_1/\lambda_2)=\big(\pi_{{\ell}_1}(\lambda_1)\otimes 
\pi_{{\ell}_2}(\lambda_2)\big)[{\cal R}]\, ,
\eea
 where $\pi_\ell$ stands for the  $2\ell+1$ dimensional representation of
the $U_q\big(\mathfrak{sl}(2)\big)$ algebra, satisfies
the Yang-Baxter equation. Let's suppose that $\ell_1=\frac{1}{2}$.
Denoting
\bea
H^{(\ell)}=\pi_\ell({\tt h})\, ,\  F^{(\ell)}=\pi_\ell({\tt e}_-)\, ,\  E^{(\ell)}=\pi_\ell({\tt e}_+)\,\in 
\,{\rm End}\big(\mathbb{C}^{(2\ell+1)}\big)
\eea
the matrix \eqref{poasop21}
is given explicitly by \cite{Kulish}
\be\label{oias89129821}
{R}_{ {\scriptscriptstyle \frac{1}{2}},\ell}(\lambda)=r
(\lambda)\ \Big(\,
q^{\frac{1}{2}+\frac{1}{2}\sigma^z\otimes H^{(\ell)}}-\lambda^2\, q^{-\frac{1}{2}-\frac{1}{2}\sigma^z\otimes H^{(\ell)}}+
\lambda\,(q-q^{-1})\,  (\sigma^+\otimes F^{(\ell)}+\sigma^-\otimes
E^{(\ell)}
\big)\,\Big)
\ee
with
 some overall factor $r(\lambda)$, which cancels out in the Yang-Baxter equation and is not
essential for our purposes.
We set $r(\lambda)=1$
and define
\bea\label{saisiusaias}
\bm{R}^{(\ell)}(q\zeta)=
\lambda^{\frac{1}{2}\sigma^z\otimes 1}\
{R}_{ {\scriptscriptstyle \frac{1}{2}},\ell}\big(q^{\frac{1}{2}}\lambda\big)\ 
\lambda^{-\frac{1}{2}\sigma^z\otimes 1}\ \ \  \ \ \ \qquad {\rm with}\ \ \ \qquad\zeta=-\lambda^2\ .
\eea
It is a $2\times 2$ matrix, 
\be\label{rmat2}
\bm{R}^{(\ell)}(q\zeta)=
\left(\begin{array}{cc} q^{\frac{1}{2}(1+H^{(\ell)})}+q^{\frac{1}{2}(1-H^{(\ell)})}\,\zeta & 
-(q-q^{-1})\,q\, \zeta\,F^{(\ell)} \\[0.2cm]
(q-q^{-1})\,E^{(\ell)} & q^{\frac{1}{2}(1-H^{(\ell)})}+q^{\frac{1}{2}(1+H^{(\ell)})}\,\zeta
\end{array}\right)\,,
\ee
whose elements act in a $2\ell+1$ dimensional linear space.
The lattice  transfer-matrix is constructed from such building blocks.
First introduce the monodromy matrix
\be\label{M-in}
{\bm M}\Big(\zeta\,\Big|\begin{array}{ll}
\ell_{N},\ell_{N-1},\,\ldots,\,\ell_1\\
\eta_N,\eta_{N-1},\,\ldots,\,\eta_1
\end{array}\!\!
\Big)=q^{-\frac{N}{2}}\ 
\bm{R}_{N}^{(\ell_N)}\big(q\zeta/\eta_N\big)\,
\bm{R}_{N-1}^{(\ell_{N-1})}\big(q\zeta/\eta_{N-1}\big)\cdots
\bm{R}_{1}^{(\ell_1)}\big(q\zeta/\eta_{1}\big)\,.
\ee
Its entries  act in the ``physical space'', which is the tensor product
\be\label{aoisd89128912}
\mathscr{V}_N=\mathbb{C}^{(2\ell_N+1)}_N\otimes\mathbb{C}^{(2\ell_{N-1}+1)}_{N-1}\otimes\ldots
\otimes \mathbb{C}^{(2\ell_1+1)}_1\ .
\ee
Taking the trace over the two dimensional  ``auxiliary space'', one obtains the transfer-matrix
\be\label{oias89128921}
\mathbb{T}(\zeta)={\rm Tr}\big[\,\omega^{\sigma^z}\bm{M}(\zeta)\,\big]\,.
\ee
Here the parameter $\omega$ can be arbitrarily chosen. It will be assumed to be a unimodular number,
so that
\be\label{oas981298}
\omega^2=\re^{2\pi\ri{\tt k}}
\ee
with some real ${\tt k}$.
Note that the similarity transformation, which is performed in \eqref {saisiusaias},
 has no effect
on the trace in the definition \eqref{oias89128921}. It  
 was done  to make apparent that $\mathbb{T}(\zeta)$ is a polynomial
in $\zeta=-\lambda^2$ of order $\zeta^N$, normalized according to
\be
\mathbb{T}(0)=\omega^{+1}\,q^{+\mathbb{S}^z}+\omega^{-1}\,q^{-\mathbb{S}^z} \,.
\ee
Also, as $\zeta\to\infty$ one has
\bea\label{normT}
\lim_{\zeta\to\infty}\,\zeta^{-N}\,\mathbb{T}(\zeta)&=&
\big(\omega^{+1}\,q^{-\mathbb{S}^z}+\omega^{-1}\,q^{+\mathbb{S}^z}\big)\,
\prod_{J=1}^N\,\eta_J^{-1}\ .\nonumber
\eea
Here $\mathbb{S}^z$  stands for the $z$ projection of the total spin operator,
\be
 \mathbb{S}^z=\tfrac{1}{2}\sum_{m=1}^N  H^{(\ell)}_m\ ,
\ee
which commutes with the transfer-matrix.

\bigskip

The Yang-Baxter equation guarantees that $\mathbb{T}(\zeta)$ commutes with
itself for different values of the spectral parameter $\zeta$. The diagonalization problem can be
solved using whatever version of the Bethe ansatz approach (e.g., quantum inverse scattering method).
In the sector with given value of $\mathbb{S}^z$,
the  Bethe ansatz equations read as \cite{Baxter:1971cs,Sogo,Kirillov}
\be\label{bae}
\prod_{J=1}^{N}
\frac{\eta_J+q^{+2\ell_J} \,\,\zeta_m}
{\eta_J+q^{-2\ell_J}\,\zeta_m }
=-\omega^2\,q^{2S^z}\,
\prod_{j=1}^M\,
\frac{\zeta_j-q^{+2}\,\zeta_m }
{\zeta_j-q^{-2}\,\zeta_m }
\,\qquad\qquad (\,m=1,2,\ldots,M\,)\,,
\ee
where $M=\sum_{J=1}^N \ell_J-S^z$.

\subsection{Main conjecture concerning the scaling}
Being a purely algebraic procedure, the 
 diagonalization of $\mathbb{T}(\zeta)$   can be performed
for arbitrary values of the inhomogeneities $\eta_J$ and the positive integers $2\ell_J$. However, to make
connection with the field theory where a fundamental symmetry is translational invariance,
we'll focus on the lattice with 
 $N$ being divisible by 
some integer $r$,
\bea\label{aoisd981221}
N=r L\,,
\eea  
while the parameters $\eta_J,\ell_J$   satisfy the $r$\,-\,site periodicity conditions
\bea\label{iisasausa}
\eta_{J+r}=\eta_{J}\,,\qquad\qquad
\ell_{J+r}=\ell_{J}\  \ \ \ \ \ \   \qquad \qquad \qquad (J=1,2,\ldots,N)\, .
\eea
\smallskip

The lattice model turns out to be critical when 
 $q$ is a unimodular number, $|q|=1$.
With a suitably defined scaling limit where $L\to\infty$,
the universal properties are described by a CFT. However,
different types of critical behaviour occur depending on the
domain of the parameter $q$.  We conjecture that 
with an appropriate definition of the scaling limit, and
a proper identification of the parameters,
the lattice transfer-matrix 
$\mathbb{T}(\zeta)$
becomes the operator ${\boldsymbol \tau}_{\scriptscriptstyle \frac{1}{2}}(\lambda)$ 
\eqref{asusuya} as $N\to\infty$. 
The scaling limit of the Baxter $Q$-operator (whose 
eigenvalues are polynomials in $\zeta$ with zeroes being the
roots of the Bethe ansatz equations \eqref{bae}) yields 
${{\mathlarger{\mathlarger{\mathlarger {\boldsymbol a}}}}}_+(\lambda)$.\footnote{%
There are two Baxter $Q$-operators ${\mathbb Q}_\pm(\zeta)$. The zeroes of the eigenvalues of ${\mathbb Q}_+$
satisfy \eqref{bae}, while those of ${\mathbb Q}_-$ obey the  system, which
is obtained from \eqref{bae} via the substitutions
$S^z\to-S^z$ and $\omega\to \omega^{-1}$
 (for details see, e.g., ref.\cite{Bazhanov:2020fbp}). In what follows we always assume that
$S^z\ge0$, and discuss the zeroes of the eigenvalues  of ${\mathbb Q}_+$, rather than $\mathbb{Q}_-$.
}
The anisotropy parameter $q$ should belong to the domain
\bea\label{oiasio98129831}
\pi \big(1-\tfrac{1}{K}\big)<\arg(q)<\pi\ \  \ \qquad \ {\rm with}\ \qquad \ \ \ \ K=\sum_{a=1}^r2\ell_a
\eea
and can be written as
\be\label{aoisoi891289}
q=-\,\re^{\frac{\ri\pi}{K}(\beta^2-1)}\,,\ \ \ \ \ \qquad \ \ {\rm where}\ \ \ \qquad \ \ 0<\beta<1\ .
\ee 
Also, the positive integers $k_a$ from the GAGM are identified as
\be
k_a=2\ell_a\,.
\ee
In taking the scaling limit, one should send  $\zeta\to 0$ such that the combination
$\zeta\,N^{\frac{2}{K}(1-\beta^2)}$ remains fixed. Up to an overall constant, it coincides with the 
spectral parameter
 $\lambda^2$ entering into
${\boldsymbol \tau}_{\scriptscriptstyle \frac{1}{2}}(\lambda)$ and
${{\mathlarger{\mathlarger{\mathlarger {\boldsymbol  a}}}}}_\pm(\lambda)$. 
The $r$ inhomogeneities $\eta_a$ are related to the $r$ parameters $z_a$
though, in general, the relation is highly non-trivial.\footnote{%
It may also be that some restrictions on the domain of the inhomogeneities  $\eta_J$  need to
be imposed in order to make connection with the GAGM. This point requires
further investigations.
}
It is important to keep in mind that the scaling limit is defined not
for the full space of states $\mathscr{V}_N$ \eqref{aoisd89128912}, but only
a certain class of so-called low energy states. 
\bigskip

A  study of the scaling limit,
among other things, requires an analysis of the algebraic structures
underlying the lattice system. 
In the case when all the $\ell_a=\frac{1}{2}$ or, equivalently,
$k_a=1$ the matrix $\mathbb{T}(\zeta)$ coincides with 
the one-row transfer-matrix of the original Baxter inhomogeneous six-vertex model \cite{Baxter:1971cs}
subject to quasi-periodic boundary conditions. The recent paper \cite{Bazhanov:2020fbp} contains
a summary of
the  known results concerning the algebraic aspects of this model.
A comprehensive study of the critical behaviour of the lattice system is
beyond the scope of this work. Here
we would merely like to provide an
 illustration of how the GAGM appears in the scaling limit.
As such
we'll stick to the case $\ell_a=\frac{1}{2}$ and rely on the results of \cite{Bazhanov:2020fbp}.
For the reader's convenience, we follow the notations of that work.

\bigskip
In the case of the homogeneous six-vertex model the transfer-matrix
commutes with the spin $\frac{1}{2}$ Heisenberg $XXZ$ Hamiltonian.
A similar important property holds true for the 
model, where the parameters $\eta_J$
satisfy  the periodicity conditions \eqref{iisasausa}  with any $r\ge 1$.
Namely, the   commuting family 
contains spin chain Hamiltonians
that are given by a sum of
terms, each of which  is built out of  local spin operators supported on
 $r+1$ consecutive sites of the lattice. 
There are $r$ such Hamiltonians and, in terms of the one-row transfer-matrix, they are expressed as
(see eq.\,(6.11) in ref.\cite{Bazhanov:2020fbp})
\be\label{Helldef1}
\mathbb{H}^{(a)}=
2\ri\,\zeta\partial_\zeta\log\big(\mathbb{T}(-q^{-1}\zeta)\big)\big|_{\zeta=\eta_a} \
-2\ri L\ \sum_{b=1}^r\big(1-q^2\,\eta_b/\eta_{a}\big)^{-1}
\,\qquad \qquad (\,a=1,2,\ldots,r\,)\ .
\ee
For the definition of the scaling limit, a central r\^{o}le belongs to  the sum
\be\label{aiosd901212}
\mathbb{H}=\sum_{a=1}^r\mathbb{H}^{(a)}\, ,
\ee
which is essentially the logarithmic derivative of a $r$ row transfer-matrix.
The ground state of this Hamiltonian
serves as the reference state from which the energy of the
excited states is counted.  In performing the scaling limit one takes $N\to\infty$ but considers only the
class of  states, whose excitation energy over the ground state energy is sufficiently low. Then as $N\to\infty$
the low energy spectrum organizes into the conformal towers, which are classified w.r.t 
to the algebra of extended conformal symmetry. In the case at hand, it is expected that 
the latter coincides with $\overline{W}_{\bm 1}^{(c,r)}\otimes {W}_{\bm 1}^{(c,r)}$.
The two $W$-algebras in the tensor product are isomorphic and describe the left and right
chiralities of the underlying CFT. 
In turn, for the low energy spectrum of 
$\mathbb{H}$ \eqref{aiosd901212}  at large $N$, one would have
\bea\label{ENjkaskj1a}
\mathbb{H}\,\asymp\, N e_\infty{\bf 1}+\frac{2\pi r}{N}\, v_{\rm F}\, \big({\bf I}_1+\overline{{\bf I}}_{1}\big)+o(N^{-1})\, .
\eea
Here ${\bf I}_1$ is the local IM in the generalized affine Gaudin model,
defined via 
\eqref{oi89129821},\,\eqref{oias81aaa212}. The operator $\overline{\bf I}_1$, corresponding to the
other chirality, is described by the similar equations.
The first term in the r.h.s. is proportional to the identity operator 
and so does not depend on the particular low energy state.
It contains the constant $e_\infty$, which stands for the specific bulk energy.
The explicit
form of $e_\infty$ is not significant in the context of the field theory.
The positive constant $v_{\rm F}$, usually referred
to as the Fermi velocity,  depends on the overall normalization of the
Hamiltonian.

\subsection{Example of the scaling limit  with $r=2$ and $\ell_1=\ell_2=\frac{1}{2}$}
The asymptotic formula  \eqref{ENjkaskj1a} involves a contribution
from both the left and right chiralities. In order to 
study the scaling limit of  a low energy state,
it is useful to employ the relations which
involve each chirality separately. 
Among these are the sum rules for the Bethe roots.
Let's illustrate them on the example  of the inhomogeneous
six-vertex model with anisotropy parameter $q=-\,\re^{\frac{\ri\pi}{2}(\beta^2-1)}$,
subject to the two site periodicity condition.
We focus on the case when $\eta_1$ and $\eta_2$
are unimodular numbers and, without loss of 
generality, set
\be\label{askjdjkA}
\eta_1=\eta_{2}^{-1}=\re^{\ri\varpi}\, .
\ee
 Also, it will be assumed that
\be\label{askjdjkB}
\tfrac{\pi}{2}\,(1-\beta^2)<\varpi< \tfrac{\pi}{2}\,(1+\beta^2)\, .
\ee
\smallskip

The number of lattice sites $N$ should be even. 
Suppose for now that $N$ is divisible by 4 and 
 $S^z=0$.
Then the ground state of the Hamiltonian $\mathbb{H}$ \eqref{aiosd901212} is a singlet.
Together with  $\zeta_n$ it is  convenient to use
\be
\theta_n=-\tfrac{1}{2}\,\log(\zeta_n)\, ,
\ee 
which are roots of  the Bethe ansatz equations written in trigonometric form.
For the ground state, they  are split into two groups
$\{\theta^{(+)}_n\}_{n=1}^{N/4}$  and $\{\theta^{(-)}_n\}_{n=1}^{N/4}$
such that $\Im m(\theta_n^{(\pm)})\approx \pm\tfrac{1}{2}\,\varpi$ (see  top 
left panel of fig.\ref{opas9012}).
We impose an ordering as 
\be
\Re e\big(\theta_1^{(\pm)}\big)\le \Re e\big(\theta_2^{(\pm)}\big)\le\ldots\le \Re e\big(\theta_{N/4}^{(\pm)}\big)\, .
\ee
The roots $\theta^{(\pm)}_n$ with $n\ll N$ or $N/4-n\ll N$ will be referred to as the left
and right edge roots, respectively, while the rest will be called the bulk roots. 
The latter, as $N\to\infty$,
 become densely distributed along the lines  $\Im m(\theta)= \pm\tfrac{1}{2}\,\varpi$.
Namely, introducing
\be
\rho_{N}^{(\pm)}\big(\theta_{n+\frac{1}{2}}^{(\pm)}\big)=\frac{2}{N\big(\theta_{n+1}^{(\pm)}-\theta_n^{(\pm)}\big)}\
\qquad\qquad  {\rm with}\qquad\qquad 
\theta_{n+\frac{1}{2}}^{(\pm)}=\tfrac{1}{2}\,\big(\theta_{n+1}^{(\pm)}+\theta_{n}^{(\pm)}\big)
\ee
one finds
\be\label{oisa8912093}
\lim_{N\to\infty}\rho_N^{(\pm)}(\theta)=\rho_\infty(\theta)\equiv
\frac{1}{2\pi(1-\beta^2)\cosh(\frac{\theta}{1-\beta^2})}\ :
\qquad\qquad \int_{-\infty}^{\infty}\rd \theta\,\rho_{\infty}(\theta)=\frac{1}{2}\  .
\ee
In contrast the edge roots develop a scaling behaviour. 
It turns out that, keeping either $n$ or $N/4-n$  fixed as $N\to\infty$, the following limits exist
\be\label{oiasoi98121}
s_n^{(\pm)} 
=\lim_{N\to\infty} N^{1-\beta^2}\,\zeta_{N/4-n}^{(\pm)} \,,
\qquad\qquad\qquad
\bar{s}_n^{(\pm)} 
=\lim_{N\to\infty} N^{1-\beta^2}\,\big(\zeta_{n}^{(\pm)}\,\big)^{-1}\ ,
\ee
where $\zeta_n^{(\pm)}=\re^{-2\theta_n^{(\pm)}}$. 
Note that $s_n^{(\pm)}$ and  $\bar{s}_n^{(\pm)}$ are complex numbers, such that
\be\label{opaspo0912}
\lim_{n\to\infty}\arg\big(s_n^{(\pm)}\big)= \pm\varpi\ ,\qquad\qquad\qquad\qquad
\lim_{n\to\infty}\arg\big(\bar{s}_n^{(\pm)}\big)= \mp \varpi\ ,
\ee
while
\be\label{opas09109212}
\big|s_n^{(\pm)}\big|^{\frac{1}{1-\beta^2}}\asymp 2\pi n + O(1)\,,\qquad\quad
\big|\bar{s}_n^{(\pm)}\big|^{\frac{1}{1-\beta^2}}\asymp 2\pi n + O(1)\qquad\quad (n\to\infty)\, .
\ee
\smallskip

A consequence of our general proposal is that, 
up to an overall factor,
$s_n^{(\pm)}$ coincide with the zeroes of the spectral determinant
 $D_{{\boldsymbol {\mathfrak j}}, \mathfrak{m}, A}(\mu)$  for the vacuum ODE \eqref{aois90121} with 
\be
r=2\,,\qquad\qquad k_1=k_2=1\,,\qquad\qquad
 {\mathfrak j}_1=\mathfrak{j}_2=\mathfrak{m}=0\,,\qquad\qquad
P=\frac{{\tt k}}{\sqrt{2}\beta}\,.
\ee
As usual, $\beta^2=\frac{\xi}{\xi+1}$
and ${A}=
\tfrac{\ri }{\beta^{-1}-\beta}\ \big(\,\sqrt{K}\, P-\tfrac{1}{2}\, \beta\, \mathfrak{m}\,\big)$
while, for the case under consideration, $K=k_1+k_2=2$.

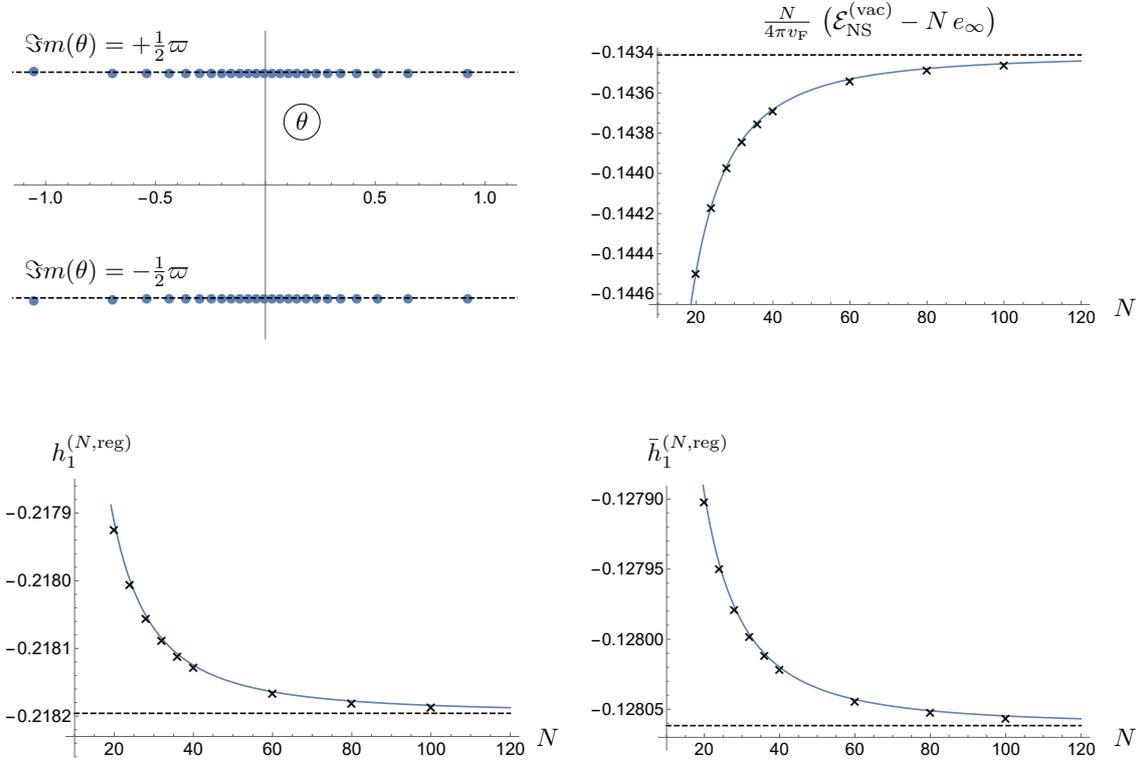
\begin{figure}
\centering
\scalebox{0.95}{
\begin{tikzpicture}
\node at (7.9,1.4) {\small $N$};
\node at (7.9,-4.6) {\small $N$};
\node at (-0.1,-4.6) {\small $N$};
\node at (-3.5,4) {\small $\theta$};
\draw (-3.5,3.99)  circle (0.25cm);
\node at (-6.2,5.05) {\small $\Im m(\theta)=+\tfrac{1}{2}\varpi$};
\node at (-6.2,1.9) {\small $\Im m(\theta)=-\tfrac{1}{2}\varpi$};
\node at (-6.4,-0.6) {\small $h_1^{(N,{\rm reg})}$};
\node at (1.85,-0.6) {\small $\bar{h}_1^{(N,{\rm reg})}$};
\node at (4.5,5.4) {\small $\tfrac{N}{4\pi v_{\rm F}}\ \big({\cal E}_{\rm NS}^{(\rm vac)}-N\,e_\infty\big)$};
\node at (-4,3.1) {\includegraphics[width=7cm]{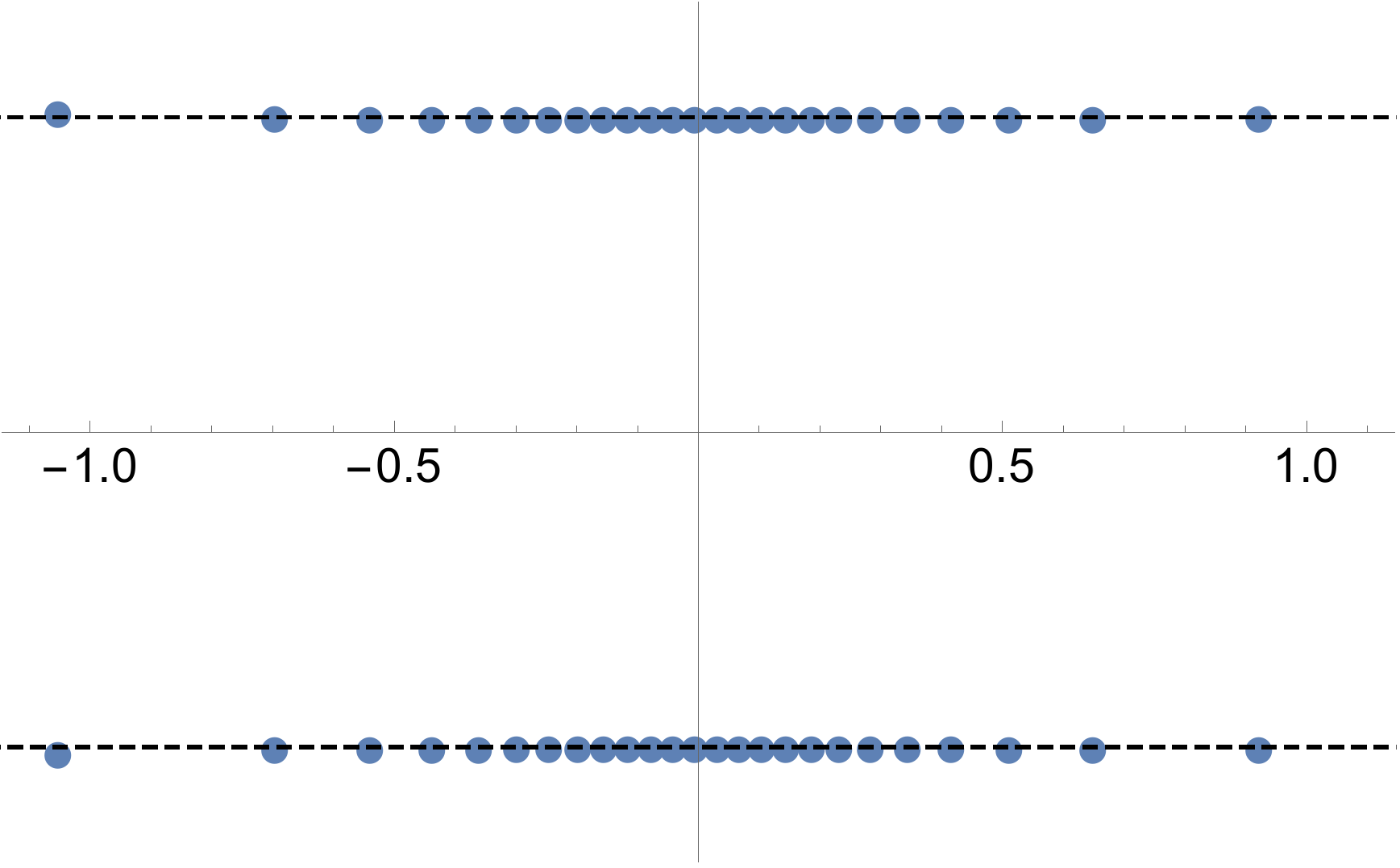}};
\node at (+4,3.1) {\includegraphics[width=7cm]{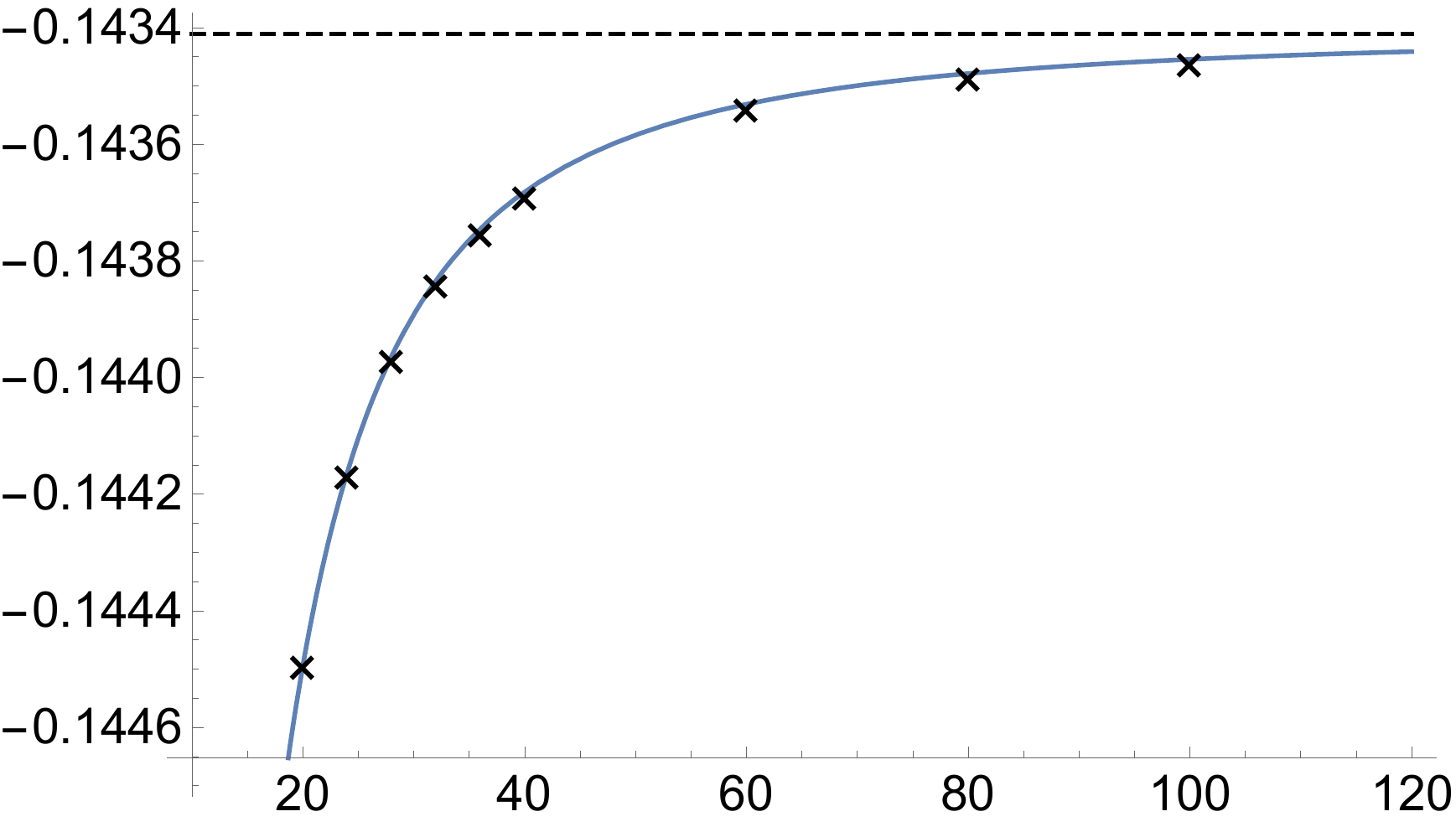}};
\node at (-4,-2.95) {\includegraphics[width=7.2cm]{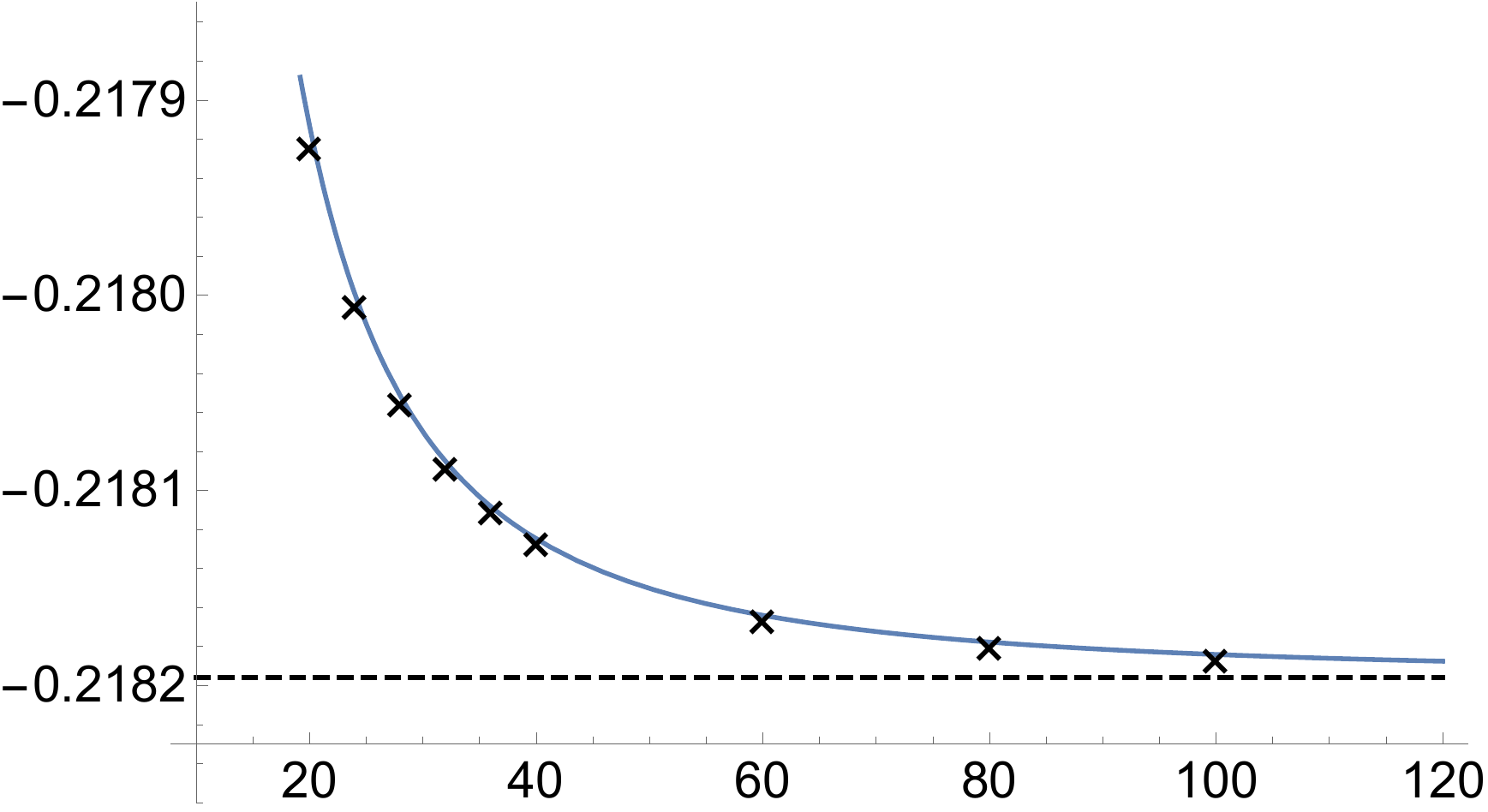}};
\node at (+4,-3) {\includegraphics[width=7cm]{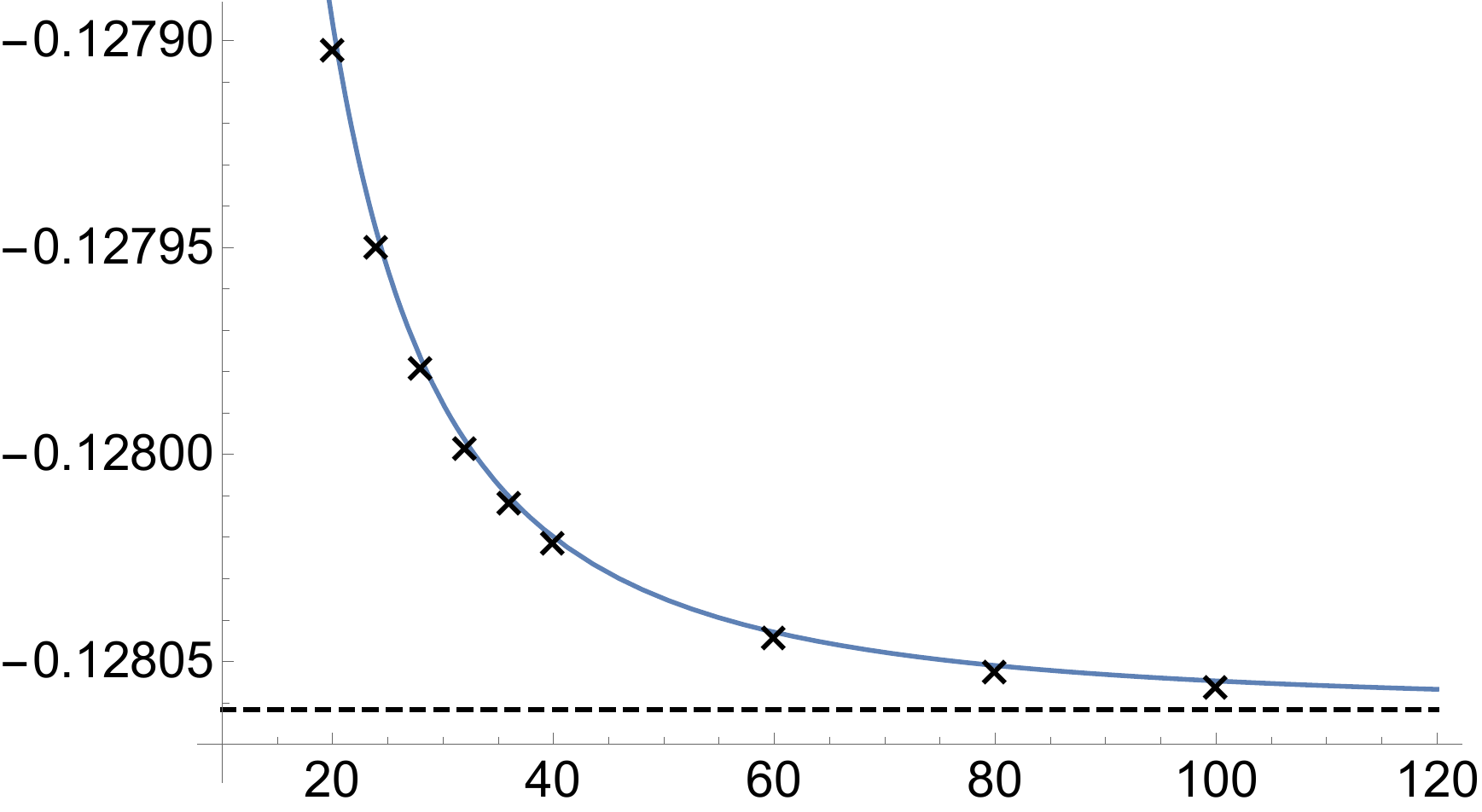}};
\end{tikzpicture}
}
\bigskip
\bigskip

\caption{\small Numerical data for the ground state of the Hamiltonian $\mathbb{H}$
\eqref{aiosd901212} for even $N/2$ and $S^z=0$.  The top left panel
depicts the  pattern of Bethe roots in the complex $\theta$ plane
with $\theta=-\frac{1}{2}\log(\zeta)$ ($N=100$). The top right panel
concerns the ground state energy ${\cal E}_{{\rm NS}}^{({\rm vac})}$. 
The black crosses were obtained through the solution of the Bethe ansatz equations for increasing $N$,
while the blue solid line represents the fit $-0.14341-0.43286/N^2$. The limiting value,
as predicted by eq.\,\eqref{ENjkaskjsss1a}, is marked by the dashed black line. The
bottom two panels illustrate the sum rules \eqref{asisiasA} and \eqref{asisiasC}. The superscript
``$\small {\rm reg}$'' indicates that a subtraction and rescaling has been made,
which makes the limit $N\to\infty$ well defined, e.g., 
$h_1^{(N,{\rm reg})}=N^{-(1-\beta^2)}\big(h_1^{(N)}-\frac{N}{2}\,
\frac{\cos(\varpi)}{\sin(\pi\beta^2/2)}
\,\big)$. Again, the crosses stand for the numerical values, which  were found by solving 
 the Bethe ansatz equations, while the solid blue line is a fit, 
$-0.21820+0.11285/N^2$ and $-0.12806+0.06623/N^2$ 
for the bottom left and bottom right panels,
respectively. The dashed line represents the limiting value according to
eqs.\,\eqref{asisiasA} and \eqref{asisiasC}. 
For the eigenvalue of the 
first non-local IM on the primary states, see \eqref{oais89czA} and \eqref{oais89czB}.
The parameters were taken to be
$\beta^2=0.43$, ${\tt k}=0.1$ and $\varpi=1.3$.
\label{opas9012}}
\end{figure}

\bigskip

The ODE  involves the parameters $z_1$, $z_2$ and we are free
to set  $z_1z_2=1$. Provided the inhomgeneities are restricted as in
eqs.\,\eqref{askjdjkA} and \eqref{askjdjkB}, 
they turn out to be unimodular numbers and it is convenient to swap them for  $x$,
\be
z_1=z_2^{-1}\,:\ \ \ \ \ \qquad x=\tfrac{1}{2}\,(z_1+z_2)\ ,\qquad\qquad -1<x<1\, .
\ee
With such reality conditions imposed on $z_1$ and $z_2$, the 
properties of the zeroes of the spectral determinant are described in
appendix \ref{hagf}.
In particular, they are split into two groups $\mu_n^{(\pm)}$, such that
\be
\lim_{n\to\infty}\arg\big(\mu_n^{(\pm)}\big)\,=\, \pm\,\alpha_0(x)
\ee
with $\alpha_0(x)$ from \eqref{ioas89128912}. Comparing the above with eq.\,\eqref{opaspo0912}
one finds that
\be
\alpha_0(x)=\varpi\, .
\ee
This allows one to relate the inhomogeneities of the lattice system $\eta_1$ and $\eta_2$ with
$z_1$ and $z_2$. Notice that the domains $\varpi\in\big(\frac{\pi}{2}\,(1-\beta^2),\frac{\pi}{2}\,\big]$
and $\varpi\in\big[\,\frac{\pi}{2},\frac{\pi}{2}\,(1+\beta^2)\big)$  map to $x\in[\,0,1)$ and $x\in(-1,0\,]$, respectively. 
Moreover, a comparison of \eqref{opas09109212}  with the asymptotic formula \eqref{oasoi192831}
yields
\be\label{oias891232}
s_{n}^{(\pm)}=f^{1-\beta^2}\,\mu_n^{(\pm)}\,.
\ee
Here $f=f(x)$,  given by \eqref{aois89128912},
  is a real and positive number.
\bigskip

Relation \eqref{oias891232} can be checked using the sum rules.
For any given solution to the Bethe ansatz equations \eqref{bae} consider
the finite sum
\be\label{oias89128932}
h^{(N)}_1=\sum_{n=1}^M(\zeta_n)^{-1}\, .
\ee
For the class of low energy states
it turns out that the following limit exists
\be\label{iosaisaias}
h^{(\infty)}_1=\lim_{N\to\infty} N^{-(1-\beta^2)}\bigg(h_1^{(N)}-\frac{N}{2}\,
\frac{\cos(\varpi)}{\sin(\frac{\pi\beta^2}{2})}
\,\bigg)\,.
\ee
Without going into details, we just mention that the existence of the
 limit comes from the Bethe ansatz equations.
Our main conjecture implies that for a low energy state,
$h_1^{(\infty)}$ appears in the first term of the Taylor expansion
of the corresponding spectral determinant. In view of \eqref{asoioi128921},
as well as the $\mu-\lambda$ relation specialized to $K=2$ (see  eq.\eqref{asususuaas}),
 the latter can be expressed in terms
of the eigenvalues of the non-local integrals of motion ${\bf H}_n^{(+)}$:
\be\label{oias981232}
\log \big(D_{{\boldsymbol {\mathfrak j}}, \mathfrak{m}, A}(\mu)\big)\,=\,
\sum_{n=1}^\infty(-1)^{n-1}\,
\big(\tfrac{2}{1-\beta^2}\big)^{-n(1-\beta^2)}\ 
\Gamma^{-2n}\big(\tfrac{1-\beta^2}{2}\big)\ 
{ H}_n^{(+)}\ \mu^n\, .
\ee
Combined with the relation \eqref{oias891232}, this yields
\be\label{asisiasA}
h_1^{({\infty})}=
\bigg(\frac{1-\beta^2}{2f}\bigg)^{1-\beta^2}
\frac{{ H}_1^{(+)}}{\Gamma^{2}\big(\tfrac{1-\beta^2}{2}\big)}\ .
\ee
Though the  sum in \eqref{oias89128932} goes over all the $\zeta_n$,
the contribution of the roots at the left edge of the distribution
becomes negligible as $N\to\infty$.
Thus the limit  \eqref{iosaisaias} encodes the scaling properties
of the right edge roots. 
The similar characteristic can be introduced for the left edge roots:
\be\label{asisiasB}
\bar{h}_1^{(\infty)}=\lim_{N\to\infty} N^{-(1-\beta^2)}\bigg(\sum_{n=1}^{M}\zeta_n-\frac{N}{2}\,
\frac{\cos(\varpi)}{\sin(\frac{\pi\beta^2}{2})}
\,\bigg)\,.
\ee
Then
\be\label{asisiasC}
\bar{h}_1^{({\infty})}=\bigg(\frac{1-\beta^2}{2f}\bigg)^{1-\beta^2}
\frac{\bar{H}_1^{(+)}}{\Gamma^{2}\big(\tfrac{1-\beta^2}{2}\big)}\ ,
\ee
where $\bar{H}_1^{(+)}$ is the eigenvalue of the non-local IM 
corresponding to the second chirality. 
\bigskip

The sum rules \eqref{asisiasA} and \eqref{asisiasC} can be applied to
the Bethe roots corresponding to the
ground state of the Hamiltonian
$\mathbb{H}$ in the sector $S^z=0$ with even $N/2$. We found that 
in the scaling limit it becomes the state 
\bea
\bar{{\bf e}}_{0,0}\big(\bar{P}^{(\rm vac)}_{\rm NS}\big)\otimes{\bf e}_{0,0}\big({P}^{(\rm vac)}_{\rm NS}\big)
\eea
with
\bea\label{ioas98128932}
{P}^{(\rm vac)}_{\rm NS}=-\bar{P}^{(\rm vac)}_{\rm NS}=\frac{{\tt k}}{\sqrt{2}\beta}\ \ \ \ \ \ \ \ \ \big(\tfrac{1}{2}
\leq {\tt k}< \tfrac{1}{2}\big)\ .
\eea
The numerical data in support of the identification is presented in fig.\ref{opas9012}.
\bigskip

Consider the formula describing the scaling limit of the Hamiltonian \eqref{ENjkaskj1a} specialized to the ground state.
The extensive part of the energy is determined  by the density of the bulk roots \eqref{oisa8912093}.
A standard computation results in the expression
\be\label{gsastatass}
e_\infty=-\frac{\cos\big(\frac{\pi \beta^2}{2}\big)}{\pi \big(1-\beta^2\big)}\ 
\int_{-\infty}^\infty\frac{\rd\theta}{\cosh\big(\frac{\theta}{1-\beta^2}\big)}
\bigg(\,\frac{1}{\cosh(\theta)-\sin\big(\frac{\pi \beta^2}{2}\big)}+
\frac{1}{\cosh(\theta-2\ri\varpi)-\sin\big(\frac{\pi \beta^2}{2}\big)}\,\bigg)
\vphantom{\sum_{a\atop b}}
\ee
As was already mentioned, the Fermi velocity depends on the overall normalization of the Hamiltonian.
With the definition \eqref{Helldef1} and \eqref{aiosd901212}, it turns out that $v_{\rm F}$ does 
not depend on the inhomogeneities and is given by
\be
v_{{\rm F}}=\frac{2}{1-\beta^2}\qquad\qquad\qquad (r=2)\ .
\ee
From eqs.\,\eqref{ioas98128932} and \eqref{ENjkaskj1a}, one has the prediction
\be\label{ENjkaskjsss1a}
N/2-{\rm even},\ S^z=0\ :\ \ \lim_{N\to\infty}
\frac{N}{4\pi v_{\rm F}}\ \big({\cal E}_{\rm NS}^{(\rm vac)}-N\,e_\infty\big)= 
-\frac{1}{6}+\frac{{\tt k}^2}{\beta^2}\, \ \ \ \ \ \   \big(\tfrac{1}{2}
\leq {\tt k}< \tfrac{1}{2}\big)
\ee
The latter has been confirmed via numerical work (see fig.\,\ref{opas9012}).

\bigskip

For odd $N/2$  with $S^z=0$ the ground state is doubly degenerate.
Again, using the sum rules, one finds that the states in the scaling limit can
be identified with (some details can be found in fig.\,\ref{fig3})
\bea\label{oas09120932}
\bar{{\bf e}}^{(-)}_{2,0}\big(\bar{P}^{(\rm vac)}_{\rm NS}\big)\otimes{\bf e}_{0,0}\big({P}^{(\rm vac)}_{\rm NS}\big)\ ,\ \ \ \ \ \ 
\bar{{\bf e}}_{0,0}\big(\bar{P}^{(\rm vac)}_{\rm NS}\big)\otimes{\bf e}^{(-)}_{2,0}\big({P}^{(\rm vac)}_{\rm NS}\big)\ .
\eea 
In this case, 
\be\label{ENjkaskjsss1b}
N/2-{\rm odd},\ S^z=0\ :\ \ 
\lim_{N\to\infty}
 \frac{ N}{4\pi v_{\rm F}}\ \big({\cal E}_{\rm NS}^{\rm (vac)}-N\,e_\infty\big)=
+\frac{1}{3}+\frac{{\tt k}^2}{\beta^2}\, \ \ \ \ \  \quad \big(\,\tfrac{1}{2}
\leq {\tt k}< \tfrac{1}{2}\,\big)
\ee
There is a doublet of first excited states, whose energy in the scaling limit
is described by the same formula. Clearly, they correspond to
\bea\label{iosa9818924}
\bar{{\bf e}}^{(+)}_{2,0}\big(\bar{P}^{(\rm vac)}_{\rm NS}\big)\otimes{\bf e}_{0,0}\big({P}^{(\rm vac)}_{\rm NS}\big)\ ,
\qquad\ \ \ \ \ \ 
\bar{{\bf e}}_{0,0}\big(\bar{P}^{(\rm vac)}_{\rm NS}\big)\otimes{\bf e}^{(+)}_{2,0}\big({P}^{(\rm vac)}_{\rm NS}\big)\ .
\eea
In making the above identifications it was assumed that $0<x<1$.
In the domain $-1<x<0$, the picture is reversed. The ground states flow to \eqref{iosa9818924},
while the first excited states become \eqref{oas09120932}.
\bigskip

For even $N/2$ and $S^z=0$ the first excited state is also two-fold degenerate.
Applying the sum rules, one finds that in the scaling limit they become 
\bea\label{oias8989132}
\bar{{\bf e}}_{1,+1}\big(\bar{P}^{(\rm vac)}_{\rm R}\big)\otimes{\bf e}_{1,+1}\big({P}^{(\rm vac)}_{\rm R}\big)\,,
\qquad\qquad
\bar{{\bf e}}_{1,-1}\big(\bar{P}^{(\rm vac)}_{\rm R}\big)\otimes{\bf e}_{1,-1}\big({P}^{(\rm vac)}_{\rm R}\big)
\eea
for $-1<x<1$. Here
\bea
{P}^{(\rm vac)}_{\rm R}=-\bar{P}^{(\rm vac)}_{\rm R}=\frac{{\tt k}-\frac{1}{2}}{\sqrt{2}\beta}\ 
\ \ \ \ \ \ \ \ \ \ \ \ \ \ \ \ \big( 0\leq{\tt k}\leq 1\big)\ .
\eea
As for the energy, one has
\be\label{ENjkaskjsssss1a}
N/2-{\rm even},\ S^z=0\ :\ \ \lim_{N\to\infty}
 \frac{ N}{4\pi v_{\rm F}}\ \big({\cal E}_{\rm R}^{(\rm vac)}-N\,e_\infty\big)=
+\frac{1}{12}+\frac{({\tt k}-\frac{1}{2})^2}{\beta^2}\, \ \ \ \ \ \quad  (
0\leq {\tt k}< 1)\ .
\ee
\smallskip

When studying the scaling limit, it is apparent how to 
assign an $N$ dependence to the ground state and perhaps the first few excited
states in the disjoint sectors of the Hilbert space. These sectors, in the
case under consideration, are distinguished by the value of the quantum number
$S^z$ and the parity of $N/2$.
However, 
for a general  lattice system there are difficulties in introducing  the $N$-dependence,
i.e., the Renormalization Group  (RG) flow
for an individual  stationary state. Of course, 
 since the Hilbert space is not isomorphic for different lattice sizes,
the problem  only makes sense for the low energy part of the spectrum. 
But even then, forming   
  individual   RG flow trajectories   is not a trivial task.
For  integrable spin chains, there is a procedure for assigning an $N$ dependence 
to a Bethe state that
relies explicitly on the  Bethe ansatz equations (for a discussion see, e.g., ref.\cite{Bazhanov:2019xvyA}).

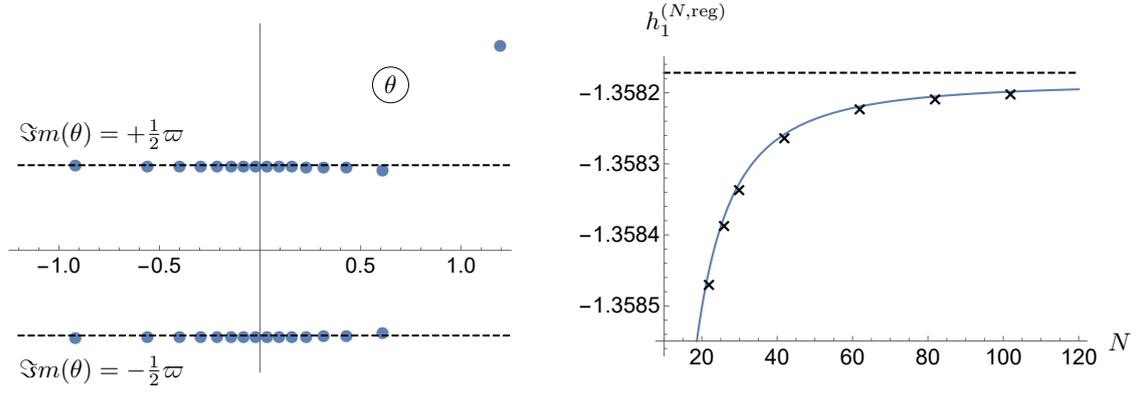
\begin{figure}
\centering
\scalebox{0.95}{
\begin{tikzpicture}
\node at (-4,0) {\includegraphics[width=7.0cm]{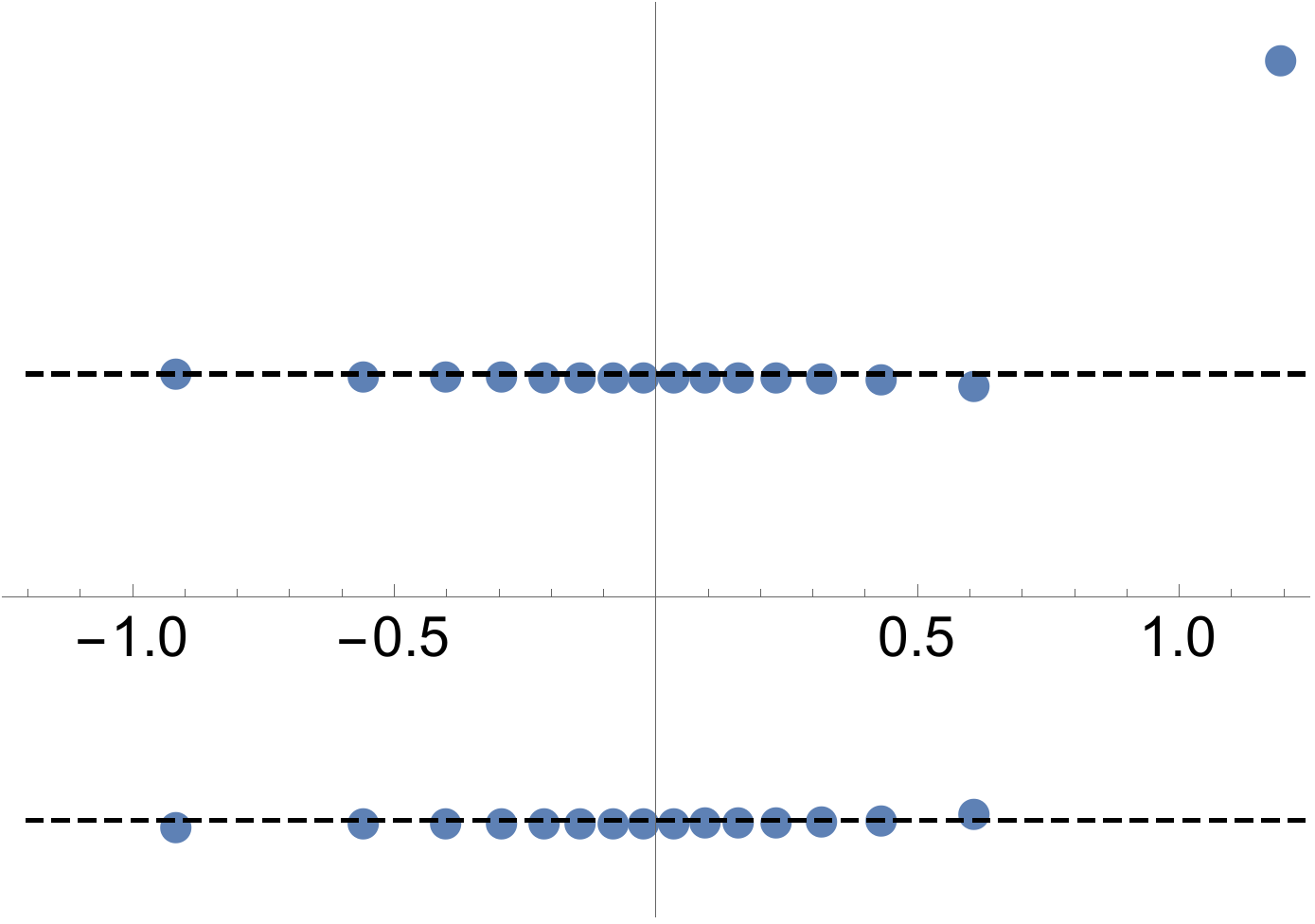}};
\node at (4,-0.2) {\includegraphics[width=7.2cm]{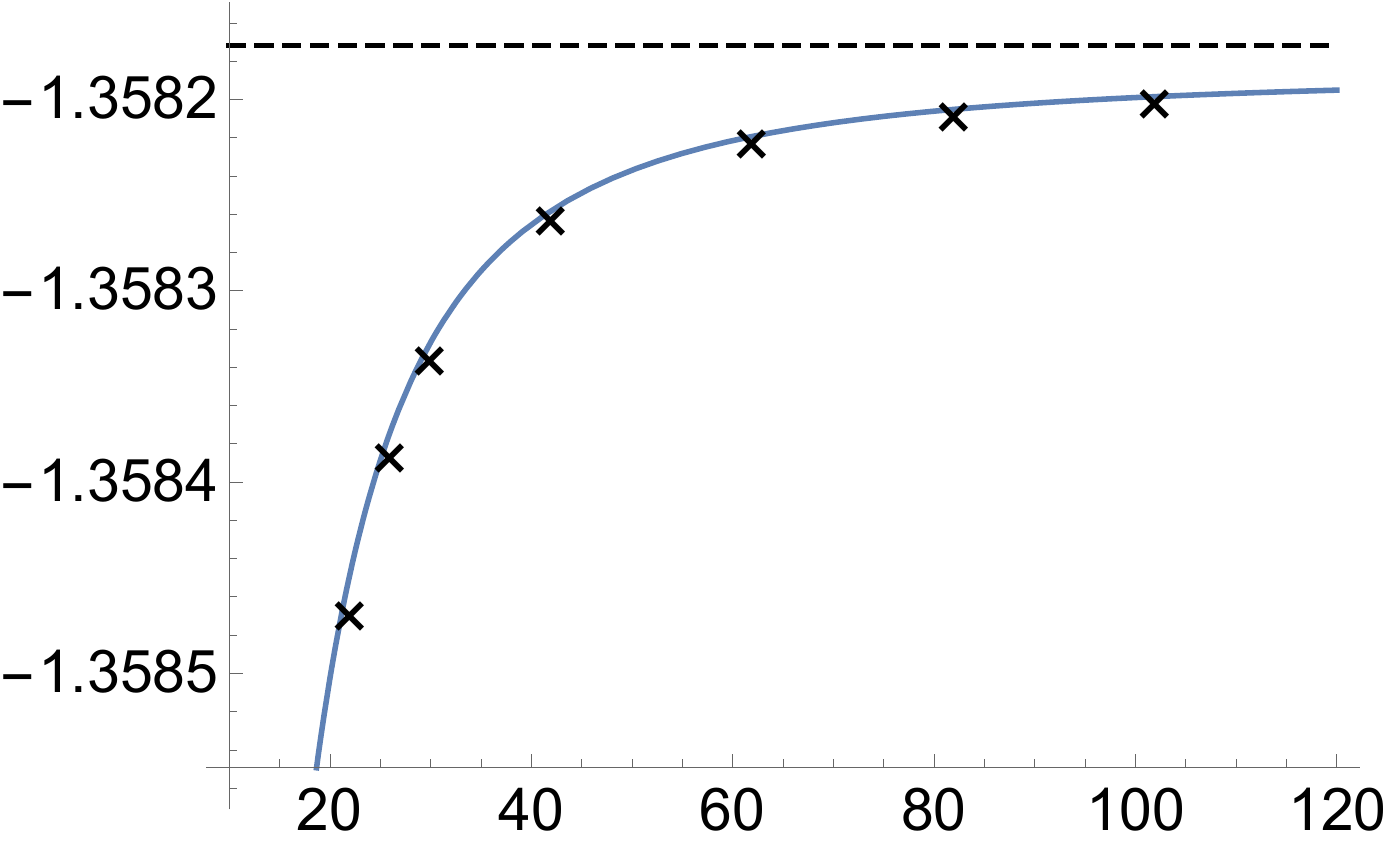}};
\node at (1.9,2.5) {\small ${h}_1^{(N,{\rm reg})}$};
\node at (-2.2,1.6) {\small $\theta$};
\draw (-2.2,1.59) circle (0.25cm);
\node at (-6.2,0.9) {\small $\Im m(\theta)=+\tfrac{1}{2}\varpi$};
\node at (-6.2,-2.4) {\small $\Im m(\theta)=-\tfrac{1}{2}\varpi$};
\node at (7.9,-2) {\small $N$};
\end{tikzpicture}
}
\caption{\small The left panel shows the typical pattern of Bethe roots 
in the complex $\theta$ plane  $\big(\theta=-\frac{1}{2}\log(\zeta)\big)$ for
the ground state of $\mathbb{H}$ with $N/2$ odd and $S^z=0$, which flows
to 
$\bar{{\bf e}}_{0,0}\big(\bar{P}^{(\rm vac)}_{\rm NS}\big)\otimes
{\bf e}^{(-)}_{2,0}\big({P}^{(\rm vac)}_{\rm NS}\big)$. 
Note that the right most root, located far above the real axis,
has $\Im m(\theta)=\frac{\pi}{2}$. 
The right panel
presents numerical data for
$h_1^{(N,{\rm reg})}=N^{-(1-\beta^2)}\big(h_1^{(N)}-\frac{N}{2}\,
\frac{\cos(\varpi)}{\sin(\pi\beta^2/2)}
\,\big)$. The  crosses were obtained from the solution
of the Bethe ansatz equations,
 while the solid blue line  represents the fit $-1.3582 - 0.12659/N^2$.
Formula \eqref{asisiasA} relates the limiting value $\lim_{N\to \infty}h_1^{(N,{\rm reg})}$
to the eigenvalue of the first  non-local IM corresponding to the state
${\bf e}^{(-)}_{2,0}\big({P}^{(\rm vac)}_{\rm NS}\big)$. This eigenvalue is
not available in analytical form. 
A numerical integration of the ODE \eqref{aois90121} corresponding  to 
${\bf e}^{(-)}_{2,0}$
leads to the prediction
$h_1^{(\infty)}=-1.35817$, which is depicted by the black dashed line.
The  parameters were set to be
$\beta^2=0.43$, ${\tt k}=0.1$ and $\varpi=1.3$.
\label{fig3}}
\end{figure}

\bigskip
In the general set-up with $S^z\ge 0$ and both even or odd $N/2$,
the low energy 
stationary states in the scaling limit  
can be classified according to the irreps of the $\overline{W}_{\bm 1}^{(c,2)}\otimes {W}_{\bm 1}^{(c,2)}$ 
algebra. They are splitted into 
the Neveu-Schwarz  and Ramond sectors.
The 
 Neveu-Schwarz states are those, which have the form
\be
\bar{ \boldsymbol e}_{\bar{\boldsymbol {\mathfrak j}}, {\mathfrak{m}},
\bar{P}_{\rm NS}}(\bar{\boldsymbol\gamma};\bar{\bm{w}})
\otimes{ \boldsymbol e}_{\boldsymbol {\mathfrak j}, {\mathfrak{m}},
P_{\rm NS}}({\boldsymbol\gamma};\bm{w})\ \  \ \ \ \  \ {\rm with}\ \ \ \  \ \ \ 
{\bar{\boldsymbol {\mathfrak j}}},{\boldsymbol {\mathfrak j}}\in\big\{(0,0), \, (\tfrac{1}{2},\tfrac{1}{2})\big\}\, ,
\quad \mathfrak{m}=0\, .
\ee
The admissible values of $P_{\rm NS}$ and ${\bar P}_{\rm NS}$ are given by
 \bea\label{89xc983}
2P_{\rm NS}
=\frac{\beta}{\sqrt{2}}\
\,S^z+\frac{\sqrt{2}}{\beta}\ \big({\tt k}+{\tt w}\big)\ ,\ \ \ \ \ \ \ 2\bar{P}_{\rm NS}=
\frac{\beta}{\sqrt{2}}\
\,S^z-\frac{\sqrt{2}}{\beta}\ \big({\tt k}+{\tt w}\big)\,,
\eea
where
\bea\label{oasoi98128932}
-{\half}\leq{\tt k}<{\half}\ ;\ \ \qquad \ {\tt w}=0,\,\pm1,\pm 2,\ldots\ .
\eea
In turn, the  Ramond states are described as 
\be
\bar{ \boldsymbol e}_{\bar{\boldsymbol {\mathfrak j}}, \mathfrak{m},
\bar{P}_{\rm R}}(\bar{\boldsymbol\gamma};\bar{\bm{w}})
\otimes{ \boldsymbol e}_{\boldsymbol {\mathfrak j}, \mathfrak{m},
P_{\rm R}}({\boldsymbol\gamma};\bm{w})\ \  \ \ \ \  \ {\rm with}\ \ \ \  \ \ \ 
{\bar{\boldsymbol {\mathfrak j}}},{\boldsymbol {\mathfrak j}}\in\big\{(\tfrac{1}{2},0), \, (0,\tfrac{1}{2})\big\}\,,
\quad \mathfrak{m}=1\,.
\ee
Here
 \bea\label{89xc983A}
2P_{\rm R}
=\frac{\beta}{\sqrt{2}}\
\,S^z+\frac{\sqrt{2}}{\beta}\ \big({\tt k}-\tfrac{1}{2}+{\tt w}\big)\ ,\ \ \ \ \ \ \ 2\bar{P}_{\rm R}=
\frac{\beta}{\sqrt{2}}\
\,S^z-\frac{\sqrt{2}}{\beta}\ \big({\tt k}-\tfrac{1}{2}+{\tt w}\big)\,,
\eea
while
\bea\label{asoi98123}
0\leq{\tt k}<1\ ;\ \ \ \ {\tt w}=0,\,\pm1,\pm 2,\ldots\ .
\eea
\smallskip

The following comment is in order here. The transfer-matrix and the Hamiltonian 
essentially depend only on the unimodular parameter $\omega^2$ \eqref{oas981298}.
Taking the logarithm results in
\be
\log(\omega^2)=2\pi\ri\,({\tt k}+{\tt w})
\ee
with ${\tt w}\in\mathbb{Z}$.
The parameter
${\tt k}$ can be brought to lie in any segment of unit length, which 
is usually referred to as the ``first Brillouin zone''. 
Formulae \eqref{oasoi98128932} and \eqref{asoi98123} mean that for 
the Neveu-Schwarz and Ramond sectors, the first Brillouin zone should be chosen
in a different way, to be
 $\big[-\frac{1}{2},\frac{1}{2}\big)$
and $[\,0,1)$,  respectively. The integer
${\tt w}$ (the winding number), which labels the  bands of the spectrum, is a 
 quantum number that appears only in the scaling limit.
It is worth reiterating that 
for  finite $N$ this quantum number  is not well defined.

\bigskip
Let $|\bm{\Psi}_N\rangle$ be a trajectory describing the RG flow of some low energy Bethe 
state. For fixed $N$ it is an eigenstate of the matrices $\mathbb{H}^{(a)}$ \eqref{Helldef1}
and we denote by ${\cal E}^{(a)}_N$ the corresponding eigenvalues.
Formula \eqref{ENjkaskj1a} implies
\bea\label{oicx98312}
\frac{N}{4\pi v_{\rm F}}\, \Big(\,{\cal E}^{(1)}_N+{\cal E}^{(2)}_N-e_\infty\,N\,\Big)\,\asymp \,
I_1^{(e)}+\bar{I}_1^{(e)}+o(1)
\eea
with $v_{\rm F}=2\,(1+\xi)$,  $e_\infty$ as in \eqref{gsastatass} and $I_1^{(e)}$
 ($\bar{I}_1^{(e)}$) stands for the
eigenvalues of  the right (left) local IM
\eqref{aoisi912898291}. In general,
\be
I_1^{(e)}=I_1^{(e,{\rm vac})}(P)+{\tt L} \,,\qquad\qquad 
\bar{I}_1^{(e)}=I_1^{(e,{\rm vac})}(\bar{P})+\bar{{\tt L}}\,,
\ee
where the vacuum eigenvalues $I_1^{(e,{\rm vac})}(P)$ correspond to the 
${W}_{\bm 1}^{(c,2)}$ primary states and
are listed in the fourth column of tab.\,\ref{tab1}, while ${\tt L}$, $\bar{\tt L}$ are non-negative
integers.
The admissible values of $P$, $\bar{P}$ are given by eqs.\,\eqref{89xc983}  and \eqref{89xc983A}
 in the Neveu-Schwarz and Ramond sectors, respectively. Notice that the relation  \eqref{oicx98312}
does not depend on the inhomogeneity parameters $\eta_a$. 
The inhomogeneous six-vertex model with $r=2$ and $\eta_1=-\eta_2=\ri$
possesses an additional global ${\cal Z}_2$ symmetry. 
This model was studied in the work \cite{IJS2}. Among other things, 
based on the modular invariance of the CFT
partition function (for ${\tt k}=0$),
the authors discuss
the gluing conditions for the right and left chiralities.
Their results  are also applicable to the more general case with the inhomogeneities as in 
eqs.\,\eqref{askjdjkA} and \eqref{askjdjkB}. 
\bigskip

Contrary to the eigenvalues of the operator $\mathbb{H}$ \eqref{aiosd901212}, i.e., 
${\cal E}^{(1)}_N+{\cal E}^{(2)}_N$ the scaling part of the eigenvalues of
each $\mathbb{H}^{(a)}$ individually
contain  an explicit dependence on the inhomogeneities.
One can obtain the asymptotic formula
\bea\label{poas9012}
\frac{N}{4\pi v_{\rm F}}\, \big(\,{\cal E}^{(1)}_N-{\cal E}^{(2)}_N\,\big)&\asymp&
\frac{\ri\, \xi(1+\xi)}{2\pi}\  f\partial_x f\, \big(\,I_1^{(e)}- \bar{I}_1^{(e)}\,\big)
\\[0.2cm]
&-&
\frac{ \xi (1+\xi)^2f^2}{2\pi \sqrt{1-x^2}}\  
\sqrt{2}\beta\, \big(\,I_1^{(o)}-\bar{I}_1^{(o)}\,\big)+o(1)\, ,\nonumber
\eea
where $\xi=\frac{\beta^2}{1-\beta^2}$ and $f=f(x)$ is given by  \eqref{aois89128912}. Note that
\bea
 f(0)=\frac{\sqrt{\pi}}{1+\xi}\ \frac{\Gamma(\frac{\xi}{2})}{\Gamma(\frac{1+\xi}{2})}\ ,\ \ \ \ \ \ \ \ \ \qquad
  \ \partial_xf(0)=0\ .
\eea
An illustration for the Bethe states, which flow
 to the tensor product of the Ramond primary states \eqref{oias8989132}, is provided 
in fig.\,\ref{fig4}.

\begin{figure}
\centering
\scalebox{0.95}{
\begin{tikzpicture}
\node at (-3,1.1) {\small $\theta$};
\draw (-3,1.09) circle (0.25cm);
\node at (-6.2,2.35) {\small $\Im m(\theta)=+\tfrac{1}{2}\varpi$};
\node at (-6.2,-1.5) {\small $\Im m(\theta)=-\tfrac{1}{2}\varpi$};
\node at (-4,0) {\includegraphics[width=7cm]{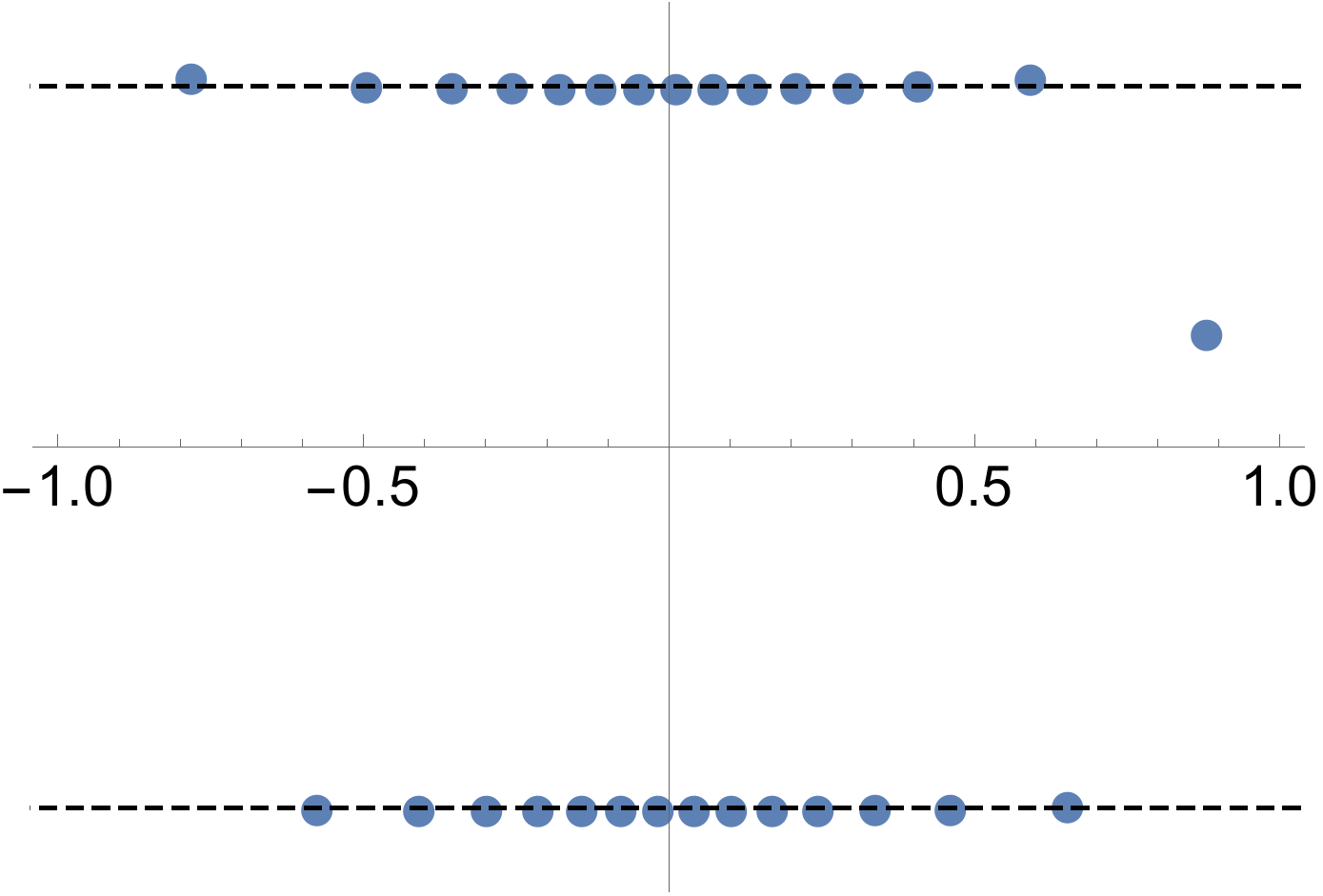}};
\node at (4,0)  {\includegraphics[width=7cm]{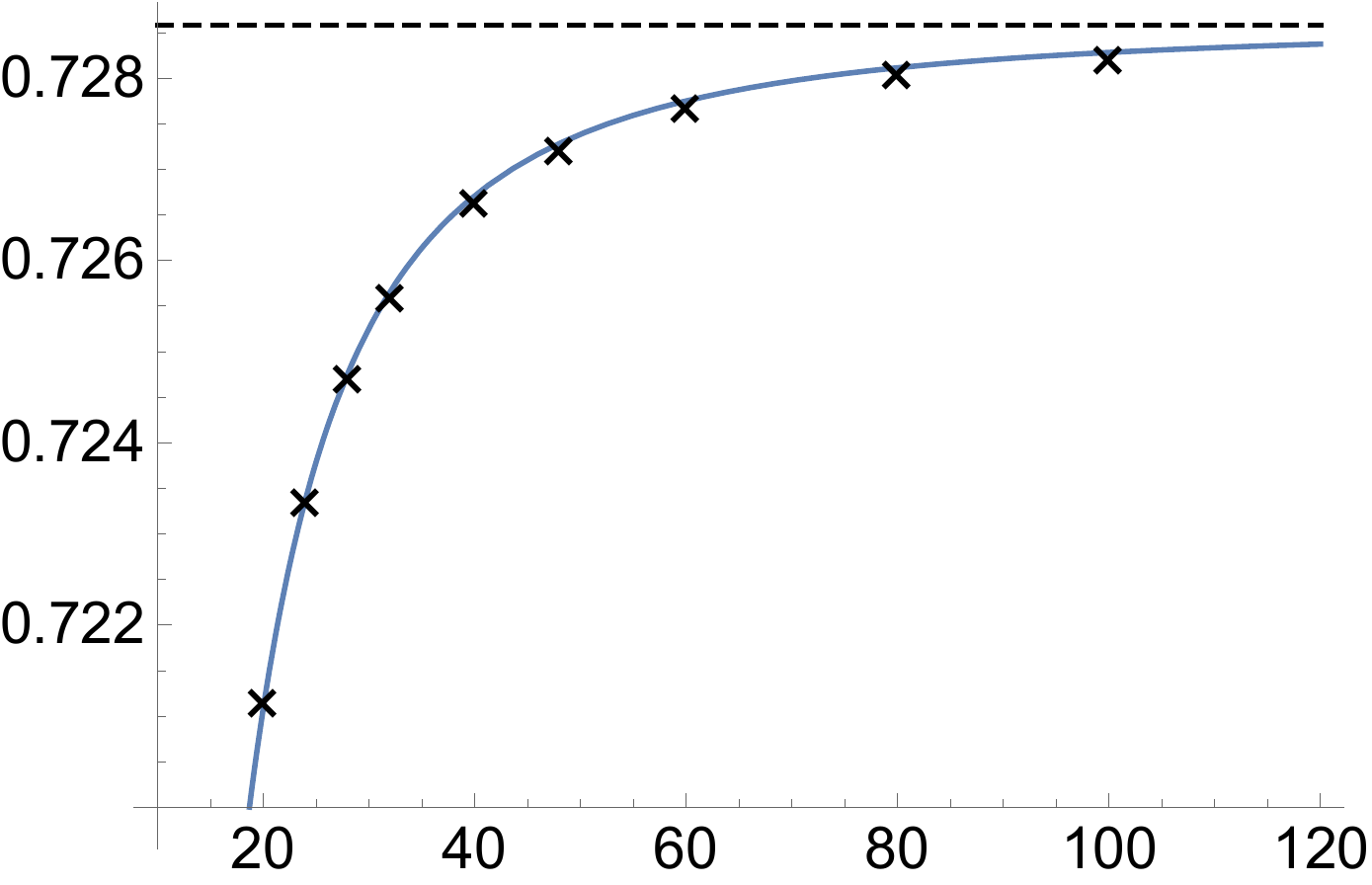}};
\node at (7.8,-1.9) {\small $N$};
\node at (4,2.55) {\small $\tfrac{N}{4\pi v_{\rm F}}\big({\cal E}^{(1)}_N - {\cal E}^{(2)}_N\big)$};
\end{tikzpicture}
}
\caption{\small
The first excited states of the Hamiltonian $\mathbb{H}$ with $N/2$ even and $S^z=0$
form a doublet. The Bethe roots for the states in the doublet are related to each other
via complex conjugation. Depicted in the left panel of the figure is the typical pattern of Bethe
roots in the complex $\theta$ plane
for the Bethe state, which flows to $\bar{{\bf e}}_{1,+1}\big(\bar{P}^{(\rm vac)}_{\rm R}\big)\otimes
{\bf e}_{1,+1}\big({P}^{(\rm vac)}_{\rm R}\big)$. The right panel compares the numerical data
for the difference ${\cal E}^{(1)}_N-{\cal E}^{(2)}_N$ with the prediction  \eqref{poas9012}. The crosses
 were computed via the Bethe ansatz, while the solid blue line is the fit
$0.72858-2.9919/N^2$. The dashed black line represents the r.h.s. of eq.\,\eqref{poas9012}.
Note that, in the case at hand, $I_1^{(e)}-\bar{I}_1^{(e)}$ is zero, while the eigenvalue of
${\bf I}_1^{(o)}$ on the Ramond primary states follows from eq.\,\eqref{aiuususausa}
and the last column of tab.\ref{tab1}. The parameters were set to be
$\beta^2=0.43$, ${\tt k}=0.1$ and $\varpi=1.8$, so that 
$P_{\rm R}^{{\rm vac})}=-\bar{P}_{\rm R}^{({\rm vac})}=-0.43133$.
\label{fig4}}
\end{figure}
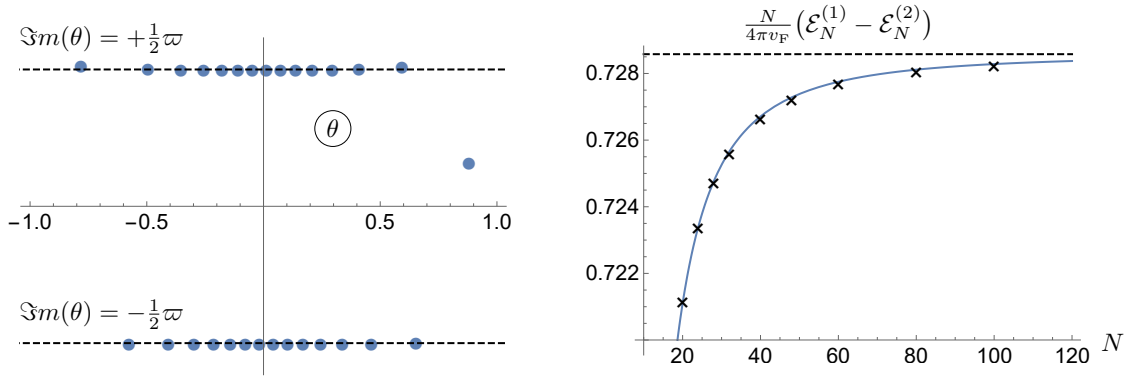

\bigskip

To determine the scaling limit of $|\bm{\Psi}_N\rangle$ one could employ formulae 
\eqref{oicx98312} and \eqref{poas9012}.
However, unlike the sum rules, they contain the contribution of
the eigenvalues of both right and left IM ${\bf I}_1^{(e,o)}$ and $\bar{{\bf I}}_1^{(e,o)}$.
Nevertheless, their computation is a much
simpler task than that of the eigenvalues of the non-local IM. 
Perhaps the most convenient way of identifying the scaling limit
of an RG trajectory involves the reflection operator. 
Its eigenvalues admit a closed expression \eqref{usaussuasa},
while the construction of the eigenstates is a straightforward algebraic procedure. 
On the other hand, the eigenvalues of the left and right reflection operators appear
separately in the large $N$ limit of certain products over the Bethe roots. 
Some related discussion can be found in sec.\,11 of ref.\cite{Bazhanov:2019xvyA}.\footnote{%
To avoid confusion, let us emphasize that the work 
\cite{Bazhanov:2019xvyA} considered the inhomogeneous six-vertex model
for $r=2$,
where $\eta_1=-\eta_2=\ri$, but in the domain of the anisotropy parameter
${\rm arg}(q)\in(0,\frac{\pi}{2})$. This is complementary to \eqref{oiasio98129831}
with $K=2$, i.e., ${\rm arg}(q)\in(\frac{\pi}{2},\pi)$.
}

\section{Outlook}
Among the  lessons we took away from carrying out this project
is that the generalized affine Gaudin model fits within the standard framework
of Yang-Baxter integrability. As a result it can be studied using 
conventional techniques such as  the quantum inverse scattering method
and its variants. This has exciting implications, since the model
shares  the same integrability  hierarchy as
 a variety of physically interesting systems.
The latter includes a large class of classically
 integrable Non-Linear Sigma Models (NLSM). For example, in refs.\cite{Vicedo:2017cge,Delduc:2018hty,Delduc:2019bcl}
the  classical affine Gaudin model  \eqref{ioas98128932a}
and its generalizations to Lie algebras of higher rank are used in
assembling  integrable NLSM.
The results of our work provide an avenue for the systematic
quantization of such theories within the ODE/IQFT correspondence,
in the spirit of ref.\cite{Bazhanov:2017nzh}.
Another application lies in the domain of Condensed Matter Physics. 
The GAGM  may be understood as an 
integrable multiparametric generalization of the 
Kondo model.
Finally, we hope that the output of this work would be useful in the context
of the representation theory of infinite dimensional algebras.

 \section*{Acknowledgment}

The authors are grateful to  V.~V. Bazhanov for collaboration at an earlier stage of
this work. They also  thank S.~Lacroix, J.~Teschner and B.~Vicedo for fruitful discussions.

\medskip
\noindent
The research of GK is funded by the Deutsche Forschungsgemeinschaft (DFG, German Research
Foundation) under Germany's Excellence Strategy -- EXC 2121 ``Quantum Universe'' -- 390833306.
Part of this work was carried out during GK's stay
at the Institut des Hautes \'{E}tudes Scientifiques. He would like to thank that institute for its support
and hospitality.

\medskip
\noindent
The research of SL is supported by the 
Rutgers New High
Energy Theory Center.

 \appendix
 
 \section{Appendix: leading  asymptotic of 
\texorpdfstring{$D_{{ \bm{\mathfrak j}}, \mathfrak{m}, A}$}{} for $\mu\to+\infty$\label{AppA}}

A few technical assumptions   are made
in order to simplify the calculation.
Namely,  that $(i)$ both $\mu$ and $\kappa=\mu^{\frac{1}{2}(1+\xi)K}$  
 are real and positive, $(ii)$ all $z_{a}\not\in(0,+\infty)$ and  $(iii)$
 $0<(1+\xi)K<2$. Later we'll discuss how the third  assumption can be relaxed.

\bigskip
Suppose that $\mu$ or, equivalently, $\kappa$
is small.
As long as  the term 
$\kappa^2{\cal P}$
can be neglected, the ODE  \eqref{aois90121AA} takes the form $\big(-\partial^2_z+t_{\tt L}(z)\big)\Psi=0$ and the solution
$\Psi^{\scriptscriptstyle{(\leftarrow)}}_{A}$ is  approximated as
\bea\label{iiiasis}
\Psi^{\scriptscriptstyle{(\leftarrow)}}_{A}(z)\approx C\  z^{\frac{1}{2}-\ri A}\ 
\frac{\prod_{m=1}^{{\tt M}_+}(z-x^{\scriptscriptstyle{(+)}}_m)}
{\prod_{a=1}^r  
           (z-z_a)^{\mathfrak{j}_a}\prod_{\alpha=1}^{\tt L}(w-w_\alpha)}\ \ \ \ \ \ \ \ \ \ \ \ \  \ \ \
            \big(\, |\kappa^2{\cal P}|\ll1\,\big)\,.
         \eea 
To ensure the asymptotic condition $\Psi^{\scriptscriptstyle{(\leftarrow)}}_{A}(z)\to z^{\frac{1}{2}-\ri A}$ for $z\to 0$, the constant $C$
 should be set to be
 \bea
 C=\frac{\prod_{a=1}^r  
           (-z_a)^{\mathfrak{j}_a}\prod_{\alpha=1}^{\tt L}(-w_\alpha)}
           {\prod_{m=1}^{{\tt M}_+}(-x^{\scriptscriptstyle{(+)}}_m)}\ .
 \eea
For $\kappa\ll 1$, 
there is a region with $|z|$ large, where
  the condition $|\kappa^2{\cal P}|\ll1$ is still satisfied.  In that domain \eqref{iiiasis} becomes
 \bea\label{iisis}
\Psi^{\scriptscriptstyle{(\leftarrow)}}_{A}(z)\approx C\  z^{\frac{1}{2}-\ri A-\frac{1}{2}\, {\mathfrak m}}\ 
\ \ \ \ \ \ \ \ \ \ \ \ \  \ \
           (1\ll |z|\ll\mu^{-1}) \ .    
         \eea 
 Here we use 
 $\mathfrak{m}=2({\mathfrak  J}+{\tt N}-{\tt M}_+)$ (see eqs.\,\eqref{ioas19829A} and \eqref{ioas19829BA}). 
On the other hand, taking into account \eqref{kausy}, for large $z$ one has
 \bea
 t_{\tt L}(z)=\frac{1}{4z^2}\ \big( 2\ri A+\mathfrak{m})^2-1\big)+O\big(z^{-3}\big)
 \eea
 so that the ODE  \eqref{aois90121AA} can be replaced by
 \bea
 \Big(-\partial^2_z+\frac{1}{4z^2}\ \big( (2\ri A+\mathfrak{m})^2-1\big)+\kappa^2\ z^{-2+(1+\xi)K}
 \Big)\ \Psi=0\ \ \ \ \ \ \ \ \  \ \  ( |z|\gg 1)\ .
 \eea
 The latter is a variant  of the Bessel equation. The solution 
 \eqref{iisis} can be continued to the whole domain $|z|\gg 1$:
 \bea
 \Psi^{\scriptscriptstyle{(\leftarrow)}}_{A}(z)\approx \sqrt{\frac{\pi}{(1+\xi)K}}\ \, 
   W_{{\boldsymbol {\mathfrak j}}, \mathfrak{m}, A}(0)\ \, \mu^{\ri  A+\frac{1}{2}\mathfrak{m}}\ z^{\frac{1}{2}}\ 
 I_{\nu}\Big(\tfrac{2\kappa}{(1+\xi)K}\ z^{\frac{1}{2}(1+\xi)K} \Big)\ ,
  \eea
where we use the notation
\bea
W_{{\boldsymbol {\mathfrak j}}, \mathfrak{m}, A}(0)\equiv
C\ \sqrt{\frac{(1+\xi)K}{\pi}}\ \ \Gamma\big(1+\nu)\ \big((1+\xi)K\big)^{\nu}
\ , \ \ \ \ \ \ \  \ \  \ \ \nu=-\frac{2\ri  A+\mathfrak{m}}{(1+\xi)K}\ .
\eea
Keeping in mind the asymptotic behaviour of the modified Bessel function at large argument,
$I_\nu(x)\asymp \re^{x}\big/\sqrt{2\pi x}$, one finds
\be
\Psi^{\scriptscriptstyle{(\leftarrow)}}_{A} (z)\asymp \frac{1}{2\sqrt{\kappa}}\ W_{{\boldsymbol {\mathfrak j}}, \mathfrak{m}, A}(0) \,
\mu^{\ri  A+\frac{1}{2}\mathfrak{m}}\  \ z^{\frac{1}{2}-\frac{1}{4} K(1+\xi)}\ 
  \exp\Big(+\tfrac{2\kappa}{(1+\xi)K}
 \ z^{\frac{1}{2}(1+\xi) K}
 \Big)\ \ \  \  {\rm as}\ \ \ z\to+\infty\,.
 \ee
 Recall that the other solution $\Psi^{\scriptscriptstyle{(\rightarrow)}} (z)$ was defined through the condition\footnote{%
The  assumption $0< (1+\xi)K <2$ is being used here.}
 \bea
 \Psi^{\scriptscriptstyle{(\rightarrow)}} (z)\asymp \frac{1}{\sqrt{\kappa}}
 \ z^{\frac{1}{2}-\frac{1}{4} (1+\xi)K}\ 
  \exp\Big(-\tfrac{2\kappa}{(1+\xi)K}
 \ z^{\frac{1}{2}(1+\xi)K}
 \Big)\ \ \  \ \ {\rm as}\ \ \ z\to+\infty\, .
 \eea
 Calculating the Wronskian  of  these two solutions at $z\to+\infty$ one arrives at the relation
 \bea\label{oi98398239823A}
 \lim_{\mu\to 0^{+}}\, \mu^{- \ri A-\frac{1}{2}\mathfrak{M}}\
{\rm Wron} \big[\,\Psi^{\scriptscriptstyle{(\rightarrow)}}, \Psi^{\scriptscriptstyle{(\leftarrow)}}_{A}\,\big]=
W_{{\boldsymbol {\mathfrak j}}, \mathfrak{m}, A}(0)
\eea
with
\be\label{oi98398239823B}
W_{{\boldsymbol {\mathfrak j}}, \mathfrak{m}, A}(0)=
\frac{(-1)^{\frac{1}{2} \mathfrak{m}}}{\sqrt{\pi}}\  \big((1+\xi)K\big)^{\frac{1}{2}-\frac{2\ri  A+ \mathfrak{m}}{(1+\xi)K}}\,
 \Gamma\big(1-\tfrac{2\ri  A+ \mathfrak{m}}{(1+\xi)K})\ \frac{\prod_{a=1}^r  
           (z_a)^{\mathfrak{j}_a}\prod_{\alpha=1}^{\tt L}w_\alpha}
           {\prod_{m=1}^{{\tt M}_+}x^{\scriptscriptstyle{(+)}}_m}\ .
\ee
A similar formula can be obtained for $W_{{\boldsymbol {\mathfrak j}}, -\mathfrak{m}, -A}(0)$.

\bigskip
Let's turn to the large $\mu$ behaviour of 
${\rm Wron} \big[\,\Psi^{\scriptscriptstyle{(\rightarrow)}}, \Psi^{\scriptscriptstyle{(\leftarrow)}}_{A}\,\big]$.
 With the Langer correction taken into account,
the WKB approximation for  $\Psi^{\scriptscriptstyle{(\rightarrow)}}$ reads as 
\be\label{WKBjkas12A}
\Psi^{\scriptscriptstyle{(\rightarrow)}}\asymp  \frac{1}{\sqrt{\kappa}}\ 
\big({\cal S}(z)\big)^{-\frac{1}{4}}
  \exp\bigg[-\frac{2\kappa}{(1+\xi)K}
 \ z^{\frac{1}{2}(1+\xi)K}  
+\kappa\int_{z}^{+\infty}\rd z\ \Big(\, \sqrt{{\cal S}(z)}- z^{\frac{1}{2} (1+\xi)K  -1}\, \Big)\bigg]\,,
\vphantom{\sum_{a\atop b}}
\ee
where
\bea
{\cal S}(z)={\cal P}(z)+
\kappa^{-2}\ \Big(t_{\tt L}(z)+\frac{1}{4z^2}\Big)\, .
\eea
The other solution $\Psi^{\scriptscriptstyle{(\leftarrow)}}_A$  is approximated by
\be\label{WKBjkas12}
\Psi^{\scriptscriptstyle{(\leftarrow)}}_{A} (z)\approx 
\ \frac{\tilde{C}}{\sqrt{\kappa}}\ z^{-\ri A}
\,\big({\cal S}(z)\big)^{-\frac{1}{4}}
  \exp\bigg[\int_{0}^{z}\rd z\ \Big(\, \kappa\sqrt{{\cal S}(z)}-\, \frac{\ri A}{z}\Big)\bigg]
\ee
with some constant $\tilde{C}$. 
Computing the Wronskian of
the  solutions $\Psi^{\scriptscriptstyle{(\rightarrow)}}$ and $\Psi^{\scriptscriptstyle{(\leftarrow)}}_A$
at $z\to 0$, yields 
\bea\label{oias982323}
\kappa^{-1}\log\bigg(
\frac{{\rm Wron}[\,\Psi^{\scriptscriptstyle{(\rightarrow)}}, \Psi^{\scriptscriptstyle{(\leftarrow)}}_{A}\,]}{
2\tilde{C}}\bigg)
&\asymp&
 \lim_{\varepsilon\to 0^+} \bigg[
-  \frac{\ri A}{\kappa}\, \log(\varepsilon)
\\[0.2cm]
&+&\int_{\varepsilon}^{+\infty}\rd z\ \Big(\, \sqrt{{\cal S}(z)}-
 z^{\frac{1}{2} (1+\xi)K  -1}\, \Big)\bigg]+O(\kappa^{-2})\nonumber
\eea
In order to determine $\tilde{C}$ we note that for large $\mu$ and
$|z|\ll1$, the function $\kappa^2{\cal P}(z)$ can be replaced by
$\kappa^2\, z^{-2+\xi K}$  (here we use the convention $\prod_{a=1}^r(-z_a)^{k_a}=1$,
see \eqref{concj128921}). In turn, the
ODE \eqref{aois90121AA} becomes
\bea
 \Big(-\partial^2_z-\frac{1}{4z^2}\ \big( 4A^2+1\big)+\kappa^2\ z^{-2+\xi K}
 \Big)\ \Psi=0\ \ \ \ \ \ \ \ \  \ \  (\, |z|\ll 1\,)\ .
 \eea
Suppose that $\Im m(A)>0$. Then
\be\label{ioas89128921}
\Psi^{\scriptscriptstyle{(\leftarrow)}}_{A} (z)\approx \Gamma(1+\rho)\ 
\big(\xi K/\kappa\big)^{\rho}
\ \ z^{\frac{1}{2}}\ 
 I_{\rho}\big(\,\tfrac{2\kappa}{\xi K}\ z^{\frac{1}{2}\xi K}\, \big)\qquad\qquad {\rm with}\qquad
\rho=-\tfrac{2\ri  A}{\xi K}\ .
\ee
In the domain where the argument of the Bessel function is large
\be
\Psi^{\scriptscriptstyle{(\leftarrow)}}_{A} (z)\approx \frac{\Gamma(1-\tfrac{2\ri  A}{\xi K}) }{2\sqrt{\pi}}\,
\bigg(\frac{\kappa}{\xi K}\bigg)^{\frac{2\ri  A}{\xi K}-\frac{1}{2}}\,
z^{\frac{1}{4}(2-\xi K)}\,\exp\bigg(\frac{2\kappa}{ \xi K}\,z^{\frac{1}{2}\xi K}\bigg)\,\qquad
\big(\mu^{-\frac{1+\xi}{\xi}}\ll z\ll1\big) \, .
\ee
Comparing the above equation with 
the WKB formula \eqref{WKBjkas12}, one finds 
\be
\tilde{C}\,=\,\frac{  \mu^{\ri \frac{\xi+1}{\xi} A}}{2\sqrt{\pi}}\ 
 \big(\xi K\big)^{\frac{1}{2}-\frac{2\ri  A}{\xi K}}
\  \Gamma\big(1-\tfrac{2\ri  A}{\xi K}\big)\, .
\ee
The relations \eqref{oias982323},\,\eqref{oi98398239823A},\,\eqref{oi98398239823B},
combined
with the definition of $D_{{\boldsymbol {\mathfrak j}}, \mathfrak{m}, A}(\mu)$ \eqref{iissisia}, yields the
large $\mu$ asymptotic formula 
\be\label{oias98138923ss}
D_{{\boldsymbol {\mathfrak j}}, \mathfrak{m}, A}
\asymp
R_{{\boldsymbol {\mathfrak j}}, \mathfrak{m}, A}
\ \mu^{\frac{\ri A}{\xi}-\frac{1}{2}\mathfrak{m}}\ 
\ \exp\Big(\mu^{\frac{1}{2}(1+\xi)K}\ q_{-1}+o(1)\Big)\ \ \ \ \ \ \qquad\ (\mu\to+\infty)\ ,
\ee
where
\bea\label{iasisaisa}
q_{-1}=\int_0^\infty\rd z\, \Big(\,\sqrt{{\cal P}(z)}-z^{\frac{1}{2}(1+\xi)K-1}\,\Big)
\eea
and
 $R_{{\boldsymbol {\mathfrak j}}, \mathfrak{m}, A}$ is given by \eqref{usaussuasa}.

\bigskip
The asymptotic coefficient $q_{-1}$  is expressed in terms of an
integral, which converges only for $0<(1+\xi) K<2$. Keeping in mind that
 $K$ is a positive integer,
it can be literally applied  to the case $K=1$ and $0<\xi<1$.
Nevertheless, the
 integral for $q_{-1}$ can be analytically continued from the domain of convergency. This is done
by replacing the integration over the positive real  semiaxis by  that over a certain  contour integral. 
To  illustrate the procedure, consider the case $r=1$ when
\bea
{\cal P}(z)=z^{-2+\xi K}(1+z)^K\ .
\eea
Then for $\xi K>0,\ \xi(K+1)<2$ the integral \eqref{iasisaisa} converges and coincides with the
Euler beta function
\be\label{isaisisi}
q_{-1}=\int_0^\infty\rd z\, \Big(\,\sqrt{{\cal P}(z)}-z^{\frac{1}{2}(1+\xi)K-1}\,\Big)=\frac{\Gamma\big(\frac{\xi K}{2}\big)\,
\Gamma\big(-\frac{(1+\xi) K}{2}\big)}{\Gamma\big(-\frac{K}{2}\big)}\ \ \ \ \qquad (r=1)\  .
\ee
The latter admits a well known integral representation,
\bea\label{opopsapospa}
\oint_{C_\zeta}\rd \zeta\ \zeta^{\alpha-1}(1-\zeta)^{\beta-1}=
\big(1-\re^{2\pi\ri\alpha}\big)\, \big(1-\re^{2\pi\ri\beta}\big)\ \frac{\Gamma(\alpha)\Gamma(\beta)}{\Gamma(\alpha+\beta)}\ ,
\eea
where
 the integration contour $C_\zeta$ 
is the Pochhammer loop  shown in fig.\ref{figa1}.
\begin{figure}
\centering
\includegraphics[width=7.cm]{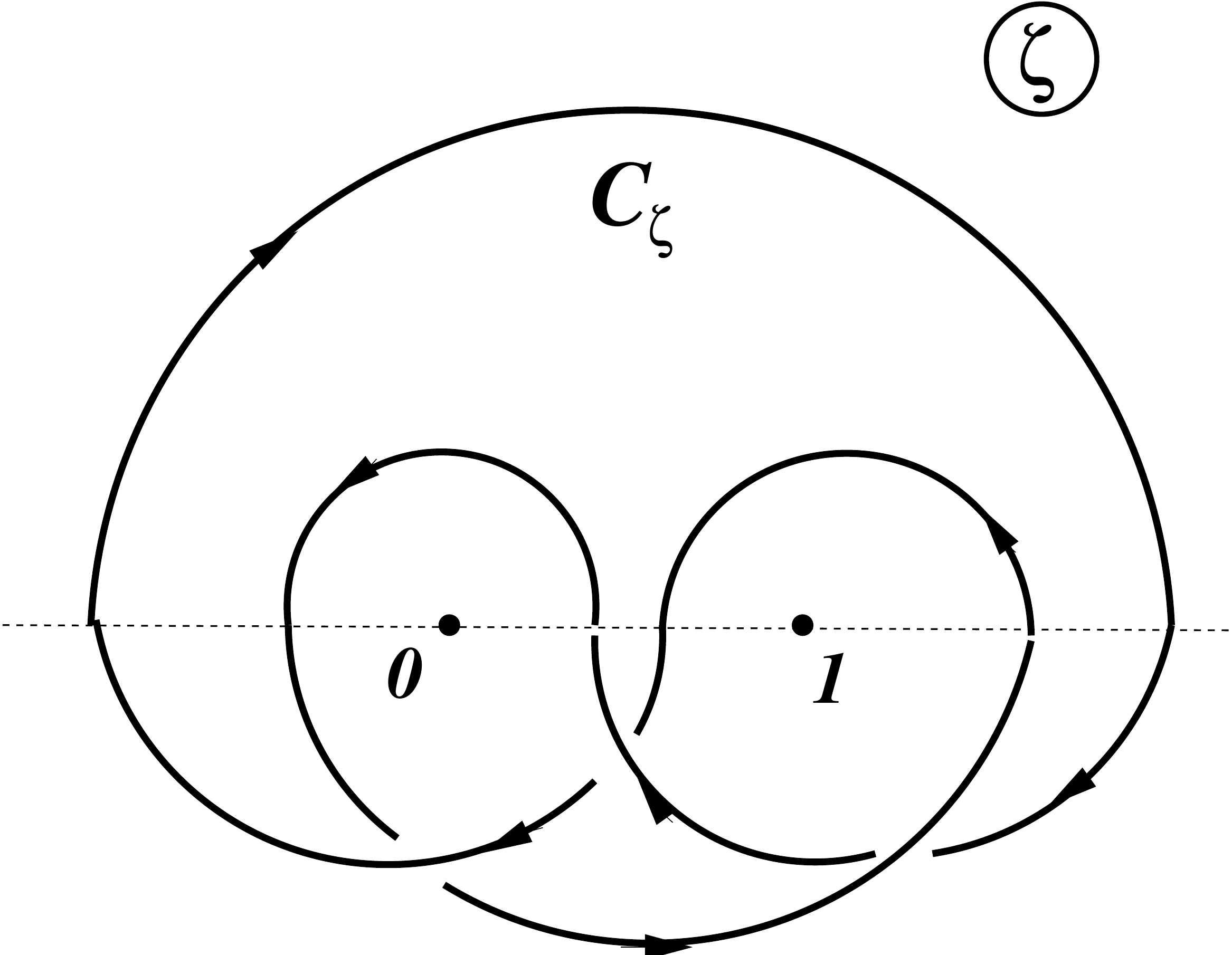}
\caption{\small  The Pochhammer loop $C_\zeta$ appearing in \eqref{opopsapospa}.
\label{figa1}}
 \end{figure}
 The branch of the integrand in \eqref{opopsapospa} should be chosen in such a way that it is real and positive
within the segment $\zeta\in(0,1)$.
Performing the change of variables,
\bea\label{usausausa}
\zeta\mapsto z=\frac{\zeta}{1-\zeta}\ ,
\eea
the coefficient $q_{-1}$ in \eqref{isaisisi} may be represented as
\bea\label{jasassay}
q_{-1}=\frac{1}{(1-\re^{2\pi\ri \nu_\infty})\ 
(1-\re^{2\pi\ri \nu_0})}\ 
\oint_{{\cal C}}\rd z\, \sqrt{{\cal P}(z)}
\eea
with
\bea\label{ksssasaysay}
\nu_0=\tfrac{1}{2}\, \xi K-1\ ,\ \ \ \  \qquad\ \nu_\infty=1-\tfrac{1}{2}\, (1+\xi) K\ .
\eea
The integration contour ${\cal C}$ is the image of $C_\zeta$ under the 
conformal map \eqref{usausausa}.
Alternatively it can be described as follows.
Consider the Riemann sphere under stereographic projection
such that  the  south and north poles are mapped to 
$z=0$ and $z=\infty$,  respectively.
Then ${\cal C}$
 is the image of
a Pochhammer  loop on the sphere
that
winds around the  two poles.
The homotopy class of that loop is schematically depicted in fig.\,\ref{figa2}.
Notice  that
$\nu_0$ and $\nu_\infty$ are the exponents, which 
 appear in the leading behaviour of $\sqrt{{\cal P}(z)}$ at $z=0$ and $z=\infty$:
\bea
\sqrt{{\cal P}(z)}\big|_{z\to 0}\sim z^{\nu_0}\ ,\ \ \ \ \ \ \ \qquad
 \sqrt{{\cal P}(z)}\big|_{z\to \infty}\sim z^{-\nu_\infty}\, .
\eea

\bigskip
For the general case with $r>1$, the analytic continuation of $q_{-1}$  is given 
by equations \eqref{jasassay} and \eqref{ksssasaysay}. However 
one should take care that the integration contour ${\cal C}$ does not 
wind around  the points $z=z_{a}$.
This can be achieved by taking ${\cal C}$ to lie inside the domain ${\cal D}$,  which is the
union of $|z|>\varepsilon^{-1}$, $|z|<\varepsilon$
and the wedge $|\arg(z)|<\delta$  such that  $z_a\not\in {\cal D}$.
Also in the case $r>1$ the
branch of $\sqrt{{\cal P}(z)}$ can not be chosen to be real for positive real $z$. Instead  it
is sufficient to require that $\sqrt{{\cal P}(z)}$ asymptotically approaches to  positive real values as $z\to+\infty$.
Under the above conditions formulae \eqref{jasassay},\,\eqref{ksssasaysay} provide an
analytical continuation of the asymptotic coefficient  $q_{-1}$.
 In turn, \eqref{oias98138923ss} becomes applicable for $\xi>0 $ and any
positive integers $k_a$, provided that
\bea
\xi K,\  (\xi+1)K\not\in 2\mathbb{Z}\ \qquad\qquad\qquad\qquad
\Big(K=\sum_{a=1}^r k_a\Big)\  .
\eea

 \section{Appendix: large $|\mu|$ asymptotic for  $r=2,\,k_1=k_2=1$\label{hagf}}
\subsection*{Leading asymptotic}
The asymptotic behaviour of the connection
coefficient, as described by \eqref{oias98138923ss},
assumes that $\mu$ is a large,  positive real number. 
A full description  of the large $|\mu|$ asymptotic  in the complex plane for generic values
of the complex parameters $z_a$ is more difficult to obtain.
Assuming  certain reality conditions imposed on $z_1$ and $z_2$,
we describe it here for 
the case  $r=2,\,k_1=k_2=1$.

\bigskip
Adopting the conventions \eqref{concj128921}, so that $z_1z_2=1$,
introduce
 \bea
 x=\tfrac{1}{2}\, (z_1+z_2)\ .
 \eea
Then
 \bea
 {\cal P}(z)=z^{2 (\xi-1)}\ \big(z^2-2x z+1\big)\ .
 \eea
 The asymptotic coefficient $q_{-1}$, being considered as a function
 of $x$,  can be expressed in terms of the Gauss
 hypergeometric function (in fact the Legendre function $P^{-1}_{-1-\xi}$)
 \be\label{uuasuasusa}
q_{-1}(x)=
\frac{ \pi}{\sin(\pi\xi)}\  (1+x)\ 
{}_2F_{1}\big(1+\xi, -\xi, 2, \tfrac{1}{2}\, (1+x)\big)\ \ \ \ \ \ \ \ \ \ \ \ \ \ \ (\xi\not= 1,2,\ldots)\ .
\ee

A breakdown of 
the leading asymptotic formula for large complex $\mu$
 is related to the lines along which the zeroes of the entire function
 $D_{{\boldsymbol {\mathfrak j}}, \mathfrak{m}, A}(\mu)$   accumulate asymptotically.
 In the case under consideration the positions of the  Stokes  lines significantly depend on the
domain of  $x$. 
Below we take $x^2$ to be a  real number.

 \bigskip
 For $-1<x<1$ (equivalently $|z_{1,2}|=1$),
 the zeroes of  $D_{{\boldsymbol {\mathfrak j}}, \mathfrak{m}, A}(\mu)$
 asymptotically approach the  two rays
 \bea\label{gsasatast}
 \arg(\mu)=\pm \alpha_0\ ,
 \eea
where  $\alpha_0=\alpha_0(x)$  is some monotonically
decreasing function of $x$ such that 
$\frac{\pi}{2(1+\xi)}<\alpha_0(x)\leq \frac{\pi}{2}$ for $0\le x <1$, while
  \bea
  \alpha_0(-x)=\pi- \alpha_0(x)\, .
 \eea
 The rays \eqref{gsasatast} divide the complex $\mu$ plane 
 into two wedges (see left panel of  fig.\ref{figa3}).
 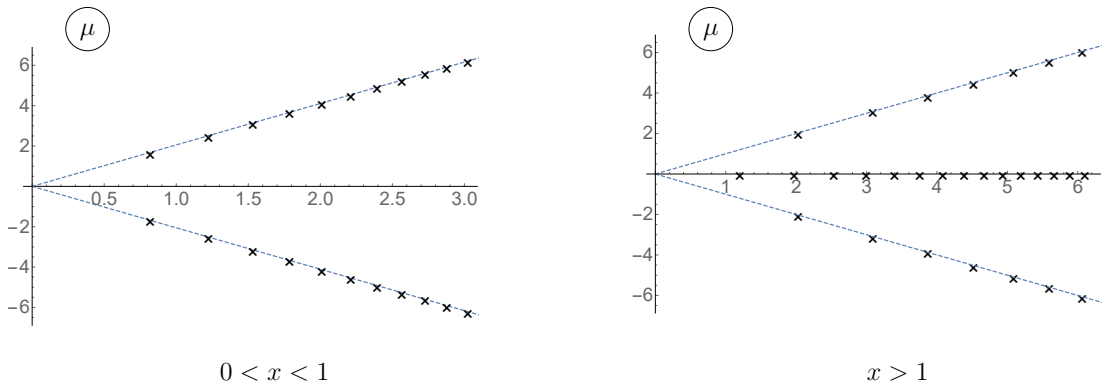
\begin{figure}
\centering
\scalebox{0.82}{
\begin{tikzpicture}
\node at (-5.5,0) {\includegraphics[width=0.5\textwidth]{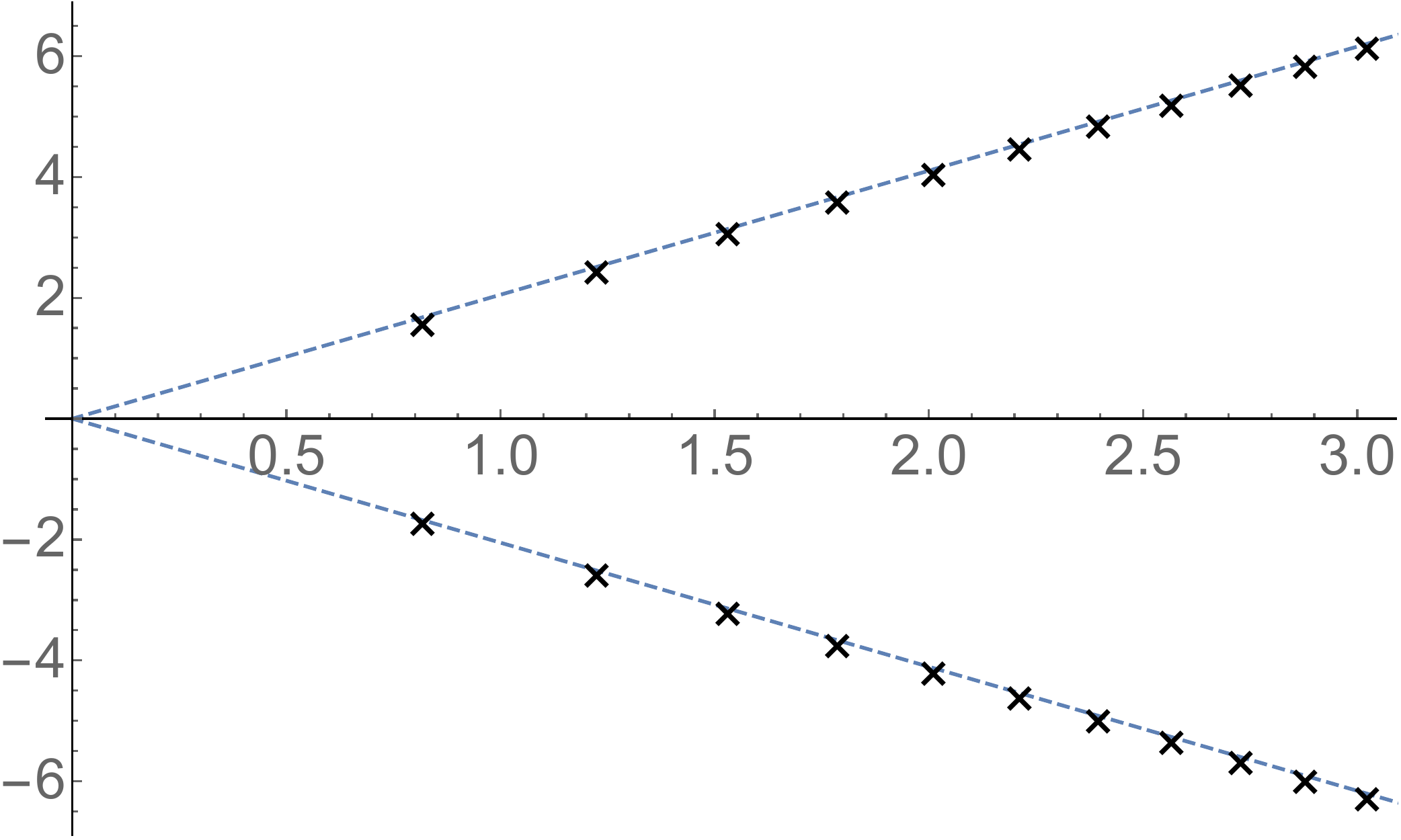}};
\node at (4.5,0.2) {\includegraphics[width=0.5\textwidth]{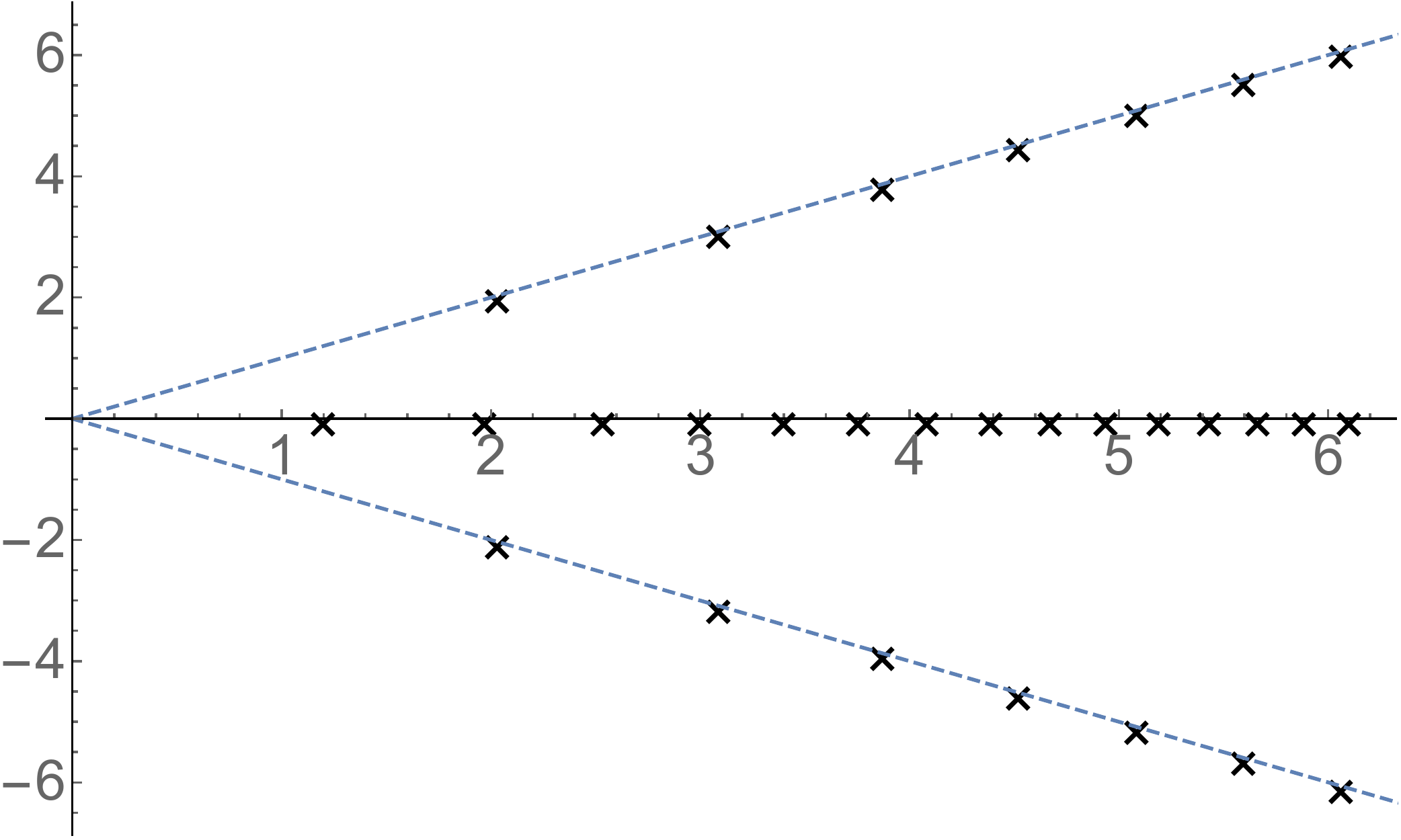}};
\node at (-5, -3) {\large $0<x<1$};
\node at (5.0, -3) {\large $x>1$};
\node at (-8,2.5) {\large $\mu$};
\node at (2.0,2.5) {\large $\mu$};
\draw (-8,2.55) circle (0.35cm);
\draw (2.0,2.55) circle (0.35cm);
\end{tikzpicture}}
\caption{\small  The typical pattern of zeroes of $D_{{\boldsymbol {\mathfrak j}}, \mathfrak{m}, A}(\mu)$
in the complex $\mu$ plane for the two
cases  $0<x<1$  and $x>1$. The numerical data was obtained 
for $\mathfrak{j}_1=\mathfrak{j}_2=\mathfrak{m}=0$, $A=\tfrac{\ri}{2}$ and $\xi=1$,
when the connection coefficient is expressed in terms of the parabolic cylinder function.
The dashed lines mark the rays
\eqref{gsasatast} and \eqref{ioasd8918921} in the left and right panels, respectively.
}
\label{figa3}
 \end{figure}
 The leading asymptotic behaviour of the connection coefficient $D_{{\boldsymbol {\mathfrak j}}, \mathfrak{m}, A}(\mu)$
 in these domains can be described through the formula:
 \be\label{oia38923ss1ABCC}
D_{{\boldsymbol {\mathfrak j}}, \mathfrak{m}, A}(\mu)\asymp
R_{{\boldsymbol {\mathfrak j}}, \mathfrak{m}, A}
\ (|\mu|\,\re^{\ri\phi_\alpha})^{\frac{\ri A}{\xi}-\frac{1}{2}\mathfrak{m}}\ 
 \exp\Big( ( |\mu| \re^{\ri\phi_\alpha} )^{1+\xi}\ q_{-1}(\sigma_\alpha x)+o(1)\Big)
\qquad\big(|\mu|\to+\infty\big)\ .
\ee
Here we use the notations
\be\label{oiasd98129821}
\sigma_\alpha=\begin{cases}
+1\ \ \ \  \ \ \ \ &{\rm for}\ \ \  |\alpha|<\alpha_0\\
-1 \ \ \  \ \ \ \ &{\rm otherwise}
\end{cases}
\ ,\ \  \ \
\phi_\alpha=\begin{cases}
\alpha\ \ \ \  \  &{\rm for}\ \ \ \ \ \   |\alpha|<\alpha_0\\
\alpha-\pi \ \ \  \  &{\rm for}\ \ \ \ \  \   \alpha_0<\alpha\leq\pi\\
\alpha+\pi \ \ \  \  &{\rm for}\ \ \ \ \ \   -\pi<\alpha<-\alpha_0
\end{cases}
\ee
where $\alpha\in (-\pi,\pi]$ stands for the argument of the complex number $\mu$, i.e., $\mu=|\mu|\,\re^{\ri\alpha}$.
\bigskip

The position of the Stokes line, i.e.,  the value of the angle $\alpha_0$, is defined through the condition
\bea
 \Re e\Big[\re^{\ri(1+\xi)\alpha_0}\ q_{-1}(x)-\re^{\ri(1+\xi)(\alpha_0-\pi)}\ q_{-1}(-x)\Big]=0\,,
\eea
which yields
\be\label{ioas89128912}
\alpha_0(x)=\frac{\pi}{2}+\frac{1}{2\ri (1+\xi)}\ \log\bigg[\frac{\re^{-\frac{\ri\pi}{2}\xi}\,
q_{-1}(x)+\re^{+\frac{\ri\pi}{2}\xi}\, q_{-1}(-x)}
{\re^{+\frac{\ri\pi}{2}\xi}\,
q_{-1}(x)+\re^{-\frac{\ri\pi}{2}\xi}\, q_{-1}(-x)}\bigg]\ \ \ \ \ \qquad  \ \ (0\leq x<1)\, .
\ee
Along the Stokes line $\arg(\mu)=\alpha_0$ the large $\mu$ asymptotic behaviour of the connection coefficient 
reads as 
\bea\label{oia38923ss1ABaaaCC}
D_{{\boldsymbol {\mathfrak j}}, \mathfrak{m}, A}(\mu)&\asymp&
R_{{\boldsymbol {\mathfrak j}}, \mathfrak{m}, A}
\ \big(|\mu|\,\re^{\ri (\alpha_0-\frac{\pi}{2})}\big)^{-\sqrt{\frac{2(1+\xi)}{\xi}}\, P}\ 
 \exp\Big( \,  \frac{g}{2 f}\, |\mu| ^{1+\xi}+o(1)\Big)\\[0.3cm]
&\times& 2\cos\Big(\, \tfrac{1}{2}\,  f\ |\mu| ^{1+\xi}-\pi 
\sqrt{\tfrac{1+\xi}{2\xi}}\ P+o(1)\Big)
\qquad\ \ \ \ \ \ \ \ \ \ \qquad\qquad\big(|\mu|\to+\infty\big)\, .\nonumber
\eea
Here 
\bea\label{aois89128912}
g&=&q^2_{-1}(x)-q^2_{-1}(-x)-2\sin(\pi\xi) \, q_{-1}(x)q_{-1}(-x)\nonumber\\[0.3cm]
f&=&\ \sqrt{q^2_{-1}(x)+q^2_{-1}(-x)+2\cos(\pi\xi) \, q_{-1}(x)q_{-1}(-x)}
\eea
and instead of $A$,  we use $P$ such that $\frac{\ri A}{\xi}-
\tfrac{1}{2}\, \mathfrak{m}=-\sqrt{\frac{2(1+\xi)}{\xi}}\ P$,
 which is assumed to be a real number. The
above asymptotic formula implies that
the position of the zeroes $\{\mu^{(\pm)}_n\}_{n=1}^\infty$  with
$\lim_{n\to\infty}\arg\big(\mu^{(\pm)}_n\big)=\pm \alpha_0$ is described at large $n$ by
\bea\label{oasoi192831}
\big|\,\mu^{(\pm)}_n\,\big|^{1+\xi}\asymp \frac{2\pi}{ f(x)}\ 
\Big(n-\tfrac{1}{2}+\sqrt{\tfrac{1+\xi}{2\xi}}\ P+
o(1)\,\Big)\ \ \ \ \ \ \qquad \qquad (n\gg 1)\,.
\eea

 \bigskip
When $x>1$ $(z_{1,2}>0)$
 there are three rays along which the zeroes  of $D_{{\boldsymbol {\mathfrak j}}, \mathfrak{m}, A}(\mu)$  accumulate,
 \bea\label{ioasd8918921}
 \arg(\mu)=0,\,\pm\, \tfrac{\pi}{2(1+\xi)}
 \eea
 (see right panel of fig.\ref{figa3}).
The asymptotics of the connection coefficient with $\mu\to\infty$ and $|\arg(\mu)|<\frac{\pi}{2(1+\xi)}$ reads as
 \be\label{oia38923ss1ABCsssC}
D_{{\boldsymbol {\mathfrak j}}, \mathfrak{m}, A}(\mu)\asymp
R_{{\boldsymbol {\mathfrak j}}, \mathfrak{m}, A}
\ (|\mu|\re^{\ri\alpha})^{\frac{\ri A}{\xi}-\frac{1}{2}\mathfrak{m}}\ 
 \exp\Big( ( |\mu| \re^{\ri\alpha} )^{1+\xi}\ q_{-1}\big( x+\ri \, \sgn(\alpha)\,\varepsilon\big)+o(1)\Big)\Big|_{\varepsilon\to +0}
\ee
(the branch of the   hypergeometric function  in \eqref{uuasuasusa} is taken in such a way that
 $q_{-1}(x)$   has a discontinuity   for  $x\in(1,+\infty)$).
 In the wedge $\frac{\pi}{2(1+\xi)}<|\arg(\mu)|\leq \pi$,   formula \eqref{oia38923ss1ABCC}
 remains applicable provided $\alpha_0$ is taken to be $\frac{\pi}{2(1+\xi)}$.

 \bigskip
 For $x<-1$ $(z_{1,2}<0)$ the zeroes of $D_{{\boldsymbol {\mathfrak j}}, \mathfrak{m}, A}(\mu)$  
accumulate along the rays
  \bea
 \arg(\mu)=\pi,\,\pm \,\big(\pi- \tfrac{\pi}{2(1+\xi)}\big)\ .
 \eea
 In this case the   large $\mu$ asymptotic behaviour  can be obtained from that  corresponding to 
 $x>1$ by simultaneously reflecting $\mu\to-\mu$ and $x\to-x$.

 \bigskip
 Finally we note that
 for pure  imaginary $x$, the zeroes  of $D_{{\boldsymbol {\mathfrak j}}, \mathfrak{m}, A}(\mu)$  accumulate along 
 the imaginary axis $\Im m(\mu)=0$.
 In this case
  \eqref{oia38923ss1ABCC},\,\eqref{oiasd98129821} remain valid  upon setting
 $\alpha_0=\frac{\pi}{2}$ therein.

\subsection*{Asymptotic coefficient $q_{1}$}
The coefficient $q_{1}$ enters into the
 subleading term in the large $\mu$ asymptotic expansion \eqref{isusuas}.
When $r=2$ and $k_1=k_2=1$ it is a
linear combination of the eigenvalues of the local IM ${\bf I}_{1}^{(1)}$ and ${\bf I}_{1}^{(2)}$,
weighted by factors, which are 
themselves  expressed
 in terms of the Gauss hypergeometric function similar to \eqref{uuasuasusa}. 
It follows
 from eqs.\eqref{aiasisia} and \eqref{jassjaasuasus} that
 \bea\label{ahsysay}
 q_{1}=\big(\, I_{1}^{(1)}\, z_1\partial_{z_1}+I_{1}^{(2)}\, z_2\partial_{z_2}\big)\, f_{1}\ ,
 \eea
 where
 \bea
f _{1}=\big[\,\big(1-\re^{2 \pi \ri \xi }\,\big)\ 
\big(1-\re^{-2 \pi\ri  \xi}\,\big)\,\big]^{-1}
\oint_{\cal C}\frac{\rd z}{\sqrt{{\cal P}}}\ \frac{1}{z^{2}}
\eea
and  the integration contour ${\cal C}$  is the Pochhammer loop that appears in eq.\eqref{aisisaasias}.
A straightforward computation gives 
  \be
  f _{1}
=
\big(\sqrt{z_1 z_2}\,\big)^{-1-\xi}\ \Phi\big(\tfrac{z_1+z_2}{2\sqrt{z_1z_2}}\big)
\ee
with
\bea\label{hasatast}
\Phi(x)=  -\frac{\pi}{\sin(\pi \xi)}\ 
{}_2F_1\big(-\xi,\ 1+\xi, 1,\tfrac{1}{2}\,(1+x)\big)\ .
\eea
The convention  $z_1 z_2=1$, which was imposed before, is not being assumed here.
The coefficient $q_{1}$ \eqref{ahsysay} is obtained
from  $f _{1}$ via differentiation w.r.t. $z_{1}$ and $z_2$, which must  be
 treated as  independent variables.
The final result can be brought to the form
\be
q_{1}=\big(\sqrt{z_1z_2}\big)^{-1-\xi}\, \Big(\, -\tfrac{1}{2}\, (1+\xi)\ 
 \Phi\big(\tfrac{z_1+z_2}{2\sqrt{z_1z_2}}\big)\, \big(I^{(1)} _{1}+I^{(2)} _{1}\big)
 +\Phi'\big(\tfrac{z_1+z_2}{4\sqrt{z_1z_2}}\big)\ \tfrac{z_1-z_2}{4\sqrt{z_1 z_2}}\ 
  \big(I^{(1)} _{1}-I^{(2)} _{1}\big)\,\Big)\vphantom{\bigg)}
\ee
where together with $\Phi(x)$ we use its derivative:
\bea\label{aisaisaisai}
\Phi'(x)=
\frac{\xi (1+\xi)}{1-x^2}\ q_{-1}(x)
\eea
with $q_{-1}$ from \eqref{uuasuasusa}.
At this point one can set $z_1z_2=1$. However
the ambiguity in the sign of the square root $\sqrt{z_1 z_2}$ requires special attention.
For instance, 
in the case of the  coefficient $q_{-1}$,
 the sign  $\sqrt{z_1z_2}=\pm1$ enters
explicitly into  the  asymptotic formulae for large complex $\mu$  (e.g. the factor $\sigma_\alpha$ in
\eqref{oia38923ss1ABCC}).
\bigskip

The local  IM,  whose eigenvalues coincide with the
linear combinations $I_1^{(e)}=-\frac{1}{4\xi} \, (I_1^{(1)}+I_1^{(2)})$ and $I_1^{(o)}=\frac{1}{4\sqrt{2\xi(1+\xi)}}\,
(I_1^{(1)}-I_1^{(2)})$, have been presented in eq.\,\eqref{aoisi912898291}.
Further developing the asymptotic series one would come up against  $q_3$ \eqref{aiasisia},
which is a certain linear combination of the eigenvalues of   ${\bf I}^{(e)}_3$ and ${\bf I}^{(o)}_3$.
These  IM are given in  eq.\,\eqref{isisaias} and their eigenvalues for 
the primary states are listed in tab.\,\ref{tab2}. Below, for completeness,  
we quote the local fields appearing in  \eqref{isisaias}
in terms of  the chiral  Bose and fermion fields.

\subsection*{Spin 4 local fields}

 The local densities for the integrals of motion ${\bf I}_{2n-1}^{(a)}$ 
are linear combinations of fields belonging to the space that was denoted by
${\cal W}^{(2n)}\subset W^{(c,2)}_{\boldsymbol 1}$.
The space ${\cal W}^{(4)}$ 
contains five spin 4 local fields, 
   $\partial S S$, $\chi_2 GS$, $G^2$, $ \partial^2 \chi_2 S$ and ${G}_{{\chi_2}} ^2$,
which are not total derivatives.
To describe them, introduce the notation
   \bea\label{iaassususa}
 G_{\pm \varphi}&=&(\partial\varphi)^2\pm \ri\rho \ \partial^2\varphi\ ,\ \ \  \\[0.2cm]
 G^2_{ \varphi}&=&(\partial\varphi)^4+(1-\rho^2)\ (\partial^2\varphi)^2+
\tfrac{1}{3}\,\partial\Big(2\ri\rho\,(\partial\varphi)^3-3\,\partial^2\varphi\,\partial\varphi-\ri\rho\,\partial^3\varphi\Big)
\nonumber
\eea
and
\bea
 G_{\chi_a}&=&\tfrac{\ri}{2}\, \chi_a\partial \chi_a\ ,\ \ \ \qquad\qquad
G_{\chi_a}^2=-\tfrac{7\ri}{12}\  \chi_a\partial^3 \chi_a+\tfrac{5\ri}{8}\ \partial
 \big( \chi_a\partial^2 \chi_a\big)\, .
 \eea
  Notice that the combination $G_{+\varphi}+G_{\chi_1}$
is the same field  as $G$ from eq.\,\eqref{ias891298321}.
  The  regular terms in the OPE \eqref{ajasusausa}  are given by
 \bea
 \partial S S&=&\ri\ (G_{+\varphi}-3 G_{-\varphi})\
  G_{\chi_1}-\tfrac{\ri}{2}\ \big(\partial^2\varphi\big)^2+ 
 \tfrac{\ri}{7}\ \big(12\rho^2-1\big)\,  G_{\chi_1}^2
  \nonumber\\[0.3cm]
 &+& \ri\,  \partial\big(-2\ri\rho\ \partial\varphi\, G_{\chi_1}-
 \tfrac{1}{28}\ (9+4\rho^2)\ \partial G_{\chi_1}
 -
   \tfrac{7}{24 }\, \partial G_{+\varphi}
  +  \tfrac{1}{24 }\,
   \partial G_{-\varphi}\,\big)\nonumber\\[0.3cm]
    G^2&=&G_{+\varphi}^2+G_{\chi_1}^2+2G_{+\varphi}G_{\chi_1}\\[0.3cm]
 GS&=& -\ri\ (\partial\varphi)^3\ \chi_1+\rho\ 
 \big( (\partial\varphi)^2\ \partial \chi_1+ \partial^2\varphi\partial\varphi\chi_1\big)
 +\ri\rho^2\  \partial^2\varphi \partial\chi_1\nonumber\\[0.3cm]
 &+&\tfrac{\ri}{4}\, \big(2\partial^3\varphi\chi_1+3\,\partial\varphi\partial^2\chi_1\big)-
 \tfrac{5}{12}\,\rho\ \partial^3\chi_1\ .\nonumber
 \eea
 For computations, it is useful to express the field $\chi_2GS$ in the form
 \bea
\chi_2GS&=&\ri\, \big(-(\partial\varphi)^3+\tfrac{1}{2}\ \partial^3\varphi\big)\, \chi_2\chi_1+
\tfrac{1}{2}\,\rho\, (\partial\varphi)^2\, \big(\chi_2\partial\chi_1+\chi_1\partial\chi_2\big)\\[0.2cm]
 &-&\tfrac{3 \ri}{4}\ \partial\varphi\, \partial\chi_2\partial\chi_1-
 \tfrac{\ri}{8}\, (3-4\rho^2)\,   \partial^2\varphi \,\partial (\chi_2 \chi_1)+\tfrac{5}{24}\,\rho\, 
 \big(\partial\chi_2\partial^2\chi_1+\partial\chi_1\partial^2\chi_2\big)\nonumber\\[0.3cm]
 &-&\tfrac{\ri}{8}\, (3-4\rho^2)\, \partial^2\varphi\,
 \big(\chi_1\partial\chi_2+\chi_2\partial\chi_1\big)+\,\partial{\cal O}_3\nonumber
 \eea
 with
 \bea
 {\cal O}_3&=&\tfrac{1}{2}\, \rho\, (\partial\varphi)^2\, \chi_2\chi_1+\ri\rho^2\ \partial\varphi\,
 \chi_2\partial\chi_1+\tfrac{\ri}{4}\, 
 (3-4\rho^2)\,  \partial\varphi \,
  \chi_2\partial\chi_1\nonumber\\[0.3cm]
 &+&\tfrac{5}{24}\,\rho\, \partial\chi_2\partial\chi_1-\tfrac{5}{12}\,\rho\, \chi_2\partial^2\chi_1\ .
 \eea
 Also, it is convenient to re-write $ \partial^2\chi_2 S=\rho\partial^2\chi_2 \partial\chi_1-
 \ri \partial\varphi\,\partial^2\chi_2\chi_1$ as  
 \bea\label{iaassususaf}
 \partial^2\chi_2 S
 &=&\ri\, \partial\varphi \, \partial\chi_2 \partial\chi_1
+ \tfrac{\ri}{2}\, \partial^2\varphi\, 
\partial(\chi_2 \chi_1)
 -\tfrac{1}{2}\,\rho\, \big(\partial\chi_2
 \partial^2\chi_1+\partial\chi_1\partial^2\chi_2\big)+
 \nonumber\\[0.3cm]
&-&\tfrac{\ri}{2}\, \partial^2\varphi\,
 \big(\chi_1\partial\chi_2+\chi_2\partial\chi_1\big)+
 \partial\,\big(\, \tfrac{1}{2}\,\rho\, \partial\chi_2\partial\chi_1
-\ri\, \partial \varphi \, \partial\chi_2 \chi_1
\big)\ .
 \eea


\begin{thebibliography}{99}

\bibitem{Gaudin}
M.~Gaudin,
\emph{Diagonalisation d'une classe d'hamiltoniens de spin},
\href{https://doi.org/10.1051/jphys:0197600370100108700}{J. Phys. France {\textbf{37}} (1976) 1087-1098}.

\bibitem{Gaudin1}
M.~Gaudin,
\emph{The Bethe wavefunction}, \href{https://doi.org/10.1017/CBO9781107053885}{Cambridge University Press, 
Cambridge (2014)}.\hfil


\bibitem{Gurco}
B.~Jur$\check{\rm c}$o,
\emph{Classical Yang-Baxter equations and quantum integrable systems},
\href{https://doi.org/10.1063/1.528305}{J. Math. Phys. {\textbf {30}}  (1989)
1289-1293}.

\bibitem{Zam}
A.~B.~Zamolodchikov,
\emph{Integrable field theory from conformal field theory},
\href{https://doi.org/10.2969/aspm/01910641}
{Adv. Stud. Pure Math. {\textbf {19}} (1989) 641-674}.

\bibitem{Feigin:2007mr}
B.~Feigin and E.~Frenkel,
\emph{Quantization of soliton systems and Langlands duality}, \href{https://doi.org/10.2969/aspm/06110185}
{Adv. Stud. Pure Math. {\textbf {61}} (2011) 185-274}
  \href{https://arxiv.org/abs/0705.2486}{{\ttfamily[arXiv:math.QA/0705.2486]}}.\hfil

\bibitem{Feigin:1994in}
B.~Feigin, E.~Frenkel and N.~Reshetikhin,
\emph{Gaudin model, Bethe ansatz and  critical level},
\href{https://link.springer.com/article/10.1007/BF02099300}{Commun. Math. Phys. \textbf{166} (1994) 27-62} 
\href{https://arxiv.org/abs/hep-th/9402022}{{\ttfamily[arXiv:hep-th/9402022]}}.\hfil



\bibitem{Frenkel:2004qy}
E.~Frenkel,
\emph{Gaudin model and opers}, in: \href{https://link.springer.com/chapter/10.1007/3-7643-7341-5_1}
{Infinite dimensional algebras and quantum integrable
systems (ed. P.~P.~Kulish et. al.)  Progr. Math. {\textbf {237}} (2005) 1-58}
\href{https://arxiv.org/abs/math/0407524}{\ttfamily[arXiv:math.QA/0407524]}.\hfil

  
\bibitem{Voros:1994}
A.~Voros, 
\emph{Exact quantization condition for anharmonic oscillators (in one
dimension)},
\href{https://iopscience.iop.org/article/10.1088/0305-4470/27/13/038/meta}{J. Phys. {\bf A}
 {\textbf{27}} (1994) 4653-4661}.
  
  

\bibitem{Dorey:1998pt} 
  P.~Dorey and R.~Tateo,
  \emph{Anharmonic oscillators, the thermodynamic Bethe ansatz, and nonlinear integral equations,}
\href{https://doi.org/10.1088/0305-4470/32/38/102}{J. Phys. {\textbf A} {\textbf {32}} (1999) L419-L425}
\href{https://arxiv.org/abs/hep-th/9812211}{{\ttfamily [arXiv:hep-th/9812211]}}.





\bibitem{Bazhanov:1998wj} 
  V.~V.~Bazhanov, S.~L.~Lukyanov and A.~B.~Zamolodchikov,
  \emph{Spectral determinants for\hfill \\  Schrodinger equation and Q operators of conformal field theory,}
\href{https://doi.org/10.1023/A:1004838616921}{J.\ Stat.\ Phys.\  {\textbf 1}{\textbf 0}{\textbf 2} (2001) 567-576}
 \href{https://arxiv.org/abs/hep-th/9812247}{{\ttfamily [arXiv:hep-th/9812247]}}.




\bibitem{Bazhanov:2003ni} 
  V.~V.~Bazhanov, S.~L.~Lukyanov and A.~B.~Zamolodchikov,
 \emph{Higher level eigenvalues of Q-operators and Schroedinger equation},
\href{https://doi.org/10.4310/ATMP.2003.v7.n4.a4}{Adv.\ Theor.\ Math.\ Phys.\  {\textbf 7} (2003) 711-725}
   \href{https://arxiv.org/abs/hep-th/0307108}{{\ttfamily[arXiv:hep-th/0307108]}}.




\bibitem{Lacroix:2018fhf}
S.~Lacroix, B.~Vicedo and C.~Young,
\emph{Affine Gaudin models and hypergeometric functions on affine opers},
\href{https://www.sciencedirect.com/science/article/pii/S0001870819302014}{Adv. Math. \textbf{350} (2019) 486-546}
 \href{https://arxiv.org/abs/1804.01480}{{\ttfamily[arXiv:math.QA/1804.01480]}}.\hfil

\bibitem{Lacroix:2018itd}
S.~Lacroix, B.~Vicedo and C.~Young,
\emph{Cubic hypergeometric integrals of motion in affine Gaudin models},
\href{https://www.intlpress.com/site/pub/pages/journals/items/atmp/content/vols/0024/0001/a005/}{Adv.\,Theor.\,Math.\,Phys.\,\textbf{24} (2020) 155-187} 
\href{https://arxiv.org/abs/1804.06751}{{\ttfamily[arXiv:math.QA/1804.06751]}}.\hfil


   \bibitem{Bazhanov:1994ft}
  V.~V.~Bazhanov, S.~L.~Lukyanov and A.~B.~Zamolodchikov,
 \emph{Integrable structure of conformal field theory, quantum KdV theory and
  thermodynamic Bethe ansatz},
\href{https://doi.org/10.1007/BF02101898}{Commun.\ Math.\ Phys.\  {\bf 177} (1996) 381-398}
\href{http://arxiv.org/abs/hep-th/9412229}{{\ttfamily [arXiv:hep-th/9412229]}}.



\bibitem{Bazhanov:1996dr}
  V.~V.~Bazhanov, S.~L.~Lukyanov and A.~B.~Zamolodchikov,
\emph{Integrable structure of conformal field theory II. Q-operator and DDV equation},
\href{https://doi.org/10.1007/s002200050240}{Commun. Math. Phys.  {\bf 190} (1997) 247-278}
  \href{http://arxiv.org/abs/hep-th/9604044}{{\ttfamily [arXiv:hep-th/9604044]}}.

\bibitem{Bazhanov:1998dq}
  V.~V.~Bazhanov, S.~L.~Lukyanov and A.~B.~Zamolodchikov,
 \emph{Integrable structure of conformal field theory III. The Yang-Baxter relation},
\href{https://doi.org/10.1007/s002200050531}{Commun. Math.\ Phys.  {\bf 200} (1999) 297-324}
\href{http://arxiv.org/abs/hep-th/9805008}{{\ttfamily [arXiv:hep-th/9805008]}}.
  
\bibitem{Baxter:1971cs} 
  R.~J.~Baxter,
 \emph{Generalized ferroelectric model on a square lattice},
%
\href{https://doi.org/10.1002/sapm197150151}
{Stud.\ Appl.\ Math.\  {\bf 50} (1971) 51-69}. 


\bibitem{Lukyanov:2006gv}
S.~L.~Lukyanov,
 \emph{Notes on parafermionic QFT's with boundary interaction},
\href{https://www.sciencedirect.com/science/article/pii/S0550321307003045}{Nucl. Phys.  {\textbf{B 784}} (2007)  151-201}
\href{http://arxiv.org/abs/hep-th/0606155}{\ttfamily[arXiv:hep-th/0606155]}.


\bibitem{Drinfeld1986}
V.~G.~Drinfel'd,
 \emph{Quantum Groups},
 \href{https://www.mathunion.org/icm/proceedings}
{Proceedings of the International Congress of
Mathematics, Berkeley 1986, {\textbf{1}}, 798-820,
Academic Press, California (1987)}
\href{https://link.springer.com/article/10.1007\%2FBF01247086}{[J. Sov. Math. {\bf 41} (1988) 898-915]}.


\bibitem{Jimbo:1985zk} 
  M.~Jimbo,
  \emph{A $q$-difference analogue of $U(\mathfrak{g})$ and the Yang-Baxter equation},
\href{https://link.springer.com/article/10.1007/BF00704588}{Lett.\ Math.\ Phys.\  {\textbf{10}} (1985) 63-69}.
  
\bibitem{Khoroshkin:1994um} 
  S.~M.~Khoroshkin, A.~A.~Stolin and V.~N.~Tolstoi,
  \emph{Gauss decomposition of trigonometric R matrices},
 \href{https://www.worldscientific.com/doi/10.1142/S0217732395001496}{Mod.\ Phys.\ Lett.\ {\textbf{A 10}} (1995)  
1375-1392}
 \href{http://arxiv.org/abs/hep-th/9404038}{\ttfamily [arXiv:hep-th/9404038]}.
  

  
  
  
\bibitem{Fateev:1985mm} 
  V.~A.~Fateev and A.~B.~Zamolodchikov,
  \emph{Nonlocal (parafermion) currents in the two-dmensional conformal quantum field 
  theory and self-dual critical points in $Z_N$-symmetric  statistical systems},
 \href{http://www.jetp.ac.ru/cgi-bin/e/index/e/62/2/p215?a=list}{Sov.\ Phys.\ JETP {\textbf{62}} (1985) 215-225}
  [Zh.\ Eksp.\ Teor.\ Fiz.\  {\bf 89}  (1985) 380].
  

\bibitem{Fateev:1982wi} 
  V.~A.~Fateev and A.~B.~Zamolodchikov,
  \emph{Self-dual solutions of the star-triangle relations in $Z_N$-models},
  \href{https://www.sciencedirect.com/science/article/pii/0375960182907368}{Phys.\ Lett.\ {\textbf {A 92}} (1982)  37-39}.
  
  
  
\bibitem{Zamolodchikov:1985wn} 
  A.~B.~Zamolodchikov,
  \emph{Infinite additional symmetries in two-dimensional conformal quantum field theory},
\href{https://doi.org/10.1007/BF01036128}{Theor.\ Math.\ Phys.\  {\bf 65} (1985) 1205-1213}
[\href{http://www.mathnet.ru/php/archive.phtml?wshow=paper&jrnid=tmf&paperid=5141&option_lang=rus}{Teor.\ Mat.\ Fiz.\  {\bf 65} (1985) 347-359}].
  
  

  
\bibitem{Fateev:1987zh}
V.~A.~Fateev and S.~L.~Lukyanov,
 \emph{The models of two-dimensional conformal quantum field theory with $Z_n$ symmetry},
\href{https://www.worldscientific.com/doi/abs/10.1142/S0217751X88000205}{Int. J. Mod. Phys. {\bf A}
 \textbf{3} (1988) 507-520}.
  
  
\bibitem{Feigin:2001yq}
B.~L.~Feigin and A.~M.~Semikhatov,
 \emph{The  $\widehat{sl}(2) \oplus \widehat{sl}(2)) /\widehat{sl}(2)$ coset theory as a
  Hamiltonian reduction of $D(2|1; \alpha)$},
\href{https://doi.org/10.1016/S0550-3213(01)00307-8}{Nucl. Phys.  \textbf{B 610} (2001) 489-530}
\href{https://arxiv.org/abs/hep-th/0102078}{{\ttfamily [arXiv:hep-th/0102078]}}.
  
\bibitem{Lukyanov:2012wq}
S.~L.~Lukyanov and A.~B.~Zamolodchikov,
\emph{Integrable boundary interaction in 3D target space: the ``pillow-brane'' model},
\href{https://www.sciencedirect.com/science/article/pii/S0550321313002708}{Nucl. Phys.  \textbf{B\ 873} (2013)  585-613}
\href{https://arxiv.org/abs/1208.5259}{{\ttfamily [arXiv:hep-th/1208.5259]}}.\hfil
  
  \bibitem{Zam1986}
  A.~B.~Zamolodchikov, {Montreal talk}, unpublished (1985).
  



   \bibitem{Wakimoto}
   M.~Wakimoto,
   \emph{Fock representations of the affine Lie algebra $A _1^{(1)}$},
 \href{https://link.springer.com/article/10.1007/BF01211068#citeas}{Commun. Math. Phys. {\bf 104} (1986) 605-609}.
 
\bibitem{Gerasimov:1989mz} 
  A.~Gerasimov, A.~Marshakov and A.~Morozov,
 \emph{Free field representation of parafermions and related coset models},
  \href{https://www.sciencedirect.com/science/article/pii/0550321389902241}{Nucl.\ Phys.\ {\bf B} {\bf 328} (1989) 664-676}.
  
  
  
  
  
\bibitem{Fateev:1995ht}
  V.~A.~Fateev,
  \emph{The duality between two-dimensional integrable field theories and sigma models},
 \href{https://www.sciencedirect.com/science/article/pii/037026939500883M}{Phys.\ Lett.\  {\bf B\ 357} (1995) 397-403}.\hfil
   
\bibitem{Fateev:1996ea} 
  V.~A.~Fateev,
  \emph{The sigma model (dual) representation for a two-parameter family of integrable quantum field theories},
   \href{https://www.sciencedirect.com/science/article/pii/0550321396002568}{Nucl.\ Phys.\ {\bf B} {\bf 473} (1996) 509-538}.


   
\bibitem{Lukyanov:2013wra}
S.~L.~Lukyanov,
\emph{ODE/IM correspondence for the Fateev model},
\href{https://link.springer.com/article/10.1007/JHEP12(2013)012}{JHEP \textbf{12} (2013) 012}\hfil\\
\href{https://arxiv.org/abs/1303.2566}{{\ttfamily [arXiv:hep-th/1303.2566]}}.
  
\bibitem{Bazhanov:2013cua}
V.~V.~Bazhanov and S.~L.~Lukyanov,
\emph{Integrable structure of Quantum Field Theory: classical flat connections versus quantum stationary states},
\href{https://link.springer.com/article/10.1007/JHEP09(2014)147}{JHEP \textbf{09}  (2014) 147}
 \href{https://arxiv.org/abs/1310.4390}{{\ttfamily [arXiv:hep-th/1310.4390]}}.

\bibitem{Gao}  
 G.~Cui, Y.~Gao, H.~H.~Rugh and L.~Tan,
 \emph{Rational maps as Schwarzian primitives},
 \href{https://arxiv.org/ct?url=https\%3A\%2F\%2Fdx.doi.org\%2F10.1007\%2Fs11425-016-5140-7&v=149cc3c5}
  {Sci. China Math. {\textbf{59}}  (2016) 1267-1284 }
 \href{https://arxiv.org/abs/1511.04246}{{\ttfamily [arXiv:math.CV/1511.04246]}}.




\bibitem{Bazhanov:2017nzh}
V.~V.~Bazhanov, G.~A.~Kotousov and S.~L.~Lukyanov,
\emph{Quantum transfer-matrices for the sausage model},
\href{https://link.springer.com/article/10.1007/JHEP01(2018)021}{JHEP \textbf{01}  (2018) 021}
\href{https://arxiv.org/abs/1706.09941}{{\ttfamily [arXiv:hep-th/1706.09941]}}.




\bibitem{Voros:1999}
A.~Voros, 
\emph{Exact resolution method for general 1D polynomial Schr\"{o}dinger equations},
\href{https://iopscience.iop.org/article/10.1088/0305-4470/32/32/311/meta}{J. Phys.
{\bf A} {\textbf{32}} (1999) 5993-6007}
\href{https://arxiv.org/abs/math-ph/9903045}{{\ttfamily [arXiv:math-ph/9903045]}}.
  
  


\bibitem{Ito:2020ueb}
K.~Ito and H.~Shu,
 \emph{TBA equations for the Schr\"odinger equation with a regular singularity},
\href{https://iopscience.iop.org/article/10.1088/1751-8121/ab96ee}{J. Phys. {\bf A} \textbf{53} (2020) 335201}
\href{https://arxiv.org/abs/1910.09406}{{\ttfamily [arXiv:hep-th/1910.09406]}}.
  
 

\bibitem{Anderson}P. W. Anderson and G. Yuval,
\emph{ Exact results in the Kondo problem: equivalence to
a classical one-dimensional Coulomb gas},
\href{https://journals.aps.org/prl/abstract/10.1103/PhysRevLett.23.89}{Phys. Rev. Lett. {\textbf{23}} (1969) 89-92}.

\bibitem{Fendley:1995kj}
P.~Fendley, F.~Lesage and H.~Saleur,
\emph{A unified framework for the Kondo problem and for an impurity in a Luttinger liquid},
\href{https://link.springer.com/article/10.1007\%2FBF02175563}{J. Stat. Phys. \textbf{85}  (1996) 211-249}
\href{https://arxiv.org/abs/cond-mat/9510055}{{\ttfamily [arXiv:cond-mat/9510055]}}.\hfil

\bibitem{Lukyanov:2003rt}
S.~L.~Lukyanov and A.~B.~Zamolodchikov,
\emph{Integrable circular brane model and Coulomb charging at large conduction},
\href{https://iopscience.iop.org/article/10.1088/1742-5468/2004/05/P05003}{J. Stat. Mech. \textbf{0405}  (2004) P05003}
\href{https://arxiv.org/abs/hep-th/0306188}{{\ttfamily [arXiv:hep-th/0306188]}}.\hfil


\bibitem{Bazhanov:2003ua}
V.~V.~Bazhanov, S.~L.~Lukyanov and A.~M.~Tsvelik,
\emph{Analytical results for the Coqblin-Schrieffer model with generalized magnetic fields},
\href{https://journals.aps.org/prb/abstract/10.1103/PhysRevB.68.094427}{Phys. Rev. \textbf{B} \textbf{68} (2003) 094427}
\hfil\\
\href{https://arxiv.org/abs/cond-mat/0305237}{{\ttfamily [arXiv:cond-mat/0305237]}}.\hfil

\bibitem{Gaiotto:2020fdr}
D.~Gaiotto, J.~H.~Lee and J.~Wu,
\emph{Integrable Kondo problems},
\href{https://link.springer.com/article/10.1007/JHEP04(2021)268}{JHEP \textbf{04} (2021) 268}\hfil \\
\href{https://arxiv.org/abs/2003.06694}{{\ttfamily [arXiv:hep-th/2003.06694]}}.



\bibitem{Gaiotto:2020dhf}
D.~Gaiotto, J.~H.~Lee, B.~Vicedo and J.~Wu,
\emph{Kondo line defects and affine Gaudin models},\hfil\\
\href{https://arxiv.org/abs/2010.07325}{{\ttfamily [arXiv:hep-th/2010.07325]}}.





















  
  
\bibitem{Zamolodchikov:1995aa} 
  A.~B.~Zamolodchikov and Al.~B.~Zamolodchikov,
  \emph{Conformal bootstrap in Liouville field theory},
\href{https://doi.org/10.1016/0550-3213(96)00351-3}{Nucl.\ Phys.\ {\bf B} {\bf 477} (1996) 577-605}
  \href{http://arxiv.org/abs/hep-th/9506136}{{\ttfamily[arXiv:hep-th/9506136]}}.










\bibitem{Kotousov:2019ygw} 
  G.~A.~Kotousov and S.~L.~Lukyanov,
  \emph{Bethe state norms for the Heisenberg spin chain in the scaling limit},
 \href{https://doi.org/10.1016/j.nuclphysb.2019.114748}{Nucl.\ Phys.\ {\bf B} {\bf 947} (2019) 114748}
 \href{https://arxiv.org/abs/1906.07081}{{\ttfamily [arXiv:hep-th/1906.07081]}}.
  
\bibitem{Kotousov:2019nvt}
G.~A.~Kotousov and S.~L.~Lukyanov,
\emph{Spectrum of the reflection operators in different integrable structures},
\href{https://doi.org/10.1007/JHEP02(2020)029}{JHEP \textbf{02} (2020) 029}
\href{https://arxiv.org/abs/1910.05947}{{\ttfamily[arXiv:hep-th/1910.05947]}}.

\bibitem{Litvinov:2020zeq}
A.~Litvinov and I.~Vilkoviskiy,
\emph{Liouville reflection operator, affine Yangian and Bethe ansatz},
\href{https://link.springer.com/article/10.1007/JHEP12(2020)100}{JHEP \textbf{12} (2020) 100}
\href{https://arxiv.org/abs/2007.00535}{{\ttfamily[arXiv:hep-th/2007.00535]}}.
  
\bibitem{Schmid1}
A.~Schmid, \emph{Diffusion and localization in a dissipative quantum system},
\href{https://doi.org/10.1103/PhysRevLett.51.1506}{Phys. Rev. Lett. {\bf 51} (1983) 1506-1509}.

\bibitem{Fendley}
P.~Fendley, A.~W.~W.~Ludwig and H.~Saleur,
\emph{Exact non-equilibrium transport through point contacts in quantum wires and fractional quantum Hall devices},
\href{https://journals.aps.org/prb/abstract/10.1103/PhysRevB.52.8934}{Phys. Rev. {\textbf{B 52}} (1995) 8934-8950}
 \href{https://arxiv.org/abs/cond-mat/9503172}{{\ttfamily [arXiv:cond-mat/9503172]}}.

\bibitem{IJS2}
Y.~Ikhlef, J.~L.~Jacobsen and H.~Saleur,
\emph{The $\mathbb{Z}_2$ staggered vertex model and its
applications}, \href{https://iopscience.iop.org/article/10.1088/1751-8113/43/22/225201}{%
J. Phys. {\bf A 43} (2010) 225201}
\href{https://arxiv.org/abs/0911.3003}{{\ttfamily [arXiv:math-ph/0911.3003]}}. 






\bibitem{Kulish}
P.~P.~Kulish and  N.~Yu.~Reshetikhin,
\emph{Quantum linear problem for the sine-Gordon equation and higher representations},
 \href{https://link.springer.com/article/10.1007\%2FBF01084171}{J. Sov. Math. {\textbf{23}}
 (1983)  2435-2441}.


\bibitem{Sogo}
K.~Sogo,
\emph{Ground state and low-lying excitations in the Heisenberg $XXZ$ chain of arbitrary spin $S$},
\href{https://www.sciencedirect.com/science/article/pii/0375960184905887}{Phys. Lett.  {\textbf{A 104}}
(1984)  51-54}.


\bibitem{Kirillov}
A.~N.~Kirillov and N.~Yu.~Reshetikhin,
\emph{Exact solution of the integrable XXZ Heisenberg model with arbitrary spin. I. The ground state and the excitation spectrum},
\href{https://iopscience.iop.org/article/10.1088/0305-4470/20/6/038/meta}{ J. Phys. {\bf A} {\textbf{20}} (1987) 1565-1585}.


\bibitem{Bazhanov:2020fbp}
V.~V.~Bazhanov, G.~A.~Kotousov, S.~M.~Koval and S.~L.~Lukyanov,
\emph{Some algebraic aspects of the inhomogeneous six-vertex model},
\href{https://www.emis.de/journals/SIGMA/2021/025/}{SIGMA \textbf{17} (2021) 025}
\href{https://arxiv.org/abs/2010.10615}{{\ttfamily [arXiv:math-ph/2010.10615]}}.\hfil



\bibitem{Bazhanov:2019xvyA} 
  V.~V.~Bazhanov, G.~A.~Kotousov, S.~M.~Koval and S.~L.~Lukyanov,
\emph{Scaling limit of the ${\cal Z}_2$ invariant inhomogeneous six-vertex model},
\href{https://doi.org/10.1016/j.nuclphysb.2021.115337}{Nucl. Phys. {\bf B 965} (2021) 115337}
\href{https://arxiv.org/abs/2010.10613}{\ttfamily [arXiv:math-ph/2010.10613]}.


\bibitem{Vicedo:2017cge}
B.~Vicedo,
\emph{On integrable field theories as dihedral affine Gaudin models},
\href{https://doi.org/10.1093/imrn/rny128}{Int. Math. Res. Not. \textbf{2020} (2020) 4513-4601}
\href{https://arxiv.org/abs/1701.04856}{\ttfamily [arXiv:hep-th/1701.04856]}.


\bibitem{Delduc:2018hty}
F.~Delduc, S.~Lacroix, M.~Magro and B.~Vicedo,
\emph{Integrable coupled $\sigma$ models},
\href{https://doi.org/10.1103/PhysRevLett.122.041601}{Phys. Rev. Lett. \textbf{122} (2019) 041601}
\href{https://arxiv.org/abs/1811.12316}{\ttfamily [arXiv:hep-th/1811.12316]}.



\bibitem{Delduc:2019bcl}
F.~Delduc, S.~Lacroix, M.~Magro and B.~Vicedo,
\emph{Assembling integrable $\sigma$-models as affine Gaudin models},
\href{https://doi.org/10.1007/JHEP06(2019)017}{JHEP \textbf{06} (2019) 017} 
\href{https://arxiv.org/abs/1903.00368}{\ttfamily [arXiv:hep-th/1903.00368]}.


\end{thebibliography}
\end{document}